\documentstyle[12pt,epsfig]{sbthesis}

\def \as {\alpha_s}

\def \MS {\overline{\rm MS}}

\def \LQCD {\Lambda_{\rm\tiny QCD}}

\def \O {\Omega}

\def \o {\omega}

\def \xic {\hat{\xi}_c}

\newcommand \Ncol {{\mathcal{N}}_C}

\newcommand \ok {{\omega}_k}

\newcommand \ol {{\omega}_l}

\newcommand \sd {\sin \delta}

\newcommand \sfi{\sin \phi}

\newcommand \cfi{\cos \phi}

\newcommand \de {\Delta \eta}

\newcommand \eh {\hat{\eta}}

\def \Tr {\mbox{Tr\,}} 

\def \be  {\begin{equation}}

\def \ee  {\end{equation}}

\def \ba  {\begin{eqnarray}}

\def \ea  {\end{eqnarray}}

\begin{document}

\pagestyle{prelim}            



%


%

\dissertation


%

\author{Carola Friederike Berger}



%

\degree{Doctor of Philosophy}



%

\department{Physics}



%

\title{Soft Gluon Exponentiation and Resummation}


%

\month{May}

\year{2003}


%

\maketitle



\makecopyright



%

\begin{approval}

 \member{Dr. George Sterman\\Advisor\\Professor, C.~N.~Yang Institute for Theoretical Physics}

 \member{Dr. John Smith\\Professor, C.~N.~Yang Institute for Theoretical Physics}

 \member{Dr. Barbara Jacak\\Professor, Department of Physics and Astronomy}

 \member{Dr. Sally Dawson\\Senior Scientist, High Energy Theory Division,\\ Brookhaven National Laboratory}

\end{approval}



%

\begin{abstract}

\vspace*{-2cm}




\vspace*{1cm}

In calculations of (semi-) inclusive events within perturbative Quantum Chromodynamics, large logarithmic corrections arise from certain kinematic
regions of interest which need to be resummed. 
When resumming soft gluon effects one
encounters quantities built out of eikonal or Wilson lines (path
ordered exponentials). In this thesis we develop a simplified method to calculate higher orders of
the singular coefficients of parton distribution functions which is based on the exponentiation of cross sections built out of eikonal lines. As an illustration of the method we determine
the previously uncalculated fermionic contribution to the
three-loop coefficient $A^{(3)}$. 

The knowledge of these
coefficients is not only important for the study of the parton
distribution functions themselves, but also for the resummation of
large logarithmic effects due to soft radiation in a variety of cross sections. 

In the second part of this thesis we study the energy flow pattern of this soft radiation in jet
events. We develop the concept of event shape-energy flow correlations that
suppress radiation from unobserved  ``minijets'' outside the region of interest and are
sensitive primarily to radiation
  from the highest-energy jets. We give analytical and numerical results at next-to-leading logarithmic order for shape/flow correlations in $e^+e^-$ dijet events. We conclude by illustrating the application of our formalism to events with hadrons in the initial state, where the shape/flow correlations can be described via matrices in the space of color exchanges.

\end{abstract}




\begin{dedication}

\begin{quote}

$[\dots] \,\,\delta \tilde{\eta} \lambda o \nu\,$ $\acute{o} \tau \iota\,$ $\kappa \alpha \grave{\iota}\,$ $\tau \tilde{\eta} \varsigma\,$ $\pi \varepsilon \rho \grave{\iota}\,$ $\phi \acute{\upsilon} \sigma \varepsilon \o \varsigma\,$ $\varepsilon \pi \iota \sigma \tau \acute{\eta} \mu \eta \varsigma\,$ $\pi \varepsilon \iota \rho \alpha \tau \acute{\varepsilon} o \nu\,$ $\delta \iota o \rho \acute{\iota} \sigma \alpha \sigma \theta \alpha \iota\,$ $\pi \rho \tilde{\o} \tau o \nu\,$ $\tau \grave{\alpha}\,$ $\pi \varepsilon \rho \grave{\iota}\,$ $\tau \grave{\alpha} \varsigma\,$ $\alpha \rho \chi \acute{\alpha} \varsigma.\,$ $\Pi \acute{\varepsilon} \phi \upsilon \kappa \varepsilon\,$ $\delta \grave{\varepsilon}\,$ $\varepsilon \kappa\,$  $\tau \tilde{\o} \nu \,$ $\gamma \nu \o \rho \iota \mu \o \tau \acute{\varepsilon} \rho \o \nu\,$ $\eta \mu \tilde{\iota} \nu\,$ $\eta\,$ $o \delta \grave{o} \varsigma\,$ $\kappa \alpha \grave{\iota}\,$ $\sigma \alpha \phi \varepsilon \sigma \tau \acute{\varepsilon} \rho \o \nu\,$ $\varepsilon \pi \grave{\iota}\,$ $\tau \grave{\alpha}\,$ $\sigma \alpha \phi \acute{\varepsilon} \sigma$- $\tau \varepsilon \rho \alpha\,$ $\tau \tilde{\eta}\,$ $\phi \acute{\upsilon} \sigma \varepsilon \iota\,$  $\kappa \alpha \grave{\iota}\,$ $\gamma \nu \o \rho \iota \mu \acute{\o} \tau \varepsilon \rho \alpha\,$ $o \upsilon\,$ $\gamma \grave{\alpha} \rho\,$ $\tau \alpha \upsilon \tau \grave{\alpha}\,$ $\eta \mu \tilde{\iota} \nu\,$ $\tau \varepsilon\,$ $\gamma \nu \acute{\o} \rho \iota \mu \alpha\,$ $\kappa \alpha \grave{\iota}\,$ $\alpha \pi \lambda \tilde{\o} \varsigma.\,$

\begin{flushright}  $\mathrm{API} \Sigma \mathrm{TOTE} \Lambda \mathrm{O} \Upsilon \Sigma\,$ $\Phi \Upsilon \Sigma \mathrm{IKH} \Sigma\,$ $\mathrm{AKPOA} \Sigma \mathrm{E} \Omega \Sigma$  A (1), 184 a 14. \end{flushright}

\vspace*{20mm}

{\renewcommand{\baselinestretch}{1.2}

\small 

[...] systematic knowledge of nature must start with an attempt to settle questions about principles. The natural course is to proceed from what is clearer and more knowable to us, to what is more knowable and clear by nature.  

\begin{flushright} Aristotle, \textit{Physics}, Book I (1), 184 a 14. \end{flushright}

\setlength{\baselineskip}{1mm} 

\renewcommand{\baselinestretch}{0.8}

}

\end{quote}

\end{dedication}



%

\tableofcontents

\listoffigures

\listoftables







%

\begin{acknowledgements}


First and foremost, I would like to thank my Ph.D. advisor George Sterman, for his advice, support, and valuable insights during my graduate studies at Stony Brook. I acknowledge a fruitful collaboration with Tibor K\'ucs, on which part of this thesis is based. Furthermore, I owe deep gratitude to Maria Elena Tejeda-Yeomans for many helpful discussions and her support in multiloop and other matters. I also thank my Master's advisor Wolfgang Schweiger for raising my interest in particle physics, and for ongoing fruitful collaborations. I am indebted to Anton Chuvakin and Chi Ming Hung for their rescue-attempts when computers tried to erase my work. I acknowledge very useful exchanges with Lilia Anguelova, Rob Appleby, John Collins, Sally Dawson, Barbara Jacak, Edward Shuryak, Jack Smith, Peter van Nieuwenhuizen, Jos Vermaseren, and Andreas Vogt. \vspace*{3mm}

Next, I want to thank all my friends, too numerous to list on this page, except the physics-related ones who are among the people listed above. I thank my family, especially my parents and my sister Petra, for their support in my attempt to learn more about the smallest building blocks of the Universe, despite their initial viewpoint that ``quark'' is some sort of cheese. \newpage

Last, but not least, I acknowledge financial support by the Austrian Ministry of Science (\"Osterreichisches Bundesministerium f\"ur Wissenschaft), the Department of Physics and Astronomy, SUNY at Stony Brook, and the U.S. National Science Foundation, grants PHY9722101 and PHY0098527.


\end{acknowledgements}


\pagestyle{body}             

\begin{singlespace}

\chapter{Prologue: Perturbative Quantum Chromodynamics}

\begin{center}
\parbox[t]{11cm}{\begin{quote}
Divergent series are the invention of the devil, and it is shameful to base on them any demonstration whatsoever.\\
\hspace*{5.7cm}
{ \raggedleft 
N. H. Abel, 1828.}
 \end{quote}}
\end{center}

The coefficients of a perturbation series in Quantum Chromodynamics (QCD) exhibit factorial growth, in other words, the series diverges. Nevertheless it is possible to construct meaningful physical observables that are calculable within perturbation theory, if the perturbative QCD series is asymptotic\footnote{Mr. Abel's statement may need to be modified to: Non-asymptotic series are the invention of the devil.}. In the following we will illustrate this for (semi-)inclusive processes \cite{Collins:1987pm,Collins:gx}. We will not discuss exclusive processes where all hadrons  in the final state are observed \cite{Brodsky:1989pv,Pire:1996bc,Stefanis:1999wy}. For exclusive processes currently experimentally accessible energies may not be high enough to make them amenable to a purely perturbative treatment, and non-perturbative effects have to be included, for example via additional parameters in effective models \cite{Berger:1998ri,Berger:1999gx,Berger:2002vc}. In the following we will denote processes where no hadronic final states  are observed by inclusive or semi-inclusive. The latter denote cross sections with some additional restrictions that do not distinguish different hadronic decompositions of the events, for example event shapes in jet cross sections.

In this introduction we will start with an overview of the basic concepts of perturbative QCD (pQCD), and the main assumptions that allow us to compare perturbative calculations with experiment. After giving a brief motivation for the work presented in this thesis we outline its contents which are based on our publications Refs. \cite{Berger:2001ns,Berger:2002ig,Berger:2003iw,Berger:2002sv}.

\section{Perturbative QCD - Basic Concepts}

We refrain here from listing the QCD Lagrangian, and other generalities of non-abelian quantum field theories. In this thesis we follow the conventions for the QCD Feynman rules listed, for example, in \cite{book}, where also a variety of useful relations regarding SU(N) and Dirac algebra can be found.  

Throughout this thesis we will use dimensional regularization \cite{'tHooft:fi}, in $n = 4 -2 \varepsilon$ dimensions, and give explicit results in the $\MS$ scheme \cite{Bardeen:1978yd}. We use Feynman gauge, unless explicitly stated otherwise.

\subsection{Asymptotic Freedom}

Here we want to point out the main feature of unbroken non-abelian, renormalizable field theories that makes them amenable to a perturbative treatment: asymptotic freedom \cite{Politzer:fx,Gross:ju}. The running coupling in asymptotically free theories vanishes at large momentum scales, as illustrated in Fig. \ref{asplot} for the strong coupling in QCD, $\as(\mu)$. This is due to the sign of the first coefficient of the beta-function (see Appendix \ref{sec:alpha} for the conventions used here), which, for QCD (SU(3)), is positive if the number of flavors is less than 33/2 = 16.5. At large scales, or equivalently, short distances, the theory is then treatable perturbatively, if long-distance effects are incoherent to short-distance effects. At long distances $> 1$ fm, which correspond to low momentum transfer of order 1 GeV or less\footnote{In the following we use natural units, for example, we set the speed of light $c$, or $\hbar$ to 1.}, confinement effects become dominant, and perturbation theory fails. 

Furthermore, if short- and long-distance effects are incoherent, we may neglect masses in the computation of short-distance effects, since  masses exhibit the same asymptotic behavior as the running coupling,
 \ba m(\mu^2) & = & m(\mu_0^2) \exp \left\{-\frac{1}{2} \int_{\mu_0^2}^{\mu^2} \frac{d \lambda^2}{\lambda^2} \left[ 1 + \gamma_m (\as(\lambda^2)) \right] \right\}, \nonumber \\ 
\lim\limits_{\mu^2 \rightarrow \infty} \frac{m(\mu^2)}{\mu^2} & = & 0. \label{mass} 
\ea
Here, $\gamma_m$ is the mass anomalous dimension, the analog of the beta-function of the running coupling. 

\subsection{Assumptions of Perturbative QCD} \label{sec:assumpt}

There are two main assumptions that go into any calculation within perturbative QCD. These assumptions have not been proven yet, but the remarkable success of pQCD seems to confirm their validity.

\subsubsection{The pQCD Series is Asymptotic}

The first of these assumptions has already been mentioned above, namely, that pQCD is an asymptotic series, despite being divergent, in the mathematical sense: In perturbation theory a physical quantity is computed as a power series in terms of the small coupling
\be
f(\as) \sim \sum\limits_{n=0}^\infty f_n \as^n,
\ee
where in field theory, and thus in QCD, one finds $n!$ growth with the order of the coefficients $f_n$ \cite{LeGuillou:nq}.
 Only at $\as = 0$ the series would equal the function, being simply a Taylor expansion. For $\as \rightarrow 0$ the series can at best be asymptotic to $f(\as)$, but does not necessarily uniquely define $f(\as)$, even if summed to all orders, irrespective of the convergence or divergence of the series. 

A series $\sum\limits_{n = 0}^\infty f_n \as^n$ is called \emph{asymptotic} to $f(\as)$ for $\as \rightarrow 0$ on a set $S$ if the remainder $R_{N+1}$ obeys 
\begin{equation} 
\left| R_{N+1} \right|  = \left| f(\as) - \sum\limits_{n = 0}^N f_n \as^n \right| \leq C_{N+1} \left|\as \right|^{N+1} \label{remain} 
\end{equation} 
for all positive integer $N$ and for all $\as$ in $S$. 
As stated above the asymptotic series does not define a unique function $f(\as)$ in general. Only under additional restrictions the series might give only one $f(\as)$. 

If the truncation error $C_N$ follows the same pattern as the coefficients $f_N$, in field theory
\begin{equation} 
C_N \sim N! a^N N^b \label{pattern}
\end{equation} 
(this follows from $\frac{C_{N+1}}{C_N} \sim \frac{1}{f_N} \sim N$ $\Rightarrow$ $C_N \sim N!$) 
the error decreases as a function of the order $N$ until order $N_* \sim \frac{1}{|a|z}$ as a short calculation of the minimum of the remainder (\ref{remain}) with the behavior (\ref{pattern}) with respect to $N$ shows, using  Stirling's formula 
\begin{equation}
\lim\limits_{n \rightarrow \infty} n! = \sqrt{2 \pi n} n^{n} e^{-n} \left[ 1 + {\mathcal{O}} \left( \frac{1}{n}\right)  \right] . \label{stirling}
\end{equation}
If we truncate the series at its minimal term $N_*$ then we get the best approximation to $f(\as)$ with an accuracy of $C_{N_*} \as^{N_{*}} \sim e^{- \frac{1}{|a| \, \as}}$. This means that the series $\sum f_n \as^n$ is not only asymptotic to $f(\as)$ but also to  
\begin{equation} 
f'(\as) = f(\as) + C e^{- \frac{1}{|a|\, \as}}, \qquad C \mbox{ real}. \label{amb} 
\end{equation} 
For $f'(\as)$ Eq. (\ref{remain}) still holds, that is, the expansions in powers of $\as$ of $f(\as)$ and $f'(\as)$, respectively, are the same even though $f(\as)$ and $f'(\as)$ are clearly two different functions. However, if $\as$ is sufficiently small, the difference between $f(\as)$ and $f'(\as)$ may be numerically small, and perturbation theory may give a well-approximated answer, up to power corrections as we will briefly mention in Sec. \ref{sec:intropower}. 

\subsubsection{Incoherence of Long- and Short-Distance Effects}

The second assumption is that properties that hold order-by-order for the asymptotic series up to power corrections in the regulated theory also hold in the full theory up to power corrections. Factorization can be proven in a sufficiently rigorous way for certain partonic quantities to any order in a regulated perturbation theory at leading power \cite{Collins:1987pm,Collins:gx}, assuming only that this regulated theory has bound states whose formation decouples from short-distance physics, and that this factorization continues to hold when the unphysical, regulator-dependent states become physical upon removing the regulator. 

However, the mechanisms that confine partons in hadrons are far from fully understood and have to be parameterized in an appropriate way in perturbative calculations. More or less heuristic argumentation, based on the parton model, suggests that these long-distance effects decouple. Colliding hadrons in the center-of-mass frame are highly Lorentz contracted, and internal interactions are time dilated. At sufficiently high energies, the interacting hadrons are in virtual states with a definite number of partons which are well separated in transverse directions. One parton in each colliding hadron then interacts incoherently at the hard scattering, interactions among partons within a hadron cannot interfere with this hard scattering because they take place at time-dilated scales. Therefore, an inclusive hadronic cross section $\sigma_{AB}$ for the process $A+B \rightarrow X$ with two hadrons in the initial state can then schematically be factorized at leading power in the hard scale, $Q$,
\be
\sigma_{AB} = \sum_{a,b} f_{a/A}(\mu) \otimes f_{b/B}(\mu) \otimes \hat{\sigma}_{ab}(\mu). \label{fact}
\ee
Here the $f_{h/H}(x)$ are parton-in-hadron distribution functions, which describe the distribution of a parton $h$ with momentum fraction $x$ in hadron $H$. These distribution functions are convoluted in terms of the momentum fractions $x$ with the partonic cross section $\hat{\sigma}_{ab}$, denoted by the symbol $\otimes$: 
\be
\left(f \otimes g \right) (x) = \int\limits_0^1 d x_1 \int\limits_0^1 d x_2 \, \delta\left( x - x_1 \, x_2 \right) f(x_1) g(x_2). \label{convoldef}
\ee
Long distance effects of hadronic distribution functions are separated from the short distance scattering by the factorization scale $\mu$. The physical cross section is of course independent of this scale. For the determination of the distribution functions experimental and/or nonperturbative input is needed, whereas the hard scattering is calculable in perturbation theory if it is infrared safe. The calculability of $\hat{\sigma}_{ab}$ follows from analyzing the partonic counterpart of 
Eq. (\ref{fact}):
\be
\sigma_{a'b'} = \sum_{a,b} f_{a/a'}(\mu) \otimes f_{b/b'}(\mu) \otimes \hat{\sigma}_{ab}(\mu). \label{factpart}
\ee
From this factorization $\hat{\sigma}_{ab}$ is calculable for infrared safe observables, as well as the parton-in-parton distribution functions, and the evolution of all these functions with the factorization scale $\mu$. Furthermore, due to the incoherence of long- and short-distance effects, parton distribution functions are universal, that is, the same functions occur in a variety of infrared safe (semi-)inclusive cross sections.

Similarly, for (semi-)inclusive cross sections without hadrons in the initial state, we assume that the observed spectra of hadrons should be mathematically similar to the calculated spectra of partons. For example in jet cross sections we assume that the distribution of experimentally observed energy deposits in the detector is calculable by studying the corresponding distribution of more or less collimated, energetic partons.

All these assumptions reduce to the assertion that power corrections in the regulated theory remain small in transition to the full theory.

\subsection{Infrared Safety}

Quantities that are dominated by the short-distance behavior of the theory are \emph{infrared (IR) safe}. For such quantities perturbation theory is applicable. In order to be IR safe a physical quantity $\tau$ in QCD has to behave in the limit of the renormalization scale $\mu \rightarrow \infty$ as
\begin{equation} 
\tau \left( \frac{Q^2}{\mu^2}, \as (\mu^2), \frac{m^2(\mu^2)}{\mu^2} \right) \stackrel{\longrightarrow}{ \mu \rightarrow \infty} \hat{\tau} \left( \frac{Q^2}{\mu^2}, \as (\mu^2)\right) + {\mathcal{O}} \left( \left(  \frac{ m^2}{\mu^2}\right)^a \right), \, a>0. \label{irsafe} 
\end{equation}
Thus $\tau$ should approach a limit as $\frac{m }{\mu } \rightarrow 0$ ($m$ represents light quark and vanishing gluon masses, $Q$ ``large'' invariants, $Q \gg \Lambda$) with $\frac{ Q}{ \mu}$ held fixed. In Chapter \ref{ch1} we will show how to identify infrared safe observables. 

Although infrared safe quantities are free of IR divergences as powers, large logarithmic corrections occur at the edge of phase space in all but fully inclusive observables, due to soft (with vanishing four-momentum) and/or collinear (parallel to primary, energetic quanta) radiation. These logarithmic corrections need to be resummed in order to provide reliable quantitative predictions. The remainder of this thesis deals with resummation of large logarithms. Another source of uncertainty in perturbative calculations are power corrections. These, however, are in the majority of cases incalculable within perturbation theory.

\subsection{Nonperturbative Effects and Power Corrections}\label{sec:intropower}

Above we have noted that Eqs. (\ref{fact})-(\ref{irsafe}) are valid up to power corrections in the hard scale $Q \gg \LQCD$. In only a few cases factorization theorems can also be proven beyond leading power. In addition, due to the at best asymptotic nature of QCD, Eq. (\ref{amb}), there will always be exponential ambiguities. These ambiguities correspond to power corrections proportional to
\be
e^{-\frac{1}{|a|\, \as(Q^2)}} \sim \left( \frac{\LQCD^2}{Q^2} \right)^{\frac{\beta_0}{4 \pi |a|}}, \label{powercorr}
\ee
using the one-loop running coupling (\ref{1as}).

Nevertheless, perturbation theory itself encodes some information about the form of these power corrections. As we have mentioned above, in field theoretic expansions one often finds factorial growth of the coefficients. This suggests to attempt summation of the series via Borel transformation, which is defined as \cite{Borel}
\be
B(t) = \sum\limits_{n = 0}^\infty f_n \frac{t^n}{n!}.
\ee
If an asymptotic series is \emph{Borel summable}, then the inverse transform, the so-called \emph{Borel integral}
\be
\tilde{f}(\as) = \frac{1}{\as} \int_0^\infty dt\, e^{-\frac{t}{\as}} B(t) \label{Binv}
\ee
uniquely determines the function $f(\as) \equiv \tilde{f}(\as)$ to which the series is asymptotic. $\tilde{f}(\as)$ is a Laplace transform (the conventional variable for a Laplace transform is $s = 1/\as$). Thus the theory of Borel summability is essentially the theory of Laplace transforms.

If the Borel transform $B(t)$ of a pQCD series has singularities on the real positive axis then the series is not uniquely Borel summable. Nevertheless, we can still define the Borel integral by moving the integration contour above or below the singularities of $B(t)$ if they are on the positive real axis. For example, consider
\be
B(t) = \frac{1}{1 - a t},
\ee
where $a$ determines the position of the singularity. Larger $a$ means smaller radius of convergence of the series. We can define the Borel integral to be the principal value which introduces an ambiguity 
\begin{displaymath}
\sim e^{-1/(|a|\, \as)}.
\end{displaymath}
This ambiguity leads according to Eq. (\ref{powercorr}) to a power correction proportional to $1/Q^{\beta_0/(2 \pi |a|)}$. 
 
Although the power of the correction can be deduced from perturbation theory, the magnitude and functional form of these power corrections cannot be inferred without additional, nonperturbative or experimental information. 
In QCD one finds $n!$ growth of perturbative coefficients, which lead to singularities of the Borel transform on the positive and negative real axis. 

One source of $n!$ growth is the factorial growth of the number of Feynman graphs with the order, which is connected to the occurrence of \emph{instantons} \cite{LeGuillou:nq,'tHooft:am,Vainshtein:1964dw}. Thus the study of instantons \cite{Schafer:1996wv}, which are solutions to the classical field equations, may provide the necessary additional information to determine the ambiguity in Eq. (\ref{amb}) stemming from instanton singularities in the Borel plane. Instantons in QCD produce singularities on the positive real axis, however, far away from the origin. 

Another source of $n!$ growth at $n$th order in the perturbative expansion is called \emph{renormalons} \cite{Beneke:1998ui,Beneke:2000kc}, classified into UV and IR renormalons, connected to the large and small loop momentum behavior, respectively. UV renormalons in QCD produce singularities in the Borel plane on the negative real axis and thus do not spoil the Borel summability. Furthermore, they are although theory-specific, process independent, analogous to UV counterterms in renormalizable field theories. IR renormalons, on the other hand, are located on the positive real axis for asymptotically free theories, in general much closer to the origin than instanton-singularities. IR renormalons therefore give rise to ambiguities of the aforementioned form (\ref{powercorr}), which are much less suppressed than instanton ambiguities. The location of IR renormalon poles is process-dependent. Renormalons are found in graphs that grow as $n!$ themselves, for example diagrams with loop insertions in the form of one or more ``bubble chains'' \cite{'tHooft:am,Lautrup:hs}.

\section{Motivation for Further Exploration}

From the above it may almost seem hopeless to attempt any calculation within perturbation theory. Nevertheless, nature itself seems to almost invite us to do perturbative calculations in QCD - the series seems asymptotic, with power corrections that are numerically small compared to the leading perturbative terms; the incoherence of long- and short-distance effects allows factorization with parton distribution functions that are universal for broad classes of observables. Thus, determined once experimentally for one observable in a class, all other observables within that class are then predictable in principle from perturbative calculations, up to power corrections.

Moreover, although much progress has been made in the development of non-perturbative techniques, or in the attempt of deriving the Standard Model from a more general theory, perturbative calculation is still the most complete and precise way to obtain quantitative predictions that can be compared to experiment. QCD processes need to be calculated as precisely as possible, in order not only to ``test'' the theory of strong interactions itself, but also to understand the background for other observables within the Standard Model and beyond, in the search for ``new physics''.

For precise quantitative predictions, it is necessary to sum the series to as high orders as possible, for the following reasons:
\begin{itemize}
\item Eq. (\ref{fact}) is in principle independent of the factorization scale $\mu$, in practice, however, fixed order calculations up to order $m$ introduce an error proportional to $\as^{m+1}$:
\ba
\mu \frac{d}{d \mu} \sigma_{AB} =  \mu \frac{d}{d \mu} \sum\limits_{n = 0}^\infty f_n(\mu) \as(\mu) = 0 \qquad \qquad & & \Rightarrow \nonumber \\
 \mu \frac{d}{d \mu} \sum\limits_{n = 0}^m f_n(\mu) \as^n(\mu) = - \mu \frac{d}{d \mu} \sum\limits_{n = m+1}^\infty f_n(\mu) \as^n(\mu). & &
\ea
Calculations to as high order as possible reduce the factorization scale uncertainty.
\item Power corrections, since they are incalculable within perturbation theory, are usually determined by experimental fits. However, numerically, at the presently calculated accuracy, power corrections are not distinguishable from higher order contributions. For example, the mean value of the thrust\footnote{A definition and discussion can be found in Chapter \ref{ch1}.}  $T$ has the following perturbative expansion (see for example \cite{Giele:2002hx} or \cite{Ellis:qj}\footnote{Resummed results can be found, for example  in \cite{Ellis:qj} which contains a collection of the results of \cite{Greco:1981kg,Tesima:1990mx,Catani:1992ua}.})
\be
\left< 1 - T \right> \approx 0.33 \,\as(Q) +  1.0 \,\as^2(Q) + c \,\as^3(Q) + \frac{\lambda}{Q} + \dots.\label{thrustaverage}
\ee
The dependence on the scale $Q$ of Eq. (\ref{thrustaverage}) with $\lambda = 1$ GeV and $c = 0$ (higher order corrections are vanishingly small) is numerically indistinguishable from $\lambda = 0.6$ GeV and $c = 3$ using the scale-dependence of the running coupling as shown in Fig. \ref{asplot}.
\item As already mentioned above, large logarithmic corrections arise that need to be resummed. In the case of the thrust, from Eq. (\ref{thrustaverage}), the average is at small values of ($1-T$), such that the corresponding logarithms of $\ln(1-T)$ are quite substantial. Resummation to high levels of logarithms also requires the calculation of the configurations that give rise to these logarithms at high orders. 
\end{itemize}

As we have already emphasized, QCD processes are present as background in the searches for new physics. A thorough understanding of this background is therefore absolutely necessary, especially the distribution of energy between energetic jets. Interjet radiation is emitted from a variety of sources, from fragments of hadrons that do not participate in the primary hard scattering which produces the jets, from multiple parton scattering, and  by soft bremsstrahlung from the primary scattering partons. All of these sources of interjet radiation are far from fully understood. 

Although an enormous amount of valuable insights has been obtained in the past thirty years of perturbative calculations within QCD, there still remains a wealth of open problems - the short list above contains only those directly connected with the content of this thesis; it would be beyond its scope to list further topics.

\section{Outline of the Thesis}
 
In this thesis we discuss two main topics - the higher order calculation of singular coefficients of partonic splitting functions, and jet event shapes, including their correlations with interjet energy flow (shape/flow correlations). The singularities in the splitting functions are due to simultaneously soft and collinear configurations. Similar configurations can be found in any quantity that contains collimated beams of particles (jets) which is not completely inclusive in the final state. It is therefore not surprising that the same coefficients appear in jet events whose discussion comprises the second part of the thesis.

\begin{figure}[htb]
\vspace*{-7mm}
\begin{center}
\epsfig{file=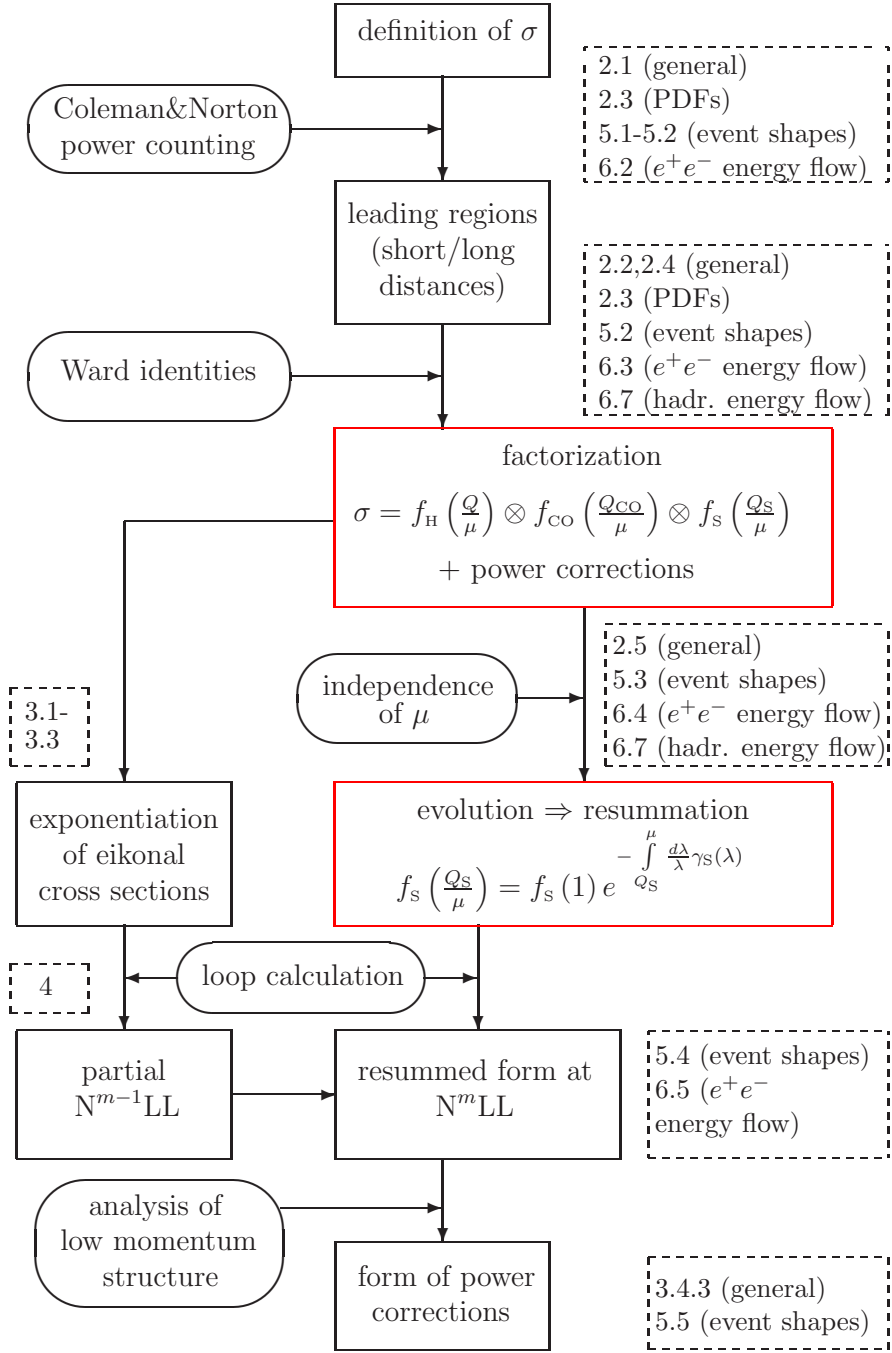,height=18cm,clip=0} \vspace*{3mm}
\caption[Outline of the thesis.]{Outline of the thesis. The boxes denote various intermediate stages in factorization and resummation procedures, the ovals describe the necessary tools. The items in the dashed boxes correspond to the sections of this thesis where corresponding descriptions and examples can be found.} \label{outline}
\end{center}
\end{figure}

The outline of this thesis which follows the successive steps in factorization and resummation procedures is illustrated schematically in Figure \ref{outline}. The starting point is in all cases the definition of a cross section or other physical observable, denoted collectively by $\sigma$ in the figure. Here $\sigma$ is either a parton distribution function, a jet event shape, or a shape/flow correlation. $\sigma$ contains in general singularities, or, in the case of IR safe observables (such as event shapes and correlations) logarithmic enhancements that need to be resummed. Resummation follows from factorization, that is, from the procedure of separating short-distance (in Fig. \ref{outline} denoted by the hard function $f_{\mbox{\tiny H}}$) from long-distance effects. In the figure we distinguish between collinear configurations, $f_{\mbox{\tiny CO}}$, which include soft/collinear radiation and soft configurations, $f_{\mbox{\tiny S}}$. These configurations have typical momentum scales $Q_i,\,i={\mbox{H,S,CO}}, Q_{\mbox{\tiny H}} = Q$.To obtain a factorized form is highly non-trivial, but once an observable is factorized, resummation is almost automatic.

\clearpage

These leading regions in momentum space are in general linked by a convolution in terms of one or more variables, depending on the observable under consideration, denoted by the symbol $\otimes$. We can disentangle this convolution (see below, Chapter \ref{sec:resum} for details) by taking for example Mellin moments, if the convolution is in terms of the variable $x$, Eq. (\ref{convoldef}), 
\ba 
\tilde{\sigma}(N) & = &  \int_0^1 dx \, x^{N-1} \sigma(x), \nonumber \\
\tilde{\sigma}(N) & = & \tilde{f}_{\mbox{\tiny H}}\left(\frac{Q}{\mu},N\right) \tilde{f}_{\mbox{\tiny CO}} \left(\frac{Q_{\mbox{\tiny CO}}}{\mu},N\right) \tilde{f}_{\mbox{\tiny S}} \left(\frac{Q_{\mbox{\tiny S}}}{\mu},N\right), \label{factmoment}
\ea
where quantities in moment space are denoted by $\tilde{\mbox{ }}$. The convolution is now a product in moment space. Eq. (\ref{factmoment}) contains potentially large ratios of the various scales intrinsic to the functions $f_i,\,i = \mbox{H,S,CO}$ to the factorization scale which give rise to the large logarithms mentioned above. These need to be resummed.

From the independence of the physical quantity $\sigma$ of the factorization scale (here in moment space)
\be
\mu \frac{d}{d\mu} \tilde{\sigma} =  0
\ee
follow the evolution equations:
\ba
\mu \frac{d}{d\mu} \ln \tilde{f}_{\mbox{\tiny H}}\left(\frac{Q}{\mu}\right) & = & - \gamma_{\mbox{\tiny H}} (\mu) \nonumber \\
 \mu \frac{d}{d\mu} \ln \tilde{f}_{\mbox{\tiny CO}}\left(\frac{Q_{\mbox{\tiny CO}}}{\mu}\right) & = & - \gamma_{\mbox{\tiny CO}} (\mu) \nonumber \\
 \mu \frac{d}{d\mu} \ln \tilde{f}_{\mbox{\tiny S}}\left(\frac{Q_{\mbox{\tiny S}}}{\mu}\right) & = & - \gamma_{\mbox{\tiny S}} (\mu) \label{evolf}
 \ea
with
\be
\gamma_{\mbox{\tiny H}} + \gamma_{\mbox{\tiny CO}} + \gamma_{\mbox{\tiny S}} = 0.
\ee
The anomalous dimensions $\gamma_i$ follow from separation of variables. The set of Eqs. (\ref{evolf}) can be solved to resum large logarithmic corrections in exponents:
\be
 \tilde{f}_{\mbox{\tiny S}}\left(\frac{Q_{\mbox{\tiny S}}}{\mu}\right) =  \tilde{f}_{\mbox{\tiny S}}\left(1\right) e^{-\int_{Q_{\mbox{\tiny S}}}^\mu \frac{d\lambda}{\lambda} \gamma_{\mbox{\tiny S}} (\lambda)}. \label{resumf}
\ee
We have evolved the soft function from its natural scale $Q_{\mbox{\tiny S}}$, where no large logarithms arise, to the factorization scale with the help of Eq. (\ref{evolf}). Calculation of the functions $f_i(1)$ and $\gamma_i$ to a specific order resums large logarithms at the $\mbox{N}^m$LL level, that is $m = 0$ denotes leading logarithmic level (LL), $m = 1$ next-to-leading logarithmic (NLL), $m = 2$ next-to-next-to-leading logarithmic, etc. On the other hand, as we will see in Chapter \ref{ch2} certain quantities exponentiate directly, not just via resummation as in (\ref{resumf}). These quantities, when calculated to order $m$ give the soft/collinear contribution to cross sections with jets at the $\mbox{N}^{m-1}$LL level.

The sections where the above is discussed in detail for parton distribution functions \cite{Berger:2002sv}, for jet shapes \cite{Berger:2002ig,Berger:2003iw}, and shape/flow correlations \cite{Berger:2001ns,Berger:2002ig,Berger:2003iw} are indicated in the dashed boxes in Figure \ref{outline}.

\chapter{Factorization, Evolution, and Resummation}
\label{ch1}

In order to factorize infrared safe, perturbatively calculable quantities from long-distance dependence it is necessary to develop means to systematically identify the latter. This chapter describes how to analyze and classify long-distance behavior, and how to separate it from short-distance contributions. For the general discussion below we follow Refs. \cite{Collins:gx,book,Sterman:1995fz} and the cited references. 

These methods were applied in our studies of the singular behavior of parton distribution functions \cite{Berger:2002sv}, and of dijet events \cite{Berger:2001ns,Berger:2002ig,Berger:2003iw}, which will be discussed in Sections \ref{sec:pdffact}  and \ref{sec:dijetfact}.

\section{Identification of Infrared Enhancements}

In Minkowski space there are two basic types of divergences remaining after ultraviolet divergences have been removed by a suitable renormalization procedure, which does not introduce new infrared singularities: \emph{soft divergences} that arise from vanishing four-momenta and \emph{collinear} ones that are associated with parallel-moving on-shell lines of finite energy. 

However, as consequences of the famous Bloch-Nordsieck  \cite{Bloch:1937pw,Yennie:ad,Grammer:1973db} (which only holds in QCD for quantities without initial-state hadrons) and Kinoshita-Lee-Nauenberg theorems \cite{Kinoshita:ur,Lee:is}, which follow from unitarity, these infrared divergences cancel between real and virtual emissions in suitably defined quan\-ti\-ties\footnote{For pedagogical reviews of these theorems see, for example, Refs. \cite{book} and \cite{Muta:vi}.}. In some important cases this cancellation is incomplete at the edge of phase space. For example for cross sections at threshold, that is, in the limit of soft and/or collinear radiation, fixed order perturbation theory is insufficient. In such cases, although no infrared divergences occur as powers, large logarithmic corrections arise that need to be resummed. 
 
While the Bloch-Nordsieck and Kinoshita-Lee-Nauenberg theorems give general arguments, it is imperative for the resummation of logarithmic corrections to identify infrared singularities at the levels of the expressions corresponding to the Feynman diagrams at arbitrarily high, but fixed, order. 
This section, which is based on Refs. \cite{Sterman:bi,Sterman:bj}, deals with the identification of infrared enhancements, while the remainder of this Chapter describe the factorization and resummation of large logarithmic corrections.

\subsection{Landau Equations} \label{sec:landau}

To see where the aforementioned singularities may come from we consider an arbitrary Feynman diagram $G\left( \{ p_s^{\mu} \} \right)$ with external momenta $ \{ p_s^{\mu} \}$ which is given by the following expression after Feynman parametrization, Eq. (\ref{feynpar}),
\begin{eqnarray}
\hspace*{-5mm} & G\left( \{ p_s^{\mu} \} \right) = & \hspace*{-3mm} \prod\limits_{\mbox{\tiny lines } i}  \int\limits_0^1 d \alpha_i \delta \left( \sum\limits_i \alpha_i - 1 \right) \prod\limits_{\mbox{\tiny loops } r} \int d^n k_r D(\alpha_i, k_r, p_s)^{-N} F(\alpha_i, k_r, p_s)  \nonumber \\ & & \label{G} \\
& D(\alpha_i, k_r, p_s) & = \sum\limits_j \alpha_j \left[ l_j^2 (p,k) - m_j^2 \right] + i \epsilon ,
\end{eqnarray}
where $F$ denotes all numerator and constant factors, and $D$ denotes the denominator. We work in $n$ dimensions, using dimensional regularization. $\alpha_j$ is the Feynman parameter of the jth line and $l_j^{\mu}$ its momentum, which is a linear function of loop momenta $\{ k_r \}$ and external momenta $\{ p_s \}$. Singularities arise in the integral \ref{G} if isolated poles cannot be avoided by contour deformation. This can happen if the pole is at one of the end-points of the integral (\emph{end-point singularity}) or if the contour is trapped between two poles (\emph{pinch singularity}).
The so-called \emph{Landau equations} summarize the conditions for the existence of pinch surfaces, which are surfaces in $(k,\alpha)$ space where $D$ vanishes \cite{Landau:1959fi,Bjorken:1959fd}.

If the poles of $D$ coalesce and therefore the contour cannot be deformed we encounter a pinch. The condition for this to occur is
\begin{equation}
\frac{\partial }{\partial k_j^\mu } D(\alpha_i, k_r, p_s) = 0
\end{equation}
at $D = 0$, because $D$ is quadratic in momenta. Considering the $\alpha_i$ we see that $D$ is only linear in each $\alpha_i$, so there are never two poles to pinch, but a pole may migrate to an end-point $\alpha_i = 0$. Or alternatively, $D$ may be independent of $\alpha_i$ at $D = 0$ if $l_i^2 - m_i^2 = 0$. 

In summarizing these conditions we arrive at the Landau equations which state that a pinch surface exists only if the following conditions hold for each point $\{ k_r^{\mu}, \alpha_i \}$ on the surface:
\begin{eqnarray}
& \mbox{either } & l_i^2 = m_i^2,  \,\, \mbox{ or }  \,\, \alpha_i = 0, \nonumber \\
& \mbox{and } \,\,\,\, & \sum\limits_{\mbox{\tiny line i in loop j}}  \eta_{i j }  \alpha_i l_i^{\mu} = 0, \label{Landau}
\end{eqnarray}
where $\eta_{i j}$ is an ``incidence matrix'' which is $\pm 1$ if the momentum $l_i$ of line $i$ flows in the same or opposite direction, respectively, as loop momentum $k_r$.

These equations can be given a physical interpretation, following Coleman and Norton \cite{colnort}. We can identify  $\alpha_i$ as the Lorentz-invariant ratio of the time of propagation to the energy for particle $i$ which is represented by the on-shell line $l_i$. So the space-time separation between the starting point and the endpoint of line $i$ is given by
\begin{eqnarray}
\Delta x_i^{\mu} & \equiv & \alpha_i l_i^{\mu} \\
& = & \Delta x_i^0 v_i^{\mu} \nonumber \label{ColNor1}
\end{eqnarray}
with
\begin{equation}
v_i^{\mu} = \left( 1, \frac{\vec{l}_i }{l_i^0 } \right) \label{ColNor2}
\end{equation}
the four-velocity of the particle. The Landau equations can then be illustrated in form of a  \emph{reduced diagram} where all off-shell lines are contracted to a point. 

Here a comment about masses is in order: The contributions of momenta near pinch surfaces are sensitive to infrared cut-offs such as quark masses, hadronic binding energies and other long-distance scales. We want to identify precisely these momentum regions, to separate them from perturbatively calculable parts which we evaluate at a large scale, $Q$. The above considerations are most relevant when $Q$ becomes very large, $Q \rightarrow \infty$. In this limit, due to Eq. (\ref{mass}), we can neglect masses since they become vanishingly small. Up and down quarks have masses of a few MeV at scales of the order of $\LQCD$, where we do not expect perturbation theory to be reliable anyway. Thus, studying the theory with all masses at their physical values is equivalent to studying the corresponding massless theory with external particles on shell. Corrections to the so identified leading behavior for infrared safe quantities, Eq. (\ref{irsafe}), will be proportional to powers of $m/Q$, where $m$ denotes any long-distance scale, including quark masses. 

\subsubsection{An Example} \label{sec:vertex}

\begin{figure}[htb]
\vspace*{-5mm}
\begin{center}
\epsfig{file=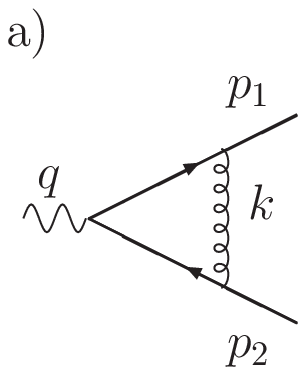,height=3.3cm,clip=0} \\
\epsfig{file=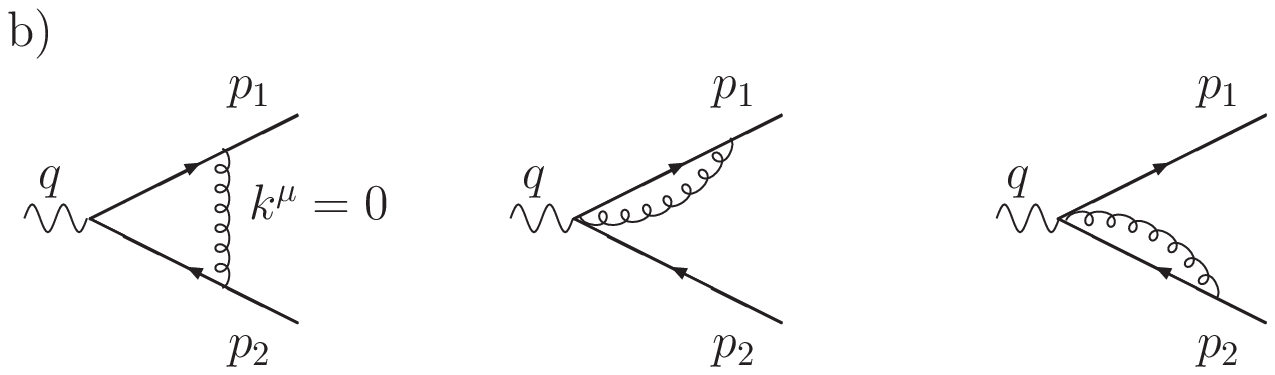,height=3.3cm,clip=0} \\
\epsfig{file=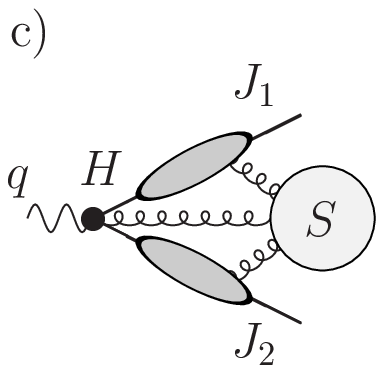,height=3.3cm,clip=0} 
\vspace*{-2mm}
\caption{The electromagnetic form factor: a) 1-loop correction, b) graphical representation of the solutions to the Landau equations, Eqs. (\ref{Landau}), in Feynman gauge, c) general reduced diagram.} \label{vertex}
\end{center}
\end{figure}

To illustrate the above, let us consider the one-loop vertex graph shown in Fig. \ref{vertex} a). As discussed above, it suffices to consider the massless limit. Its momentum structure in $n = 4 -2 \varepsilon$ dimensional regularization in Feynman gauge is given by
\be
V_\mu \left(p_1,p_2\right) = \int \frac{d^n k}{(2 \pi)^n} \frac{\bar{u}(p_1) \gamma^\alpha \left( \!\not\! p_1 + \!\not\! k \right) \gamma_\mu \left( - \!\not\! p_2 + \!\not\! k  \right)  \gamma_\alpha v(p_2)}{ \left[ (p_1 + k)^2 + i \epsilon\right] \left[(p_2 - k)^2  + i \epsilon \right]  (k^2 + i \epsilon)}, \label{vertexeq}
\ee
where we omitted all prefactors unnecessary for the argument that follows. 
The Landau equation for this expression is
\be
\alpha_1 (p_1 + k)^\mu  +  \alpha_2 (p_2 - k)^\mu + \alpha_3 k^\mu = 0,
\ee
which is not modified by the numerator. The solutions to this equation and the second condition of Eqs. (\ref{Landau}) which give non-vanishing contributions when taking the numerator into account, are
\ba
k^\mu = 0,\, \alpha_1 = \alpha_2 \!\! & \!\!  = \!\! & \!\!0, \\
\alpha_1 (p_1 - k)^\mu + \alpha_3 k^\mu \!\!& \!\!= \!\!& \!\!0,\,\, \alpha_2 = 0,\,\, k^2 = p_1 \cdot k = 0, \\
\alpha_2 (p_2 + k)^\mu + \alpha_3 k^\mu \!\!& = \!\!& \!\!0,\,\, \alpha_1 = 0,\,\, k^2 = p_2 \cdot k = 0.
\ea
These solutions are depicted graphically in Fig. \ref{vertex} b). In the first solution the radiated gluon $k$ is soft, in the other two solutions it is collinear to either of the outgoing quarks. 

As one can see by inspection of Eq. (\ref{vertexeq}), this scaling behavior is not changed by making the following approximations as $k^\mu \rightarrow 0$:
\begin{itemize}
\item We neglect $k^\mu$ compared to $p^\mu$ in numerator factors, and
\item we neglect $k^2$ compared to $p_1 \cdot k$ and $p_2 \cdot k$ in denominators.
\end{itemize}
The resulting expression for (\ref{vertexeq}) is
\be
V_\mu \left(p_1,p_2\right) \approx  - p_1 \cdot p_2 \bar{u}(p_1) \gamma_\mu v(p_2) \int \frac{d^n k}{(2 \pi)^n}
\frac{1}{\left( p_1 \cdot k + i \epsilon \right)\left( - p_2 \cdot k + i \epsilon \right) (k^2 + i \epsilon)}.
\ee
We observe that this integral is logarithmically divergent as $k^\mu \rightarrow 0$. Furthermore, for $p_1 \cdot p_2 \rightarrow \infty$ it approaches a constant value, in other words, the composite vertex exhibits the same asymptotic behavior as the elementary (Born) vertex. This behavior is characteristic of theories with vector particles, as we will show explicitly by infrared power counting in the next section. The set of approximations above is called \emph{eikonal approximation}, closely connected to path-ordered exponentials, also called Wilson lines. We will have to say more about this connection below. 
The eikonal approximation leads to the eikonal Feynman rules listed in Appendix \ref{sec:eikfeyn}. Especially the second approximation, the neglect of $k^2$ compared to $p \cdot k$ in denominators, is nontrivial in Minkowski space. In Section \ref{sec:Glauber} we will study this issue in more detail.

Following Coleman and Norton, we can generalize the above to arbitrarily high, but fixed orders. The resulting pinch surfaces for the electromagnetic form factor are shown in Fig. \ref{vertex} c) in reduced diagram notation.  Only short-distance effects contribute to the hard scattering, $H$, where the two primary outgoing partons are produced. $H$ is therefore contracted to a (slightly extended) point in the reduced diagram. Once the primary partons are produced, only collinear and/or soft radiation can be emitted, since they travel away from the hard scattering at the speed of light, if massless. They can never meet again at a point in space-time. We label the soft radiation $S$ in the reduced diagram, and the collinear configurations, the ``jets'' are labelled $J_i, \, i = 1,2$.

\subsection{Power Counting} \label{sec:powercount}

Eqs. (\ref{Landau}) are only necessary conditions for infrared divergences. In many cases, integration contours in perturbative integrals may pass through pinch surfaces without producing significant contributions in the limit of large momentum transfer. In such cases infrared safety (\ref{irsafe}) is not violated. Infrared power counting \cite{Sterman:bi} gives the potential degree of divergence of the pinch surfaces under consideration. Before becoming more explicit, let us just remark that in Feynman gauge individual diagrams in general have much worse scaling behavior than the gauge invariant contribution after summing over all diagrams, since in the sum unphysical contributions cancel. 

In the following we use light-cone coordinates, our conventions are given in Eq. (\ref{lccoord}).
The following momentum configurations, scaled relative to the large momentum scale in the problem, $Q$, are possible:
\begin{itemize} 
\item Soft momenta that scale as $k^\mu \sim \lambda\, Q, \, \lambda \ll 1$, in all components. 
\item Soft momenta with components that scale in the strongly ordered form $k^+ \sim \sigma \, Q, k^- \sim \lambda \, Q, k_\perp^2 \sim \lambda \, (Q)^2$, or  $k^- \sim \sigma\, Q, k^+ \sim  \lambda\, Q, k_\perp^2 \sim \lambda\, (Q)^2$, respectively, where $\sigma \ll \lambda \ll 1$. These are the so-called Glauber or Coulomb momenta \cite{Bodwin:1981fv,Collins:ta}. Glauber momenta have to be considered separately.
\item Momenta collinear to the momenta of initial or final state particles. Momenta collinear to particles moving in the plus direction scale as $k^+ \sim  Q, k^- \sim \lambda\, Q, k_\perp^2 \sim \lambda\, (Q)^2$, whereas momenta collinear to the minus direction behave as $k^- \sim  Q, k^+ \sim\lambda\, Q, k_\perp^2 \sim \lambda\, (Q)^2$. 
\item Hard momenta that are far off-shell, and thus scale as $\sim \, Q$ in all components.  
\end{itemize} 

Real momenta contributing to the final state have the same scaling behavior as purely virtual momenta since 
\begin{itemize}
\item for a jet momentum crossing the cut we obtain
\begin{displaymath}
\int d k^+ dk^- d k_\perp^2 \delta_{+}\left( 2 k^+ k^- - k_\perp^2 \right) = \int \frac{d k^+}{2 k^+} d k_\perp^2 \sim \lambda Q^2,
\end{displaymath}
which is the same scaling behavior found for a virtual jet line
\begin{displaymath}
\int d k^+ dk^- d k_\perp^2\frac{1}{k^2}  \sim  \frac{\lambda^2 Q^4}{\lambda Q^2} = \lambda Q^2.
\end{displaymath}
\item Similarly, we find for real soft momenta
\begin{displaymath}
\int d k^+ dk^- d k_\perp^2 \delta_+\left( 2 k^+ k^- - k_\perp^2 \right) \sim \lambda^2 Q^2,
\end{displaymath}
which coincides with the behavior of virtual soft momenta.
\end{itemize}
In the remainder of this section we scale all momenta implicitly by $Q$ which we drop from here on, that is
\be
k^\mu \rightarrow k^\mu/Q.
\ee

\subsubsection{Vertex Suppression Factors}

Let us now study how numerator factors change the scaling behavior of momentum lines and loops. 
These numerator suppression factors are different in covariant and physical gauges. Let us consider a fermion-gluon-fermion vertex in Feynman and in axial gauge as representative examples. This vertex is given by
\be
\left( \!\not\! p + \!\not\! k \right) \gamma^\mu \!\not\! p = - \gamma^\mu \left( \!\not\! p + \!\not\! k \right) \!\not\! p  +  2 \left(  p + k \right)_\mu \!\not\! p.  \label{ffv}
\ee
The first term in Eq. (\ref{ffv}) scales as $\sim \lambda^{1/2}$ if $k$ is part of the jet. The contribution of the second term, however, is different in covariant and physical gauges, respectively. In physical gauges, this term does not contribute, due to 
\be
\lim\limits_{k^2 \rightarrow 0} k^2 k^\mu D_{\mu \nu} (k,\xi,\kappa) = 0, \label{axialid}
\ee
where $D_{\mu \nu} (k,\xi,\kappa)$ is the gluon propagator from a Lagrangian with gauge fixing term $\left[-1/2 \kappa (\xi \cdot {\mathcal{A}})^2\right]$, ${\mathcal{A}}$ being the gauge potential, modulo color factors,
\be
D_{\mu \nu} (k,\xi,\kappa) = \left[- g_{\mu \nu} + \frac{\xi_\mu k_\nu + k_\mu \xi_{\nu}}{\xi \cdot k} - \xi^2 \left(1 + \frac{k^2}{\kappa \xi^2}\right) \frac{k_\mu k_\nu}{(\xi \cdot k)^2} \right] \frac{1}{k^2 + i \epsilon}. \label{gluonpropaxial}
\ee
Eq. (\ref{axialid}), however, does not hold in covariant gauges with gauge fixing $\left[-1/2 \kappa (\partial \cdot {\mathcal{A}})^2\right]$. \emph{Scalar polarized} gluons, that is, gluons that are contracted into their own momenta at jet-vertices, are then unsuppressed. Analogous considerations apply to other three-point vertices in jets. 

All jet three-point vertices  contribute a numerator suppression factor of $\lambda^{1/2}$, unless one of the attached gluons is scalar polarized and attaches to a hard part, in a covariant gauge, or soft. This is due to the above observation, and because we obtain a numerator suppression-factor proportional to $\lambda$ in any gauge when analyzing purely soft three-point gluon-gluon or ghost-gluon vertices in a similar manner. Four-point vertices do not suppress the scaling behavior. 

Contracted vertices have the same scaling behavior as the elementary vertices analyzed above. In contracted vertices in reduced diagrams internal lines which are off-shell by $\sim 1$ have been shrunk to a point, following (\ref{ColNor1}) and (\ref{ColNor2}). The behavior of contracted vertices can be analyzed by decomposing each vertex into its most general Lorentz structure, and by using Ward identities. 

For example, the most general decompositions in Feynman and physical gauges for the gluon two-point one-particle irreducible Green function are
\ba
\Pi_{\mu \nu}^{\rm Feynman} (p) & = & p^2 g_{\mu \nu}  f_1 + p_\mu p_\nu f_2, \\
\Pi_{\mu \nu}^{\rm physical}(p,\xi) & = & \frac{(p \cdot \xi)^2}{\xi^2} g_{\mu \nu} f'_1 + p_\mu p_\nu f'_2 + \nonumber \\
& & \quad + (p_\mu \xi_\nu + \xi_\mu p_\nu) \frac{(p \cdot \xi)}{\xi^2} f'_3 + \frac{\xi_\mu \xi_\nu}{\xi^2} \frac{(p \cdot \xi)^2}{\xi^2} f'_4,
\ea
where the $f_i,\,f'_i$ are dimensionless functions of contracted momenta with scaling behavior $\sim 1$. Upon insertion of these composite propagators into a diagram they are contracted with elementary gluon propagators. For the Feynman gauge expression it is immediately obvious that the combination (gluon jet line-$\Pi^{\rm Feynman}$-gluon jet line) has the same scaling behavior as an elementary gluon jet line. In the physical gauge we observe that the terms proportional to $f'_3$ and $f'_4$ drop out, using Eq. (\ref{axialid}). The term with $f'_2$ gives at least one factor $p^2$ upon contraction with a gluon jet line, and thus the same scaling contribution arises as for elementary propagators. The term with $f'_1$, however, seems to spoil this behavior. It can be shown using the techniques of \cite{Sterman:bi,Sen:sd} that this term is absent in $\Pi^{\rm physical}$. Therefore, the composite gluon jet line has also in a physical gauge the same scaling behavior as an elementary jet line. Similar considerations apply to all other propagators and vertices in QCD, further details can be found in Refs. \cite{Sterman:bi,Sen:sd}. 

\subsubsection{Summary}

\begin{itemize}
\item Every internal jet line scales as $\sim \lambda^{-1}$. Every bosonic soft momentum contributes as $\sim \lambda^{-2}$, fermionic soft momenta are proportional to $\sim \lambda^{-1}$. Jet loops scale as $\sim \lambda^2$, whereas soft loops behave proportional to $\sim \lambda^4$. This leads to the following superficial degrees $\omega$ of infrared (IR) divergence 
\ba
\omega_S & = & 4 L_S - 2 N^b_S - N^f_S + t_S,  \label{softdiv}\\
\omega_J & = & 2 L_J - N_J + t_J, \label{jetdiv}
\ea
where the subscripts denote soft ($S$) or jet ($J$) pinch surfaces. $L_i$ are the number of loops in $i = S,\,J$, $N_i$ are the number of lines therein, where the superscripts $b$ and $f$ label bosonic and fermionic lines, respectively, 
\be
N = N^b + N^f. 
\ee
We can count the degree of IR divergence separately for each pinch surface, if we carefully take into account momenta which link the surfaces in order to avoid double counting. $t_i$ denote numerator suppression factors which are summarized below.
\item Soft three-point vertices suppress the scaling by $\lambda$. Therefore the soft suppression factor is given by
\be
t_S = v^{(3)}_S, \label{softnum}
\ee
where $v^{(3)}_S$ is the number of soft three-point vertices.
\item On the other hand, jet three-point vertices give a suppression of $\lambda^{1/2}$, unless the gluons involved are scalar polarized in covariant gauges. This leads to a suppression factor $t_J$ for jets in physical gauges:
\be
t_J \geq \mbox{ max } \left\{ \frac{1}{2} \left[ v^{(3)}_J - s_J \right],0 \right\}, \label{physnum}
\ee
where $v^{(3)}_J$ is the number of jet 3-point vertices, and $s_J$ denotes the number of soft lines attached to the jet. In covariant gauges this is modified to
\be
t_J \geq \mbox{ max } \left\{ \frac{1}{2} \left[ v^{(3)}_J - s_J - l_J \right],0 \right\}, \label{covnum}
\ee
due to scalar polarized gluons $l_J$, linking jet lines and the hard scattering.
\end{itemize}

Furthermore, we will need the following useful identities: the relation between the number of $i$-point vertices $v^{(i)}$, the number of internal momentum lines $N$, and the number of external lines $E$,
\be
2 N + E = \sum_i i v^{(i)}, \label{linevertex}
\ee
and the Euler identity
\be
L = N - \sum_i v^{(i)} +1, \label{euler}
\ee
where $L$ is the number of loops.  
We will use these rules below.

In this section we have described how to identify regions in momentum space that give leading contributions, and how to determine the degree of infrared divergence by counting powers in each of these regions. Before going on to actually apply the techniques developed above, we will first explain how to factorize the leading regions which we expect to be linked by soft and/or scalar polarized gluons because of the numerator suppression factors Eqs. (\ref{physnum}) and (\ref{covnum}), respectively.
 
\section{Factorization} \label{sec:factor}

We are now going to show how to factorize infrared safe quantities from long-distance behavior, and how to refactorize these leading regions, which will allow us to resum large logarithmic corrections. In the following we will denote both procedures by factorization. The final results can in many cases be rewritten in terms of gauge-independent functions which reproduce the leading behavior in any gauge. 

\subsection{Ward Identities}

\begin{figure}[htb] 
\vspace*{-5mm}
\begin{center}
\epsfig{file=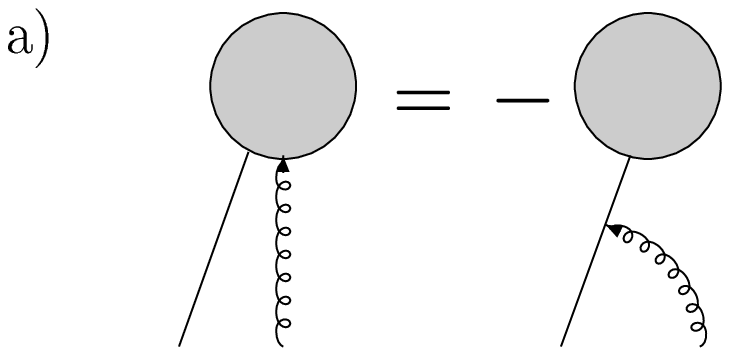,height=2.7cm,clip=0} 
\hspace*{1cm}
\epsfig{file=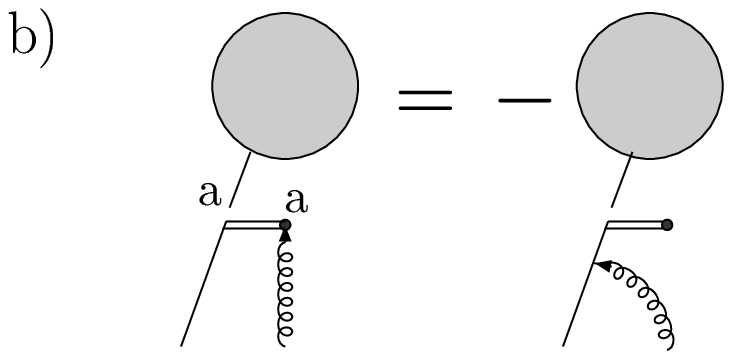,height=2.7cm,clip=0} \\
\vspace*{3mm}
\epsfig{file=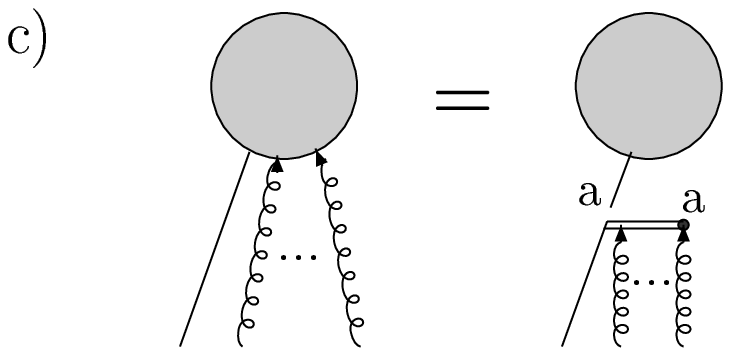,height=2.7cm,clip=0}
\caption[a) Ward identity for a scalar polarized gluon. b) Identity for a single longitudinally polarized gluon attaching to an eikonal line. c) Resulting identity after iterative application of Figs. a) and b).]{a) Ward identity for a scalar polarized gluon. b) Identity for a single longitudinally polarized gluon attaching to an eikonal line. c) Resulting identity after iterative application of Figs. a) and b). Repeated gauge-group indices are summed over. }  \label{wardeik}    
\end{center}    
\vspace*{-6mm}
\end{figure}

The factorization of scalar polarized gluons from hard contributions is a matter of straightforward application of the Ward identity shown in Fig. \ref{wardeik} c) for scalar polarized gluons.  
The grey blob denotes a hard part. 

Fig. \ref{wardeik} c) follows from the Ward identity shown in Fig. \ref{wardeik} a), and the identity for scalar polarized gluons attaching to an eikonal line in Fig. \ref{wardeik} b). 
Fig. \ref{wardeik} a) is the graphical representation (in momentum space) of the following equation \cite{'tHooft:fh,'tHooft:1972ue}
\be
\left< N \left| \kappa \,\partial \cdot A^b (x) \right| M \right> = 0, \label{wardeq}
\ee 
where $M$ and $N$ are physical states, and $A_\mu^b$ is a nonabelian gauge field carrying color $b$. By physical states we denote states involving on-shell fermions and gauge particles with physical polarizations; ghosts are not included. Throughout this thesis, gluons with arrows are scalar polarized. Eq. (\ref{wardeq}) can be proven, for example, by taking the BRST variation \cite{Becchi:1974md,Becchi:1975nq,Tyutin} of the Green function $\left<0 \left| T\, b^b(x) \prod_i \Psi (y_i) \prod_j \bar{\Psi}(z_j) \right| 0 \right>$. The sum of BRST variations of all of the fields is 0, since the QCD Lagrangian is BRST invariant. Then we use the reduction formula to relate the truncated Green function to the transition matrix element above. All variations of the quark and antiquark fields $\Psi$ and $\bar{\Psi}$  vanish due to truncation, and since 
\be
\bar{u}(p)(\!\not\! p - m )= 0,\,\quad (\!\not\! p - m) u(p) = 0\, . \label{quark}
\ee
 In the following we set all masses to zero. The remaining variation of the antighost $b^b$ gives Eq. (\ref{wardeq}), where now the fields $\Psi$ and $\bar{\Psi}$ are on-shell, denoted by $M$ and $N$. Similar considerations apply to gluons.
Eq. (\ref{wardeq}) says that the sum of all possible attachments of a scalar polarized gluon to a matrix element vanishes. From this follows Fig. \ref{wardeik} a), since, by definition, we do not include the graph into the hard function where the gluon attaches to the physically polarized parton (quark or gluon), shown on the right-hand side of Fig. \ref{wardeik} a). 

The eikonal identity in Fig. \ref{wardeik} b) follows from the eikonal Feynman rules in Fig. \ref{Frules}. The attachment of the unphysical gluon to the fermion line is equivalent to its attachment to an eikonal or Wilson line, or path-ordered exponential $\Phi$, in direction $\beta$ opposite to the fermion line's momentum:
\ba
\Phi_\beta^{(\rm f)} (0,\eta;0) & \equiv & P e^{i g_s \int_0^{\eta} d \xi \beta \cdot {{\mathcal{A}}^{(\rm f)}} (\xi \beta^\mu)} \nonumber \\
& = & 1 + P \sum\limits_{m = 1}^\infty \prod\limits_{i = 1}^m \int \frac{d^n k_i}{(2 \pi)^n} g_s \beta \cdot {\mathcal{A}^{(\rm f)}}(k^\mu_i) \frac{1}{\beta \cdot \sum\limits_{j = 1}^i k_j + i \varepsilon}. \quad \label{path}
\ea
The exponent is the resulting phase rotation on a particle of flavor $(\rm f)$ due to unphysical, scalar polarized, gluons. Here $P$ denotes path ordering, $g_s$ is the strong coupling, and $\mathcal{A}^{(\rm f)}$ is the vector potential in representation $(\rm f)$. In the second line of (\ref{path}) we have expanded the ordered exponential in momentum space. The resulting Feynman rules are precisely the ones mentioned above, of the eikonal approximation, listed in Appendix \ref{sec:eikfeyn}. The eikonal propagators are the result of the path ordering, because
\be
\int\limits_{-\infty}^\infty dz\, e^{i k z} \theta(z) = \frac{i}{k+i \epsilon}.
\ee
 We will represent eikonal lines graphically as double lines. Fig. \ref{wardeik} b) displays the following equality for $p^2 = k^2 = 0$, $k$ collinear to $p$,
\ba
- g_s T^b \bar{u}(p) \frac{\beta^\mu k_\mu}{- \beta \cdot k} & = & g_s T^b \bar{u}(p) \gamma^\mu \frac{1}{\!\not\! p + \!\not\!k} k_\mu \nonumber \\
& = & g_s T^b \bar{u}(p) \frac{\!\not\!k + \!\not\!p - \!\not\!p}{\!\not\! p + \!\not\!k} = g_s T^b \bar{u}(p).
\ea 
where we have used Eq. (\ref{quark}) in the last equality on the right-hand side. $p$ is the quark momentum which flows into the final state, and $k$ the gluon's momentum. $T^b$ is a SU(N) generator in the fundamental representation. Since the right-hand sides of Figures \ref{wardeik} a) and b) are the same (the ``empty'' eikonal line carries no momentum), the left-hand sides are the same. Repeated application of this identity results in Fig. \ref{wardeik} c) \cite{Collins:1985ue,Collins:ig}. Note that the color factors are included in the Ward identity, resulting in the appropriate color factor for the attachments of the gluons as shown in Fig. c). We have succeeded in decoupling unphysical gluons from physical processes, their only effect being phase rotations on the factorized physical momentum lines. In completely inclusive cross sections these phase rotations cancel due to the unitarity of Wilson lines \cite{Collins:ta}:
\be
\Phi_\beta^{(\rm f)\, \dagger} \Phi_\beta^{(\rm f)} = \mathbf{1}.
\ee

\subsection{Glauber/Coulomb Gluons and Infrared Safety} \label{sec:Glauber}

The factorization of soft gluons, which are not necessarily scalar polarized, from jet lines does not follow immediately from the Ward identity Fig. \ref{wardeik}. To achieve factorization in this case, we have to make the following two approximations: the neglect of non-scalar gluon polarizations, and the eikonal approximation discussed above in Sec. \ref{sec:vertex}, which consists of neglecting soft momenta $k^\mu$ in the numerator compared to jet-momenta $p^\mu$ and $k^2$ compared to $p \cdot k$ in the denominator. 
After making these approximations, which are referred to by the term \emph{soft approximation}, we can factor soft momenta from collinear jet momenta with the help of Fig. \ref{wardeik} c).

We start by decomposing each soft gluon propagator $D_{\mu \nu}$ with momentum $k^\mu$ into a scalar-polarized contribution and a remainder \cite{Grammer:1973db,Libby:1978qf}:
\ba
D_{\mu \nu} \left(k \right) & = & G_{\mu \nu} \left(k, \beta \right) + K_{\mu \nu} \left(k,\beta \right),  \label{kgdec}
\ea
 where we define
\ba
K_{\mu \nu} \left(k \right) & \equiv & D_{\mu \rho} \left(k \right) \frac{\beta^\rho \, k_{\nu}}{\beta \cdot k + i \epsilon}, \nonumber \\
G_{\mu \nu} \left(k,\beta \right) & \equiv & D_{\mu \nu} \left(k \right) - D_{\mu \rho} \left(k \right) \frac{\beta^\rho \, k_{\nu}}{\beta \cdot k + i \epsilon}. \label{kgdef}
\ea 
$\beta$ is chosen to flow in the direction opposite to $p$, that is, if the jet flows into the plus-direction, $\beta$ points into the minus-direction, and vice versa. 
 The sign of the $i \epsilon$-prescription is chosen in such a way as to not introduce new pinch singularities near the ones produced by the soft gluons under consideration. For momenta to the right of the cut in initial-state jets we also have $+i\epsilon$, since the sign of the momentum flow to the right of the cut is reversed relative to the jet momentum $p$. That is, we have for momenta to the left of the cut $(p+k)^2 + i \epsilon \approx 2 p \cdot k + i \epsilon$, whereas to the right of the cut we obtain $(p-k)^2 - i \epsilon \approx -2 p \cdot k - i \epsilon$.

From Eq. (\ref{kgdef}) we see, that for the scalar polarized $K$-gluons the identity shown in Fig. \ref{wardeik} c) is immediately applicable, leading to the desired factorized form. So it remains to be shown that the $G$-gluons do not give leading contributions. Power-counting, as we have demonstrated in the previous section, shows that only 3-point vertices are relevant for the coupling of soft gluons to jets. 

\begin{figure}[hbp] 
\begin{center}
\epsfig{file=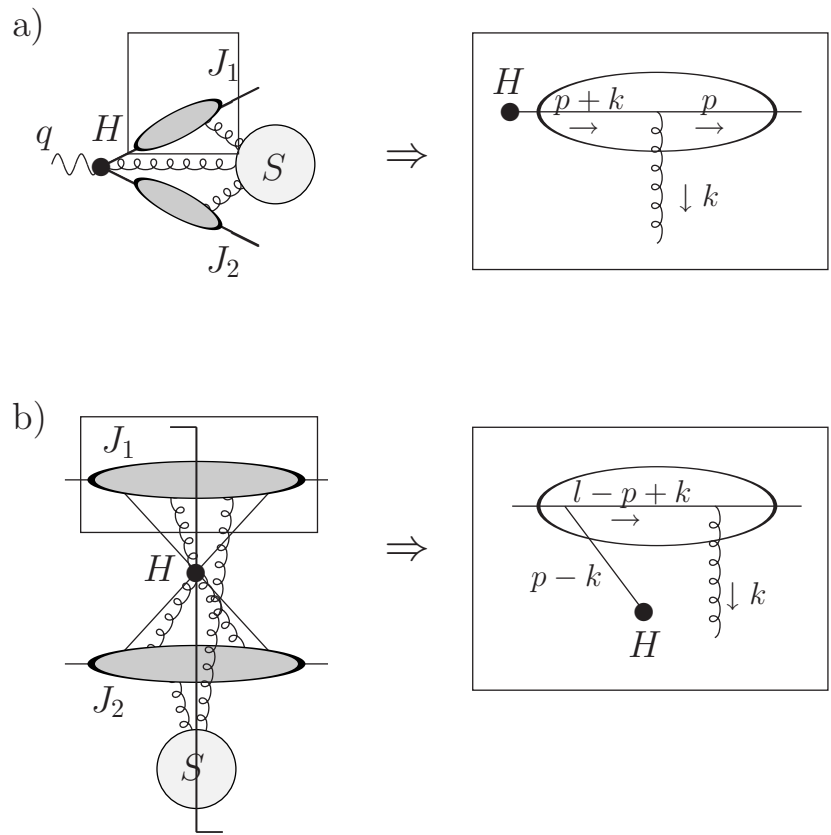,height=12.0cm,clip=0} 
\caption[Illustration of momentum flow for two jet-configurations: a) a final-state jet in $e^+e^-$ annihilation, b) a jet which radiates into initial and final states in the Drell-Yan process. The reduced diagrams in cut-diagram notation for the two processes are shown on the left, the jet-configurations which we study as examples are shown in the boxes.]{Illustration of momentum flow for two jet-configurations: a) a final-state jet in $e^+e^-$ annihilation, b) a jet which radiates into initial and final states in the Drell-Yan process. The reduced diagrams in cut-diagram notation for the two processes are shown on the left, the jet-configurations which we study as examples are shown in the boxes. The vertical line (cut) represents the final state. }  \label{Glaubfig}    
\end{center}    
\end{figure}

First, let us consider a 3-point vertex in a purely initial or purely final state jet, as shown in Fig. \ref{Glaubfig} a). Such jets, as indicated in the left part of Fig. \ref{Glaubfig} a) occur for example in $e^+e^-$ annihilation, as we have seen in Sec. \ref{sec:vertex} above. The soft $G$-gluon with propagator $G_{\mu \nu}(k,\beta)$ couples to a fermion jet-line with momentum $p^\mu$ in a jet moving in the plus-direction:
\be
\frac{\not \! k + \not \! p}{(k+p)^2+i \epsilon} \gamma^\nu \frac{\not \! p}{p^2+i\epsilon} G_{\mu \nu}  \left(k,\beta  \right) \approx 2 \frac{\not \! p}{\left((k+p)^2+i \epsilon\right)\left(p^2+i\epsilon\right) } p^\nu 
G_{\mu \nu}  \left(k,\beta \right) \approx 0, \label{gapprox}
\ee
because $p^\nu \approx p^+ \beta^\nu$.  Corrections are proportional to $\lambda Q, \lambda \ll 1$, as follows from the power counting described in Sect. \ref{sec:powercount} when neglecting $k$, and because $ G_{\mu \nu}  \left(k,\beta  \right) \beta^\nu = 0$. An analogous observation holds for the coupling of $G$-gluons to jet-lines via triple-gluon vertices. In (\ref{gapprox}) we neglect all terms of order $\lambda Q$ in the numerator, including the momentum $k$, because we assume that the denominator also scales as $\sim \lambda Q$. This approximation is only valid for soft gluons not in the Glauber or Coulomb region, where the denominator behaves as $\sim \lambda^2 Q$. If the soft momenta are not pinched in this Glauber/Coulomb region we can deform the integration contours over these momenta away from this dangerous region, into a purely soft region where the above approximations are applicable. This is straightforward to show for purely virtual  initial or final state jet configurations, as displayed in Fig. \ref{Glaubfig} a)\footnote{Initial and final states are defined with respect to the hard scattering.}.

Consider again the 3-point vertex in Eq. (\ref{gapprox}), where now the gluon with momentum $k$ is in the Glauber/Coulomb region. If $|p^+ k^-|$ is not dominant over $|2 p_\perp \cdot k_\perp + k_\perp^2|$ in the denominator our approximation fails. The poles of the participating denominators are in the $k^-$ complex plane at
\ba
k^- & = & \frac{k_\perp^2 - i \epsilon}{2 k^+}, \nonumber \\
k^- & = & \frac{(k_\perp+p_\perp)^2 - i \epsilon}{2 (k^+ + p^+)} - p^- .
\ea
As long as the jet-line $p$ carries positive plus momentum, we see that the $k^-$-poles are not pinched in the Glauber region. In this case we can deform the contour away from this region into the purely soft region, where $ |p^+ k^-| \gg |2 k_\perp \cdot p_\perp + k_\perp^2|$. In a reduced diagram, which represents a physical process, there must be at every vertex at least one line whose plus momentum  flows into the vertex, and at least one line whose plus momentum flows out of the vertex. Thus, at every vertex,  we can always find a momentum $p$ for which the above observation holds, and we can always choose the flow of $k$ through the jet along such lines. If the soft gluon momenta $k$, which we want to decouple, connect only to a purely virtual or final state jet, this observation remains true throughout the jet.  An analogous argument applies to the right of the cut.

For other jet configurations, where the jet appears in both initial and final states, the situation is significantly more complicated. Here pinches in the Glauber/Coulomb region occur on a diagram by diagram basis, and cancel only in sufficiently inclusive cross sections. Let us illustrate this at the example of the inclusive Drell-Yan cross section shown in Fig. \ref{Glaubfig} b) in cut-diagram notation, $A + B \rightarrow X$, where $A$ and $B$ are hadrons, and $X$ is all radiation into the final state. In a cut diagram, the amplitude is displayed to the left of the cut (the vertical line), the complex conjugate amplitude is shown to the right. 

\clearpage

\begin{figure}[hbt] 
\begin{center}
\epsfig{file=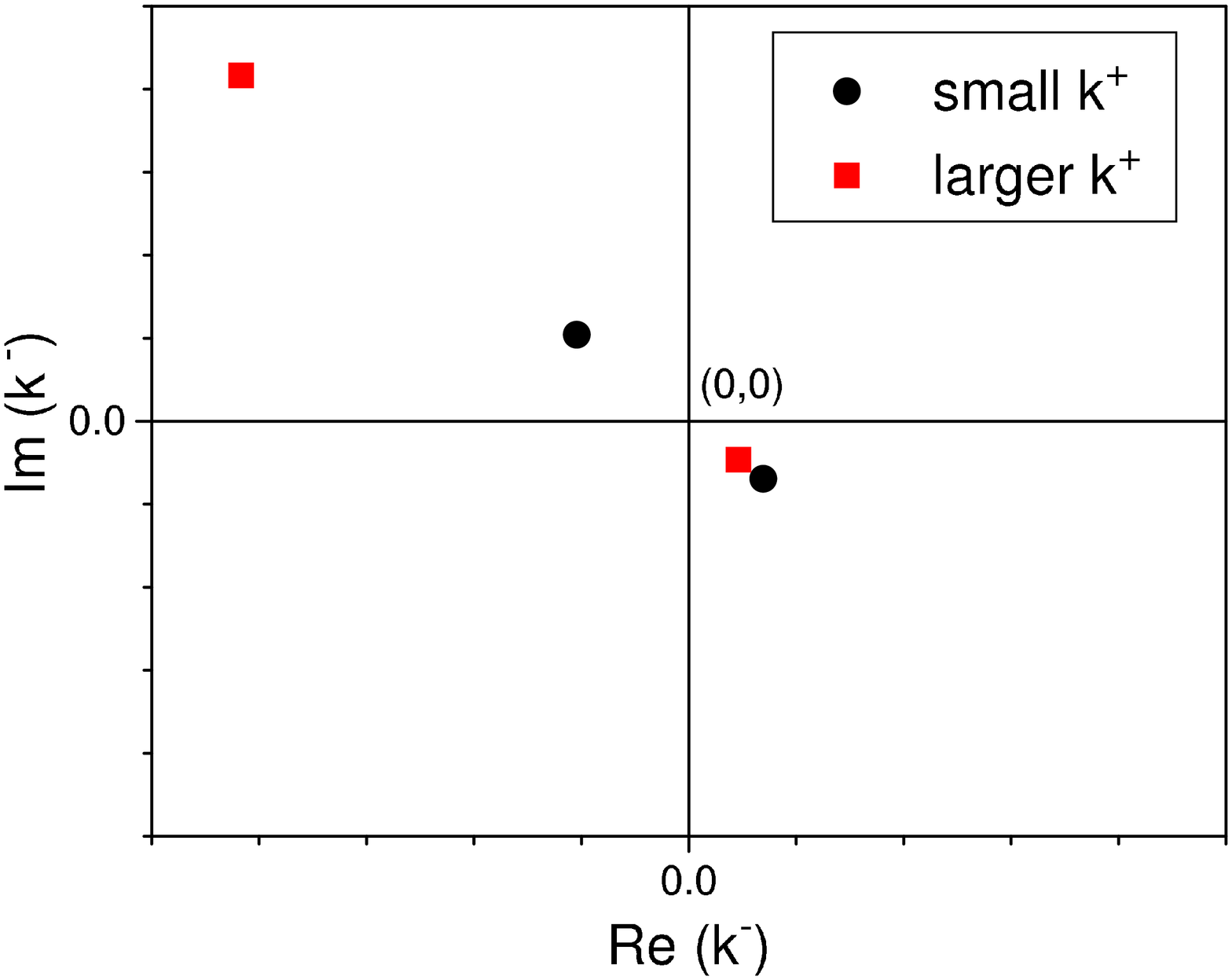,height=7.7cm,angle=270,clip=0} 
\caption{Pole structure of configuration Fig. \ref{Glaubfig} b) in the complex plane, according to Eq. (\ref{DYpole}) for small and larger $k^+$. The scale is arbitrary. }  \label{pole}    
\end{center}    
\end{figure}

In Fig. \ref{Glaubfig} b), a soft gluon with momentum $k$ connects the active jet line $p - k$ which participates in the hard scattering, and the spectator jet line $l - p + k$ which flows into the final state after emitting the soft gluon. The relevant denominators have poles in the $k^-$ complex plane at
\ba
k^- & = & \frac{k_\perp^2 - i \epsilon}{2 k^+}, \nonumber \\
k^- & = & \frac{-(p_\perp - k_\perp)^2 + i \epsilon}{2 (p^+ - k^+)} + p^-, \nonumber \\
k^- & = & \frac{(l_\perp - p_\perp + k_\perp)^2 - i \epsilon}{2 (l^+ - p^+ + k^+)} + p^- - l^-. \label{DYpole}
\ea
The pole-structure for $k^-$ in the complex plane is shown in Fig. \ref{pole}. For fixed, small $k^+$ the contour is pinched between the poles of lines $p - k$ and $l - p + k$ which are in the Glauber/Coulomb region. This is the case for the pole-configuration shown as filled circles in the figure. If $k^+$ is not pinched, however, then its contour can be deformed such that the two poles move away from each other, as illustrated by the poles drawn as squares. Then the $k^-$ contour is not trapped near the Glau\-ber/Cou\-lomb region, and the soft approximation is applicable. On the other hand, if also $k^+$ is pinched by a configuration similar to Fig. \ref{Glaubfig} b) in the other jet, then the contour cannot be deformed, and the soft approximation fails. This occurs in general if there are at least two initial state hadrons or jets whose spectator fragments proceed into the final state. 
  
A careful analysis \cite{Collins:1985ue,Collins:ig,Bodwin:1984hc,Collins:1998ps} shows that these pinches in the Glau\-ber/Cou\-lomb region cancel in the sum over states for the inclusive Drell-Yan cross section, and in general, in sufficiently inclusive cross sections. Let us sketch the proof of cancellation to all orders, following \cite{Collins:ig}.

\subsubsection{Cancellation of Pinches in the Glauber/Coulomb Region}

The cancellation is best seen in light-cone ordered perturbation theory (LCOPT). 
The rules for LCOPT can be found in Appendix \ref{app:LCOPT} \cite{Chang:1968bh,Kogut:1969xa,Brodsky:1973kb}, they can be derived by performing the minus integrals. LCOPT is similar to old-fashioned, or time-ordered, perturbation theory, but ordered along the light-cone, $x^+$, rather than in $x^0$. In a LCOPT diagram all internal lines are on-shell, in contrast to a covariant Feynman diagram. A covariant Feynman diagram is comprised of one or more LCOPT diagrams.

In LCOPT the jet shown in Fig. \ref{Glaubfig} b) at a specific cut $(C)$ can be written schematically as a sum over $x^+$-orderings $T$ of its vertices:
\be
J^{(C)} = \sum_T I'_T \left(k^\alpha_l \right)^{*} \otimes F_T^{(C)} \left(k^\alpha_l \right) \otimes I_T \left(k^\alpha_l \right). \label{IFI}
\ee
The factor $I_T$ collects all initial state interactions to the left of the cut, ${I'_T}^*$ contains all initial state interactions to the right of the cut, and $F_T^{(C)}$ collects all final state interactions consistent with cut $(C)$. The functions are linked by soft gluons $\left\{k^\alpha_l \right\}$, as indicated by the symbol $\otimes$.  Following the rules in Appendix \ref{app:LCOPT}, the factors $I_T, {I'_T}^*$ and $F_T^{(C)}$ are given by
\ba
I_T \left(k^\alpha_l \right) \!\! & = & \!\!\prod\limits_{\mbox{ \tiny states } \xi < H} \left[ \sum\limits_{\mbox{\tiny vertices } l < \xi} \left( k^-_l + i \epsilon \right) - \sum\limits_{\mbox{\tiny lines } j \in \xi} \frac{q_{j,\,\perp}^2}{2 q_j^+} \right]^{-1}, \label{init} \\
I'_T \left(k^\alpha_l \right)^{*} \!\!& = &\!\! \prod\limits_{\mbox{ \tiny states } H' < \xi} \left[- \sum\limits_{\mbox{\tiny vertices } \xi < l} \left( k^-_l + i \epsilon \right) - \sum\limits_{\mbox{\tiny lines } j \in \xi} \frac{q_{j,\,\perp}^2}{2 q_j^+} \right]^{-1},  \label{init2} \\
F_T^{(C)} \left(k^\alpha_l \right) \!\!& = & \!\!\int\limits_{-\infty}^\infty \frac{d p_J^-}{2 \pi} \prod\limits_{\mbox{ \tiny states } C < \xi < H'} \left[- p_J^- + \sum\limits_{\mbox{\tiny vertices } l < \xi} \left( k^-_l - i \epsilon \right) - \sum\limits_{\mbox{\tiny lines } j \in \xi} \frac{q_{j,\,\perp}^2}{2 q_j^+} \right]^{-1} \nonumber \\
& & \times \, 2 \pi \delta \left( - p_J^- + \sum\limits_{\mbox{\tiny vertices } l < C}  k^-_l - \sum\limits_{\mbox{\tiny lines } j \in \xi} \frac{q_{j,\,\perp}^2}{2 q_j^+} \right) \nonumber \\
& & \times \,  \prod\limits_{\mbox{ \tiny states } H < \xi < C} \left[- p_J^- + \sum\limits_{\mbox{\tiny vertices } l < \xi} \left( k^-_l + i \epsilon \right) - \sum\limits_{\mbox{\tiny lines } j \in \xi} \frac{q_{j,\,\perp}^2}{2 q_j^+} \right]^{-1} \, . \label{final}
\ea
The vertices and lines are ordered with respect to $x^+$. $p_J^-$ is the minus momentum leaving the jet $J$ at the hard vertex $H$, and $p^-_J - \sum k^-_l$ flows back into the vertex at $H'$. Eqs. (\ref{init})-(\ref{final}) exhibit the same pinching of poles as the simple example, Eq. (\ref{DYpole}).

For a given $x^+$-ordering of vertices we can sum over all cuts $(C)$, if the remainder of the factorized cross section (that is, the other jet or jets) is independent of the choice of which of the vertices where soft gluons join the jet $J$ are to the left or to the right of the cut $(C)$. Let us assume this independence for the moment. Then the cancellation of final state interactions, $F_T^{(C)}$, is a matter of straightforward application of the unitarity relation:
\ba
\sum\limits_{c = 1}^N \Bigg\{  \prod\limits_{n = c+1}^N \frac{1}{-p_J^- -D_n - i \epsilon} \!\!&\!\! 2 \pi \!\!& \!\! \delta\left(-p_J^- - D_c \right) \prod\limits_{n = 1}^{c-1} \frac{1}{-p_J^- -D_n + i \epsilon} \Bigg\} \nonumber \\
& & \hspace*{-20mm} =\, i \prod\limits_{n = 1}^N \frac{1}{-p_J^- -D_n + i \epsilon} - i \prod\limits_{n = 1}^N \frac{1}{-p_J^- -D_n - i \epsilon}, \label{unit}
\ea
and the fact that the $p_J^-$ integral vanishes for any $N > 1$.
Applying Eq. (\ref{unit}) to (\ref{IFI}), we obtain
\be
\sum_{C} J^{(C)} = \sum_{C} \sum_T I'_T \left(k^\alpha_l \right)^{*} \otimes I_T \left(k^\alpha_l \right), \label{endglaub}
\ee
where now only initial states are kept.

Thus the Glauber/Coulomb pinches disappear, and the soft approximation is applicable. Two subtleties remain to be discussed for a complete proof: why the remainder of the graph is independent of the choice of which of the soft gluon fields are to the left or to the right of the final state cut, and why all transverse components of soft momenta can safely be neglected in Eq. (\ref{endglaub}). 

The independence of the remainder of the graph of soft gluon arrangements can be proven in exactly the same way as the above cancellation of Glauber/Coulomb pinches, by using LCOPT in the form (\ref{IFI}) for the remainder, and applying (\ref{unit}). Details can be found in \cite{Collins:ig}. 

The neglect of transverse components may not seem obvious from the LCOPT-forms (\ref{init}) and (\ref{init2}), since it can happen that $\sum_l k_l^- = 0$, that is, no minus-momentum flows into a specific set of vertices. As stated above, the sum of all LCOPT-expressions of a given graph and performing all internal minus-integrals of this graph give equivalent expressions, however, the two forms differ by partial fraction manipulations.
In the latter form these apparent divergences as $\sum_j q_{j,\,\perp}^2 \rightarrow 0$ when $\sum_l k_l^- = 0$,  are absent. Therefore, the transverse components can be neglected, and our overview of the proof of cancellation of Glauber/Coulomb pinches is complete. For further information and a nontrivial example we refer to Refs. \cite{Collins:ig,Collins:1998ps}.

\subsection{Summary}

We have succeeded in identifying and disentangling the regions in momentum space that give leading contributions by means of power counting and application of Ward identities. This results in a product of hard scattering, jet, and soft functions. These functions, although separated in momentum space by a factorization scale, are still at least additionally linked by eikonal lines. The eikonal lines replace in each of the functions the momenta of the partons that do not give leading contributions to the particular region. 
In most cases, we can construct operators for each of the functions which reproduce exactly their leading behavior. Examples will be given below. 

The above arguments are valid at arbitrary orders in the perturbative expansion.  In Refs. \cite{Sen:sd,Collins:1981uk} a recursive algorithm, the so-called ``tulip-garden formalism'',  was developed to systematically disentangle the leading contributions of any given diagram. This algorithm is similar to Zimmermann's forest formula for ultraviolet divergences \cite{Zimmermann:1969jj}. The result of applying the tulip-garden formalism to all diagrams up to a given order in perturbative QCD is of course equivalent to the result of the approach described above. We refer the interested reader to the literature \cite{Sen:sd,Collins:1981uk} for further information.

We emphasize that a careful treatment of Glauber/Coulomb gluons is necessary for a complete proof of factorization in processes where two or more initial-state jets contribute to the final state.

\section[Refactorization of Nonsinglet Partonic Splitting Functions]{Refactorization of Nonsinglet Partonic \\ Splitting Functions} \label{sec:pdffact}

Let us now apply the above to determine the leading behavior of parton-in-parton distribution functions in the limit $x\rightarrow 1$, that is, in the limit that the emerging parton carries nearly all the momentum of the original parton. In the introduction, Sec. \ref{sec:assumpt}, we have introduced parton distribution functions (PDFs) $f_{a/A}(x)$ which describe the distribution of parton $a$ in hadron $A$.

\subsection{Definition of Parton-in-Parton Distribution Functions}

Hadronic distribution functions $f_{a/A}(x)$ are incalculable within perturbation theory. However, their evolution is perturbatively calculable from the renormalization group equation \cite{Gribov:ri,Lipatov:qm,Altarelli:1977zs,Dokshitzer:sg}
\be
\mu \frac{d}{d \mu} f_{a/A} (x,\mu) = \sum_b \int\limits_{x}^1 \frac{d \zeta}{\zeta} P_{a b}
\left(\zeta,\alpha_s(\mu)\right) f_{b/A} \left(\frac{x}{\zeta},\mu \right), \label{evol}
\ee
where $P_{a b}$ is the evolution kernel or splitting function, and $\mu$ denotes the factorization scale, usually taken equal to the renormalization scale. This follows from the factorization assumption, Eq. (\ref{fact}), which enables us to write a large class of  physical cross sections as convolutions of these 
PDFs with perturbatively calculable short-distance functions. Eq. (\ref{evol}) follows from the independence of the physical cross section of the factorization scale $\mu$, as seen in Eq. (\ref{evolf}).
Similar considerations apply to partonic cross sections. There the parton-in-parton distribution functions  $f_{f_i/f_j}$ describe the probability of finding parton $f_i$ in parton $f_j$. The evolutionary behavior of the partonic PDFs obeys the same equation, (\ref{evol}), as for the hadronic PDFs. Thus the splitting functions can be computed in perturbation theory.

Up to non-leading corrections which vanish as $x \rightarrow 1$ we can neglect flavor mixing, that is, we deal with non-singlet distributions:
\be
f_{NS}^{\pm} = f^\pm_{q_a/q} - f^\pm_{q_b/q},\quad  f^\pm_{q_i/q} = f_{q_i/q} \pm f_{\bar{q}_i/q}.
\ee
Factorization allows us to define PDFs in terms of nonlocal operators.  At leading power one can define \cite{Collins:gx,Collins:1981uw} the following function\footnote{Recently there has been some discussion about the correct behavior of transverse-momentum dependent PDFs in some noncovariant gauges \cite{Brodsky:2002ue,Ji:2002aa}. This question does not arise in this thesis since we will work in Feynman gauge and consider only distribution functions where all transverse-momentum dependence has been integrated out.}
\ba
f_{q_i/q} (x) & =  & \frac{1}{4 \pi} \int d y^- e^{-i x p^+ y^-} \left< p \left| \bar{\psi}_i (0,y^-,0_\perp) \gamma^+  \right. \right. \nonumber \\
& & \qquad \qquad \qquad \qquad \quad \times \, \left. \left. P e^{i g_s \int_0^{y^-} d z^- {\mathcal{A}}^{(q)\, +} (0,z^-,0_\perp)} \psi_i(0) \right| p \right> \nonumber \\
& = &  \frac{1}{4 \pi} \int d y^- e^{-i x p^+ y^-} \sum_n \left< p \left| \bar{\Psi}_i (0,y^-,0_\perp)\right| n \right>
\gamma^+ \left< n \left|  \Psi_i(0)  \right| p \right>, \,\,\,\,\,\,\,\,\,\,\,\,\,\label{pdfdef}
\ea
which describes the distribution of a quark $q_i$, created by the operator $\bar{\psi}_i$, in a quark $q$ with momentum $p$. We use light-cone coordinates where our conventions are as in (\ref{lccoord}). The operators are separated by a light-like distance, and are joined with a path-ordered exponential, denoted by $P$, to achieve gauge-invariance. This exponential describes the emission of arbitrarily many  gluons of polarization in the plus-direction. ${\mathcal{A}}^{(q)}$ is the vector potential in the fundamental representation. In the second line we have inserted a complete set of states, have used the identity
\be
P e^{i g_s \int_0^{\eta} d \xi \beta \cdot {\mathcal{A}} (\xi n^\mu )}  = \left[P e^{i g_s \int_0^{\infty} d \xi \beta \cdot {\mathcal{A}} ( (\xi+\eta) \beta^\mu )}\right]^\dagger \left[ P e^{i g_s \int_0^{\infty} d \xi \beta \cdot {\mathcal{A}} (\xi \beta^\mu )} \right],  \label{phaseop}
\ee
and have defined
\be
\Psi_i(y) = P e^{i g_s \int_0^{\infty} d \xi \beta \cdot {\mathcal{A}} (y + \xi \beta^\mu )}  \psi_i(y),
\ee
with the light-like vector $\beta^\mu$ chosen in the minus-direction, $\beta^+ = \beta_\perp = 0$. As we have seen in the previous section, the occurrence of the path-ordered exponential follows from the factorization of unphysical gluons from the physical, short-distance, cross section.  Gluon parton distribution functions can be constructed analogously \cite{Collins:gx,Collins:1981uw}, with the vector potentials in the adjoint representation, and appropriate operators for the creation of gluons.

In the following we will be interested in the limit $x \rightarrow 1$, since there large logarithmic corrections arise due to soft-gluon radiation.  We will show below that the factorized form of a perturbative non-singlet parton-in-parton distribution function contains a cross section built out only of eikonal lines that absorbs all collinear and infrared singular behavior as $x \rightarrow 1$. An equivalent observation was made by Korchemsky \cite{Korchemsky:1988si}, who related the flavor-diagonal splitting function $P_{ff}$ to the cusp anomalous dimension of a Wilson loop.

\subsection{Power Counting as $x \rightarrow 1$} 

\begin{figure}
\begin{center}
\parbox{5.5cm}{\vspace*{-1.55cm} \epsfig{file=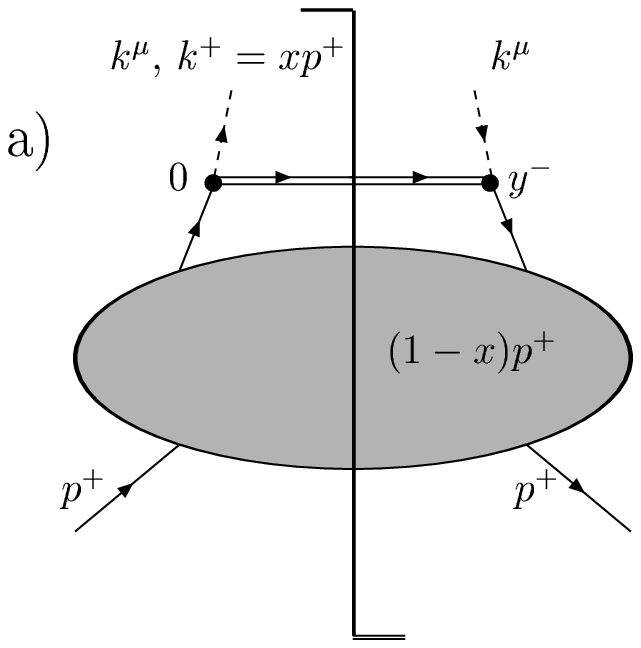,height=5.3cm,clip=0} }
\hspace*{1mm} 
\parbox{6cm}{\epsfig{file=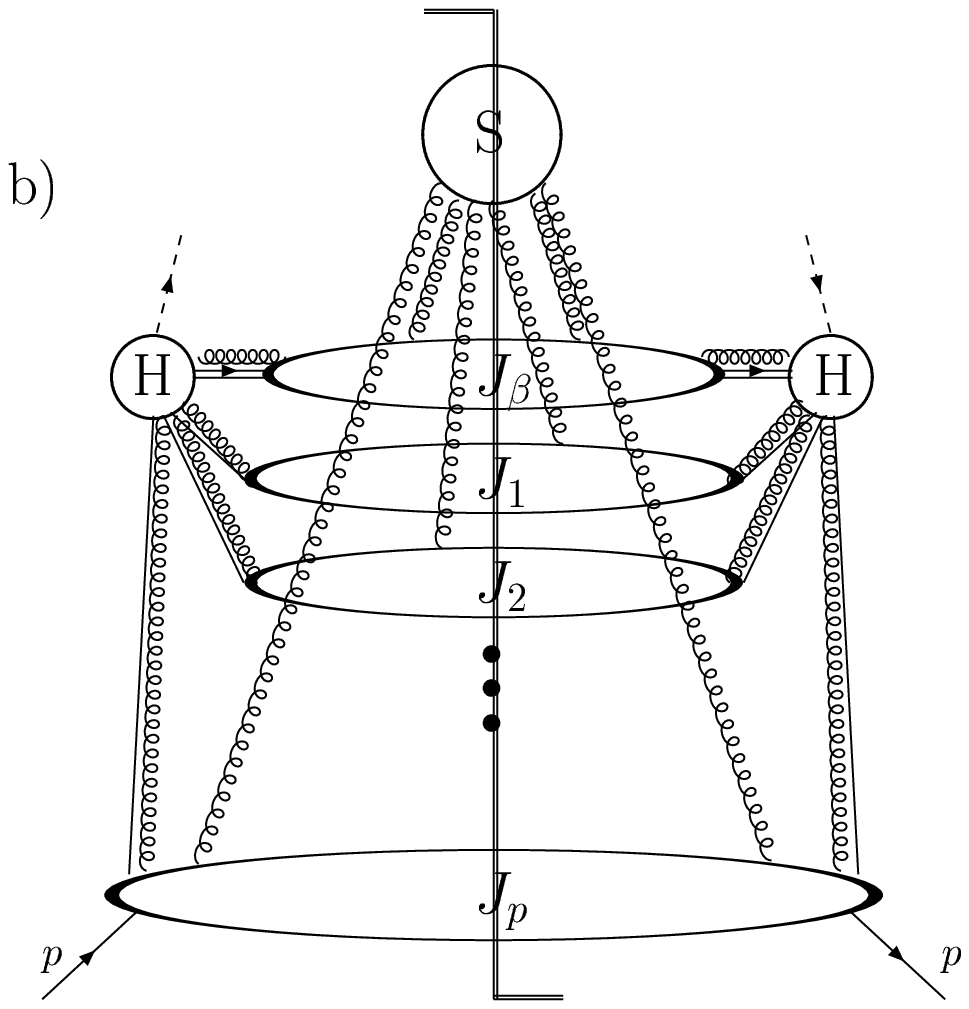,height=7.9cm,clip=0}}
\vspace*{2mm}
\\
\vspace*{1mm}
\epsfig{file=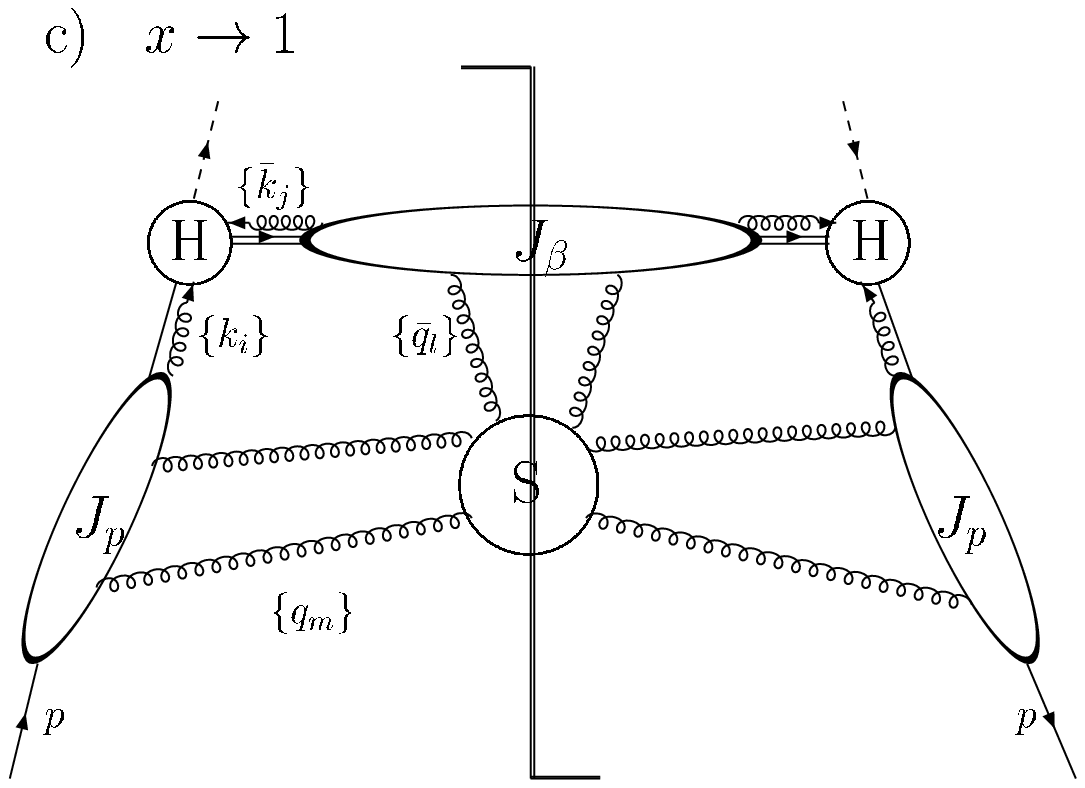,height=6.2cm,clip=0}
\caption[a) Graphical representation of a parton-in-parton distribution function, Eq. (\ref{pdfdef}), and b) its factorized form for arbitrary momentum fraction $x$, drawn as a reduced diagram. c) The reduced diagram for the PDF in the limit $x \rightarrow 1$.]{a) Graphical representation of a parton-in-parton distribution function, Eq. (\ref{pdfdef}), and b) its factorized form for arbitrary momentum fraction $x$, drawn as a reduced diagram. c) The reduced diagram for the PDF in the limit $x \rightarrow 1$. Here we have suppressed the labels $L$ and $R$  for purely virtual contributions to the left and to the right of the cut, respectively, compared to the notation in the text. The cut represents the final state.} \label{partondef}
\end{center}
\end{figure}

We start with the definition Eq. (\ref{pdfdef}) of a perturbative parton-in-parton distribution function, which is shown in Fig. \ref{partondef} a) in cut-diagram notation.  We pick the incoming momentum to flow in the plus direction, 
\be
p^\mu = (p^+,0,0_\perp).
\ee 
Because the minus and transverse momenta in (\ref{pdfdef}) are integrated over, they can flow freely through the eikonal line, whereas no plus momentum flows across the cut in the eikonal line. 

The regions that can give leading contributions are shown in Fig. \ref{partondef} b) in form of a reduced diagram, following the discussion in Sec. \ref{sec:landau}. 
We can have a jet $J_p$ collinear to the incoming momentum $p^+$, as well as an arbitrary number of jets $J_i$ emerging from the hard scattering. Furthermore, we can have momenta collinear to the eikonal moving in the minus direction, $\beta^-$, represented by $J_\beta$ in the figure. The jets can be connected by arbitrarily many soft gluons, $S$. Here and below, unless explicitly stated otherwise, we use the term ``soft'' for both soft and Glauber momenta.

A further simplification of the leading behavior occurs when we perform the limit to $x \rightarrow 1$, in which we are interested here. Jet lines having a finite amount of plus momentum in the final state become soft, and only virtual contributions can have large plus momenta. This does not affect the jet collinear to the eikonal, since it is moving in the minus direction. Thus we arrive at the leading regions depicted in Fig. \ref{partondef} c), with hard scatterings $H_L$ and $H_R$, which are the only vertices where finite amounts of momentum can be transferred, virtual jets $J_{p,\,L}$ and $J_{p,\,R}$ collinear to the incoming momentum $p^+$, a jet $J_\beta$ collinear to the eikonal $\beta^-$, connected via soft momenta, $S$. Here and below the subscripts $L$ and $R$, respectively, indicate that the momenta and functions are purely virtual, located to the left or to the right of the cut.

Let us now determine the degree of infrared divergence $\omega(f)$ of Fig. \ref{partondef} c) using the tools developed in Sec. \ref{sec:powercount}. The degree of divergence of the various regions is additive,
\be
\omega (f) = \o_{J_{p,\,L}} + \o_{J_{p,\,R}} + \o_{J_{\beta}} + \o_S.
\ee

The degree of divergence of the soft function is given by
\be
\omega_S = 4 (E^b + E^f -1) - 2 E^b - E^f + \omega'_S, \label{soft1}
\ee
where $\o'_S$ denotes the degree of divergence of the internal part of the soft function only, without the lines attaching to the jets, which are denoted by $E^b$ and $E^f$, for bosonic and fermionic lines, respectively. The first term in Eq. (\ref{soft1}) comes from $E - 1$ loop integrations over these attachments, the second and third term stem from the denominators of these loops. $\o'_S$ is found from Eqs. (\ref{softdiv}) with (\ref{softnum}), (\ref{linevertex}), and (\ref{euler}):
\be
\o'_S = 4 - E^b - \frac{3}{2} E^f.
\ee
Putting everything together, we arrive at
\be
\o_S = E^b + \frac{3}{2} E^f. \label{dodsoft}
\ee
This result can also be obtained by simple dimensional analysis of the soft function since soft momenta have the same scaling behavior in all components.

The degree of divergences of the jet functions can be found similarly, using Eqs. (\ref{jetdiv}), (\ref{physnum}) or (\ref{covnum}), (\ref{euler}), and slight modifications of (\ref{linevertex}) for jet $J_{p,\,L/R}$:
\be
2 N_{J_{p,\,L/R}} + E_{J_{p,\,L/R}} + 1 = \sum_i i v^{(i)}_{J_{p,\,L/R}} + l_{J_{p,\,L/R}} + p_{J_{p,\,L/R}},
\ee
where the number of external lines counts only attachments to the soft function, but not to the hard scattering. These have to be added separately, where $l_{J_{p,\,L/R}}$ and $p_{J_{p,\,L/R}}$ denote scalar polarized gluons and physical lines (scalars, fermions, physically polarized gluons), respectively, which connect the jet (to the left or to the right of the cut) and the hard part.
Similarly, the modifications for jet $J_\beta$ result in 
\be
2 N_{J_\beta} + E_{J_\beta} = \sum_i i v^{(i)}_{J_\beta} + l_{J_\beta} + p_{J_\beta}.
\ee
After a bit of algebra we find in Feynman gauge:
\ba
 \o_{J_{p,\,L/R}} & \geq & \frac{1}{2} \left[ - 1 - E^f_{J_{p,\,L/R}} - E^b_{J_{p,\,L/R}} + p_{J_{p,\,L/R}} - s_{J_{p,\,L/R}} \right.  \label{dodjetp} \\ 
& & \quad \left.  + \left(s_{J_{p,\,L/R}} + l_{J_{p,\,L/R}} - v^{(3)}_{J_{p,\,L/R}} \right) 
\theta \left( s_{J_{p,\,L/R}} + l_{J_{p,\,L/R}} - v^{(3)}_{J_{p,\,L/R}} \right) \right],\nonumber   \\
\o_{J_\beta} & \geq & - 2 - \frac{1}{2} E^f_{J_\beta} - \frac{1}{2} E^b_{J_\beta} + \frac{1}{2} p_{J_\beta} - \frac{1}{2} s_{J_\beta} \nonumber \\
& & \quad  + \left(s_{J_{\beta}} + l_{J_{\beta}} - v^{(3)}_{J_{\beta}} \right) 
\theta \left( s_{J_{\beta}} + l_{J_{\beta}} - v^{(3)}_{J_{\beta}} \right). \label{dodjetb}
\ea

Adding Eqs. (\ref{dodsoft}), (\ref{dodjetp}), and (\ref{dodjetb}), we arrive at
\ba
\o(f) & \geq & \left\{ \frac{1}{2} \left( E^b_{J_{p,\,L}} +  E^b_{J_{p,\,R}} \right) + E^f_{J_{p,\,L}} +  E^f_{J_{p,\,R}} \nonumber  \right. \\
& & \quad  + \frac{1}{2} \left( p_{J_{p,\,L}} + p_{J_{p,\,R}} - s_{J_{p,\,L}} - s_{J_{p,\,R}} \right) - 1 \nonumber  \\
 & & \quad  + \frac{1}{2} \left[  \left(s_{J_{p,\,L}} + l_{J_{p,\,L}} - v^{(3)}_{J_{p,\,L}} \right) 
\theta \left( s_{J_{p,\,L}} + l_{J_{p,\,L}} - v^{(3)}_{J_{p,\,L}} \right) \nonumber \right. \\
& & \qquad \,\,\, \left. \left. +  \left(s_{J_{p,\,R}} + l_{J_{p,\,R}} - v^{(3)}_{J_{p,\,R}} \right) 
\theta \left( s_{J_{p,\,R}} + l_{J_{p,\,R}} - v^{(3)}_{J_{p,\,R}} \right) \right] \right\} \nonumber \\
& & \quad + \left\{ E^f_{J_\beta} + \frac{1}{2} E^b_{J_\beta} - 2 + \frac{1}{2} p_{J_\beta} - \frac{1}{2} s_{J_\beta} \nonumber \right. \\
& & \quad  + \left. \left(s_{J_{\beta}} + l_{J_{\beta}} - v^{(3)}_{J_{\beta}} \right) 
\theta \left( s_{J_{\beta}} + l_{J_{\beta}} - v^{(3)}_{J_{\beta}} \right) \right\}.
\ea
From this lengthy expression we see that the maximum degree of divergence is
\be
\o(f) \geq -1.
\ee
$\o(f) = -1$ corresponds to a divergence proportional to $1/(1-x)$.

To get this maximum degree of IR divergence the following conditions have to be fulfilled (compare to Fig. \ref{partondef} c) ):
\begin{itemize}
\item No soft vectors directly attach $H_L$ or $H_R$ with $S$.
\item The jets and the soft part can only be connected through soft gluons, denoted by the sets $\left\{q_{L,\,m_L}\right\}$, $\left\{q_{R,\,m_R}\right\}$, and $\left\{\bar{q}_l\right\}$.
\item Arbitrarily many scalar-polarized gluons, $\left\{k_{L,\,i_L}\right\}$, $\left\{k_{R,\,i_R}\right\}$, $\{\bar{k}_{L,\,j_L}\}$ and $\{\bar{k}_{R,\,j_R}\}$ , attach the jets with $H_L$ and $H_R$, respectively. The barred momenta are associated with $J_\beta$, the unbarred with the $J_p$'s.
\item Exactly one scalar, fermion, or physically polarized gluon with momentum $k_L^\mu - \sum\limits_{i_L} k_{L,\,i_L}^\mu$, $k_R^\mu - \sum\limits_{i_R} k_{R,\,i_R}^\mu$,  $\bar{k}_L^\mu - \sum\limits_{j_L} \bar{k}_{L,\,j_L}^\mu$ or  $\bar{k}_R^\mu - \sum\limits_{j_R} \bar{k}_{R,\,j_R}^\mu$, respectively, connects each of the jets with the hard parts.
The momenta $k_L^{\nu_L}$ ($k_R^{\nu_R}$) and $\bar{k}_L^{\rho_L}$ ($\bar{k}_R^{\rho_R}$) denote the \emph{total} momenta flowing into $H_L$ ($H_R$) from $J_{p,\,L}$ ($J_{p,\,R}$) and $J_\beta$, respectively, that is, they are the sum of the scalar, fermion or physically polarized gluon momenta, and the scalar-polarized gluon momenta.

In an individual diagram we can have only scalar-polarized gluons connecting the jets with the hard parts, and no scalar, fermion or physically polarized gluon. However, the sum of these configurations vanishes after application of the Ward identity shown in Fig. \ref{wardeik} a).
\item The number of soft and scalar-polarized vector lines emerging from a particular jet is less or equal to the number of 3-point vertices in that jet.
\end{itemize}

In summary, we have found that the regions in momentum space which give leading contributions may in Feynman gauge be represented as:
\ba
f_{f}^{x \rightarrow 1}\;(x)\!\! &\!\!\! = \!\!& \!\!\! \sum\limits_{C_\beta,C_S} \int\frac{d^n k_L}{(2 \pi)^n} \int \frac{d^n k_R}{(2 \pi)^n}\int\frac{d^n \bar{k}_L}{(2 \pi)^n}  \int\frac{d^n \bar{k}_R}{(2 \pi)^n} \nonumber \\
& &
\prod\limits_{i_L,i_R,j_L,j_R} \int \frac{d^n k_{L,\,i_L}}{(2 \pi)^n}  \frac{d^n k_{R,\,i_R}}{(2 \pi)^n}  \frac{d^n \bar{k}_{L,\,j_L}}{(2 \pi)^n}  \frac{d^n \bar{k}_{R,\,j_R}}{(2 \pi)^n}\nonumber \\
& &  \prod\limits_{m_L,m_R,l} \int \frac{d^n q_{L,\,m_L}}{(2 \pi)^n} \frac{d^n q_{R,\,m_R}}{(2 \pi)^n} \frac{d^n \bar{q}_l}{(2 \pi)^n} \;  S^{(C_S)} \left( \{q_{L,\,{m_L}}^{\gamma_{k_L}} \}; \{q_{R,\,{m_R}}^{\gamma_{k_R}} \};\{\bar{q}_l^{\delta_l}\} \right) \nonumber \\
& & \,\, \times \;  H_L \left( k_L^{\nu_L} , \{k_{L,\,i_L}^{\alpha_{i_L}} \}; \bar{k}_L^{\rho_L}, \{\bar{k}_{L,\,j_L}^{\eta_{j_L}} \} \right)\;
H_R \left( k_R^{\nu_R} , \{k_{R,\,i_R}^{\alpha_{i_R}} \}; \bar{k}_R^{\rho_R}, \{\bar{k}_{R,\,j_R}^{\eta_{j_R}} \} \right) \nonumber \\
& & \,\, \times \;  J_{p,\,L} \left(k_L^{\nu_L}, \{k_{L,\,i_L}^{\alpha_{i_L}} \};  \{q_{L,\,{m_L}}^{\gamma_{m_L}} \} \right) \;  J_{p,\,R} \left(k_R^{\nu_R}, \{k_{R,\,i_R}^{\alpha_{i_R}} \};  \{q_{R,\,{m_R}}^{\gamma_{m_R}} \} \right) \nonumber \\
& & \,\, \times \; J_\beta^{(C_\beta)} \left(\bar{k}_L^{\rho_L}, \{\bar{k}_{L,\,j_L}^{\eta_{j_L}} \}; \bar{k}_R^{\rho_R}, \{\bar{k}_{R,\,j_R}^{\eta_{j_R}} \};\{\bar{q}_l^{\delta_l}\} \right) \nonumber   \\
& &\,\, \times \; \delta^n\left(\sum_{m_L} q^\mu_{L,\,m_L} + \sum_{m_R} q^\mu_{R,\,m_R} +  \sum_l \bar{q}^\mu_{l} \right) \;  \delta^n\left(k^\mu_L + \bar{k}^\mu_L - x p^\mu \right) \; \; \nonumber \\
& & \,\, \times \;\delta^n\left(k^\tau_R + \bar{k}^\tau_R - x p^\tau \right) \nonumber \\
& & \,\, \times \; \delta^n\left( k_L^\mu -\sum_{m_L} q^\mu_{L,\,m_L} - p^\mu \right) \; \delta^n\left( k_R^\tau - \sum_{m_R} q^\tau_{R,\,m_R}  - p^\tau \right)\;.  \label{nonfactor}
\ea
The sum in Eq. (\ref{nonfactor}) runs over all cuts of jet $J_\beta$, $C_\beta$,  and of the soft function $S$, $C_S$, which are consistent with the constraints from the delta-functions due to momentum conservation.
The functions in (\ref{nonfactor}) are still connected with each other by scalar polarized or soft gluons. This obscures the independent evolution of the functions. In the next subsection we will show how to simplify this result.

\subsection{Refactorization}

We will show here that the scalar polarized gluons decouple via the help of the Ward identity, Fig. \ref{wardeik} c), and that we can disentangle the jets and the soft part, which are connected by soft gluon exchanges, via the soft approximation as described in Sec. \ref{sec:factor}. In the following we will work in Feynman gauge throughout.

\subsubsection{Decoupling of the Hard Part}

Starting from Eq. (\ref{nonfactor}), we use the fact that the leading contributions come from regions where the gluons carrying momenta $\left\{k_{L,\,{i_L}}\right\}$, $\left\{k_{R,\,{i_R}}\right\}$, $\{\bar{k}_{L,\,{j_L}}\}$ and $\{\bar{k}_{R,\,{j_R}}\}$ are scalar polarized. Thus $H_L$, $H_R$, $J_{p,\,L}$, $J_{p,\,R}$, and $J_\beta$ have the following structure:
\ba
H_L \! \!& = \!&\! \left( \prod\limits_{i_L} \xi^{\nu_{i_L}} \right) \left( \prod\limits_{j_L} \beta^{\rho_{j_L}} \right) \nonumber \\
&  &\times \;
  H_{L,\, \{\nu_{i_L} \},\{\rho_{j_L}\} } \left(k_L^+\; \xi^{\nu_L}, \left\{k_{L,\,i_L}^+ \xi^{\alpha_{i_L}}\right\}; \bar{k}_L^- \; \beta^{\rho_L}, \{\bar{k}_{L,\,{i_L}}^- \beta^{\eta_{j_L}} \} \right), \,\,\,\,\,\,\,\,
 \\
J_{p,\,L} & = & \left( \prod\limits_{i_L} \beta^{\nu_{i_L}} \right)  J_{p,\,L \, \{\nu_{i_L} \} } \left(  k_L^{\nu_L}, \{k_{L,\,i_L}^{\alpha_{i_L}} \}; \{q_{L,\,{k_L}}^{\gamma_{m_L}} \} \right);
\\
J_\beta & = & \left( \prod\limits_{j_L} \xi^{\rho_{j_L}} \right) \left( \prod\limits_{j_R} \xi^{\rho_{j_R}} \right) \nonumber \\
& & \times \;
J_{\beta\, \{\rho_{j_L}\},\,\{\rho_{j_R}\}} \left(  \bar{k}_L^{\rho_L}, \{\bar{k}_{L,\,j_L}^{\eta_{j_L}} \};  \bar{k}_R^{\rho_R}, \{\bar{k}_{R,\,j_R}^{\eta_{j_R}} \} ;\{\bar{q}_l^{\delta_l}\}\right),
\ea
where, as above, for the functions to the right of the cut, we replace the subscripts $L$ with $R$. In these relations the vectors
\ba
\xi^\mu & = & \delta_+^\mu \, , \nonumber \\
\beta^\mu & = & \delta_{-}^\mu \,  \label{lightdef}
\ea
are the light-like vectors parallel to $p^\mu$ and parallel to the direction of the eikonal, respectively.

\begin{figure}
\begin{center}
\epsfig{file=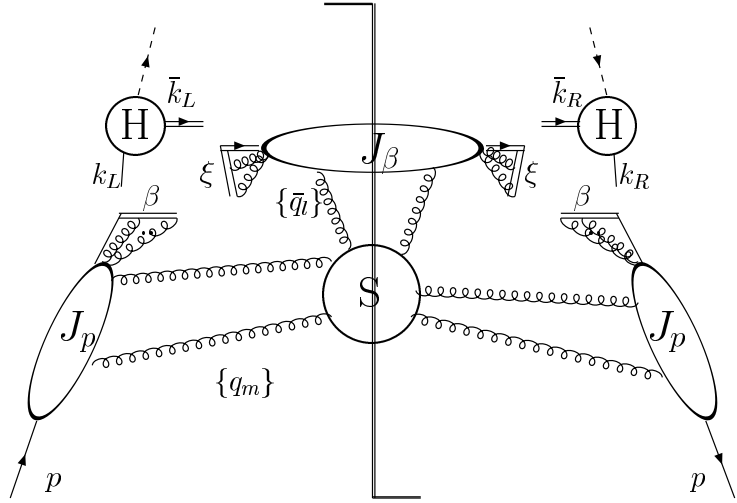,height=7.0cm,clip=0}
\caption{Parton distribution function for $x \rightarrow 1$ with the jet functions factorized from the hard part, graphical representation of Eq. (\ref{factform1}).} \label{partonfact1}
\end{center}
\end{figure}

We now use the identity depicted in Fig. \ref{wardeik} c) as described above for all scalar polarized gluons in the sets $\left\{k_{L,\,{i_L}}\right\}$, $\left\{k_{R,\,{i_R}}\right\}$, $\{\bar{k}_{L,\,{j_L}}\}$ and $\{\bar{k}_{R,\,{j_R}}\}$, to decouple the jet functions from the hard function. This decoupling occurs in Feynman gauge only after summation over the full gauge-invariant set of graphs which contribute to the reduced diagram Fig. \ref{partondef} c). The result is shown in Fig. \ref{partonfact1}. The products over the vectors $\beta$ and $\xi$ are replaced by eikonal factors $\mathcal{E}$, which we can group with the jets. Furthermore, the hard scatterings, by definition far off-shell, become  independent of $x$ up to corrections which vanish for $x \rightarrow 1$. Eq. (\ref{nonfactor}) then becomes
\ba
f_{f}^{x\rightarrow 1}\;(x) \!\!\! &\! = \!\! & \!\! H_L(p,\mu;\beta,\xi)\; H_R(p,\mu;\beta,\xi) \sum\limits_{C_\beta,C_S}  \prod\limits_{m_L,m_R,l}\int \frac{d^n q_{L,\,m_L}}{(2 \pi)^n} \frac{d^n q_{R,\,m_R}}{(2 \pi)^n} \frac{d^n \bar{q}_l}{(2 \pi)^n} \;
  \nonumber \\
& \times & \;  \tilde{J}_{p,\,L}(p,\mu;\beta; \{q_{L,\,{m_L}}^{\gamma_{m_L}} \} ) \; \tilde{J}_{p,\,R}(p,\mu;\beta; \{q_{R,\,{m_R}}^{\gamma_{m_R}} \})
 \;  \nonumber \\
& \times & \; \int dy \int dz\; S^{(C_S)} \left( y p,\mu; \{q_{L,\,{m_L}}^{\gamma_{m_L}} \}; \{q_{R,\,{m_R}}^{\gamma_{m_R}} \};\{\bar{q}_l^{\delta_l}\} \right) \; \nonumber
 \\
&  & \,\,  \times \; \tilde{J}^{(C_\beta)}_{\beta}(z p, \mu;\xi;\{\bar{q}_l^{\delta_l}\}) \nonumber \\
& \times & \; \delta^n\left(\sum_{m_L} q^\mu_{L,\,m_L} + \sum_{m_R} q^\mu_{R,\,m_R} +  \sum_l \bar{q}^\mu_{l} \right) \; \delta(1-x-y-z) \;, \,\,\,  \label{factform1}
\ea
where $\mu$ denotes the renormalization scale, which we set equal to the factorization scale, for simplicity. Corrections are subleading by a power of $1-x$. We define the functions $\tilde{J}_{p,\,L}$, $\tilde{J}_{p,\,R}$, and $\tilde{J}_\beta$ as follows:
\ba
\tilde{J}_{p,\,L} \!\!\!& \!\!=\!\!\! &\!\! \int\frac{d^n k_L}{(2 \pi)^n} \prod\limits_{i_L}  \int \frac{d^n k_{L,\,i_L}}{(2 \pi)^n}\; { \mathcal{E}} \left(\beta, \{k_{L,\,i_L}^+\} \right)^{\{\nu_{i_L}\}} \;  \nonumber \\
& & \; \times \; J_{p,\,L \, \{\nu_{i_L} \} } \left(  k_L^{\nu_L}, \{k_{L,\,i_L}^{\alpha_{i_L}} \}; \{q_{L,\,{k_L}}^{\gamma_{m_L}} \} \right) \; \delta^n \left(k_L^\mu - \sum_{m_K} q^\mu_{L,\,m_L} - p^\mu\right), \hspace*{10mm} \label{Jp}  \\
\tilde{J}^{(C_\beta)}_\beta & = & \int\frac{d^n \bar{k}_L}{(2 \pi)^n}  \int\frac{d^n \bar{k}_R}{(2 \pi)^n}  \prod\limits_{j_L,j_R} \int \frac{d^n \bar{k}_{L,\,j_L}}{(2 \pi)^n} \frac{d^n \bar{k}_{R,\,j_R}}{(2 \pi)^n} \;  \nonumber
 \\
& & \; \times \; {\mathcal{E}} \left(\xi, \{\bar{k}_{L,\,j_L}^-\} \right)^{\{\rho_{j_L}\}} \; {\mathcal{E}}^{*} \left(\xi, \{\bar{k}_{R,\,j_R}^-\} \right)^{\{\rho_{j_R}\}}  \; \nonumber \\
& & \; \times \; J^{(C_\beta)}_{\beta\, \{\rho_{j_L}\},\,\{\rho_{j_R}\}} \left( zp; \bar{k}_L^{\rho_L}, \{\bar{k}_{L,\,j_L}^{\eta_{j_L}} \}; \bar{k}_R^{\rho_R}, \{\bar{k}_{R,\,j_R}^{\eta_{j_R}} \} ; \{\bar{q}_l^{\delta_l}\}\right) \;,
\ea
and $\tilde{J}_{p,\,R}$ is defined analogously to $\tilde{J}_{p,\,L}$, with the subscripts $L$ replaced by $R$, and with a complex conjugate eikonal line, since it is to the right of the cut. In (\ref{factform1}) the total plus momentum flowing across the cut is restricted to be $(1-x) p^+$, and flows through the soft function and/or the eikonal jet. The plus momenta flowing across the cuts $C_S$ and $C_\beta$, denoted by $y p^+$ and $z p^+$,  respectively, are therefore restricted to be $(1-x) p^+$ via the delta-function in Eq. (\ref{factform1}). Above we have factorized the hard part from the remaining functions, which are still linked via soft momenta.

\subsubsection{Fully Factorized Form} \label{softapprox}

Here we will use the soft approximation to factorize the jets $J_p$ from the soft function, which, by power counting, are connected only through soft gluons. We could factorize the eikonal jet $J_\beta$ from $S$ in an analogous way, but we choose not to do so here because eventually we will combine all soft and eikonal functions to form an eikonal cross section. From the arguments given in Sec. \ref{sec:Glauber}, we see that in our case Glauber gluons do not pose a problem, since there is only one initial-state jet, which in addition, does not proceed into the final state due to the restriction that $x \rightarrow 1$. 

The application of the soft approximation is therefore straightforward, and we arrive
arrive at the factorized form of the parton distribution function as $x \rightarrow 1$:
\ba
f_{f}^{x\rightarrow 1}\;(x) &\! = & \! H_L(p,\mu;\beta,\xi)\; H_R(p,\mu;\beta,\xi)  \bar{J}_{p,\,L}(p,\mu;\beta; \xi ) \; \bar{J}_{p,\,R}(p,\mu;\beta; \xi)  \nonumber \\
& \times & \;  \sum\limits_{C_\beta,C_S}
\prod\limits_{m_L,m_R,l} \int \frac{d^n q_{L,\,m_L}}{(2 \pi)^n} \frac{d^n q_{R,\,m_R}}{(2 \pi)^n} \frac{d^n \bar{q}_l}{(2 \pi)^n} \; \nonumber \\
& \times & \; 
{\mathcal{E}} \left(\xi, \{q_{L,\,m_L}^-\} \right)^{\{\gamma_{m_L}\}} {\mathcal{E}}^{*} \left(\xi, \{q_{R,\,m_R}^-\} \right)^{\{\gamma_{m_R}\}}
\nonumber \\
& \times & \; \int dy \int dz \; S^{(C_S)} \left( y p,\mu; \{q_{L,\,{m_L}}^{\gamma_{m_L}} \}; \{q_{R,\,{m_R}}^{\gamma_{m_R}} \};\{\bar{q}_l^{\delta_l}\} \right) \;  \nonumber
  \\
  & \times & \; \tilde{J}^{(C_\beta)}_{\beta}(z p, \mu;\xi;\{\bar{q}_l^{\delta_l}\}) \nonumber \\
& \times &  \; \delta^n\left(\sum_{m_L} q^\mu_{L,\,m_L} + \sum_{m_R} q^\mu_{R,\,m_R} +  \sum_l \bar{q}^\mu_{l} \right) \; \delta(1-x-y-z) \;,\,\,\,\,\,\,\,\,\,\label{factform2}
\ea
where we have grouped the eikonal factors stemming from the soft approximation with the soft function and the eikonal jet. We define
\be
\tilde{J}_{p,\,L}(p,\mu;\beta; \{q_{L,\,{m_L}}^{\gamma_{m_L}} \} ) =  {\mathcal{E}} \left(\xi, \{q_{L,\,m_L}^-\} \right)^{\{\gamma_{m_L}\}} \;\bar{J}_{p,\,L}(p,\mu;\beta; \xi ), \,\,\, \label{bardef}
\ee
and analogously for the jet to the right of the cut, $J_{p,\,R}$, with a complex conjugate eikonal.

\subsubsection{Fully Factorized Form with an Eikonal Cross Section}

Although in Eq. (\ref{factform2}) the various functions are clearly separated in their momentum dependence, the parton distribution function is not quite in the desired form yet. We want to write the PDF in terms of a color singlet eikonal cross section, built from ordered exponentials:
\ba
\sigma^{(\mbox{\tiny eik})}_{a a}\left(\frac{(1-x)p^+}{\mu},\alpha_s(\mu),\varepsilon\right)\! \!\! &\!\! =\!\! &\!\! \frac{p^+}{\mbox{Tr } \mathbf{1}} \int \frac{d y^-}{2 \pi} e^{i(1-x)p^+ y^-} \label{eiksigdef} \\
& \times & \mbox{ Tr } \left< 0 \left| \bar{\mbox{T}} \left[ {\mathcal{W}}^{(aa)} (0,y^-,0_\perp)^\dagger \right] \; \mbox{T} \left[ {\mathcal{W}}^{(a a)} (0) \right] \right| 0 \right>, \nonumber
\ea
where the product of two non-Abelian phase operators (Wilson lines) in the representation $a$, for quarks, is defined as follows:
\ba
{\mathcal{W}}^{(aa)}(x) & = &   \Phi^{(a)}_\beta (\infty,0;x)\; \Phi^{(a)}_\xi (0,-\infty;x),  \\
\Phi^{(f)}_\beta (\lambda_2,\lambda_1;x) & = & P e^{-i g \int_{\lambda_1}^{\lambda_2} d \lambda \beta \cdot {{\mathcal{A}}^{(f)}} (\lambda \beta + x )},
\ea
where the light-like velocities $\xi$ and $\beta$ are defined in (\ref{lightdef}), and where ${\mathcal{A}}^{(f)}$ is the vector potential in the representation of a parton with flavor $f$.  The trace  in (\ref{eiksigdef}) is over color indices. The lowest order of the eikonal cross section is normalized to $\delta(1-x)$.
This eikonal cross section has ultraviolet divergences which have to be renormalized, as indicated by the renormalization scale $\mu$. Furthermore, the delta-function for the soft momenta in Eq. (\ref{factform2}) constrains the momentum of the final state in $\sigma^{(\mbox{\tiny eik})}$ to be $(1-x)p^+$.

We can factorize the eikonal cross section (\ref{eiksigdef}) in a manner analogous to the full parton distribution function, and obtain
\ba
\sigma^{(\mbox{\tiny eik})}_{aa}\left(1-x\right) & = & \sum\limits_{C_\beta,C_S}
\prod\limits_{m_L,m_R,l} \int \frac{d^n q_{L,\,m_L}}{(2 \pi)^n} \frac{d^n q_{R,\,m_R}}{(2 \pi)^n} \frac{d^n \bar{q}_l}{(2 \pi)^n} \; \nonumber \\
& \times & \;
\tilde{J}^{(\mbox{\tiny eik})}_{p,\,L}(\xi,\mu;\beta; \{q_{L,\,{m_L}}^{\gamma_{m_L}} \} )
 \; \tilde{J}^{(\mbox{\tiny eik})}_{p,\,R}(\xi,\mu;\beta; \{q_{R,\,{m_R}}^{\gamma_{m_R}} \} )\; \nonumber \\
& \times & \; \int dy \int dz\; S^{(C_S)} \left(yp,\mu; \{q_{L,\,{m_L}}^{\gamma_{m_L}} \}; \{q_{R,\,{m_R}}^{\gamma_{m_R}} \};\{\bar{q}_l^{\delta_l}\} \right)
\; \nonumber \\
& \times & \; 
\tilde{J}^{(C_\beta)}_{\beta}(zp,\mu;\xi;
\{\bar{q}_l^{\delta_l}\})
\;   \label{eikfact}
  \\
& \times & \;
\delta^n\left(\sum_{m_L} q^\mu_{L,\,m_L} + \sum_{m_R} q^\mu_{R,\,m_R} +  \sum_l \bar{q}^\mu_{l} \right) \; \delta(1-x-y-z).   \nonumber
\ea
The eikonal jets $\tilde{J}^{(\mbox{\tiny eik})}_{p,\,L}$, $\tilde{J}^{(\mbox{\tiny eik})}_{p,\,R}$ moving collinear to the momentum $p$, are defined analogously to Eq. (\ref{Jp}), with the fermion line carrying momentum $p$ replaced by an eikonal line in representation $a$ with velocity $\beta$. We can define analogous to Eq. (\ref{bardef})
\be
\tilde{J}^{(\mbox{\tiny eik})}_{p,\,L}(\xi,\mu;\beta; \{q_{L,\,{m_L}}^{\gamma_{m_L}} \} ) =  {\mathcal{E}} \left(\xi, \{q_{L,\,m_L}^-\} \right)^{\{\gamma_{m_L}\}} \;\bar{J}^{(\mbox{\tiny eik})}_{p,\,L}(\mu;\beta; \xi ),  \label{bardefeik}
\ee
and similarly for the jet to the right of the cut, $J_{p,\,R}^{(\mbox{\tiny eik})}$, with a complex conjugate eikonal. In the following we suppress the index $a$ for better readability.

\begin{figure}[htb]
\begin{center}
\epsfig{file=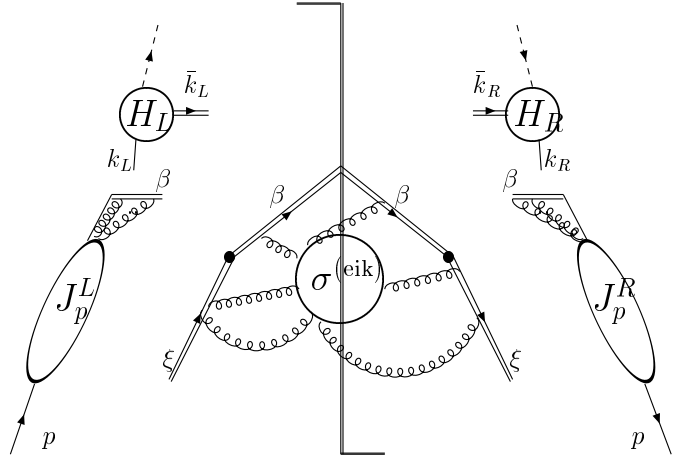,height=7.0cm,clip=0}
\caption[Parton distribution function for $x \rightarrow 1$, factorized into hard scatterings, an eikonal cross section, and purely virtual jet-remainders, as derived in Eq. (\ref{finalform}).]{Parton distribution function for $x \rightarrow 1$, factorized into hard scatterings, an eikonal cross section, and purely virtual jet-remainders, as derived in Eq. (\ref{finalform}). The virtual jet functions are normalized by their eikonal analogs, as in Eq. (\ref{jetvirdef}).} \label{partonfactend}
\end{center}
\end{figure}

Combining Eqs. (\ref{factform2}), (\ref{eikfact}), and (\ref{bardefeik}),  we arrive at the final form of the factorized parton distribution function, shown in Fig. \ref{partonfactend},
\ba
f_{f}^{x\rightarrow 1}\;(x) & = &  H_L\left( p, \mu \right)\; H_R\left(p, \mu \right) \; J_{p,\,L}^R (p,\mu)\; J_{p,\,R}^R (p,\mu)\; \sigma^{(\mbox{\tiny eik})} \left((1-x)p,\mu\right), \nonumber \\
& & \label{finalform}
\ea
suppressing the dependence on the lightlike vectors (\ref{lightdef}).
The purely virtual jet-remainders are defined by
\be
J_{p,\,L}^R (p,\mu) = \frac{\bar{J}_{p,\,L}(p,\mu;\beta; \xi )  }
{ \bar{J}^{(\mbox{\tiny eik})}_{p,\,L}(\mu;\beta; \xi )}\;. \label{jetvirdef}
\ee

In Chapter \ref{ch2} we will show that the eikonal cross section exponentiates, where the resulting exponent can be given a simple, recursive definition. We will use this exponentiation in Chapter \ref{ch4} to study the renormalization properties of eikonal cross sections. These studies lead to a powerful method for the calculation of the singular contribution to the splitting functions as introduced in Eq. (\ref{evol}). We will illustrate this method with the calculation of the fermionic contribution proportional to $1/(1-x)$ at three loops. 

Let us now turn to the second topic of this thesis, resummation of large logarithmic corrections to dijet event   shapes. These topics are linked by the exponentiation of soft gluon radiation, as will become apparent shortly.

\section{Factorization of the Thrust Cross Section in $e^+e^-$ Annihilation} \label{sec:dijetfact}

\subsection{Jet Event Shapes} \label{sec:thrust}

Jet cross sections measure the probability of producing jet-like final states, where most of the radiation is collimated. Jet cross sections are infrared safe \cite{Sterman:1977wj}. In the following we will study the  characteristics of hadronic final states in  $e^+e^-$ annihilation more generally by weighting the final states with \emph{shape variables} ${S}(p_1,\dots,p_n)$. These shape variables are functions of the final state momenta $p_i,\,i = 1,\dots,n$, and characterize the shape of an event, whether it is pencil-like, spherical, etc.. They provide information about the distribution of radiation, information which is complementary to the information obtained by computing inclusive or threshold jet cross sections. Event shapes also provide important tests of perturbative QCD. The shape of the distributions is a direct test of the QCD matrix elements, and the strong coupling $\as$ can be determined very precisely through the normalization of the cross sections.  

A weighted cross section $d\sigma/d{\mathcal{S}}$ at fixed shape variable ${\mathcal{S}}$ is given by
\be
\frac{d\sigma}{d {\mathcal{S}}} = \frac{1}{2 Q^2} \sum_n \int\limits_{\mbox{PS}_n} \left| M( p_1,\dots,p_n ) \right|^2 \delta \left(  S(p_1,\dots,p_n) - {\mathcal{S}} \right),   
\ee
where $Q$ is the center-of-mass (c.m.) energy, and  $M$ are the matrix elements integrated over the $n$-particle phase space $\mbox{PS}_n$ available for radiation to the momenta $p_i$.
Such cross sections are infrared safe if the weight is insensitive to collinear and/or soft radiation,
\be
S \left(p_1,\dots,p_i,\dots,p_{n-1},\alpha \,p_i \right) = S \left(p_1,\dots,p_i + \alpha \,p_i ,\dots,p_{n-1} \right).
\ee

A prominent example of an infrared safe shape function is the \emph{thrust} $T$ in $e^+ e^-$ annihilation \cite{Farhi:1977sg}:
\be
T = \max\limits_{\hat{n}} \frac{\sum_i \left| \vec{p}_i \cdot \hat{n}\right| }{\sum_j \left| \vec{p}_j \right|} = \frac{1}{\sqrt{s}} \max\limits_{\hat{n}} \sum_i \omega_i \,\left|\cos \theta_{i\hat{n}}\right|, \label{thrust}
\ee
where $\hat{n}$ is an arbitrary unit vector, whose direction is called the ``thrust axis'' when $T$ is maximal. In the second equality we have expressed the thrust in terms of the energies $\o_i$ of radiated particles $i$ and their angles with respect to the thrust axis $\hat{n}$. $\sqrt{s} = Q$ denotes the c.m. energy. The second definition in terms of angles and energies is equivalent to the first definition in terms of three-momenta only at the massless level, which we are considering for the main part of this thesis. The thrust measures how pencil-like a two-jet event is. For two jets perfectly back-to-back, its value is 1, at the three-jet boundary it assumes a value of 2/3, and for completely spherical events it takes a value of $1/2$.

Another well-known example of an infrared safe shape observable is the \emph{jet-broadening} $B$ in $e^+ e^-$ dijet events \cite{Catani:1992jc},
\be
B = \sum_{c=1}^2  \left[\frac{1}{2} \, \frac{1}{ \sqrt{s}} \sum_{i \in \Omega_c} \omega_i \left|\sin \theta_i  \right| \right] = \sum_{c=1}^2  \left[ \frac{1}{2}\, \frac{1}{\sqrt{s}} \sum_{i \in \Omega_c} \left| k_{\perp,\,i}\right| \, \right] , \label{broad}
\ee
where all angles and the transverse components of the final state momenta, $k_{\perp,\,i}$, are taken relative to the thrust axis. Instead of minimizing the axis as above, the event's thrust axis is found first, and then the phase space is divided into two hemispheres, $\Omega_1$ and $\Omega_2$, by the plane perpendicular to the thrust axis. The jet broadening is again $1$ for pencil-like configurations, but $\pi/4$ for spherical ones.

Many other event shape functions can be found in the literature, see for example Refs. \cite{Catani:1992ua,Ellis:qj}, and references therein. In the following we will discuss the factorization of the thrust cross section in the two-jet limit.  For $T \rightarrow 1 $, as we will see below, large logarithmic enhancements occur, of the form $\ln (1-T)$, which we will resum in the next section. In general, any weighted, differential cross section $d \sigma/d {\mathcal{S}}$ has at $n$th order in perturbation theory logarithmic enhancements proportional to $\as^n (\ln^m {\mathcal{S}})/{\mathcal{S}}$, where $m \leq 2 n -1$.  

In Chapter \ref{ch5} we will introduce a generalized shape function that interpolates continuously between the thrust, Eq. (\ref{thrust}) and the jet broadening, Eq. (\ref{broad}) \cite{Berger:2002ig,Berger:2003iw}.

\subsection{Leading Regions and Factorization for the Thrust} \label{sec:powerthrust}

 The role of an event shape at its limiting value with regard to power counting is to constrain the final state radiation to physical configurations which contribute to that value. It is at this edge of phase space that   large logarithmic corrections occur which need to be resummed. At other values of the event shape, fixed order perturbation theory suffices. In order to resum these large corrections we need to identify the regions in momentum space where these logarithms originate. 
The procedure is quite analogous to the previous argumentation on parton distribution functions.  

Following Coleman and Norton once more, the leading contributions in momentum space for the thrust cross section as $T \rightarrow 1$ are given by the reduced diagram shown in Fig. \ref{vertex} c), for the electromagnetic form factor. Due to the requirement that $T \approx 1$ the final state is restricted to contain exactly two very narrow jets. Thus the discussion of leading regions reduces to the one given in Sec. \ref{sec:vertex}.

The degree of infrared divergence $\o(T)$ is, as above, given by the incoherent contributions of  jet and soft  regions
\be
\o(T) = \sum\limits_{c = 1}^2 \o_{J_c} + \o_S.
\ee
The degree of IR divergence of the soft function is the same as in Eq. (\ref{dodsoft}) since it can be found by  dimensional analysis. The degrees of IR divergence of the jet functions are easily found, using Eqs.  (\ref{jetdiv}), (\ref{physnum}) or (\ref{covnum}), (\ref{euler}), and (\ref{linevertex}). The final result is in Feynman gauge
\ba
\o(T) & \geq & \sum\limits_{c = 1}^2 \left\{ \frac{1}{2} \left( p_{J_c} - 1 \right) + N_{J_c}^f + \frac{1}{2} \left( N_{J_c}^b - s_{J_c} \right) \right. \nonumber \\
& & \qquad \left. + \frac{1}{2} \left( s_{J_c} + l_{J_c} - v^{(3)}_{J_c} \right) \theta\left( s_{J_c} + l_{J_c} - v^{(3)}_{J_c} \right) \right\},
\ea
with the same notation as in Sec. \ref{sec:powercount}.

The maximal degree of divergence in this case is logarithmic in $1 - T$, $\o(T \rightarrow 1) = 0$, if and only if:
\begin{itemize}
\item No soft vectors directly attach the hard scattering with the soft function.
\item The jets and the soft part can only be connected through soft gluons.
\item Exactly one scalar, fermion, or physically polarized gluon, respectively, connects each of the jets with the hard part.  
\item Additionally, only scalar polarized gluons can connect the jets with the hard scattering.
\item The number of soft and scalar-polarized vector lines emerging from a particular jet is less or equal to the number of 3-point vertices in that jet.
\end{itemize}
As announced above, the IR behavior of the thrust cross section is proportional to $\ln (1-T)$, more precisely, at order $n$ in the perturbative expansion, maximally proportional to $\as^n \ln^{2n} (1-T)$. Simultaneously collinear and soft configurations give two logarithms per loop, only collinear or only soft gluons contribute at the level of one logarithm per loop, as one can see in the simple example discussed above in Sec. \ref{sec:vertex}. Here we will consider the differential cross section, whose behavior is therefore $\as^n [\ln^m {(1-T)}]/{(1-T)}$, where $m \leq 2 n -1$. $m = 2 n - 1$ is referred to as leading logarithmic (LL) behavior, $m = 2 n -2$ is called next-to-leading logarithmic (NLL) and so forth.

\begin{figure}[htb]
\begin{center}
\epsfig{file=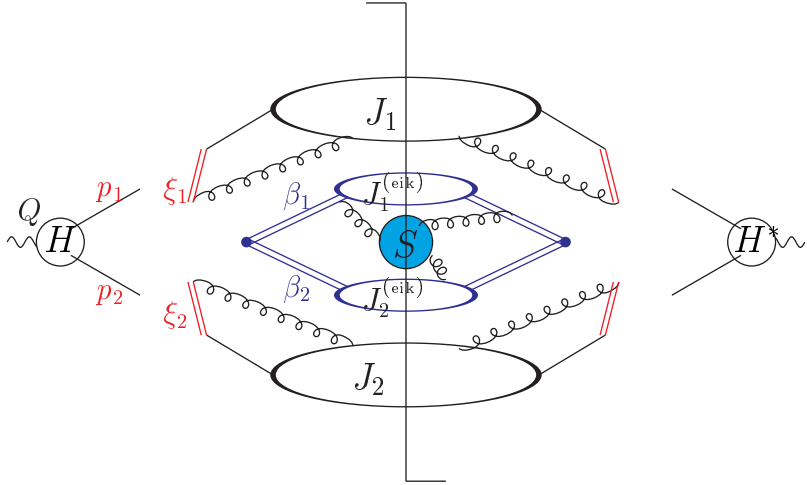,height=7cm,clip=0}
\caption{Factorized cross section (\ref{thrustfact})
after the application of Ward identities. The vertical line
denotes the final state cut.} \label{factorized}
\end{center}
\end{figure}

The factorization is straightforward, using the Ward identities and the decomposition of gluon propagators discussed ins Section \ref{sec:factor}, Glauber configurations do not pose a problem here, as we have seen in Sec. \ref{sec:Glauber}. The resulting cross section is linked via a convolution in $1-T$
\ba
\frac{d \sigma(T,s)}{d (1-T) } & = & \sigma_0 \,H \left(\frac{\sqrt{s}}{\mu},\frac{p_{J_c} \cdot \xic}{\mu}, \as(\mu) \right) \prod\limits_{c = 1}^2 \int d \tau_{J_c}   J_c \left( \frac{p_{J_c} \cdot \xic}{\mu}, \tau_{J_c} \frac{\sqrt{s}}{\mu}, \as(\mu) \right) \nonumber \\
& & \!\!\! \times \, \int  d \tau_{s} \, S \left( \tau_{s} \frac{\sqrt{s}}{\mu},\hat{\beta}_c \cdot \xic,  \as(\mu) \right) \delta \left(1-T - \sum_{c = 1}^2 \tau_{J_c} - \tau_s \right), \label{thrustfact}
\ea
since the leading contributions are incoherent and thus additive to the weight in the elastic limit, up to corrections that vanish as $\tau^2$ for small $\tau$, where
\be
\tau \equiv 1 - T. \label{taudef}
\ee
 Eq. (\ref{thrustfact}) is illustrated in Fig. \ref{factorized}.
$\mu$ is the factorization scale which we set in the following equal to the renormalization scale.  
The arguments of the various dimensionless functions in Eq. (\ref{thrustfact}) follow from dimensional considerations. $\sigma_0$ is the dimensionful Born cross section, which we separated such that the hard scattering $H$ begins at $1 + {\mathcal{O}}(\as)$. $\xic^\mu = \xi_c^\mu/|\xi_c|$ are the normalized eikonal vectors that arise in the course of the factorization. The physical cross section on the left hand side is of course independent of the factorization scale and of the $\xic$. Due to the soft approximation the soft function $S$ cannot depend on the magnitude of the jet momenta $p_{J_c}$, only on their normalized directions, denoted by $\hat{\beta}_c,\,c = 1,2$. Explicit definitions in form of (nonlocal) operators of the various functions above will be discussed in Chapter \ref{ch5}. In Eq. (\ref{thrustfact}) we have neglected recoil effects, which, in principle also link the various functions. In Chapter \ref{ch5} we will provide the justification for this approximation. 

Following the power-counting arguments above, the dimensionless jet and soft functions above begin at $1/\tau_i, \, i = s, J_c$, and are multiplied in higher orders by logarithms of their arguments. These logarithms stem from the expansion of  kinematic combinations such as $1/\varepsilon \,\, 1/(\tau_i \sqrt{s})^{1+\varepsilon}$:
\be
\frac{1}{\varepsilon} \frac{1}{(\tau_i )^{1+\varepsilon}} = \frac{1}{\varepsilon^2} \delta(\tau_i  ) + \frac{1}{\varepsilon} \left[ \frac{1}{\tau_i }\right]_+ + \left[ \frac{\ln (\tau_i )}{\tau_i }\right]_+ + {\mathcal{O}}(\varepsilon)\, , \label{logexpand}
\ee
when working in $n = 4 - 2 \varepsilon$ dimensions. The plus-distribution above is defined as
\be
\int_z^1 dx \,f(x) \left[ g(x) \right]_{+} = \int_z^1 dx\, g(x) \left[f(x)- f(1) \right] - f(1) \int_0^z dx\, g(x). \label{plusdef}
\ee

\section{Evolution and Resummation at the Example of the Thrust} \label{sec:resum}

The following discussion of resummation of large logarithms in $1- T$ is based on Refs. \cite{Sterman:1995fz,Contopanagos:1996nh}. 
The natural scale for the hard scattering is $\sqrt{s}/2$, such that it contains no large ratios. Setting $\mu =  \sqrt{s}/2$ in Eq. (\ref{thrustfact}) we obtain
\ba
\frac{d \sigma(T,s)}{d (1-T) } \!\!&\!\! =\!\! &\!\! \sigma_0 H \left(\frac{2\,p_{J_c} \cdot \xic}{\sqrt{s}}, \as\left(\frac{\sqrt{s}}{2}\right) \right) \prod\limits_{c = 1}^2 \int d \tau_{J_c}  J_c \left( \frac{2\,p_{J_c} \cdot \xic}{\sqrt{s}},2 \tau_{J_c}, \as\left(\frac{\sqrt{s}}{2}\right) \right) \nonumber \\
& & \,\,\, \times \, \int d \tau_{s}\, S \left( 2 \tau_{s}, \hat{\beta}_c \cdot \xic, \as\left(\frac{\sqrt{s}}{2}\right) \right) \delta \left(1-T - \sum_c \tau_{J_c} - \tau_s \right). \label{thrustfact2}
\ea
As $T \rightarrow 1$, the $\tau_i$ are restricted to be very small by the delta-function. In this limit the  logarithms of the $\tau_i$ in the soft and jet functions become large and fixed-order calculations become inadequate. Resummation of these logarithms is needed to provide reliable predictions.

Since the functions in the factorized thrust are still linked by a convolution in the variables $\tau_i$ it is advantageous to take moments to disentangle this convolution:
\be
\tilde{\sigma}(N,s) \equiv \int_0^1 d T\, T^{N-1} \frac{d \sigma(T,s)}{d (1-T)} = \int_0^\infty d \tau e^{-N \,\tau}  \frac{d \sigma(\tau,s)}{d \tau} + {\mathcal{O}}\left(\frac{1}{N}\right)\, . \label{thrustmoment}
\ee
The first definition in (\ref{thrustmoment}) is the Mellin transform, which is equivalent to the second definition, the Laplace transform, at large $N$, because then $e^{-N \, \tau} \approx (1-\tau)^N$ (recall the definition of $\tau$, Eq. (\ref{taudef})). Tildes denote quantities in moment space in this chapter. 

The factorized thrust becomes a simple product in moment space,
\ba
\tilde{\sigma}(N,s) & = & \sigma_0  H \left(\frac{\sqrt{s}}{\mu},\frac{p_{J_c} \cdot \xic}{\mu}, \as(\mu) \right)   \tilde{S} \left( \frac{\sqrt{s}}{\mu N}, \hat{\beta}_c \cdot \xic, \as(\mu) \right) \nonumber \\
& & \qquad \times \, \prod\limits_{c = 1}^2 \tilde{J}_c \left( \frac{p_{J_c} \cdot \xic}{\mu},  \frac{\sqrt{s}}{ \mu N}, \as(\mu) \right) \label{thrustmomentb} \\
& = & \sigma_0 H \left(\frac{2\,p_{J_c} \cdot \xic}{\sqrt{s}}, \as\left(\frac{\sqrt{s}}{2}\right) \right) 
\tilde{S} \left( \frac{2}{N}, \hat{\beta}_c \cdot \xic, \as\left(\frac{\sqrt{s}}{2}\right) \right) \nonumber \\
& & \qquad \times \prod\limits_{c = 1}^2 \tilde{J}_c \left( \frac{2 p_{J_c} \cdot \xic}{\sqrt{s}},  \frac{2}{N},\as\left(\frac{\sqrt{s}}{2}\right) \right), \label{thrustmoment2}
\ea
where
\be
\tilde{S} \left( \frac{\sqrt{s}}{\mu N}, \as(\mu) \right) \equiv \int_0^{\infty} d \tau_s \, e^{- N \tau_s} S \left( \tau_{s} \frac{\sqrt{s}}{\mu}, \as(\mu) \right),
\ee
and analogously for the jet functions.
In  Eq. (\ref{thrustmoment2})  we have set $\mu = \sqrt{s}/2$. 
The large logarithms of the $\tau_i$ are transformed into large logarithms of $N$ in moment space. Specifically, logarithms of the form displayed in Eq. (\ref{logexpand}) transform under moments as
\be
\int_0^1 d \tau (1- \tau)^{N-1} \left[ \frac{\ln^m (\tau)}{\tau}\right]_+ = \frac{(-1)^{m+1}}{m+1} \ln^{m+1} N + {\mathcal{O}}\left(\frac{1}{N}\right).
\ee 

The factorized cross section (\ref{thrustfact2}) or (\ref{thrustmoment}) already provides all the information necessary for the resummation. The physical cross section is independent of the factorization scale
\be
\frac{d}{d \ln \mu} \, \frac{d \sigma(T,s)}{d (1-T) } = \frac{d}{d \ln \mu} \tilde{\sigma}(N,s) = 0, \label{thrustRGE}
\ee
and of the choice of the eikonal directions $\xic$,
\be
\frac{\partial}{\partial \ln \left(p_{J_c} \cdot \xic \right)} \, \frac{d \sigma(T,s)}{d (1-T) } = \frac{\partial}{\partial \ln \left(p_{J_c} \cdot \xic \right)} \tilde{\sigma}(N,s) = 0. \label{thrustxi}
\ee
These conditions are exactly fulfilled only if the cross section is calculated to all orders in perturbation theory. Upon truncation at order $m$, the relations (\ref{thrustRGE}) and (\ref{thrustxi}) are fulfilled only up to the same order, with corrections proportional to $\as^{m+1}$.
In the remainder of this section we explore the consequences of the renormalization group conditions (\ref{thrustRGE}) and (\ref{thrustxi}) \cite{Contopanagos:1996nh}.

\subsection{Resummation of Single Logarithms} \label{sec:single}

The renormalization group equation (\ref{thrustRGE}) organizes all single logarithms in the soft function, and some of the single logarithms in the jet functions. The jet functions contain, as we will see shortly, double logarithms, due to emissions that are simultaneously soft and collinear. Applying Eq. (\ref{thrustRGE}) to the factorized thrust, Eq. (\ref{thrustmoment}), we derive the following consistency conditions since all functions are multiplicatively renormalizable:
\ba
 \frac{d}{d \ln \mu} \ln \tilde{S} \left( \frac{\sqrt{s}}{\mu N},\hat{\beta}_c \cdot \xic,  \as(\mu) \right) & = & - \gamma_s(\as (\mu)), \label{softRGE} \\
 \frac{d}{d \ln \mu} \ln \tilde{J}_c \left( \frac{p_{J_c} \cdot \xic}{\mu},  \frac{\sqrt{s}}{\mu N}, \as(\mu) \right) & = & - \gamma_{J_c} (\as(\mu)), \label{jetRGE} \\
\frac{d}{d \ln \mu} \ln H \left(\frac{\sqrt{s}}{\mu},\frac{p_{J_c} \cdot \xic}{\mu}, \as(\mu) \right) & = & - \gamma_H (\as(\mu)), \label{HRGE}
\ea 
with
\be
\gamma_s + \sum_{c=1}^2 \gamma_{J_c} + \gamma_H = 0.
\ee
The anomalous dimensions $\gamma_d,\, d = s,J_c,H$ can depend on variables held in common between at least two of the functions. Because each function is infrared safe, while ultraviolet divergences are present only in virtual diagrams, the anomalous dimensions cannot depend on $N$. This leaves as arguments of the $\gamma_d$ only the coupling $\as(\mu)$ and the dimensionless ratio $(2 p_{J_c} \cdot \xic)/\sqrt{s}$. Since the dependence on $p_{J_c} \cdot \xic$ will be studied with the help of Eq. (\ref{thrustxi}) below, we suppress this argument for now. 

The solutions to Eqs. (\ref{softRGE}) and (\ref{jetRGE}) are given by
\ba
\tilde{S} \left( \frac{\sqrt{s}}{\mu N}, \hat{\beta}_c \cdot \xic, \as(\mu) \right) & = & \tilde{S} \left( \frac{\sqrt{s}}{\mu_0 N},\hat{\beta}_c \cdot \xic,  \as(\mu_0) \right) e^{-\int\limits_{\mu_0}^\mu \frac{d \lambda}{\lambda} \gamma_s (\as(\lambda))}\!\!\!\!, \label{softsol} \\
\tilde{J}_c \left( \frac{p_{J_c} \cdot \xic}{\mu},  \frac{\sqrt{s}}{\mu N}, \as(\mu) \right) & = & 
\tilde{J}_c \left( \frac{p_{J_c} \cdot \xic}{\mu_0},  \frac{\sqrt{s}}{\mu_0 N}, \as(\mu_0) \right)
e^{-\int\limits_{\mu_0}^\mu \frac{d \lambda}{\lambda} \gamma_{J_c} (\as(\lambda))}\!\!\!\! . \label{jetsol}
\ea
Setting $\mu_0 = \sqrt{s}/N$,  we avoid logarithms of $N$ in the soft function in Eq. (\ref{thrustmoment2}):
\ba
\tilde{\sigma}(N,s) & = & \sigma_0 H \left(\frac{2\,p_{J_c} \cdot \xic}{\sqrt{s}}, \as\left(\frac{\sqrt{s}}{2}\right) \right)  \nonumber \\
& & \,\,\,\, \times \,
\tilde{S} \left( 1,\hat{\beta}_c \cdot \xic,  \as\left(\frac{\sqrt{s}}{N}\right) \right) e^{-\int\limits_{\sqrt{s}/N}^{\sqrt{s}/2} \frac{d \lambda}{\lambda} \gamma_s (\as(\lambda))}  \nonumber \\
& & \,\,\,\, \times \prod\limits_{c = 1}^2 \tilde{J}_c \left( \frac{2 N p_{J_c} \cdot \xic}{\sqrt{s}}, 1 ,\as\left(\frac{\sqrt{s}}{N}\right) \right) e^{-\int\limits_{\sqrt{s}/N}^{\sqrt{s}/2} \frac{d \lambda}{\lambda} \gamma_{J_c} (\as(\lambda))}. \label{thrustmoment3}
\ea

\subsection{Resummation of Sudakov Double Logarithms} \label{sec:jetdouble}

The remaining unorganized large logarithms of $N$ in Eq. (\ref{thrustmoment3}) reside in the jet functions. $p_{J_c} \cdot \xic$ is of the order of the hard scale, $\sim \sqrt{s}$. The requirement that the cross section be independent of $p_{J_c}\cdot \hat{\xi}_c$, Eq. (\ref{thrustxi}), implies that the jet, soft and hard functions obey equations analogous to (\ref{softRGE})-(\ref{HRGE}), again in terms of the variables
that they hold in common \cite{Contopanagos:1996nh}.  The same results may be derived following the method of Collins and Soper \cite{Collins:1981uk}, by defining the jets in an axial gauge, and then studying their variations under boosts. The latter method  provides an explicit construction of the functions that control the variation which is needed for explicit calculations. We will discuss both ways below.

\subsubsection{$p_{J_c}\cdot \hat{\xi}_c$-Dependence in Axial Gauge}

Since the definition of the jets can be made gauge invariant, we can derive the evolution with $\frac{\partial}{\partial \ln \left(p_{J_c} \cdot \xic\right)}$ in any gauge. The derivation is most easily done in axial gauge with the gauge vector $\xic$. The variation with respect to $p_{J_c}^\mu$ is equivalent to the variation with respect to $\xic^\mu$:
\be
\frac{\partial }{\partial \ln p_{J_c}^\mu} J_c = \frac{\partial }{\partial \ln \xic^\mu} J_c.
\ee
We will derive the effect of the variation with respect to $\xi_1 \equiv \xi$ on jet 1, which we choose to move in the plus direction, 
\be
p_{J_1}^\mu = (p^+,0,0_\perp).
\ee 
The derivation for jet 2 is analogous. The final cross section obeys Eq. (\ref{thrustxi}) for each $\xi_c,\, c = 1,\,2$ separately.

\begin{figure}[htb]    
\begin{center}    
\epsfig{file=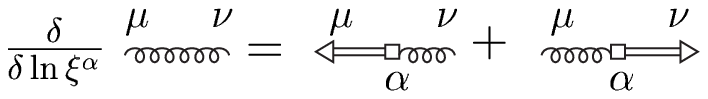,height=2.2cm,clip=0}    
\caption{Graphical illustration of the variation of the gluon propagator with respect to the gauge fixing vector $\xi$, Eq. (\ref{variation}).} \label{varfig}    
\end{center}    
\end{figure}

The only dependence on the gauge-vector $\xi$ in axial gauge is in the gluon propagator, given by Eq. (\ref{gluonpropaxial}) with $\kappa \rightarrow \infty$,  which allows us to trace out the $\xi$-dependence relatively easily. We choose as variation a boost in the plus-direction, which leaves $\xi^2$ invariant, with $\xi_\perp = 0$. A straightforward calculation gives
\be
\frac{\delta}{\delta \, \ln \hat{\xi}^\alpha} D_{\mu \nu} = - \frac{ \, k_\mu}{k \cdot \hat{\xi}} \left( \frac{1}{\hat{\xi} \cdot \bar{v} } \bar{v}^\alpha + v_\perp^\alpha \right)  D_{\alpha \nu} + \left\{ \mu \leftrightarrow \nu \right\}, \label{variation}
\ee
where $\bar{v}$ is a unit vector in the minus-direction $\bar{v}^\mu = (0^+,1,0_\perp)$, and $v_\perp$ is a unit vector in the perpendicular direction, $v_\perp^\mu = (0^+,0^-,1_\perp)$. This variation is shown graphically in Fig. \ref{varfig}, where the box denotes $i  \left( \frac{\xi^2}{\xi \cdot \bar{v} } \bar{v}^\alpha + v_\perp^\alpha \right)$.

\begin{figure}[htb]    
\begin{center}    
a)\hspace*{12.9cm} $\mbox{ } $ \\
\epsfig{file=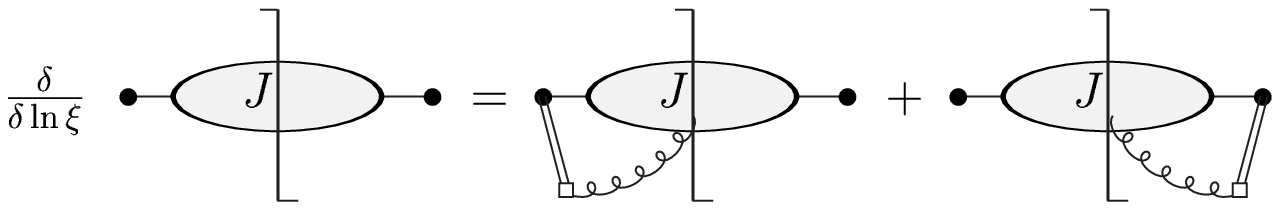,height=1.63cm,clip=0} \vspace*{2mm}   \\
\vspace*{4mm}
b)\hspace*{12.9cm} $\mbox{ } $ \\
\epsfig{file=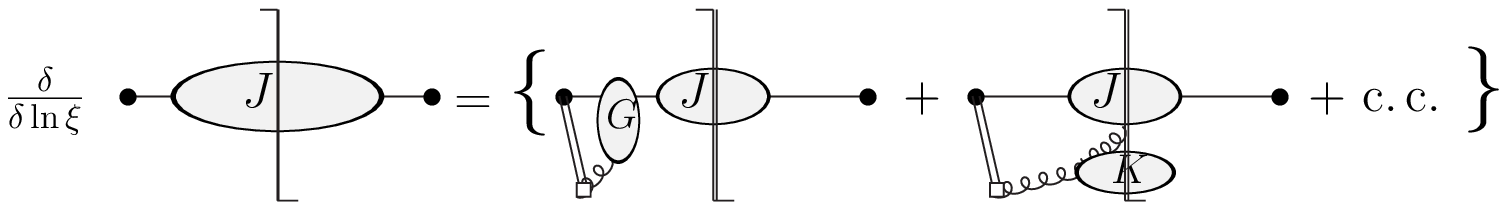,height=2.53cm,clip=0}   \\
\caption[a) Graphical illustration of the variation of the jet function with respect to the gauge fixing vector $\xi$. b) Factorization of the variation into a soft and an ultraviolet part, Eq. (\ref{prekg}).]{a) Graphical illustration of the variation of the jet function with respect to the gauge fixing vector $\xi$. b) Factorization of the variation into a soft and an ultraviolet part, Eq. (\ref{prekg}). c.\,c. denotes the complex conjugate contribution.} \label{jetvar}    
\end{center}    
\end{figure}

Applying this to the jet function in axial gauge we obtain Fig. \ref{jetvar} a). We will provide an explicit construction of the jet function in Chapter \ref{ch5}. Here it suffices to know that the jet can only have one external physical line on each side of the cut, which follows from power-counting. In the axial gauge $\xi \cdot {\mathcal{A}} = 0$ the eikonal lines from factoring unphysically polarized gluons connecting the jet with the hard part are not present. With the help of power counting, as presented in \ref{sec:powercount}, we find that the effect of the variation with respect to $\xi$ on $J_c$ can either be far off-shell or soft to give a leading contribution. Collinear configurations are subleading. Only virtual diagrams can contribute to the off-shell part, in the following denoted by $G$. $G$ is therefore an overall factor. The soft part, denoted by $K$, can be factored from the jet via the Ward identity \ref{wardeik} c), linked only by an overall convolution in the momentum $q$ flowing through the jet and $K$.
\ba
 \frac{\partial }{\partial \ln \hat{\xi}^\mu} J_1\left(p_{J_1}^\mu,\tau_{J_1}, \mu \right) & = & 
\int \frac{d^n q}{(2 \pi)^n} K_1'\left(q^\mu,\tau_{J_1}, \mu',\mu \right) J_1\left(p_{J_1}^\mu-q^\mu,\tau_{J_1},\mu \right) \nonumber \\
& & + G_1\left(\frac{ p_{J_1} \cdot \hat{\xi}}{\mu'},\mu'\right)  J_1\left(p_{J_1}^\mu,\tau_{J_1}, \mu  \right), \label{prekg}
\ea
where $\mu'$ is the scale separating hard contributions in $G$ from soft ones in $K$. The jet, of course, is independent of $\mu'$. Eq. (\ref{prekg}) is displayed in Fig. \ref{jetvar} b). Both $K'_1$ and $J_1$ on the right-hand-side contribute to the weight $\tau_{J_1}$.

We can reduce the convolution in the four-momentum $q^\mu$ further, since $q^\mu$ is soft, 
\be
\ln \frac{ \left(p_{J_1}-q\right) \cdot \hat{\xi}}{\mu' } \sim \ln \frac{ p_{J_1} \cdot \hat{\xi}}{\mu'}. 
\ee
We can therefore neglect $q^\mu$ in $J_1$ to leading accuracy. But we cannot neglect its contribution to $\tau_{J_1}$. Thus the convolution in (\ref{prekg}) reduces to a convolution in terms of $\tau$:
\ba
 \frac{\partial }{\partial \ln \hat{\xi}} J_1\left(\frac{p_{J_1} \cdot \hat{\xi} }{\mu },\tau_{J_1} \frac{\sqrt{s}}{\mu}, \as(\mu) \right) \! & \!\!=\!\! & \!\!
\int d \tau_K \, K_1\left(\tau_K \frac{\sqrt{s}}{\mu}, \frac{\mu'}{\mu},\as(\mu') \right) \nonumber \\
& & \hspace*{-25mm} \times \int d \tau'_1 \, J_1\left(\frac{p_{J_1} \cdot \hat{\xi} }{\mu }, \tau'_{1} \frac{\sqrt{s}}{\mu}, \as(\mu) \right) \delta\left( \tau_{J_1} - \tau'_1 - \tau_K \right) \nonumber \\
& & \hspace*{-35mm} + G_1\left(\frac{p_{J_1} \cdot \hat{\xi} }{\mu'},\as(\mu')\right)  J_1\left(\frac{p_{J_1} \cdot \hat{\xi}  }{\mu }, \tau_{J_1} \frac{\sqrt{s}}{\mu}, \as(\mu)  \right). \label{kga0} 
\ea
Taking moments, and setting $\mu' = \mu$ (corrections are non-leading), we arrive for both jets at 
\ba
   \frac{\partial }{\partial \ln \left(p_{J_c} \cdot \xic\right)}
\ln\ \tilde{J}_c \left( \frac{p_{J_c} \cdot \xic}{\mu},
\frac{\sqrt{s}}{\mu N} ,\as(\mu)
\right)
& \ & \nonumber
\\
&\ & \hspace{-55mm} =
   \tilde{K}_c\left(\frac{\sqrt{s}}{\mu\,
N},\as(\mu)
\right)   +  G_c\left(\frac{p_{J_c} \cdot \xic}{\mu},\as(\mu)\right)   \, .
   \label{KGend}
\ea
We note that although the derivation of Eq. (\ref{KGend}) has been performed in axial gauge, the result is gauge invariant, since the jet function can be given a gauge-invariant definition, as we will show in Chapter \ref{ch5}.

\subsubsection{$p_{J_c}\cdot \hat{\xi}_c$-Dependence from General Considerations}

Eq. (\ref{KGend}) can also be derived in a manner similar to the derivation of Eqs. (\ref{softRGE})-(\ref{HRGE}), by considering the variables held in common by at least two of the functions \cite{Contopanagos:1996nh}. Applying Eq. (\ref{thrustxi}) to Eq. (\ref{thrustmoment}) we obtain for jet~1
\ba
  \frac{\partial }{\partial \ln \left(p_{J_1} \cdot \hat{\xi}_1\right)}
\ln\ \tilde{J}_1 \left( \frac{p_{J_1} \cdot \hat{\xi}_1}{\mu},
\frac{\sqrt{s}}{\mu N} ,\as(\mu)
\right) &  & \nonumber \\
 & & \hspace*{-65mm} = -  \frac{\partial }{\partial \ln \left(p_{J_1} \cdot \hat{\xi}_1\right)} \ln H \left(\frac{\sqrt{s}}{\mu},\frac{p_{J_1} \cdot \hat{\xi}_1}{\mu}, \frac{p_{J_2} \cdot \hat{\xi}_2}{\mu},\as(\mu) \right) \nonumber \\
& & \hspace*{-61mm} -  \frac{\partial }{\partial \ln \left(p_{J_1} \cdot \hat{\xi}_1\right)} \ln \tilde{S} \left( \frac{\sqrt{s}}{\mu N}, \hat{\beta}_1 \cdot \hat{\xi}_1, \hat{\beta}_2 \cdot \hat{\xi}_2,  \as(\mu) \right)\,.
\ea
The logarithmic derivative of the jet can depend additively on a function containing the hard scale $p_{J_1} \cdot \hat{\xi}_1 \sim \sqrt{s}/2$, or on a function containing the soft scale $\sqrt{s}/N$. These functions can contain all arguments that the jet and the hard function, or the jet and the soft function hold in common, respectively. Aside from the hard and the soft scales, the only other common variable is the running coupling, $\as(\mu)$. These considerations result again in Eq. (\ref{KGend}).

\subsubsection{The Resummed Jet Functions} 

From the above considerations, $\tilde{K}_c+G_c$ are renormalized
additively, and satisfy \cite{Collins:1981uk}
\ba
\mu \frac{d}{d \mu}\
\tilde{K}_c\left(\frac{\sqrt{s}}{\mu\, N},\as(\mu) \right) & = & - \gamma_{K_c}
\left(\as(\mu)\right),
\nonumber\\
\mu \frac{d}{d \mu}G_c\left(\frac{p_{J_c} \cdot \xic }{\mu},\as(\mu)\right)
  & = &  \gamma_{K_c}
\left(\as(\mu)\right) \, .
\label{Gevol}
\ea
$\gamma_{K_c}$ is the
universal Sudakov anomalous dimension \cite{Sen:sd,Collins:1981uk,Korchemsky:wg}. 

This Sudakov anomalous dimension is the anomalous dimension of the soft-collinear functions $\tilde{K}_c$, which is built out solely of eikonal lines. Therefore, as we will also see by explicit calculation below, the anomalous dimension of $\tilde{K}_c$, and the anomalous dimension of the eikonal cross section in Eq. (\ref{eiksigdef}) are the same, since the eikonal anomalous dimension for lightlike lines is independent of the directions of the scattering eikonals.  The anomalous dimension of (\ref{eiksigdef}) controls the most singular evolutionary behavior of parton-in-parton distribution functions as $x \rightarrow 1$ \cite{Berger:2002sv,Korchemsky:1988si}, which is, as in Eq. (\ref{Gevol}), due to simultaneously soft and collinear radiation. By analogous reasoning, the same anomalous dimension appears in a variety of other jet-related processes \cite{Sterman:2002qn}.   

With the help of these evolution equations, the terms $\tilde{K}_c$ and $G_c$
in Eq. (\ref{KGend}) can be reexpressed as
\cite{Collins:1984kg}
\ba
   \tilde{K}_c\left(\frac{\sqrt{s}}{\mu\,N},\as(\mu)
\right)   +  G_c\left(\frac{p_{J_c} \cdot \xic}{\mu},\as(\mu)\right)
&\ &  \nonumber \\
&\ & \hspace{-80mm} =
\tilde{K}_c\left(\frac{1}{c_1},\as\left(c_1 \,
\frac{\sqrt{s}}{ N} \right)
\right)
+  G_c\left(\frac{1}{c_2},\as\left(c_2 \, p_{J_c} \cdot \xic \right) \right)
\ - \!\!\! \int\limits_{c_1 {\sqrt{s}}/ N }^{ c_2\,
p_{J_c} \cdot \xic }
\frac{d  \lambda'}{\lambda'} \gamma_{K_c}\left(\as\left(\lambda'\right)
\right)
\nonumber \\
&\ & \hspace{-80mm} =
   - B_c\left(c_1,c_2,  \as\left(c_2 \, p_{J_c} \cdot \xic \right) \right)
-
2 \int\limits_{c_1 {\sqrt{s}}/{ N} }^{ c_2 \,
p_{J_c}
\cdot \xic }
\frac{d  \lambda'}{\lambda'} A_c\left(c_1, \as\left(\lambda'\right)
\right)\, ,
\label{ABabbr}
\ea
where in the second equality we have shifted the argument of
the running coupling in $\tilde{K}_c$, and have introduced the notation
\ba
B_c\left(c_1,c_2, \as\left(\mu \right) \right)
& \equiv & -
\tilde{K}_c\left(\frac{1}{c_1}, \as\left(\mu \right)  \right) -
G_c\left(\frac{1}{c_2} ,\as\left(\mu \right)\right),
\nonumber \\
2 A_c\left( c_1,  \as\left(\mu \right) \right) & \equiv &  \gamma_{K_c}
\left(\as(\mu) \right) + \beta(g(\mu)) \frac{\partial}{\partial
g(\mu)} \tilde{K}_c\left(\frac{1}{c_1},\as(\mu)\right).
\label{ABdef} \quad
\ea
The choice of the constants $c_1$ and $c_2$ is a matter of convenience. They reflect the freedom in separating soft/collinear ($K_c$) from collinear ($G_c$) contributions. If the functions in (\ref{ABabbr}) were calculated to all orders the jet evolution would be independent of their choice. Residual dependence on the $c_i$ is due to fixed logarithmic resummation.   

The solution to Eq. (\ref{KGend}) with $\mu = \mu_0$ is
\ba
\tilde{J}_c \left( \frac{p_{J_c} \cdot \xic }{\mu_0},
  \frac{\sqrt{s}}{ \mu_0 N} \, ,\as(\mu_0)
\right)
&=&
\tilde{J}_c \left( \frac{\sqrt{s}}{\mu_0 N},
  \frac{\sqrt{s}}{\mu_0 N} ,\as(\mu_0)
\right)   \label{orgsol}  \\
&\ & \hspace{-57mm}
   \times \, \exp \left\{\! -\int\limits_{\sqrt{s}/N }^{p_{J_c} \cdot
\xic}
   \frac{d \lambda}{\lambda} \left[B_c\left(c_1,c_2,
\as\left(c_2 \lambda \right) \right)  + \!  2 \int\limits_{c_1 \frac{s
}{ 2 \lambda N  } }^{c_2\, \lambda}\frac{d \lambda'}{\lambda'} A_c\left(
c_1,\as\left(\lambda'\right) \right) \right] \right\}\, .\,\,
\nonumber
\ea

After combining Eqs.\ (\ref{jetsol}) and (\ref{orgsol}),
the choice $\mu_0 = \sqrt{s}/N$
  allows us to control
all large logarithms in
the jet functions simultaneously:
\ba
\tilde{J}_c \left( \frac{p_{J_c} \cdot \xic}{\mu},
\frac{\sqrt{s}}{\mu N} ,\as(\mu)
\right)
\!\!\!&=& \!\!\!
\tilde{J}_c \left(
1, 1,\as\left(\frac{\sqrt{s}}{N} \right) \right)
\exp \left\{- \!\!\!\! \int\limits_{\sqrt{s}/N}^\mu
\frac{d\lambda}{\lambda} \gamma_{J_c} \left(\as(\lambda)\right) \right\}
\,  \nonumber  \\
&\ & \hspace{-52mm}
   \times \, \exp \left\{ -\int\limits_{\sqrt{s}/N }^{p_{J_c}
\cdot \xic}
   \frac{d \lambda}{\lambda} \left[ B_c\left(c_1,c_2, \as\left(c_2 \lambda
\right) \right) + 2 \int\limits_{c_1 s/(2 \lambda N) }^{c_2\, \lambda}\frac{d \lambda'}{\lambda'} A_c\left(
c_1,\as\left(\lambda'\right) \right) \right] \right\}\, .
 \nonumber \\
 & & \label{jetxiend}
\ea

\subsection{The Resummed Thrust}

Putting everything together, we arrive at the fully resummed form of the thrust,
\ba
\tilde{\sigma}(N,s) & = & \sigma_0 H \left(1, \as\left(\frac{\sqrt{s}}{2}\right) \right)  \nonumber \\
&  & \hspace*{-22mm} \times 
\tilde{S} \left( 1,  \as\left(\frac{\sqrt{s}}{N}\right) \right) \exp \left\{-\int\limits_{\sqrt{s}/N}^{\sqrt{s}/2} \frac{d \lambda}{\lambda} \gamma_s (\as(\lambda)) \right\}   \nonumber \\
&  &  \hspace*{-22mm} \times
\prod\limits_{c = 1}^2 \tilde{J}_c \left(
1, 1,\as\left(\frac{\sqrt{s}}{ N} \right) \right) \exp \left\{ -\int\limits_{\sqrt{s}/N }^{\sqrt{s}/2}
   \frac{d \lambda}{\lambda} \left[ \gamma_{J_c} (\as(\lambda)) + B_c\left(c_1,c_2, \as\left(c_2 \lambda
\right) \right)  \right. \right. \nonumber \\
& & \qquad \qquad \qquad \qquad \left. \left. \, + \, 2 \int\limits_{c_1 s/(2 \lambda N) }^{c_2\, \lambda}\frac{d \lambda'}{\lambda'} A_c\left(
c_1,\as\left(\lambda'\right) \right) \right] \right\}\, . \label{thrustend}
\ea
We have  set $\mu = \sqrt{s}/2$ and identified
\be
p_{J_c} \cdot \xic = \frac{\sqrt{s}}{2}, \label{pxi}
\ee
since the cross section is independent of the choice of $\xi_c$. Eq. (\ref{thrustend}) contains double logarithms from the integral with $A_c$, and single logarithms from the remaining exponents. At NLL it suffices to evaluate $A_c$ at two loops, and the remaining functions at one loop. 

In general, since the term with $A_c$ contains one more integral compared to the remaining exponents, this term has to be evaluated to one order higher for consistency. A simplified method to calculate this function at higher orders is therefore desirable. At NNLL the calculation of the Sudakov anomalous dimension is required at three loops. In the next two chapters, Ch.s \ref{ch2}-\ref{ch4}, we develop the necessary tools to perform this calculation at higher loops, before returning to the discussion of event shapes. In Chapter \ref{ch4} the method is illustrated with the fermionic contribution to the three-loop anomalous dimension.

\chapter{Nonabelian Eikonal Exponentiation}
\label{ch2}

As we have shown in the previous chapter, soft radiation at wide angles from the hard scattering directions decouples, and is equally well described by radiation from path ordered exponentials. These eikonal lines replace each of the partons involved in the hard scattering. 
We have shown how to resum large logarithmic corrections in exponentials by solving renormalization group equations. For certain quantities that are built out only of eikonal lines, as we will show here, this exponentiation occurs directly by reordering of diagrams. 

This observation, first made by Sterman \cite{Sterman:jc}, then proved by
Gatheral \cite{Gatheral:cz}, and Frenkel and Taylor \cite{Frenkel:pz}, overcomes the following difficulties which one faces in perturbative calculations at higher order: 
Although perturbative calculations within the eikonal approximation are significantly less complex than calculations within the full theory, we pay the price of introducing new infrared divergences. Of course, these infrared divergences cancel in the infrared safe physical observable under consideration. Also, the number of graphs at each order is in general\footnote{Self-energies of light-like eikonal lines vanish in Feynman gauge.} the same as in the full theory, and can be quite significant at higher orders in the perturbative expansion.

Nonabelian eikonal exponentiation remedies these difficulties. This theorem states that a cross section $X$ with two eikonal lines in a nonabelian theory exponentiates,
\begin{equation}
\sigma^{(\rm eik)} \equiv X = e^{Y}, \label{eq1}
\end{equation}
where $Y$ can be given a simple recursive definition.
In Section \ref{twoexp} we will recall the proof of Eq. (\ref{eq1}) \cite{Gatheral:cz,Frenkel:pz}, for the sake of completeness, including a few illustrative examples which will be used in Chapter \ref{ch4}. The exponent $Y$ in (\ref{eq1}) has the following properties:
\begin{enumerate}
\item $Y$ is a subset of the diagrams contributing to $X$, which we will call ``webs'' in the following, since, as we will see below, their lines
`` [...] are all nested [...] in a spider's web pattern'' \cite{Sterman:jc}.
\item The color weights of the diagrams in $Y$ are in general different from those in $X$.
\item For Eq. (\ref{eq1}) to hold, the phase-space region should be symmetric in the real gluon momenta.
\end{enumerate}

Below we will outline the arguments necessary to prove this theorem.  The proof relies on the recursive definition of color-weights and on the iterated application of a  well-known eikonal identity. In the exponentiated form, the number of graphs to be calculated at each order is significantly reduced. Furthermore, IR and UV subdivergences cancel in the exponent at each order as we will show in Section \ref{wardidproof}, before extending the proof to three eikonal quantities, and listing a few implications of this exponentiation, including the form of power corrections for the thrust in Sec. \ref{sec:webpower}. Sections \ref{twoexp} and \ref{wardidproof} were published in \cite{Berger:2002sv}.

\section{Proof of Exponentiation for Cross Sections with Two Eikonal Lines} \label{twoexp}

\subsection{Some Terminology} \label{defs}

In order to specify which subset of diagrams of the original perturbation series $X$ contributes to the exponent $Y$ we need to introduce some terminology.

\begin{figure}[htb]
\begin{center}
\epsfig{file=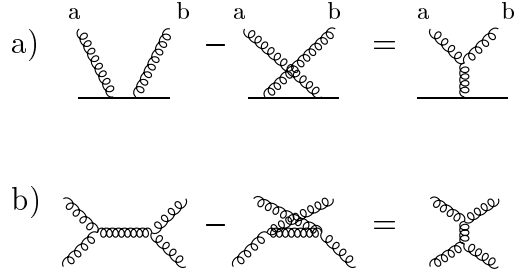,height=5cm,clip=0}
\caption{Graphical representations of the commutation relations between a) $T^a_{ij}$s, and b)  $i\, f_{ijk}$s (Jacobi identity), Eq. (\ref{jacid}).} \label{commrel}
\end{center}
\end{figure}

Each diagram will be decomposed into its color part and its Feynman integral in the eikonal approximation. The eikonal Feynman rules can be found in Appendix \ref{sec:eikfeyn}. The color part can be represented graphically in a diagram which is similar to an ordinary Feynman diagram, but the vertices represent the color part of the Feynman rules, i.\,e. the vertices are just the $T^a_{ij}$s and $i\, f_{ijk}$s, for quark or gluon, respectively, and the lines are $\delta_{ij}$s. In addition, all soft lines have to be drawn inside the (cut) eikonal loop for reasons which will become clear shortly. Certain color diagrams are related to each other by use of the commutation relations of the $T^a_{ij}$s and $i\, f_{ijk}$s (Jacobi identity) which are graphically represented in Fig. \ref{commrel}.
\ba
\left[ T^a, T^b \right] & = & i f_{abc} T^c \nonumber \\
f_{ilm} f_{mjk} &+ & f_{jlm} f_{imk} + f_{klm} f_{ijm} = 0. \label{jacid}
\ea

\begin{figure}[hbt]
 \begin{center}
\epsfig{file=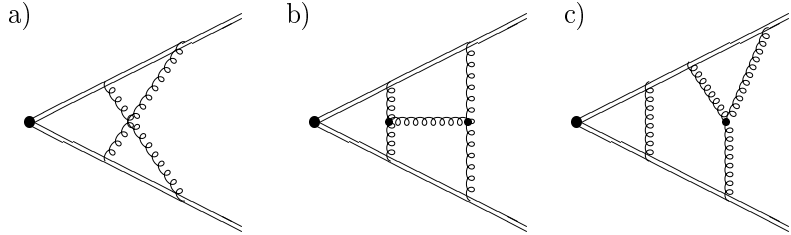,height=4cm,clip=0}
\caption{Examples of a) a web (order 2), b) a c-web (web of order 3), and c) a diagram which is not a web (consisting of a product of an order-1 web and of an order-2 web).} \label{webexample}
\end{center}
\end{figure}

As mentioned above, the diagrams contributing to $Y$ will be called ``webs'' \cite{Sterman:jc}. Originally \cite{Gatheral:cz},  a web was defined as a set of gluon lines which cannot be partitioned without cutting at least one of its lines. As already stated above, all soft lines are to be drawn inside the eikonal loop(s). However, at ${\mathcal{O}}\left(\alpha_s^3\right)$ new types of diagrams arise which Frenkel and Taylor \cite{Frenkel:pz} called connected webs (``c-webs''). c-webs are not included in the original definition for the following reason: If one cuts the horizontal gluon line of the c-web drawn in Fig. \ref{webexample} one would get two webs consisting of three-point vertices since real and virtual gluon lines are treated on equal footing in a color-weight diagram. Below we will refer to webs and c-webs just as webs.

The definitions given by Gatheral, and Frenkel and Taylor can be unified by the following \emph{definition}:
\vspace*{1mm}

A web is a (sub)diagram consisting of soft gluon lines connecting two eikonal lines which cannot be partitioned into webs of lower order by cutting all eikonal lines exactly once. Stated differently, webs are two-eikonal irreducible diagrams. The \emph{order of a web} is defined to be equal to the powers  of $\alpha_s$ it contains, e.\,g. a web of ${\mathcal{O}}\left(\alpha_s^2\right)$ will be called a web of order 2. Diagrammatic examples are shown in Fig. \ref{webexample}. A web has a color factor $\overline{C}$ and a Feynman integral part $\mathcal{F}$. $\mathcal{F}$ contains only those eikonal propagators which are \emph{internal} to the web. The color weight is in general different from the one which one would get from the usual Feynman rules.
\vspace*{1mm}

The color weight of  a web of order $m$ is recursively defined as
\begin{eqnarray}
\overline{C}(W^{(m)}) & \equiv & \frac{1}{\mbox{Tr } \mathbf{1}} C(W^{(m)}) - \sum_d  \prod\limits_{n_i} \overline{C} (W_{n_i}^{(i)}), \nonumber  \\
\overline{C}(W^{(1)}) & \equiv & \frac{1}{\mbox{Tr } \mathbf{1}} C(W^{(1)}) ,
\label{colwe}
\end{eqnarray}
where $C(W^{(m)})$ is the ordinary color factor, $\frac{1}{\mbox{\tiny Tr } \mathbf{1}}$ is the usual normalization of the lowest order, $C^{(0)} = \mbox{Tr } \mathbf{1}$, $\sum_d$ is the sum over the set of all non-trivial decompositions $d$ of $W^{(m)}$ into webs of order $i < m$, and $\prod\limits_{n_i}$  denotes the product of all webs $n_i$ of order $i$ ($1 \leq i < m$) in a particular decomposition $d$. The set of all non-trivial \emph{decompositions} of a given web can be obtained by successively disentangling crossed gluon lines in the web by repeated application of the color identities given in  Fig. \ref{commrel}.

In \cite{Gatheral:cz} Gatheral showed that webs in the original definition have what he called ``maximally nonabelian'' color weights $\sim \alpha_s^m C_F C_A^{m-1}$, where the $C_i$s are the Casimir factors in the fundamental and adjoint representation, respectively. This statement, however, is misleading at orders  $\geq \alpha_s^3$ \cite{Frenkel:pz}. We will see an example in Ch. \ref{ch4}, in the calculation of the $N_f$ term contributing to the coefficient $A$ at three loops.

\begin{figure}[htb]
\begin{center}
\epsfig{file=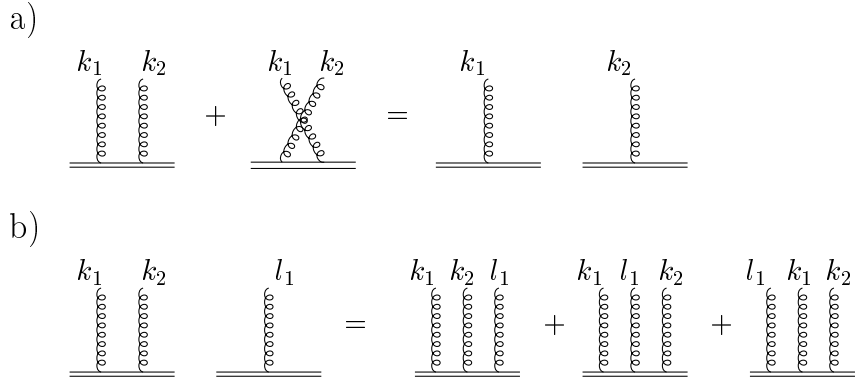,height=5cm,clip=0}
\caption{a) Eikonal identity for 2 gluons, and b) an illustration of the generalized eikonal identity Eq. (\ref{gen2eikid})  for two webs. } \label{rules}
\end{center}
\end{figure}

In the following subsection we will clarify  the above definitions in an example which shows how to factorize eikonal Feynman diagrams into sums of products of webs. This then leads directly to exponentiation. The recursive definition of the color weights of the webs ensures the factorization of the color parts. For the factorization of the Feynman eikonal integrals $\mathcal{F}$ we will make repeated use of the eikonal identity  \cite{Levy:1969cr}
\begin{equation}
 \frac{1}{p \cdot k_1} \frac{1}{p \cdot (k_1 + k_2)} +  \frac{1}{p \cdot k_2} \frac{1}{p \cdot (k_1 + k_2)} = \frac{1}{p \cdot k_1} \frac{1}{p \cdot k_2} \label{eikid},
\end{equation}
illustrated in Fig. \ref{rules} a). This identity can be extended to an arbitrary number of soft gluons in a straightforward way by repeated application of Eq. (\ref{eikid}): For two webs $W_1$ and $W_2$ with gluon legs $k_i \,(i = 1,\dots,m)$ and $l_j\,(j=1,\dots, n)$ attached to an eikonal line with velocity $p$ the generalized identity reads
\begin{eqnarray}
{{\mathcal{F}}}(W_1) {{\mathcal{F}}}(W_2) & \sim  &  \frac{1}{p \cdot k_1 \, p \cdot (k_1 + k_2) \dots p \cdot (k_1 + \dots + k_m)} \nonumber \\
& & \,\,\, \times \,  \frac{1}{p \cdot l_1 \, p \cdot (l_1 + l_2) \dots p \cdot (l_1 + \dots + l_n)} \nonumber \\
& = & \sum\limits_{\mbox{\tiny perms} (n,m)} F, \label{gen2eikid}
\end{eqnarray}
where the sum is over all Feynman diagrams $F$ obtained by permuting the $n+m$ gluon lines such that the order of the $k_i$, and $l_j$, respectively, \emph{within} each web is not changed. A simple example is shown in Fig. \ref{rules} b). The extension to more than two webs follows by repeating the above argument:
\begin{equation}
\sum\limits_{F \,  \mbox{\tiny in } d} F = \prod\limits_{n_i} {\mathcal{F}} \left( W_{n_{i}}^{(i)} \right). \label{geneikid}
\end{equation}

\subsection{An Example}    \label{examplesubsect}

\begin{figure}
\begin{center}
\epsfig{file=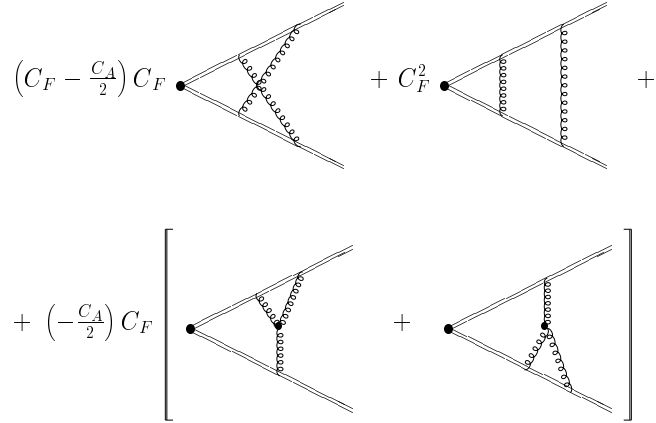,height=6.3cm,clip=0}
\caption[Diagrams contributing at ${\mathcal{O}}(\alpha_s^2)$ (excluding self-energies).]{Diagrams contributing at ${\mathcal{O}}(\alpha_s^2)$ (excluding self-energies). The color factors are given for eikonal lines in the fundamental representation, omitting the overall normalization $\frac{1}{\mbox{Tr } \mathbf{1}}$ everywhere.} \label{order2}
\end{center}
\end{figure}

\begin{figure}
\begin{center}
\epsfig{file=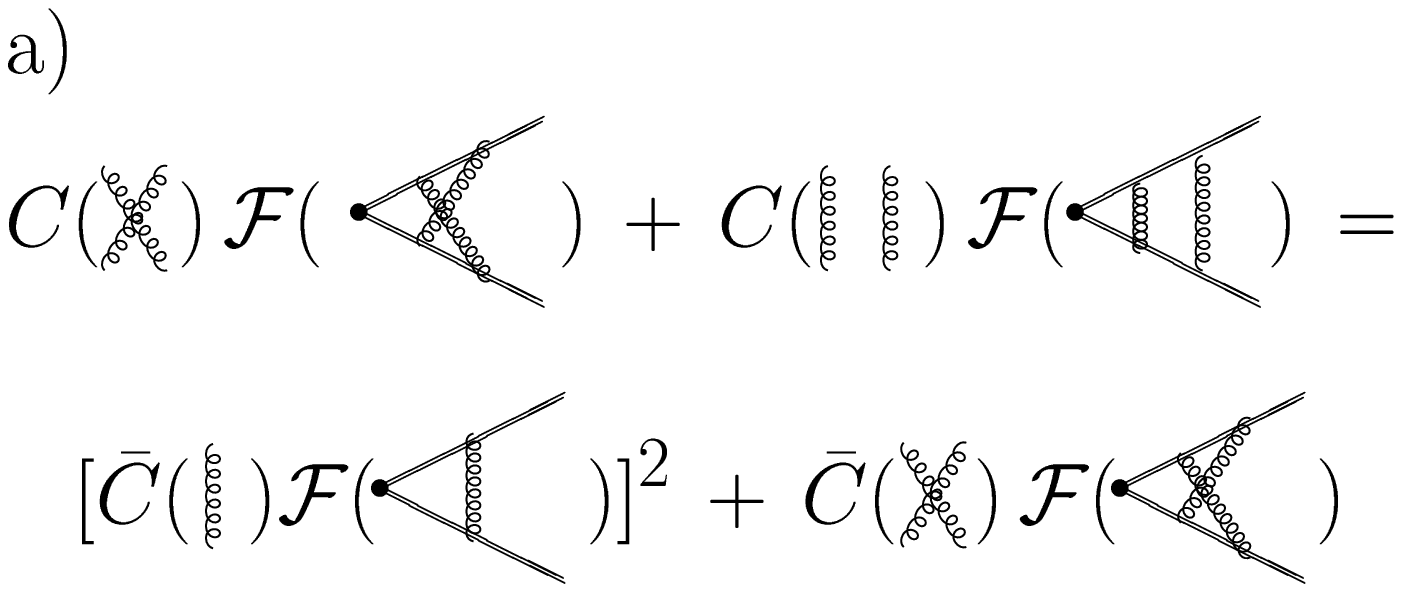,height=3.6cm,clip=0}
\\
\hspace*{-9mm} \epsfig{file=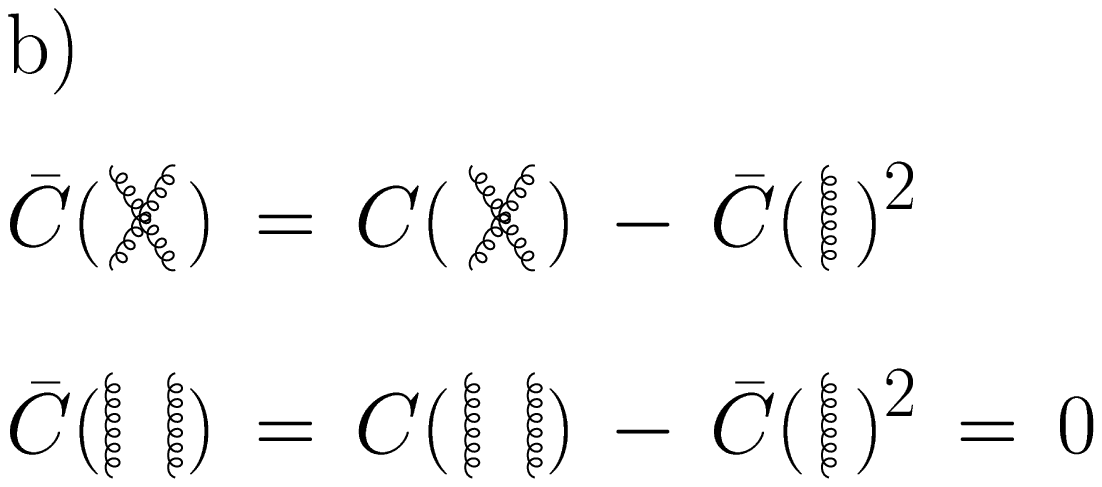,height=3.3cm,clip=0}
\caption{a) Rearrangement of the first two terms of Fig. \ref{order2} using Eq. (\ref{eikid}), and b) Eq. (\ref{colwe}).} \label{rearr}
\end{center}
\end{figure}

We will show by induction that the terms in the perturbation series $X$ in Eq. (\ref{eq1}), normalized by the zeroth order contribution, can be reorganized into a sum of products of webs which can be rewritten as $\exp(Y)$. Therefore it is necessary and also instructive to start with the first nontrivial example, diagrams of ${\mathcal{O}}(\alpha_s^2)$. At this order we have the Feynman diagrams shown in Fig. \ref{order2}, for quark or antiquark eikonal lines, excluding eikonal and gluon self-energies. Eikonal self-energies vanish if we work in Feynman gauge. The sum of diagrams at a given order in $\alpha_s$ is gauge invariant, of course. The contribution of the first two terms in Fig. \ref{order2} can be rearranged by applying Eqs. (\ref{colwe}) and (\ref{eikid}) as shown in Fig. \ref{rearr}.

\begin{figure}[htb]
\begin{center}
\epsfig{file=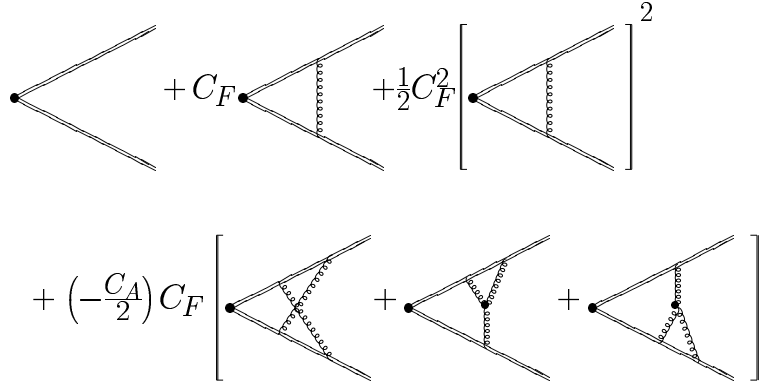,height=5.4cm,clip=0}
\caption{Graphical expression of the exponentiation up to webs of order 2 in Feynman gauge. The color weights are given for (anti)quark eikonal lines. } \label{expgraph}
\end{center}
\end{figure}

Thus we arrive at the following series obtained by rearranging the expansion of $X$ up to ${\mathcal{O}}(\alpha_s^2)$:
\begin{eqnarray}
X & = & {\mathbf{1}} + \sum\limits_{\stackrel{\mbox{\tiny all webs }}{\mbox{\tiny of order } 1}}  \overline{C} \left(W^{(1)} \right) {\mathcal{F}} \left(W^{(1)} \right) + \nonumber \\
& + & \frac{1}{2 !} \bigg(  \sum\limits_{\stackrel{\mbox{\tiny all webs }}{\mbox{\tiny of order } 1}}  \overline{C} \left(W^{(1)} \right) {\mathcal{F}} \left(W^{(1)} \right) \bigg)^2 +  \sum\limits_{\stackrel{\mbox{\tiny all webs }}{\mbox{\tiny of order } 2}}  \overline{C} \left(W^{(2)} \right) {\mathcal{F}} \left(W^{(2)} \right) + \dots, \nonumber \\
& & \mbox{ }  \label{exporder2}
\end{eqnarray}
which is illustrated in Fig. \ref{expgraph}. The combinatorial factor $\frac{1}{2!}$ is necessary to avoid overcounting since two webs with the same structure are indistinguishable if the integration measure is symmetric in the real gluon momenta.

\subsection{Exponentiation}

Looking at the above example it is now clear that any Feynman diagram with two eikonal lines can be expressed as a sum of products of webs by applying Eqs. (\ref{colwe}) and (\ref{geneikid}) repeatedly. By induction we arrive at the following equation for the set of  Feynman diagrams of ${\mathcal{O}}(\alpha_s^n)$, $F^{(n)}$:
\begin{equation}
F^{(n)} = \sum\limits_{\left\{ n_i \right\} } \delta_{n \sum_i i n_i} \prod_i \frac{1}{n_i !} \bigg( \sum\limits_{\stackrel{\mbox{\tiny all webs}}{\mbox{\tiny of order }i}} \overline{C}(W^{(i)}) {\mathcal{F}}(W^{(i)}) \bigg)^{n_i}, \label{nfac}
\end{equation}
where $i$ labels the order of the webs and the sum is over all sets $\left\{ n_i \right\},\, 0 \leq n_i < \infty$ such that $\sum_i i \, n_i = n$. For example, at ${\mathcal{O}} (\alpha_s^3)$ we can have $n_1 = 3$ webs of order 1 ($\{3,0,\dots\}$), or $n_1 = 1$ webs of order 1 and $n_2 = 1$ webs of order 2 ($\{1,1,0,\dots\}$), or $n_3 = 1$ webs of order 3 ($\{0,0,1,0,\dots\}$). The combinatorial factor of $\frac{1}{n!}$ is needed to avoid overcounting because of  property 3.) of $X$,  stated in the introduction to this section, namely that the integration measure is symmetric in the real gluon momenta, for example ${\mathcal{F}}^{(1)}(k_1) {\mathcal{F}}^{(2)}(k_2, k_3) = {\mathcal{F}}^{(1)}(k_2) {\mathcal{F}}^{(2)}(k_1, k_3)$, which means that webs of the same structure are indistinguishable. Were property 3.) not fulfilled, the perturbation series would not exponentiate.

We now rearrange the original perturbation series given in powers of $\alpha_s^n$
\begin{equation}
X = \sum\limits_{n=0}^\infty F^{(n)}, \label{ser}
\end{equation}
\begin{eqnarray}
X & = & \sum\limits_{n=0}^\infty \sum\limits_{\left\{ n_i \right\} } \delta_{n \sum_i i n_i}
\prod_i \frac{1}{n_i !} \bigg( \sum\limits_{\stackrel{\mbox{\tiny all webs}}{\mbox{\tiny of order }i}} \overline{C}(W^{(i)}) {\mathcal{F}}(W^{(i)}) \bigg)^{n_i} \nonumber \\
& = & \sum\limits_{\stackrel{\mbox{\tiny all possible }}{ \{n_i\}}}
\prod_i \frac{1}{n_i !} \bigg( \sum\limits_{\stackrel{\mbox{\tiny all webs}}{\mbox{\tiny of order }i}} \overline{C}(W^{(i)}) {\mathcal{F}}(W^{(i)}) \bigg)^{n_i}    \nonumber
\\
& = & \prod_i \left\{ \sum_{n_i} \frac{1}{n_i !} \bigg( \sum\limits_{\stackrel{\mbox{\tiny all webs}}{\mbox{\tiny of order }i}} \overline{C}(W^{(i)}) {\mathcal{F}}(W^{(i)}) \bigg)^{n_i} \right\} \nonumber \\
&  = & \prod\limits_{i} \exp \bigg( \sum\limits_{\stackrel{\mbox{\tiny all webs}}{\mbox{\tiny of order }i}} \overline{C}(W^{(i)}) {\mathcal{F}}(W^{(i)}) \bigg),
\end{eqnarray}
where we have used the fact that for any function $f(n_i,i)$
\be
\sum\limits_{\stackrel{\mbox{\tiny all possible }}{ \{n_i\}}}  \prod_i f(n_i,i) = \prod_i \sum_{n_i} f(n_i,i)
\ee
which is easy to see by comparing the expansions of the left and the right hand sides.

So the series exponentiates
\begin{equation}
X = e^Y, \quad Y \equiv \sum\limits_{i} \bigg( \sum\limits_{\stackrel{\mbox{\tiny all webs}}{\mbox{\tiny of order }i}} \overline{C}(W^{(i)}) {\mathcal{F}}(W^{(i)}) \bigg). \label{exp2eq}
\end{equation}
This completes the proof that eikonal cross sections with two eikonal lines can be written as an exponent of an infinite sum of webs.

\section{Cancellation of Subdivergences in the Exponent}  \label{wardidproof}

Gatheral, Frenkel, and Taylor showed in  \cite{Frenkel:1983di} by explicit fixed-order calculations that
infrared/collinear subdivergences cancel in the exponent. Here we will outline the proof of this
cancellation, as well as of the cancellation of UV subdivergences involving the eikonal vertex,
to all orders with the help of the identities in Fig. \ref{wardeik} in the soft approximation.
The remaining UV subdivergences are removed via ordinary QCD counterterms, and thus an additional
investigation of the renormalizability of the eikonal vertex is unnecessary.

 To show the absence of subdivergences, let us rewrite
 Eq. (\ref{nfac}) as
\be
\sum\limits_{\mbox{\tiny order } n} \bar{C}^{(n)} {\mathcal{F}}^{(n)} = F^{(n)}_{\mbox{\tiny conv}} + F^{(n)}_{\mbox{\tiny div}}
- \sum\limits_{\left\{ n_i \right\},\, i < n } 
\prod_i \frac{1}{n_i !} \bigg( \sum\limits_{\stackrel{\mbox{\tiny all webs}}{\mbox{\tiny of order }i < n}} \overline{C}(W^{(i)}) {\mathcal{F}}(W^{(i)}) \bigg)^{n_i}. \label{eqsubfact}
\ee
Eq. (\ref{eqsubfact}) means, that the sum of all webs at order $n$ are given by the original perturbation series at that
 order where all lower-order webs have been subtracted out. The original perturbation series can be classified
 into terms without subdivergences, denoted by $F^{(n)}_{\mbox{\tiny conv}}$, and terms which contain
 subdivergences, $F^{(n)}_{\mbox{\tiny div}}$.

\begin{figure}[hbt]
\begin{center}
\epsfig{file=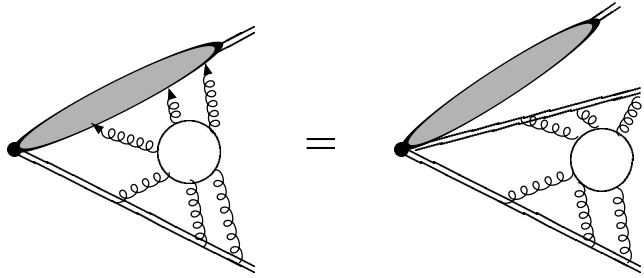,height=4.2cm,clip=0}
\caption{Factorization of jet-like collinear configurations, represented by the grey oval, from the eikonal cross section with the help of the soft approximation. } \label{subfact}
\end{center}
\end{figure}

In eikonal cross sections, infrared/collinear divergences stem from the same momentum configuration as UV divergences.
Since eikonal cross sections are scaleless, when a line becomes collinear to an eikonal it can carry infinite
momentum in a light-like direction.
 But in this case we can employ the soft approximation described in Sect. \ref{sec:Glauber} to factorize
 these jet-like configurations from the rest of the eikonal cross section. The reasoning follows Section
 \ref{sec:Glauber}, and we arrive at the equality shown in Fig. \ref{subfact}. The grey oval in the figure
 represents a specific jet-like configuration, collinear to one of the eikonal lines.  The displayed equality
 states that the sum of all webs at a given order, where this jet-like configuration is connected to the rest
 of the eikonal cross section by soft gluons, can be expressed in the factored form shown on the right-hand side.
 As in Section \ref{sec:Glauber}, remainders  are non-leading. Due to the definition of the color weights
 (\ref{colwe}), the right-hand side does not constitute a web of the same order, but rather a product of webs
 of lower orders. The contribution shown on the left-hand side of Fig. \ref{subfact} is a contribution to
 $F^{(n)}_{\mbox{\tiny div}}$ of Eq. (\ref{eqsubfact}). In (\ref{eqsubfact}), however, we subtract out all products of webs of lower orders, thus cancelling the divergent contributions because of the equality shown in Fig. \ref{subfact}.
Using the equality in Fig. \ref{subfact} and Eq. (\ref{eqsubfact}) recursively for every IR/collinear
subdivergence, we see that the sum of webs at a given order is free of such subdivergences.

To summarize,
 the collinear configuration does not contribute at order $n$ after summing over all relevant webs at that order, because this collinear configuration has already been taken into account at a lower order $< n$. The only possible collinear and UV vertex divergence can occur in the final, overall integral. Of course, in the original perturbative expansion $X$ of the eikonal cross section in Eq. (\ref{ser}) these collinear and UV subdivergent configurations contribute, but in the exponent $Y$ of Eq. (\ref{exp2eq}) they only appear as overall divergences.

\section{Exponentiation for Quantities with Three Eikonal Lines} \label{sec:threeexp}

Above we have presented the necessary ingredients to show that this exponentiation generalizes to three eikonal lines. The proof is analogous to the proof for two eikonal lines, because the vertex structure is again a singlet in color space - $T^{a}_{i j}$ for $q (i) \, \bar{q}(j) \, g (a)$ or $i f_{a b c}$ for $g (a) \, g(b) \, g(c)$, respectively, where $q$ and $g$ denote a quark and a gluon, respectively, the $i,j,a,b,c$ are color indices. Soft gluons do not change this basic color flow. This is not true, however, for more than three eikonal lines.
 
To be specific, let us consider a process involving a $q(i)$-, a $\bar{q}(j)$-, and a $g(a)$-eikonal line, for example $q \, \bar{q} \rightarrow g \, \gamma  + \mbox{ soft gluons}$. We want to show that 
\begin{equation} 
X' = \exp Y'. \label{threeexp} 
\end{equation}
 
The properties of $Y'$, the recursive definition of the color weights Eq. (\ref{colwe}), and the application of the eikonal identity (\ref{eikid}) are exactly analogous to the simpler case of two eikonal lines, with the normalization
\begin{equation} 
C(W^{(0)}) =  T^a_{ij} T^a_{ji} \label{c0qqg} 
\end{equation} 
instead of $\mbox{Tr } \mathbf{1}$ in Eq. (\ref{colwe}):
\begin{eqnarray}
\overline{C'}(W^{(m)}) & \equiv & \frac{1}{C(W^{(0)})} C(W^{(m)}) - \sum_d  \prod\limits_{n_i} \overline{C'} (W_{n_i}^{(i)}), \nonumber  \\
\overline{C'}(W^{(1)}) & \equiv & \frac{1}{C(W^{(0)})} C(W^{(1)}) ,
\label{colwe2}
\end{eqnarray}
The only difference from the 2-eikonal case is the number and complexity  of webs contributing at a given order. 
 
\begin{figure}[htb] 
\hspace*{-5mm} \epsfig{file=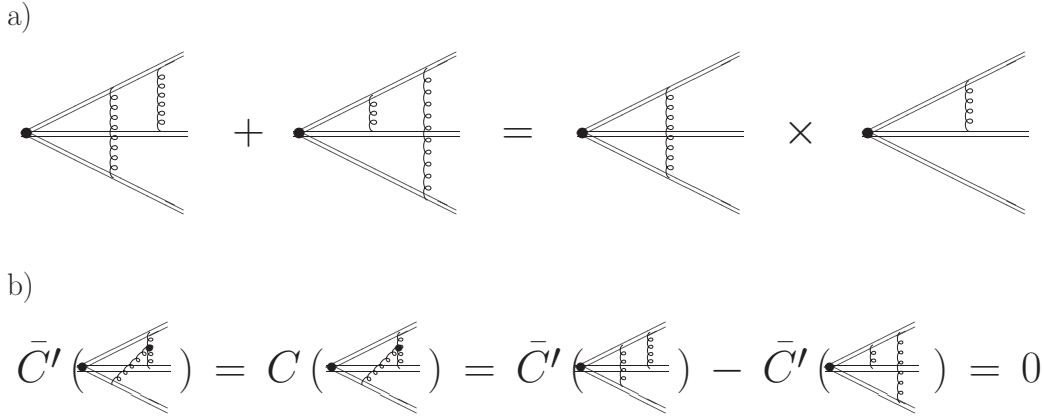,height=6cm,clip=0}
\caption{a) Factorization of contributions at order 2 into webs of order 1. \newline b) Diagram with color weight 0, and decomposition into diagrams of Fig. \ref{factor} a) using Fig. \ref{commrel}. } \label{factor} 
\end{figure}

The proof of exponentiation resulting in Eq. (\ref{nfac}) and then in Eq. (\ref{threeexp}) is again by induction on the order. The first web where all three eikonal lines could in principle be connected occurs at order 2. 
Any contribution at order 2 can be factored into contributions of order 1 with the help of the eikonal identity Eq. (\ref{eikid}), graphical examples are shown in Fig. \ref{factor}. These terms which factor into order-1 webs connecting only two eikonal lines are exactly the terms that stem from 
\begin{displaymath}
\frac{1}{2 !} \bigg( \sum\limits_{\stackrel{\mbox{\tiny all webs}}{\mbox{\tiny of order }1}} \overline{C'}(W^{(1)}) {\mathcal{F}}(W^{(1)}) \bigg)^2.
\end{displaymath} 
 There are no webs at order 2 that connect all three lines, since the color weight $\overline{C'}$ of all the diagrams in Fig. \ref{factor} b) is 0. This also applies to the case when all lines are gluon lines. In that case we apply the Jacobi identity Fig. \ref{commrel} b) instead of \ref{commrel} a) to arrive at a color weight $\overline{C'} = 0$. 
 
At order 3 or higher, this observation does not remain true. Then all three eikonal lines can be connected. Nevertheless, the combinatorics remains the same, since the definition of the color weights is not modified for color singlet configurations except for the overall normalization (\ref{c0qqg}). 

By induction we arrive at Eq. (\ref{nfac}) with webs of a slightly more complicated structure. 
Summation of Eq. (\ref{nfac}) over all powers of  $\alpha_s^n$ results again in exponentiation, and the full cross section consisting of three eikonal lines can be written as
\begin{equation} 
X' = e^{Y'}, \quad Y' \equiv \sum\limits_{i} \bigg( \sum\limits_{\stackrel{\mbox{\tiny all webs}}{\mbox{\tiny of order }i}} \overline{C'}(W^{(i)}) {\mathcal{F}}(W^{(i)}) \bigg). \label{yprime} 
\end{equation} 
 
The case of three gluon eikonal lines, $g(a)\, g(b) \, g(c)$, is analogous to the one considered above, with \begin{equation} 
C(W^{(0)}) = i f_{a b c} i f_{c b a} 
\end{equation} 
instead of (\ref{c0qqg}). 

This completes the proof of exponentiation for three eikonal cross sections which are free of subdivergences, following the same argumentation as for two eikonal cross sections, given in Sec. \ref{wardidproof}. 
 
\section{Implications for Physical Observables} \label{sec:powercorr}

As we have seen in Chapter \ref{ch1}, eikonal cross sections contain all information about soft radiation, and their anomalous dimensions control double logarithms that arise from soft-collinear emission. The exponentiation of webs directly implies this exponentiation of double logarithmic behavior. In addition, important consequences on the behavior of power corrections arise, as was also observed, for example, in Refs. \cite{Laenen:2000ij,Korchemsky:1994is}.

\subsection{Properties of Webs} \label{sec:webprop}

The properties proved in the Section \ref{wardidproof}, and the invariance under rescalings of the light-like eikonal momenta give the following constraints on the possible dependence of webs on the web-momentum $k$ and the eikonal momenta, denoted in the following by $\beta_i,\, i = 1,2$ or $i = 1,2,3$: 
\begin{itemize} 
\item Invariance under rescalings requires that each eikonal momentum appears with  the same power in the numerator as in the denominator of the argument. 
\item After regulating the one overall UV divergence the webs obey
\be
\mu \frac{d}{d \mu} W(k,\beta_i,\mu,\as(\mu),\varepsilon) = 0. \label{muindep}
\ee
Here $\mu$ is the renormalization scale, and $\varepsilon$ the dimensional regulator.
\item The fact that they have at most one overall IR divergence, coupled with a single overall collinear divergence requires that the factors $k \cdot \beta_i$ occur only once for each $\beta_i$. 
\end{itemize} 

From the above properties we deduce for $n=2$ eikonal lines
\begin{equation} 
W \left(k, \beta_1,\beta_2,  \mu, \alpha_s(\mu), \varepsilon \right) = W \left( k^2, \frac{(k \cdot \beta_1) (k \cdot \beta_2)}{\beta_1 \cdot \beta_2}, \mu^2,  \alpha_s(\mu), \varepsilon \right). \label{arg2} 
\end{equation} 
The last property allows us to rewrite this as 
\begin{equation} 
 W \left( k^2, \frac{(k \cdot \beta_1) (k \cdot \beta_2)}{\beta_1 \cdot \beta_2}, \mu^2,  \alpha_s(\mu), \varepsilon \right) = W\left(k^2, k^2 + k_{\perp}^2, \mu^2, \alpha_s(\mu), \varepsilon \right), \label{arg2perp} 
\end{equation} 
where $k_\perp$ is the momentum transverse to the plane spanned by $\beta_1$ and $\beta_2$. It is easy to derive this form using a Sudakov parametrization for $k$, 
\be
k^\mu = x \beta_1^\mu + \frac{k^2 + k^2_\perp}{2 x} \beta_2^\mu + k_\perp^\mu. 
\ee
We have reduced the maximum number of independent parameters from four, $k^\mu$, to two, $k^2$ and $k^2_\perp$. 

These observations do not remain true for the case of three eikonal lines. Nevertheless, our argumentation on two-eikonal webs also applies to a subset of three-eikonal cross sections, which we will call in the following ``degenerate''. In degenerate webs only two out of the three eikonals are connected. As we have seen in Sec. \ref{sec:threeexp}, these are the only contributions up to, and including, next-to-leading order. Nondegeneracy begins at order 3. For degenerate webs, the dependence splits into the sum 
\be
W = W_{ab} + W_{ac} + W_{bc},
\ee
where $a,b, c$ are the colors carried by the three eikonal lines, in the fundamental or adjoint representation, respectively. 

\subsection{Leading Logarithms of the Thrust from Eikonal Exponentiation} \label{sec:thrustexp}

We now return to the study of the thrust cross section to illustrate the consequences of the above observations. In the previous chapter we have exponentiated large logarithms by solving renormalization group equations. The final result for the thrust cross section in moment space is given in Eq. (\ref{thrustend}). 
However, as we have seen in this chapter, eikonal cross sections exponentiate directly. Here we will demonstrate how the exponentiation of eikonal cross sections and the properties listed in Section \ref{sec:webprop} lead to the same form as Eq. (\ref{thrustend}). 

The thrust is related to the minus momentum, from Eq. (\ref{thrust}),
\be
\tau = 1 - T =  \frac{\sqrt{2}}{Q} \sum_i k_i^-. \label{tauminus}
\ee  

The factorized thrust contains also an eikonal cross section, as displayed in Fig. \ref{factorized}. This eikonal cross section exponentiates. 
Taking moments with respect to $\tau = 1-T$, we obtain with Eq. (\ref{arg2perp}),
\ba
\tilde{\sigma}^{(\mbox{\tiny eik})}_T (N) &  =  & \exp \Bigg\{ 2 \int\limits_0^{Q^2} \frac{d^{2-2\varepsilon} k_\perp}{\Omega_{1-2 \varepsilon}} \int\limits_0^{Q^2 - k_\perp^2} d k^2 \int\limits_{k_\perp^2/(\sqrt{2}Q)}^{|k_\perp|/\sqrt{2}} \frac{d k^-}{2 k^-} \left(e^{-N/Q \sqrt{2} k^-  } - 1 \right) \nonumber \\
& & \qquad \times \, W\left(k^2, k^2 + k_\perp^2,\mu^2,\as(\mu),\varepsilon \right) \Bigg\}.
\ea
 The factor of 2 is due to adding the complex conjugate contributions.
Here $\Omega_{1-2 \varepsilon} = 2 \pi^{1-\varepsilon}/\Gamma(1- \varepsilon)$ is the dimensionally continued transverse angular volume, and the limits of the integrals correspond to the one-particle phase space \cite{Catani:1992ua}.
We have normalized the virtual contributions to 1 for $N = 0$, and the single-gluon emission in $\MS$ scheme to
\be
W^{(1),\,\mbox{\tiny real}}(k) = \frac{2 C_F \as}{\pi} \mu^{2 \varepsilon} \frac{1}{k_\perp^2} \delta_+(k^2).
\ee

Using the equivalence of the minus momentum to $\tau$, Eq. (\ref{tauminus}) we obtain
\ba
\tilde{\sigma}^{(\mbox{\tiny eik})}_T (N) & = & \exp \Bigg\{ \int\limits_{0}^{Q^2} \frac{d k_\perp^2}{k_\perp^{2\varepsilon}} \int\limits_{k_\perp^2/Q^2}^{k_\perp/Q} \frac{d \tau}{\tau} \left(e^{-N \tau } - 1 \right)    \nonumber \\
&  & \qquad \times \int\limits_0^{Q^2 - k_\perp^2} dk^2 \, W\left(k^2, k^2 + k_\perp^2,\mu^2,\as(\mu),\varepsilon \right)  \Bigg\}, \label{powert0}
\ea
where we have used the azimuthal symmetry of the webs, Eq. (\ref{arg2perp}).

From the considerations in the previous two sections we know that there are no internal divergences. Moreover, for fixed $k_\perp$ the webs do not require overall UV regularization, and the convergence occurs on a scale set by $k_\perp$, independent of $N$ or $Q$. Thus we can formally expand the integral over $k^2$ in inverse powers of $Q^2$ \cite{Laenen:2000ij}
\be
 \int\limits_0^{Q^2 - k_\perp^2} dk^2 \, W\left(k^2, k^2 + k_\perp^2,\mu^2,\as(\mu),\varepsilon \right) = \frac{2 A\left(\as(k_\perp^2) \right)}{(k_\perp^2)^{1-2\varepsilon}} + {\mathcal{O}}\left(\frac{k_\perp^{2\varepsilon}}{Q^2} \right). \label{webexpand}
\ee
We have used the independence of the webs of the renormalization scale, Eq. (\ref{muindep}), to set the scale of the running coupling to $k_\perp^2$.
We will confirm the form (\ref{webexpand}) by explicit calculation in Chapter \ref{ch4} below. 

Eq. (\ref{powert0}) can then be written as
\be
\tilde{\sigma}^{(\mbox{\tiny eik})}_T (N) =  \exp \left\{ 2 \int\limits_{0}^{Q^2} \frac{d k_\perp^2}{k_\perp^2} A\left(\as(k_\perp^2) \right)   \int\limits_{k_\perp^2/Q^2}^{k_\perp/Q}  \frac{d \tau}{\tau} \left(e^{- N \tau} - 1 \right)   \right\}. \qquad \label{powert1}
\ee
We have neglected all terms that vanish as $k_\perp \rightarrow 0$. 

The form (\ref{powert1}), obtained directly from the exponentiation of webs, is identical to the term with $A_c$ in Eq. (\ref{thrustend}) \cite{Korchemsky:1999kt}. To see this, we use the relation
\be
e^{-x/y} - 1 \approx - \theta \left( x - y e^{-\gamma_E} \right), \label{thetarel}
\ee
which is valid to next-to-leading logarithmic order. Upon relabelling $k_\perp \rightarrow \lambda'$ we obtain
\be
\tilde{\sigma}^{(\mbox{\tiny eik})}_T (N) =  \exp \left\{- 2 \int\limits_{0}^{Q} \frac{d \lambda'}{\lambda'} A\left(\as(\lambda') \right)   \int\limits_{{\lambda'}^2/Q^2}^{\lambda'/Q}  \frac{d \tau}{\tau} \theta\left(e^{\gamma_E} N \tau - 1 \right)   \right\}. \qquad \label{powert2}
\ee
Then we change variables,  
\be
\lambda = \frac{\tau Q^2}{2 \lambda'},
\ee
and exchange orders of integration. We arrive at
\be
\tilde{\sigma}^{(\mbox{\tiny eik})}_T (N) = \exp \left\{ - 2 \int\limits_{Q/(2N)}^{Q/2} \frac{d \lambda}{\lambda} \int\limits_{e^{-\gamma_E} Q^2/(2 N \lambda)}^{2 \lambda} \frac{d \lambda'}{\lambda'} A\left(\as(\lambda') \right) \right\}.
\ee
We have reproduced the term in Eq. (\ref{thrustend}) with the double integral, with 
\ba
c_1  & = & e^{-\gamma_E}, \nonumber \\
c_2  & = & 2. \label{cipick}
\ea
The remaining terms in (\ref{thrustend}) with a single integral, on the other hand, account for the difference between the eikonal cross section and the full partonic cross section for the thrust \cite{Laenen:2000ij}. This is  analogous to the jet-remainders in Eq. (\ref{finalform}) that match the eikonal cross section to the full parton distribution function as $x \rightarrow 1$.

\subsection{Power Corrections} \label{sec:webpower}

As we have seen in the introduction, due to the asymptotic nature of the perturbation series, pQCD calculations can only be accurate up to power corrections $\sim  1/Q^p$, where $Q$ is the hard scale in the problem. In general, mean values $\left< {\mathcal{O}} \right>$ or integrated cross sections  depend on hadronization corrections and other non-perturbative effects in a more trivial way than differential observables $d \sigma/(d {\mathcal{O}})$:
\ba
\frac{1}{\sigma_{\mbox{\tiny tot}}} \frac{d \sigma}{d {\mathcal{O}}} & = & \frac{d \sigma_{\mbox{\tiny PT}}}{d {\mathcal{O}}} + f_{\mbox{\tiny hadr}} (Q^{-p}, {\mathcal{O}} ) ,\\
\left< {\mathcal{O}} \right>  & = & \left< {\mathcal{O}} \right>_{\mbox{\tiny PT}} + \frac{\lambda_p}{Q^p}. \label{lamp}
\ea
Here the subscript PT is the perturbatively calculable part, and $f_{\mbox{\tiny hadr}}$  is a nonperturbative function, related to the nonperturbative parameter $\lambda_p$, \begin{samepage}$\lambda_p/Q^p = \int d {\mathcal{O}}\, {\mathcal{O}} f_{\mbox{\tiny hadr}}$.\end{samepage} The power $p$ is a measure of the sensitivity of the observable ${\mathcal{O}}$ to confinement physics.

However, perturbation theory itself contains information about the form of these non-perturbative corrections. Sensitivity of an observable to long-distance behavior is adjustable by studying its Laplace transform, Eq. (\ref{thrustmoment}). The value of $N$ controls the influence of long-distance effects. For large $N$, non-perturbative corrections become important.

The form for the double logarithmic terms derived from eikonal exponentiation, Eq. (\ref{powert1}), allows us to deduce the form of power corrections. Using (\ref{powert1}) with the one-loop running coupling  at scale $k_\perp^2$ reexpressed in terms of the running coupling at the hard scale $Q^2$ via Eq. (\ref{asreexp}), we arrive at
\ba
\tilde{\sigma}^{(\mbox{\tiny eik})}_T (N) & = & \exp \Bigg\{  2\, \int\limits_0^{Q^2} \frac{d k_\perp^2}{k_\perp^2} A\Bigg( \frac{\as(Q^2)}{1+ \frac{\beta_0}{4 \pi} \as(Q^2) \ln \frac{k_\perp^2}{Q^2}} \Bigg) \nonumber \\
& & \qquad \times \,  \int\limits_{k_\perp^2/Q^2}^{k_\perp/Q} \frac{d \tau}{\tau} \left( e^{- N \tau} - 1 \right) \Bigg\}. \label{powertp}
\ea
Since $N$ is conjugate to $\tau$, and since we are interested in the infrared region for small $k_\perp^2$, we can expand the exponential in the exponent. This results in
\ba
\tilde{\sigma}^{(\mbox{\tiny eik})}_T (N) & = & \exp \Bigg\{ 2\, \sum_{n = 1}^\infty \frac{1}{n\, n!} \left(-N\right)^n \int\limits_0^{Q^2} \frac{d k_\perp^2}{k_\perp^2}  A\Bigg(\frac{\as(Q^2)}{1+ \frac{\beta_0}{4 \pi} \as(Q^2) \ln \frac{k_\perp^2}{Q^2}} \Bigg) \nonumber \\
& & \hspace*{4cm} \times \, \left(\frac{k_\perp}{Q}\right)^n \left[ 1 -  \left(\frac{k_\perp}{Q}\right)^n  \right]  \Bigg\}, \label{powertpb}
\ea
after performing the integration over $\tau$.
We now change variables 
\be
t_n \equiv n\, t = \frac{n}{2} \as(Q^2) \ln \frac{Q^2}{k_\perp^2}, \label{varchange}
\ee
and obtain
\ba
\tilde{\sigma}^{(\mbox{\tiny eik})}_T (N) \!\!& =\!\! & \exp \Bigg\{2\, \sum_{n = 1}^\infty  \frac{1}{n^2\, n!} \frac{1}{\as(Q^2)} \left(-N \right)^n \int\limits_0^{\infty} d t_n\, A\Bigg(\frac{\as(Q^2)}{1 -\frac{\beta_0}{2 \pi\,n} t_n} \Bigg)  e^{-\frac{t_n}{\as(Q^2)}}  \nonumber \\
& & \hspace*{4cm} \times \, \left[ 1- e^{-\frac{t_n}{\as(Q^2)}} \right] \Bigg\}.\label{powertp2}
\ea
The integral over $t$ has the form of the Borel integral introduced in the introduction, Eq. (\ref{Binv}). The Borel integral (\ref{powertp2}) has singularities at 
\be
t_n = \frac{2 \pi\, n}{\beta_0}, \quad \mbox{ or equivalently, } \quad t = \frac{2 \pi}{\beta_0},
\ee
leading with Eq. (\ref{powercorr}) to an ambiguity proportional to
\be
e^{-\frac{2 \pi \,n}{\beta_0 \, \as(Q^2)}} \sim \left(\frac{\LQCD}{Q}\right)^n. 
\ee
The crucial factor in the exponent that results in power corrections proportional to $\sim 1/Q$ comes from the upper limit in (\ref{powertp2}), and is therefore due to radiation at wide angles. In summary, the thrust behaves as
\be 
\ln \tilde{\sigma}_T(N,Q)  = \ln \tilde{\sigma}_{T,\,\mbox{\tiny PT}}(N,Q) + \ln \tilde{\sigma}_{T}^{\mbox{\tiny power}} \left(\frac{N}{Q} \right) + {\mathcal{O}}\left( \frac{N}{Q^2} \right). \label{thrustpower}
\ee
The last term is power-suppressed by $1/Q$ relative to the leading non-per\-tur\-ba\-tive corrections. This term comes from the lower limit of the integral in (\ref{powertp}), as can be seen in (\ref{powertp2}), corresponding to radiation close to the jets.

In the above arguments we have used the one-loop running coupling. But the occurrence of poles in the Borel integral is not connected to the specific form of the one-loop coupling. That is, the ambiguity above is not connected to the Landau pole. Above we have only used the relation between the coupling evaluated at two different scales, independent of $\LQCD$. Although this relation is nonlinear at higher orders (see Eq. (\ref{2as})), conclusions similar to the above are reached in studies of couplings without a simple Landau pole \cite{Grunberg:1995vx,Dokshitzer:1995af,Peris:1996in}, as also the derivation of the result (\ref{thrustpower}) for the thrust with a variety of methods shows \cite{Korchemsky:1999kt,Webber:1994cp,Dokshitzer:1995zt,Akhoury:1995fb,Dokshitzer:1997ew,Gardi:1999dq,Korchemsky:2000kp,Gardi:2001ny,Gardi:2002bg}. 

Above, we have shown that eikonal cross sections exponentiate, and have illustrated the consequences of this exponentiation. The exponentiation simplifies the explicit calculation of the term $A$ significantly at higher orders. We now turn to the   calculation of $A$, that is, the anomalous dimension of an eikonal cross section, also called the cusp anomalous dimension of a Wilson loop.

\chapter{Higher Orders in $A \left(\as \right)/\left[1-x\right]_+$ of Non-Singlet Partonic Splitting Functions}
\label{ch4}

In the previous chapters we have developed the tools necessary for the computation of eikonal diagrams at higher orders. We have seen in Chapters \ref{ch1} and \ref{ch2} that soft-collinear emission is controlled by the anomalous dimensions of eikonal cross sections. In this Chapter we will develop a simplified method to calculate this anomalous dimension at higher orders which is equivalent to the singular part of the partonic splitting functions, $P_{ff}$ (see Eq. (\ref{evol})).

To leading power in $N$  the moments of the  partonic splitting functions, 
\be
\gamma_{ff}(N) = \int_0^1 dx x^{N-1} P_{ff}(x),
\ee
take the simple form \cite{Korchemsky:1988si,Albino:2000cp}
\begin{equation}
\gamma_{ff}(N,\alpha_s) =  A_f (\alpha_s) \ln N +  B_f (\alpha_s) + {\mathcal{O}}\left(\frac{1}{N} \right),\label{pffN}
\ee
or in $x$-space,
\begin{equation}
P_{ff} (x,\alpha_s ) = A_f (\alpha_s) \left[\frac{1}{1-x}\right]_{+} + B_f (\alpha_s) \delta(1-x) + {\mathcal{O}}\left(\left[1-x \right]^0 \right), \label{pff}
\end{equation}
with the plus distribution as defined in Eq. (\ref{plusdef}).
The term with the plus distribution represents the cancellation of a single overall infrared divergence. The coefficient of $\ln N$ in Eq. (\ref{pffN}) can be expanded in the strong coupling,
\begin{equation}
A_f (\alpha_s) = \sum_n \left( \frac{ \alpha_s}{\pi} \right)^n A_f^{(n)}. \label{aaa}
\end{equation}

As we have seen above, the exact knowledge of the terms $A^{(n)}$ is important for $x \rightarrow 1$ (large $N$), since there large logarithmic corrections arise due to soft-gluon radiation. These corrections need to be resummed in order to be able to make reliable predictions within perturbation theory. The knowledge of the coefficients $A^{(3)}$ and $B^{(2)}$ is required at the next-to-next-to-leading logarithmic (NNLL) level \cite{Vogt:2000ci} (compare to Eq. (\ref{thrustend})).

The anomalous dimensions $\gamma_{ff}(N)$ are currently known to two loops \cite{Gross:ju,Gross:cs,Floratos:1977au,Gonzalez-Arroyo:df,Curci:1980uw,Floratos:1981hs,Kodaira:1981nh}, and a general formula for the $\alpha_s^n N_f^{n-1}$-terms of $\gamma_{ff}(N)$ was computed by Gracey \cite{Gracey:nn}. From the known exact values for some specific moments and the behavior at small $x$ \cite{Larin:1993vu,Blumlein:1995jp,Larin:1996wd,Retey:2000nq} a numerical parametrization for the coefficient $A^{(3)}$ was obtained in \cite{Vogt:2000ci,vanNeerven:2001pe,vanNeerven:2001tc}, although, for the above reasons, the exact knowledge of this term is desirable. A calculation at the three-loop level by Moch, Vermaseren, and Vogt of the splitting functions via the operator product expansion (OPE) in moment space will be completed in the near future \cite{Vermaseren:2002rn}. Their results for the fermionic contributions are now available \cite{Moch:2002sn}. However, the method presented here, although only applicable for the calculation of the coefficients $A$, not of the complete $x$-dependence, is complementary to the OPE method in two ways: we calculate only virtual diagrams, and furthermore, it is much less computationally intensive, thus a computation of the four- or even higher loop coefficients may be feasible. The work presented in this chapter was published in \cite{Berger:2002sv}.

In Chapter \ref{ch1} we have related the anomalous dimension of PDFs to the anomalous dimension of an eikonal cross section which exponentiates. A similar observation was made by Korchemsky \cite{Korchemsky:1988si}, who related the anomalous dimension of PDFs, Eq. (\ref{pff}), with the cusp anomalous dimension of a Wilson loop. His work was performed in a noncovariant axial gauge, whereas here  we will use Feynman gauge throughout. Korchemsky's observation was used in \cite{Korchemsky:1992xv} for the calculation of the two-loop coefficient $A^{(2)}$, which was done in Feynman gauge.  The work of Ref. \cite{Korchemsky:1992xv} was also based on the renormalization properties and exponentiation of Wilson loops (see \cite{Korchemsky:wg,Polyakov:ca,Arefeva:1980zd,Gervais:1979fv,Dotsenko:1979wb,Brandt:1981kf,Korchemskaya:1992je} and references therein). This approach is related to ours. However, the additional observations we make result in several advantages.  The number of diagrams contributing at each order is decreased by working with light-like eikonals. Furthermore, we can restrict ourselves only to virtual graphs. With the help of Ward identities we have shown explicitly the absence of infrared (IR) subdivergences and the cancellation of ultraviolet (UV) subdivergences at the eikonal vertex, leaving only the usual QCD UV divergences.

Below we first relate the renormalization properties of PDFs with those of webs, then summarize the method in light-cone ordered perturbation theory. We rederive as examples the one- and two-loop coefficients $A^{(1)}$ and $A^{(2)}$. In Section \ref{sect3loop} we derive a formula for the coefficients of $A^{(n)}$ proportional to $N_f^{n-1}$, which agrees with the corresponding contribution computed by Gracey \cite{Gracey:nn} using an effective theory. We end by illustrating the steps necessary for the complete calculation of the 3-loop coefficient $A^{(3)}$. The IR structure of $A^{(3)}$ is explored for the graphs contributing at $\alpha_s^3 N_f$, which we calculate exactly.

\section[Renormalization of Parton Distribution Functions and of Webs]{Renormalization of Parton Distribution \\ Functions and of Webs}

\subsection{Renormalization of Parton Distribution Functions}

As was shown in \cite{Collins:1981uw}, the parton distribution functions,  defined in their unrenormalized form in  terms of nonlocal operators (\ref{pdfdef}), obey the evolution  equation (\ref{evol}), where the kernel $P_{ab}$ is found from the usual relation \cite{book,Sterman:1995fz,Collins:1981uw}
\be
P_{ff}(\alpha_s,x) = A_f(\alpha_s) \left[ \frac{1}{1-x} \right]_+  +  \dots =  - \frac{1}{2} g_s \frac{\partial }{\partial g_s} \ln Z^A_1 \left[ \frac{1}{1-x} \right]_+ + \dots, \label{prela}
\ee
where $\ln\, Z^A_1$ denotes the $\frac{1}{\varepsilon}$-pole of the counterterm which multiplies the plus-distribution, plus scheme dependent constants, if we work in a minimal subtraction scheme with dimensional regularization. Above we only exhibit the term that is singular as $x \rightarrow 1$, since it is this term which we want to extract from the renormalization of our factorized form, Eq. (\ref{finalform}).

From Eq. (\ref{finalform}) we observe that only the eikonal cross section can contribute to the $A$-term proportional to a plus-distribution. This is because the hard functions are off-shell by ${\mathcal{O}}\left(x p^+\right)$, and the jet-remainders are purely virtual, thus cannot contain plus-distributions. Therefore, their renormalization has to be proportional to $B_f \, \delta(1-x)$, as was observed in \cite{Laenen:2000ij}.

It is thus the renormalization of a color singlet eikonal vertex which we have to study, in order to compute the coefficients $A^{(n)}$ in (\ref{aaa}). 

\subsection{Renormalization of Webs}

For definiteness, we pick the incoming line  $\xi$ moving in the plus direction, and the outgoing eikonal $\beta$ in the minus direction, and since quantities built from eikonal lines are scaleless, we can scale the eikonal velocities to 1. 
\ba
\xi & = & \left( 1,0,0_\perp\right) \nonumber \\
\beta & = & \left( 0,1,0_\perp \right) \label{betaframe}
\ea
in light-cone coordinates. This choice will simplify the calculations considerably, as we will see below. 

With the considerations in Sec. \ref{sec:webprop}, we can write the contributions from virtual webs of order $n$ to the eikonal cross section as
\ba
 2 \int \frac{d^{2 - 2\varepsilon} k_\perp}{(2 \pi)^{1-2 \varepsilon}} \int\limits_0^\infty \frac{d k^+}{2 k^+}  \int dk^2 
 W_{aa}^{(n)} \left(k^2,k^2 + k_\perp^2,  \alpha_s(\mu^2),\varepsilon \right) & & \nonumber \\
& & \hspace*{-99mm} = 2\; \bar{C}_a^{(n)} \left( \frac{\alpha_s(\mu^2)}{\pi} \right)^n \left(\mu^2\right)^{l \varepsilon} \left(4 \pi\right)^{l \varepsilon} K(\varepsilon)  \int \frac{d^{2 - 2\varepsilon} k_\perp}{(k_\perp^2)^{1+(l-1)\varepsilon}} \int\limits_0^\infty \frac{d k^+}{k^+},   \label{methodeq}
\ea
where $l$ is an integer $\leq n$, and $K$ contains numerical factors (including factors of $\pi$) and is, in general, a function of $\varepsilon$ due to the regulation of infrared and UV (sub)divergences.  Above, on the left hand side, all internal momenta have been integrated over, as well as $k^-$,  and internal UV divergences have been renormalized. The integration over $k^2$ results in terms $\sim \frac{1}{(k_\perp^2)^{(l-1)\varepsilon}}$. For graphs including (local) counterterms $l < n$, whereas for graphs with $n$ loops $l = n$. Both virtual webs and their complex conjugates contribute to the overall factor of 2. The structure of the integral over $k^+$ follows from boost invariance.  In Eq. (\ref{methodeq}), this integral is divergent, but these divergences cancel against the corresponding real contributions, and therefore do not affect the anomalous dimension of the eikonal vertex. The $k^+$-integral plays the role of $\frac{dx}{1-x}$ for the full  parton-in-parton distribution functions (cf. Eq. (\ref{eiksigdef})), after combining real and virtual graphs. It suffices to consider only virtual graphs, since real and virtual graphs built out of eikonal lines have the same IR singularity structure, which, due to the scalelessness of virtual graphs, is equivalent to the UV structure:
\be
\frac{1}{\varepsilon} + \frac{1}{(-\varepsilon)} = 0. 
\ee
In other words, the coefficients of the UV poles are equal to those of the IR poles. 

The final scaleless $k_\perp$ integral provides the $n$-loop UV counterterm which contributes to the anomalous dimension $P_{ff}$, Eq. (\ref{pff}). To isolate the UV pole we temporarily introduce a mass
\be
\int \frac{d^{2 - 2 \varepsilon} k_\perp}{(k_\perp^2+m^2)^{1+(l-1)\varepsilon}} = \pi^{1-\varepsilon} \frac{\Gamma(l \varepsilon)}{\Gamma(1 + (l-1)\varepsilon)} \left(m^2 \right)^{-l \varepsilon}. \label{scalepart}
\ee
The counterterm is then given, as usual, by minus the pole terms after expanding in $\varepsilon$. After summing over the contributions of all webs at a given order and their counterterms for subdivergences, all nonlocal terms ($\sim \ln \frac{\mu^2}{m^2}$) cancel as well as UV vertex counterterms and IR divergences, and we obtain the $n$-loop counterterm contributing at $x \rightarrow 1$, which can be written as a series in $\varepsilon$:
\be
Z^{(n)\,A} = \sum\limits_{m = 1}^n \frac{1}{\varepsilon^m} \left( \frac{\alpha_s(\mu^2)}{\pi} \right)^n a_m^{(n)} \int\limits_0^\infty \frac{d k^+}{k^+},
\ee
with purely numerical coefficients $a_m^{(n)}$. Because webs exponentiate, the counterterm for UV divergences in the perturbative expansion of a non-singlet parton distribution  is given by
\be
Z^A = \exp \left\{ \sum_{m = 1}^\infty \sum_{n = 1}^m \frac{1}{\varepsilon^m} \left( \frac{\alpha_s(\mu^2)}{\pi} \right)^n a_m^{(n)} \, \int\limits_0^\infty \frac{d k^+}{k^+}  \right\}  \label{counterterm}
\ee
in the limit $x \rightarrow 1$, as indicated by the superscript $A$. As noted above, the notation $\int_0^\infty dk^+/k^+$ is equivalent to $[1/(1-x)]_+$.
Now it is trivial to extract the contribution to $P_{ff}$. From (\ref{prela}) and (\ref{counterterm}) we get
\be
A_f^{(n)} = - n a_1^{(n)}. \label{aendeq}
\ee

As emphasized above, internal UV divergences, including the usual QCD divergences and divergences at the eikonal vertex, have to be renormalized. Further complications arise because collinear/IR divergences cancel only after summing over all diagrams at a given order, so an individual diagram has in general UV singularities multiplying IR/collinear singularities. Our method to resolve these technical problems is most transparent in light-cone ordered perturbation theory  \cite{Chang:1968bh,Kogut:1969xa,Brodsky:1973kb} (see Appendix \ref{app:LCOPT}), which is equivalent to performing all minus integrals of all loops, because it allows us to identify UV divergent loops in eikonal diagrams more easily.

\section{Summary of the Method in LCOPT}

The method can be summarized as follows, details will be given below:
\begin{enumerate}
\item We start with the expressions in LCOPT, as introduced in Section \ref{sec:Glauber} and Appendix \ref{app:LCOPT}, for the set of webs at a given order with a fixed coupling. The number of web-diagrams is much less than the number of all possible diagrams at a given order. Moreover, since we work in Feynman gauge, the number of possible webs is further reduced. For example, at order 2, as we will see below, only three diagrams contribute, aside from gluon self-energies.
\item Ultraviolet divergent internal $k_{\perp,\,i}$-integrals are regularized via dimensional regularization, with $\varepsilon > 0$. At this stage we do not yet encounter IR/collinear singularities since all integrals over transverse momenta are performed at fixed plus momenta.
\item We add the necessary QCD counterterms and the counterterms for the eikonal vertex which has to be renormalized as a composite operator. As we showed in Section \ref{wardidproof}, the sum of the latter cancels because of the recursive definition of webs and a Ward identity. However, in the intermediate stages the vertex counterterms are necessary to make individual diagrams UV finite.
\item After elimination of the UV singularities we dimensionally continue to $\varepsilon < 0$ to regulate the IR/collinear plus-integrals. It follows from the rules for LCOPT that all internal plus momenta are bounded by the total $k^+$ flowing into the minus eikonal. Therefore, the integrals over these internal plus momenta give no UV subdivergences.
\item When we sum over the set of diagrams at a given order the IR divergent parts cancel, as well as the UV counterterms for the vertex, thus also the internally UV divergent vertex parts  cancel.
\item The final scaleless $k_\perp$ integral provides the UV counterterm contributing to the anomalous dimension (see Eqs. (\ref{methodeq})-(\ref{aendeq}) ).
\end{enumerate}

We start by writing down all light-cone ordered diagrams of a given covariant Feynman diagram. All momenta in crossed gluon ladders have to be chosen independent of each other, such that they all flow through the eikonal vertex, since we seek the anomalous dimension for this vertex. Because $\xi$ has no minus-component (cf. Eq. (\ref{betaframe})), we have $q^- = 0$ in Eq. (\ref{denom}) of the appendix when applying the Feynman rules for LCOPT in our case. This can be depicted graphically by contracting all propagators on the minus-eikonal (here $\beta$) to a point, which coincides with the eikonal vertex. Two-loop examples can be found in Fig. \ref{2loopfig}. Sometimes, numerators stemming from triple-gluon vertices or quark propagators cancel the corresponding propagators on the plus-eikonal ($\xi \cdot k = k^-$). Graphically, this can again be described by contracting these propagators to a point. Then we can read off easily from the various light-cone ordered diagrams the analytical expressions, whose $k_{\perp,\,i}$-integrals we perform, and renormalize.

We need QCD counterterms and counterterms for the effective vertex. More specifically, by QCD counterterms we mean the usual gluon self-energy counterterms, as well as the counterterms for triple-gluon vertices and eikonal-gluon-eikonal vertices. The latter are UV divergent in any covariant diagram, however, this is not necessarily the case for all LCOPT diagrams found from a covariant diagram. Examples will be given below. Self-energies of the light-like eikonal lines vanish in Feynman gauge. Both types of counterterms are found via the (recursive) BPHZ-formalism \cite{Zimmermann:1969jj,Bogoliubov:gp,Hepp:1966eg}, and the subdivergences are identified with the help of naive power-counting on a graph-by-graph basis.

\subsection{Remark About Eikonal Integrals}

The usual methods for treating loop-integrals (see, for example, Appendix \ref{app2} and \cite{Tejeda-Yeomans:2002eh}) aim at reducing combinations of propagators to complete squares. This is not possible in the eikonal approximation where the propagator-denominators have been linearized to give Fig. \ref{Frules}.

 The linear occurrence of loop-momenta in eikonal propagators suggests a different strategy. It is advantageous to perform the first integrations by contour methods. In the case of two light-like eikonal lines, in a frame where these momenta  are back-to-back, contour integration over either minus- or plus-momenta in all internal eikonal propagators simplifies the expressions significantly. As mentioned before, the integration over all minus-momenta is equivalent to the expressions obtained via LCOPT (see Appendix \ref{app:LCOPT}),  up to partial fraction manipulations. The $n-2$ dimensional perpendicular integrals can then be performed via Feynman parametrization, Eq. (\ref{feynpar}). This results in integrals from 0 to 1 over Feynman parameters and ratios of plus-momenta, which can be expressed in closed form in terms of Beta- and generalized hypergeometric functions. For more than two eikonal lines, or when the eikonals are not light-like,  it is  best to perform all internal  energy ($k^0$) integrals via contour integration, and then proceed as above. 

The discussion in the previous paragraph applies only to loops with eikonal propagators. Loops that involve no eikonal propagators and thus contain no linearized momenta, are best treated with the methods described in the previous two sections. This also avoids potential difficulties with so-called ``z-graphs'', where fermions flow backwards \cite{Brodsky:1973kb}. For example, the integrals over all loop momenta in the fermion bubbles and triangles in Table \ref{graphtab} are performed first by usual $n$-dimensional integration, then the results are inserted into the remainder of the graph which is then treated via contour integration.

We will now illustrate our method by the rederivation of the 1- and 2-loop $A$-coefficients.

\section{Calculation of the 1- and 2-loop Coefficients $A^{(1)}_f$, $A^{(2)}_f$} \label{sectexample}

The well-known \cite{Gross:ju,Gross:cs,Floratos:1977au,Gonzalez-Arroyo:df,Curci:1980uw,Floratos:1981hs,Kodaira:1981nh} coefficients of the collinear parts of the splitting functions to one and two-loop order are given by
\begin{eqnarray}
A_a^{(1)} & = & C_a,  \label{a1number} \\
A_a^{(2)} & = & \frac{1}{2} C_a K \equiv \frac{1}{2} C_a \left[ C_A \left( \frac{67}{18} - \frac{\pi^2}{6} \right) - \frac{10}{9} T_F N_f \right], \label{a2number}
\end{eqnarray}
where $C_q = C_F, \, C_g = C_A$, $N_f$ is the number of fermions, and $T_F$ determines the normalization of the generators of the fundamental fermion representation, $T_F = \frac{1}{2}$. We will now apply our method to the recalculation of these coefficients.

The only web at order 1 in Feynman gauge is a single gluon exchanged between the two eikonal lines. The color weight is  $\overline{C}^{(1)} = C_a$ by definition (\ref{colwe}), and the web has no internal momenta. Straightforwardly we obtain
\be
2 \int  \frac{d^n k}{(2 \pi)^n} W_{aa}^{(1)}  =  \overline{C}^{(1)}  \left( \frac{\alpha_s(\mu^2)}{\pi} \right) \left( \frac{\mu^2}{m^2} \right)^\varepsilon (4 \pi)^\varepsilon\; \Gamma(\varepsilon)  \int\limits_0^\infty \frac{d k^+}{k^+}.  \label{1loopcalc}
\ee
So at lowest order we get from Eq. (\ref{aendeq})  $A_a^{(1)} = \overline{C}^{(1)} = C_a$, as in (\ref{a1number}).

At order 2 we have the webs shown in Fig. \ref{2loops}, where we rotated the eikonal lines in the figure compared to the diagrams shown in Section \ref{examplesubsect}, to make the connection to Figs. \ref{partondef} a) and \ref{partonfactend} more evident. The original color factors are (compare to Fig. \ref{order2})
\ba
C(W_b) & = & \left(C_F - \frac{C_A}{2} \right) C_a, \nonumber \\
C(W_c) & = & C(W_d) = - \frac{C_A}{2} C_a \label{2loopcolwe}
\ea
for eikonal lines in the $a$-representation. The respective color weights of the webs c) and d) are the same as the original color factors, since they do not have decompositions (these diagrams are ``maximally nonabelian''). The decomposition of diagram b) was shown as an example in Section \ref{examplesubsect}, which resulted in a color weight
\be
\bar{C}(W_b) =  - \frac{C_A}{2} C_a. \label{colweb}
\ee

The contribution of Fig. \ref{2loops} a) is easily found from Eq. (\ref{1loopcalc}) and the well-known finite terms (see e.\,g. \cite{book,Sterman:1995fz}) after renormalization of the of the gluon-self energy in the $\MS$ scheme, which is an example of what we called a QCD renormalization in the previous subsection:
\be
A_a^{(2),\,a)} = \frac{29}{36} C_A C_a + \frac{1}{18} C_A C_a - \frac{5}{9} T_F N_f C_a = \left(\frac{31}{36} C_A -
\frac{5}{9} T_F N_f \right) C_a. \label{2loopgluon}
\ee
 The first term in the first equality in Eq. (\ref{2loopgluon}) stems from the gluon loop, the second term from the ghost loop. The last term is obviously the fermion loop contribution found from the expression for the graph,
\ba
 2 \int  \frac{d^n k}{(2 \pi)^n} W_{aa,\,a)N_f}^{(2)} &  = & - T_F N_f C_a  \left( \frac{\alpha_s(\mu^2)}{\pi} \right)^2  \left( \frac{\mu^2}{m^2} \right)^{2 \varepsilon} (4 \pi)^{2\varepsilon} \frac{\Gamma(2 \varepsilon)}{\varepsilon} \nonumber \\
& & \qquad \times 2 B(2-\varepsilon,2-\varepsilon) \int\limits_0^\infty \frac{d k^+}{k^+}, \label{bubblenf}
\ea
and its counterterm.

\begin{figure}[hbt]
\vspace*{5mm}
\begin{center}
\epsfig{file=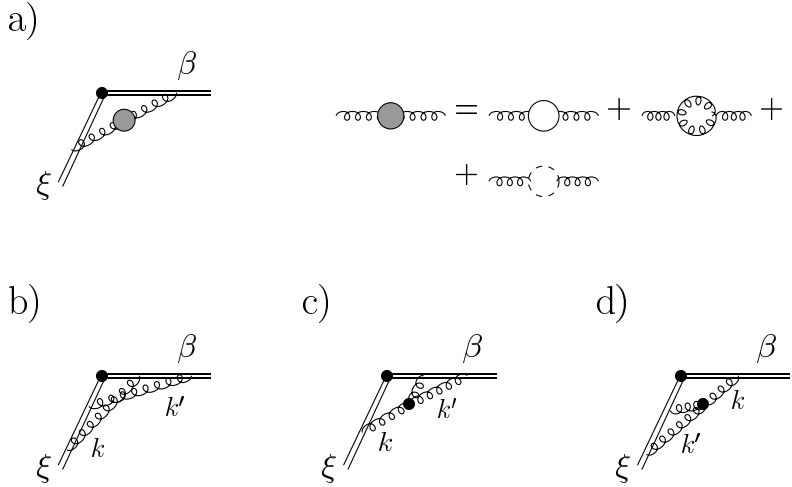,height=6.8cm,clip=0} \vspace*{6mm}
\caption{Webs contributing to $A^{(2)}_f$ (compare to Fig. \ref{expgraph}): a) web of order 1 with 1-loop gluon self-energy inserted, b) the ``crossed ladder", c) and d) graphs with a triple gluon vertex.} \label{2loops}
\end{center}
\end{figure}

The LCOPT diagrams obtained from the webs \ref{2loops} b)-d) are shown in Fig. \ref{2loopfig}. We see that due to the numerator $(2 k'^- - k^-)$ in the triple-gluon vertex, web d) contains two orderings on the light-cone; the factors of $2$ and $(-1)$ next to the eikonal vertices in the figure come from this numerator. Furthermore,  for web b) it is important to route the momenta in the crossed ladder independently of each other, such that both of them flow through the vertex, to separate the subdivergence associated with the upper loop ($k'$) from the overall UV divergence.

\begin{figure}
\vspace*{7mm}
\begin{center}
\epsfig{file=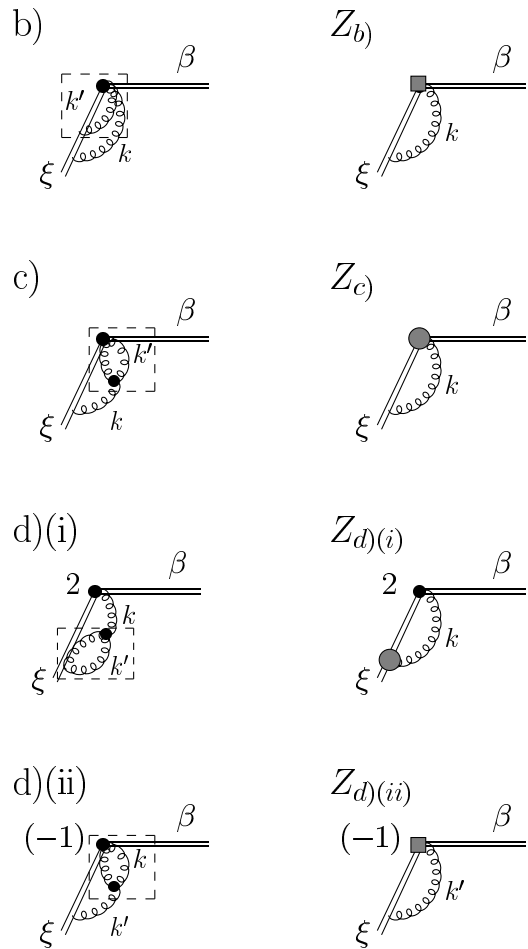,height=13cm,clip=0}
\vspace*{7mm}
\caption[LCOPT diagrams obtained from Fig. \ref{2loops} b)-d).]{LCOPT diagrams obtained from Fig. \ref{2loops} b)-d). The subgraphs in the dashed boxes are UV divergent 1-loop subgraphs, whose counterterms are shown in the second column. The grey boxes denote the eikonal vertex counterterms, whereas the grey blobs are the triple-gluon-vertex counterterms, shown in Fig. \ref{3gcounter}. } \label{2loopfig}
\end{center}
\end{figure}

Now we determine the divergent 1-loop subgraphs for each web by naively counting the powers of transverse momentum components in numerators and denominators. The UV divergent subgraphs are marked with boxes in Fig. \ref{2loopfig}. We see that for web c) and the first term of web d) we need a QCD counterterm for the triple-gluon vertex, whereas for the webs a) and d)(ii) we require vertex counterterms, as shown in the second column of Fig. \ref{2loopfig}. Web d)(ii) is an example for a LCOPT graph with a triple-gluon vertex which does not need QCD renormalization, in contrast to loop-corrections to 3-gluon-vertices in every covariant diagram. Due to the factor of (-1) in web d)(ii) the two vertex counterterms cancel each other, as announced above. The QCD counterterm, as shown in Fig. \ref{3gcounter}, is in the $\MS$ scheme for quark eikonal lines $\beta$ given by
\be
Z^{a\,\mu}_{\mbox{\tiny 3-g},\,ij} = - \frac{C_A}{2} T^a_{ij} \frac{\alpha_s}{\pi} \, g \, \beta^{\mu}  \frac{1}{2} \left( \frac{1}{\varepsilon} - \ln \frac{e^{\gamma_E}}{4 \pi}  \right), \label{3gcount}
\ee
where $g$ is the QCD coupling, $\alpha_s = g^2/(4\pi)$, and $\varepsilon > 0$.

The next step, after adding the appropriate counterterms to the respective graphs, is to perform the plus-momentum integrals. To do so, we dimensionally continue to $\varepsilon < 0$, that is, to $n > 4$ dimensions. The results for the webs b)-d) and the counterterms for UV subdivergences, denoted by $Z$ (omitting the vertex counterterms which cancel each other) are:
\ba
2 \int  \frac{d^n k}{(2 \pi)^n} W_{aa,\, b)}^{(2)} & = & - \overline{C}^{(2)} \left( \frac{\alpha_s(\mu^2)}{\pi} \right)^2 \left(\frac{\mu^2}{m^2}\right)^{2 \varepsilon} (4 \pi)^{2 \varepsilon}  \, \nonumber \\
& & \quad \times \,  \frac{1}{2}\, \frac{\Gamma(2 \varepsilon)}{\varepsilon} B(1+\varepsilon,-\varepsilon) \int\limits_0^\infty \frac{d k^+}{k^+},  \label{twoloopa} \\
2 \int  \frac{d^n k}{(2 \pi)^n} W_{aa,\, c)}^{(2)} & = & 2 \int  \frac{d^n k}{(2 \pi)^n} W_{aa,\, d)}^{(2)} = \overline{C}^{(2)} \left( \frac{\alpha_s(\mu^2)}{\pi} \right)^2  \left(\frac{\mu^2}{m^2}\right)^{2 \varepsilon} (4 \pi)^{2 \varepsilon}  \, \nonumber \\
& & \quad  \times \, \frac{1}{4} \frac{\Gamma(2 \varepsilon)}{\varepsilon} \, \left\{ B(1-\varepsilon,-\varepsilon) - 2 B(1 - \varepsilon,1-\varepsilon)\right\} \int\limits_0^\infty \frac{d k^+}{k^+}, \nonumber \\
& & \\
2 \int  \frac{d^n k}{(2 \pi)^n} Z_{c)} & = & 2 \int  \frac{d^n k}{(2 \pi)^n} Z_{d)(i)} = \overline{C}^{(2)} \left( \frac{\alpha_s(\mu^2)}{\pi} \right)^2 \left(\frac{\mu^2}{m^2}\right)^{\varepsilon} (4 \pi)^{\varepsilon}  \, \nonumber \\
& & \quad   \times \, \frac{1}{2}\,  \Gamma( \varepsilon) \left( \frac{1}{\varepsilon} - \ln \frac{e^{\gamma_E}}{4 \pi} \right) \int\limits_0^\infty \frac{d k^+}{k^+}. \nonumber \\
& & \label{twoloopz}
\ea
 The color weight, as stated in Eqs. (\ref{2loopcolwe}) and (\ref{colweb}), is $\overline{C}^{(2)} = - \frac{C_A}{2} C_a$ for all diagrams. We notice that diagram d) gives the same contribution as its upside-down counterpart c), as expected, but only after adding different types of counterterms.

\begin{figure}[hbt]
\begin{center}
\epsfig{file=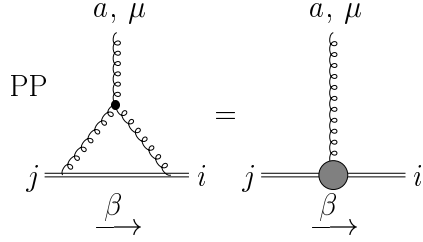,height=4.2cm,clip=0}
\caption{QCD counterterm for the triple-gluon vertex, where $PP$ denotes the pole part (omitting scheme-dependent constants).}  \label{3gcounter}
\end{center}
\end{figure}

After summing over the contribution of the webs b)-d) and the counterterms, we see that the  infrared poles $1/(-\varepsilon)$ in the Beta-functions cancel, as well as the vertex counterterms, leaving us with
\be
A_a^{(2),\,b)-d)} = \frac{C_A}{2} C_a \left( 2 - \frac{\pi^2}{6}\right) \label{remaincont}
\ee
according to Eq. (\ref{aendeq}).
The contributions of all diagrams, (\ref{2loopgluon}) and (\ref{remaincont}), result in the 2-loop coefficient (\ref{a2number}), as announced.

\newpage
\section{Higher Loops} \label{sect3loop}

\subsection{$N_f^{n-1}$-Terms in $A^{(n)}$}

It is relatively straightforward to obtain a general formula for the $N_f^{n-1}$-contribution to the $n$-loop coefficient $A^{(n)}$, since the only graphs involved are one-loop webs with $n-m-1$ fermion bubbles and $m$ counterterms for the fermion bubbles inserted into the gluon propagator. It is therefore a matter of simple combinatorics to obtain the $\alpha_s^n N_f^{n-1}$ contribution (compare to the one-loop expression Eq. (\ref{bubblenf}) ): 
\ba
2 \int \frac{d^{4-2 \varepsilon} k}{(2 \pi)^{4-2 \varepsilon}} W_{aa,\, N_f^{n-1}}^{(n)} & =  & 2\; C_a T_F^{n-1} N_f^{n-1}  \left( \frac{\alpha_s(\mu^2)}{\pi} \right)^n \int\limits_0^\infty \frac{d k^+}{k^+} \nonumber \\
& & \hspace*{-7mm} \times \, \sum\limits_{m=0}^{n-1} \left( \begin{array}{c} n-1 \\ m \end{array} \right) (-1)^{n-m-1}\left(\frac{\mu^2}{m^2}\right)^{(n-m)  \varepsilon}
   (4 \pi)^{(n-m) \varepsilon} \; \nonumber
\\ & \times & 2^{n-m-2} \frac{\Gamma\left( (n-m) \varepsilon \right)}{\Gamma\left( 1+(n-m-1) \varepsilon \right)}
 \nonumber \\
& \times & \left[ \Gamma(\varepsilon) B(2-\varepsilon,2-\varepsilon) \right]^{n-m-1} \left[ \frac{1}{3}
\left( \frac{1}{\varepsilon} - \ln \frac{e^{\gamma_E}}{4 \pi} \right) \right]^m . \nonumber \\
& &
\ea
The $\frac{1}{\varepsilon}$-pole in the expansion of the $\Gamma$- and Beta-functions in the sum is the contribution to the anomalous dimension (cf. Eq. (\ref{aendeq}) ). The contributions up to $\alpha_s^6$ are given in Table \ref{table}. They coincide with the corresponding values (the $\ln N$-terms, or equivalently, the $S_1(N)$-terms) calculated by Gracey in \cite{Gracey:nn}\footnote{Note the different overall normalization of the anomalous dimension there.}.

\begin{table}
\begin{center}
\begin{tabular}{|c|l|}
\hline
$n$ & $C_a (T_F N_f)^{n-1}$-term in $A_a^{(n)}$  \\ \hline
2 &  $-\frac{5}{9}$ \\
3 &  $-\frac{1}{27}$  \\
4 &  $-\frac{1}{81} + \frac{2}{27}  \zeta(3)$ \\
5 &  $-\frac{1}{243} - \frac{10}{243} \zeta(3) + \frac{\pi^4}{2430}$ \\
6 &  $-\frac{1}{729} - \frac{2}{729} \zeta(3) - \frac{\pi^4}{4374} + \frac{2}{81} \zeta(5)$  \\
\hline
\end{tabular}
\end{center}
\caption{$\alpha_s(\mu^2)^n N_f^{n-1} \left[ \frac{1}{1-x} \right]_+$-contributions to the anomalous dimension $P_{ff}$. The expansion of $A_f$ is performed in terms of $\alpha_s/\pi$ (cf. Eq. (\ref{aaa}) ).} \label{table}
\end{table}

\subsection{Towards the Three-Loop Coefficient $A^{(3)}_f$}

Vogt \cite{Vogt:2000ci,vanNeerven:2001pe,vanNeerven:2001tc} obtained a numerical parametrization of the $A^{(3)}$ from the known integer moments of the splitting function\footnote{Note the expansion in $\left(\frac{\alpha_s}{4 \pi}\right)$ there, whereas we expand in terms of $\left(\frac{\alpha_s}{\pi}\right)$ in Eq. (\ref{aaa}).}:

\be
A^{(3)}_f = \left[(13.81 \pm 0.14) - (4.31 \pm 0.02) T_F N_f - \frac{1}{27}T_F^2 N_f^2\right]\, C_F. \label{vogtresult}
\ee
We obtained the term proportional to $N_f^2$ in the previous subsection, as listed in Table \ref{table}. Now we will go on to compute the term proportional to $N_f$. All intermediate expressions are simple enough to be handled by the general algebraic computer program \textit{Mathematica} \cite{mathematica}. For the calculation of the full $A^{(3)}$ or even higher loops, however, an implementation of the algorithm into a more specialized computer algebra program such as FORM \cite{Vermaseren:2000nd} may be desirable.

The diagrams contributing to this term and their QCD counterterms are listed in Table \ref{graphtab}, labelled in analogy to the two-loop case. We only have to compute the contributions from the $g_{\mu \nu}$ part of the dressed gluon propagator, since the longitudinal parts $\sim k_\mu k_\nu$ cancel due to the Ward identity shown in Fig. \ref{subfact}. This cancellation has been verified explicitly.

The contributions
to the set a) are easily computed to be
\be
A_f^{(3),\,a)} = \frac{1}{18} C_A T_F N_f C_F. \label{A3a}
\ee
The contributions to the $N_f$-part of the two-loop gluon self-energy inserted into a one-loop web (set g) ) give:
\be
A_f^{(3),\,g)} = - \left[ C_A \left( \frac{509}{864} + \frac{1}{2} \zeta(3) \right) - C_F \left( - \frac{55}{48}+ \zeta(3) \right)\right] T_F N_f C_F. \label{A3g}
\ee
To compute the two-loop gluon self-energy, the occurring tensor integrals have been reduced to simple scalar one- and two-loop master integrals using the relations discussed in Appendix \ref{app:reduce}.
We checked our calculations of the set g) against previous computations of the two-loop gluon self-energy in Feynman gauge, see for example \cite{Braaten:1981dv,Davydychev:1997vh}. Note that this contribution has a term $\sim C_F^2$, which is not ``maximally non-abelian''. The results of \cite{Braaten:1981dv,Davydychev:1997vh} include the scalar polarized terms of the gluon propagator, which is dressed with a fermion bubble. Since these terms in the two-loop gluon self-energy, as stated above, cancel against the scalar polarized parts in the remaining webs, Eq. (\ref{A3g}) does not contain these contributions.

The expressions for the two-loop webs with a one-loop bubble-insertion are found easily from the corresponding two-loop expressions Eqs. (\ref{twoloopa})-(\ref{twoloopz}), taking into account the proper multiples of $\varepsilon$ in the Gamma- and Beta-functions due to the bubbles.
The calculation of the triangles e) and f) is a bit more nontrivial. The resulting contributions can be found in the table. The results for e) and f) have been expanded in terms of $\varepsilon$ and Beta-functions using various identities tabulated in \cite{polylog,Devoto:1983tc,gradshteyn}.

Since the infrared structure of the graphs is modified by the bubbles, which effectively raise the powers of the corresponding gluon propagators by $\varepsilon$ to a non-integer value, the upside-down counterparts do not give the same contributions. This asymmetry is not surprising, since we compute the coefficients collinear to the plus eikonal, thus introducing an asymmetry in how we treat the eikonal lines and the gluons attaching to them. However, we find that the sum of graphs in set d) gives the same contribution as the sum of graphs in set c), as can be seen from the tabulated expressions.

The individual diagrams b)-f) have at most three UV (QCD) divergences and one IR/collinear divergence, in addition to the overall scaleless $k^+$-integral.
We observe that the diagrams
with a one-loop counterterm for the fermion bubble and the one-loop counterterms for the triangle graphs have the same IR structure as the two-loop webs. Thus their IR divergences cancel separately from the rest of the diagrams. This implies that the collinear divergences have to cancel within the set of remaining diagrams, that is, within the set of webs with bubbles and the triangles. Moreover, we observe that the infrared divergences cancel within certain subsets of these graphs. Namely, they cancel separately between graphs b)(1), c)(1), and d)(1), between graphs b)(2), c)(2), and d)(2), as well as between c)(3), d)(3), e) and f).

Summing over all contributions from graphs b)-f) we arrive at
\be
A_f^{(3),\,b)-f)} = - \left(\frac{125}{288} - \frac{5 \pi^2}{54} + \frac{2 \zeta(3)}{3} \right) T_F N_f C_A C_F. \label{A3b}
\ee

We performed several checks of our computations. The infrared structure described above is one check of the results listed in Table \ref{graphtab}. Another check is the cancellation of non-local logarithms $\sim \log M$. Furthermore, the values of the $1/\varepsilon^3$- and $1/\varepsilon^2$-poles can be predicted from the one- and two-loop calculations performed in Section \ref{sectexample} \cite{'tHooft:1973mm}. The sum of all diagrams contributing at $\alpha_s^3 N_f$ has the following structure:
\ba
2 \int \frac{d^n k}{(2\pi)^n} W^{(3)}_{aa,\,N_f} & = & \left\{ - \frac{11}{54} C_A C_F T_F \frac{1}{\varepsilon^3} + \right. \label{poles} \\
& & +\, \left.
 \left[ \left( \frac{167}{324} -\frac{\pi^2}{108} \right) C_A + \frac{1}{12} C_F
  \right] C_F T_F  \frac{1}{\varepsilon^2} \right. 
\nonumber \\
& & + \, \left. \frac{1}{3}  A_{N_f}^{(3)} \frac{1}{\varepsilon}   \right\}  N_f \left( \frac{\alpha_s}{\pi} \right)^3 \int\limits_0^\infty \frac{d k^+}{k^+}. \nonumber
\ea
 The predictions of the higher poles in Eq. (\ref{poles})  coincide with the poles obtained from the expansion of the calculated expressions listed in the table.

Adding (\ref{A3a}), (\ref{A3g}), and (\ref{A3b}) we obtain the term proportional to $N_f$ contributing to the three-loop coefficient $A^{(3)}$:
\ba
A^{(3)}_{N_f} & = &  - \left[ C_A \left( \frac{209}{216} - \frac{5 \pi^2}{54} + \frac{7 \zeta(3)}{6} \right)  - C_F \left(- \frac{55}{48}+ \zeta(3) \right)\right] T_F N_f C_F \nonumber \\
& = & -4.293\, T_F N_f C_F,
\ea
which agrees with the numerical prediction in Eq. (\ref{vogtresult}). The same result was obtained in Ref. \cite{Moch:2002sn}, which was published simultaneously to our result \cite{Berger:2002sv}.

\begin{table}[hbt]
 Table \ref{graphtab}:
\begin{center}
\begin{tabular}{|l|c|l|}
\hline
\hspace*{1cm} Web & Factor  & \hspace*{1cm} Contribution \\ \hline
 & & \\
\raisebox{3mm}{a)} \raisebox{-3mm}{ \epsfig{file=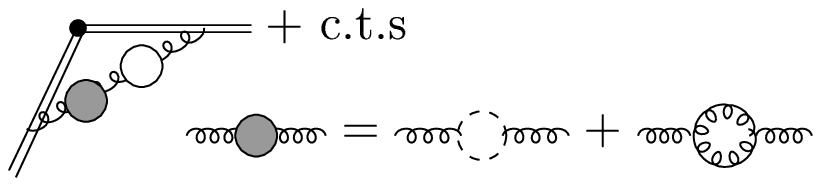,height=1cm,clip=0}}   & 2  & see Eq. (\ref{A3a}) \\
 & & \\ \hline\hline
 & & \\
 \raisebox{3mm}{b)(1)}\hspace*{4mm} \raisebox{-3mm}{ \epsfig{file=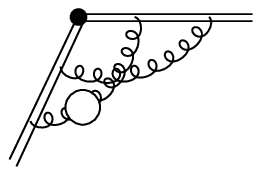,height=1cm,clip=0}} & 2 & $- K M^{3 \varepsilon} \frac{\Gamma(3 \varepsilon)}{2 \varepsilon^2} B(2-\varepsilon,2-\varepsilon) B(1+\varepsilon,-\varepsilon)$ \\
 & & \\
 \raisebox{3mm}{b)(2)}\hspace*{4mm} \raisebox{-3mm}{ \epsfig{file=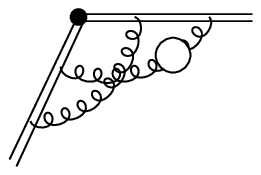,height=1cm,clip=0}} & 2 & $- K M^{3 \varepsilon} \frac{\Gamma(3 \varepsilon)}{4 \varepsilon^2} B(2-\varepsilon,2-\varepsilon)$ \\
& & $\qquad \times \, B(1+2\varepsilon,-2\varepsilon)$ \\
 & & \\ \hline
 & & \\
 \raisebox{3mm}{b)(C1)}\hspace*{1.8mm} \raisebox{-3mm}{ \epsfig{file=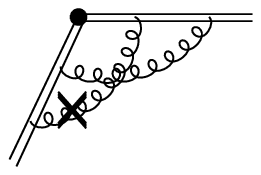,height=1cm,clip=0}} & 4 & $+ K M^{2 \varepsilon} \frac{\Gamma(2 \varepsilon)}{12 \varepsilon} N_{\varepsilon} B(1+\varepsilon,-\varepsilon)$ \\
 & & \\
\hline\hline
 & &\\
 \raisebox{3mm}{c)(1)}\hspace*{4mm} \raisebox{-3mm}{ \epsfig{file=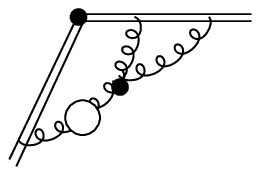,height=1cm,clip=0}} & 2 & $+ K M^{3 \varepsilon} \frac{1}{4} \frac{\Gamma(3 \varepsilon)}{e^2} \frac{\left(\Gamma(1+\varepsilon)\right)^2}{\Gamma(1+2 \varepsilon)} B(2-\varepsilon,2-\varepsilon)$ \\
 & & $\quad \times \left\{ B(1-\varepsilon,-\varepsilon) - 2 B(1 - \varepsilon,1-\varepsilon)\right\}$ \\
 & & \\
  \raisebox{3mm}{c)(2)}\hspace*{4mm} \raisebox{-3mm}{ \epsfig{file=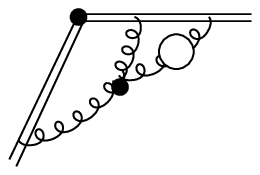,height=1cm,clip=0}} & 2 & $+ K M^{3 \varepsilon} \frac{1}{4} \frac{\Gamma(3 \varepsilon)}{2 \varepsilon^2} B(2-\varepsilon,2-\varepsilon)$ \\
 & & $\quad \times \left\{ B(1-\varepsilon,-2 \varepsilon) - 2 B(1 - \varepsilon,1-2\varepsilon)\right\}$ \\
 & & \\
 \raisebox{3mm}{c)(3)}\hspace*{1.8mm} \raisebox{-3mm}{ \epsfig{file=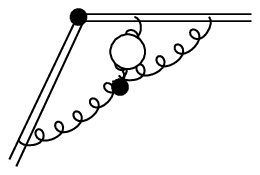,height=1cm,clip=0}} & 2 & $+ K M^{3 \varepsilon} \frac{1}{4} \frac{\Gamma(3 \varepsilon)}{2 \varepsilon^2} B(2-\varepsilon,2-\varepsilon)$ \\
 & & $\quad \times \left\{ B(1-2\varepsilon,- \varepsilon) - 2 B(1 - \varepsilon,1-2\varepsilon)\right\}$ \\
\hline
\end{tabular}
\end{center}
\end{table}

\begin{table}
Continuation of Table \ref{graphtab}:
\begin{center}
\begin{tabular}{|l|c|l|} \hline
\hspace*{1cm} Web & Factor  & Contribution  \\ \hline
 & &\\
 \raisebox{3mm}{d)(1)}\hspace*{4mm} \raisebox{-3mm}{ \epsfig{file=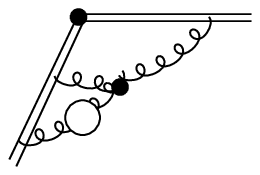,height=1cm,clip=0}} & 2 & $+ K M^{3 \varepsilon} \frac{1}{4} \frac{\Gamma(3 \varepsilon)}{\varepsilon^2}  B(2-\varepsilon,2-\varepsilon)$ \\
 & & $\quad \times \left\{ \frac{\left(\Gamma(1+\varepsilon)\right)^2}{\Gamma(1+2 \varepsilon)} B(1-\varepsilon,-\varepsilon) - B(1 - 2 \varepsilon,1-\varepsilon)\right\}$ \\
 & & \\
  \raisebox{3mm}{d)(2)}\hspace*{4mm} \raisebox{-3mm}{ \epsfig{file=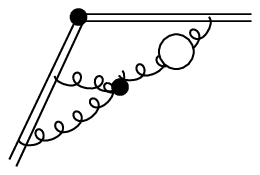,height=1cm,clip=0}} & 2 & $+ K M^{3 \varepsilon} \frac{1}{4} \frac{\Gamma(3 \varepsilon)}{2 \varepsilon^2} B(2-\varepsilon,2-\varepsilon)$ \\
 & & $\quad \times \left\{ B(1-\varepsilon,-2 \varepsilon)-4 \frac{\left(\Gamma(1+\varepsilon)\right)^2}{\Gamma(1+2 \varepsilon)} B(1 - \varepsilon,1-2\varepsilon)\right\}$ \\
 & & \\
 \raisebox{3mm}{d)(3)}\hspace*{1.8mm} \raisebox{-3mm}{ \epsfig{file=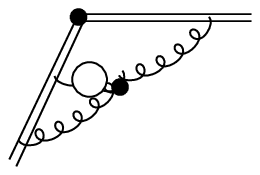,height=1cm,clip=0}} & 2 & $+ K M^{3 \varepsilon} \frac{1}{4} \frac{\Gamma(3 \varepsilon)}{2 \varepsilon^2} B(2-\varepsilon,2-\varepsilon)$ \\
 & & $\quad \times \left\{ B(1-2\varepsilon,- \varepsilon) - 2 B(1 - \varepsilon,1-2\varepsilon)\right\}$ \\
\hline
 &  &\\
\raisebox{3mm}{c)d)(C1)}\raisebox{-3mm}{ \epsfig{file=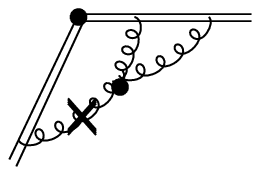,height=1cm,clip=0}} & 12 & $- K M^{2 \varepsilon} \frac{1}{24} \frac{\Gamma(2 \varepsilon)}{ \varepsilon}  N_{\varepsilon} \left\{ B(1-\varepsilon,-\varepsilon) - 2 B(1 - \varepsilon,1-\varepsilon)\right\}$ \\
 & & \\
\raisebox{3mm}{c)d)(C2)}\raisebox{-3mm}{ \epsfig{file=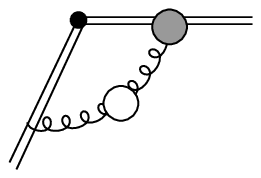,height=1cm,clip=0}} & 4 & $+K M^{2 \varepsilon} \frac{\Gamma(2 \varepsilon)}{2 \varepsilon} N_{\varepsilon} B(2-\varepsilon,2-\varepsilon) $ \\
 & & \\
\hline
 & & \\
\raisebox{3mm}{c)d)(C3)}\raisebox{-3mm}{ \epsfig{file=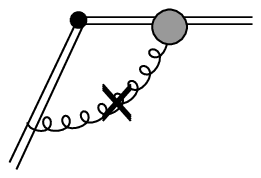,height=1cm,clip=0}} & 4 & $-K M^{ \varepsilon} \frac{\Gamma(\varepsilon)}{12} N_{\varepsilon}^2$ \\
 & & \\
\raisebox{3mm}{c)d)(C4)}\raisebox{-3mm}{ \epsfig{file=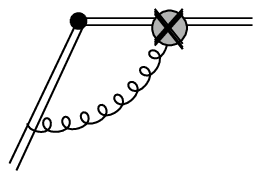,height=1cm,clip=0}} & 8 & $-K M^\varepsilon \frac{\Gamma(\varepsilon)}{2} \left[ \frac{1}{12}  N_{\varepsilon}^2 - \frac{1}{18} N_{\varepsilon} \right] $ \\
 & & \\
\hline\hline
 & & \\
\raisebox{3mm}{e)} \hspace*{4mm} \raisebox{-3mm}{ \epsfig{file=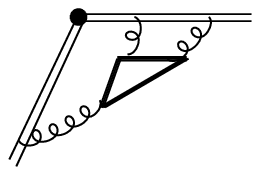,height=1cm,clip=0}} & 2 & $- K M^{3 \varepsilon} \frac{\Gamma(3 \varepsilon)}{8 \varepsilon^2 } B(2-\varepsilon,2-\varepsilon) \left\{ B(3-2\varepsilon,-\varepsilon) \right. $ \\
 & & $\quad \left. - B(1-\varepsilon,2-2\varepsilon) + \frac{2 \pi^2}{3} \varepsilon + \left(4 \zeta(3)-2\right) \varepsilon^2  \right\} $ \\
\hline
 & & \\
\raisebox{3mm}{f)} \hspace*{4mm} \raisebox{-3mm}{ \epsfig{file=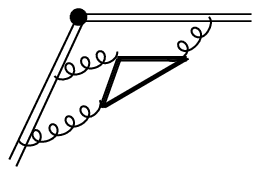,height=1cm,clip=0}} & 2 & $- K M^{3 \varepsilon} \frac{\Gamma(3 \varepsilon)}{8 \varepsilon^2} B(2-\varepsilon,2-\varepsilon)\left\{ B(3-2\varepsilon,-\varepsilon)\right. $ \\
 & & $\quad \left. - B(1-\varepsilon,2-2\varepsilon) - \frac{\pi^2}{3} \varepsilon + \left(10 \zeta(3) -2 \right) \varepsilon^2 \right\}$ \\
\hline
\end{tabular}
\end{center}
\end{table}

\begin{table}
Continuation of Table \ref{graphtab}:
\begin{center}
\begin{tabular}{|l|c|l|} \hline
\hspace*{1cm} Web & Factor  & Contribution  \\ \hline
 & & \\
\raisebox{3mm}{e)f)(C1)}\raisebox{-3mm}{ \epsfig{file=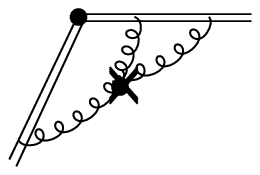,height=1cm,clip=0}} & 4 & $+ K M^{2 \varepsilon} \frac{1}{24} \frac{\Gamma(2 \varepsilon)}{ \varepsilon} N_{\varepsilon} \left\{ B(1-\varepsilon,-\varepsilon) \right.$ \\
& & $\qquad -\left. 2 B(1 - \varepsilon,1-\varepsilon)\right\} $ \\
 & & \\ \hline
 & & \\
\raisebox{3mm}{e)f)(C2)}\raisebox{-3mm}{ \epsfig{file=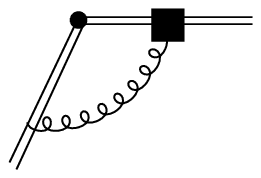,height=1cm,clip=0}} & 4 & $+ K M^\varepsilon \frac{\Gamma(\varepsilon)}{2} \frac{1}{12} \left( N_{\varepsilon}^2 - \frac{5}{12}  N_{\varepsilon} \right)$  \\
 & & \\ \hline \hline
& & \\
\raisebox{3mm}{g) }\raisebox{-3mm}{ \epsfig{file=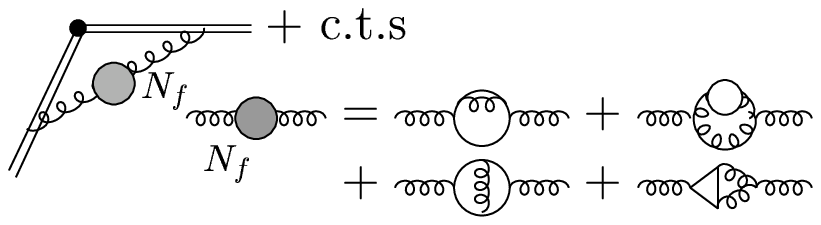,height=1.3cm,clip=0}} & 2 & see Eq. (\ref{A3g}) \hspace*{2cm} \\
 & & \\
\hline\hline
\end{tabular}
\end{center}
We introduced the following abbreviations:
\ba
K & \equiv & \frac{C_A}{2} T_F N_f C_F \left( \frac{\alpha_s}{\pi}\right)^3 \int\limits_0^\infty \frac{dk^+}{k^+}, \\
M & \equiv & \left(\frac{\mu^2}{m^2} \right) \left(4 \pi \right), \\
N_{\varepsilon} & \equiv & \frac{1}{\varepsilon} - \ln \frac{e^{\gamma_E}}{4 \pi} .
\ea
\caption[Webs contributing to the $N_f$-term of the three-loop coefficient $A^{(3)}$ and their counterterms (c.t.s).]{Webs contributing to the $N_f$-term of the three-loop coefficient $A^{(3)}$ and their counterterms (c.t.s), labelled (C). The cross denotes the counterterm for the fermion bubble. Similarly, the cross in the triple-gluon vertex denotes the counterterm for the fermion triangle. The grey blob represents the counterterm Fig. \ref{3gcounter}, the grey blob with a cross is the 2-loop counterterm for the triple-gluon vertex with a fermion bubble inside. And finally, the black box denotes the fermion part of the 2-loop counterterm for the triple-gluon vertex.
 We omitted vertex counterterms which cancel. We refrain from drawing all counterterms which give the same contribution. Instead, we indicate multiple contributions in the column ``factor''. A factor of 2 is due to the two complex conjugate contributions, and has already been taken into account in Eqs. (\ref{A3a}) and (\ref{A3g}). } \label{graphtab}
\end{table}

\clearpage

\section{Summary and Outlook}

We have developed and proved a method for the calculation of the coefficients proportional to $\left[\frac{1}{1-x}\right]_+$ of the non-singlet parton splitting functions, whose knowledge, for example, is important for NNLL resummations. The method is based on the factorization properties of the splitting functions, and on the exponentiation of eikonal cross sections.

We have illustrated the method with the rederivation of the 1- and 2-loop coefficients $A^{(1)}$ and $A^{(2)}$, as well as the $N_f^{n-1}$ terms at order $n$. We have presented the result for the term proportional to $N_f$ at three loops, which coincides with the approximate result obtained by Vogt \cite{Vogt:2000ci}. The full splitting functions at three loops are currently being computed by Moch, Vermaseren, and Vogt \cite{Vermaseren:2002rn} with the help of the operator product expansion.  Their results for the $N_f$-term \cite{Moch:2002sn} provide a further check of our calculations.

Although a calculation via the OPE provides the complete $N$, or equivalently, $x$-dependence of the splitting functions, it involves a large number of diagrams and complex expressions at higher orders. A computation at three loops is a formidable task, and it seems unlikely that higher loop calculations will be completed in the near future. However, for certain observables, large logarithms due to soft and/or collinear radiation become numerically important, and need to be resummed to as high a level in logarithms as possible. Our method, although limited to only the computation of $A$, has the advantage that higher-order computations are much less complex than within conventional methods, because the number of graphs is greatly reduced, and the expressions involved are relatively simple in LCOPT. Moreover, a fully computerized implementation of the algorithm should be straightforward. Therefore, the computation of the coefficients $A$ at four or even higher loops may be within reach.

Let us now return to the discussion of event shapes, where the resummation of large logarithmic enhancements has lead to the same anomalous dimension discussed above. This anomalous dimension controls the double logarithmic behavior, due to coherent radiation. Soft radiation, on the other hand, is emitted incoherently at wide angles. In Sec. \ref{sec:resum} we have reviewed the tools to resum coherent and incoherent radiation to all orders. We will now go on to apply this formalism to a generalized class of event shapes.

\chapter{Dijet Event Shapes}
\label{ch5}

The agreement of
theoretical predictions  with experiment for jet cross sections is often
impressive.
This is especially so for inclusive jet cross sections at
high $p_T$, using fixed-order factorized perturbation theory
and parton distribution functions \cite{Bethke:1998ja,Seidel:2002cd,Gonzalez:2002gc}.  A
good deal is also known about the substructure of jets,
through the theoretical and  experimental study
of multiplicity distributions and fragmentation functions \cite{Dokshitzer:nm},
and of event shapes \cite{Farhi:1977sg,Georgi:1977sf,Basham:1978bw,DeRujula:1978yh,Fox:1978vu,Parisi:1978eg,Donoghue:1979vi,Ellis:nc,Sterman:1979uw}.   We have discussed the example of the thrust in Chapter \ref{ch1}.
Event shape
distributions \cite{Catani:1992jc,Catani:1992ua,Catani:1991kz,Dokshitzer:1998kz} in
particular offer a bridge between the perturbative, short-distance and
the nonperturbative, long-distance dynamics  of QCD \cite{Korchemsky:1994is,Korchemsky:1999kt,Dokshitzer:1997ew,Gardi:2002bg,Korchemsky:1995zm,Belitsky:2001ij}, as we have seen in Section \ref{sec:webpower} for the thrust. 

In the following we will introduce a general class of inclusive event shapes in $e^+e^-$ dijet events. This chapter is based on our publications \cite{Berger:2002ig,Berger:2003iw}, and includes some as yet unpublished material. After introducing the general event shape we factorize the corresponding cross section in order to resum large logarithmic corrections. In Sec. \ref{sec:resevent} we give explicit analytical results at next-to-next-to-leading logarithmic order. We conclude this chapter by deriving the form of power corrections to the generalized event shape in Sec. \ref{sec:powerevent}. 

\section{A Generalized Event Shape} \label{sec:eventshape}

Schematically we consider the following cross section where the final state radiation into all of phase space is weighted with weight functions $\bar f$,
\be
e^+ + e^- \rightarrow J_1(p_{J_1}, \bar{f}_{\bar{\O}_1}) + J_2(p_{J_2},
\bar{f}_{\bar{\O}_2})\, . \label{crossdef1}
\ee
$J_1$ and $J_2$ are two jets with momenta $p_{J_c}, c = 1, 2$ at center of mass (c.m.) energy $Q = \sqrt{s} \gg \LQCD$.

To impose the two-jet condition on the states of Eq.\ (\ref{crossdef1}) we
choose weights that suppress states with substantial radiation into
$\bar{\O}$
away from the  jet axes.
We now introduce a class of event shapes $\bar{f}$, related
to the thrust, that
enforce the two-jet condition in a natural way.

These event shapes interpolate between and extend the familiar
thrust \cite{Farhi:1977sg} and jet broadening \cite{Catani:1992jc,Dokshitzer:1998kz},
through an adjustable parameter $a$.
For each state $N$ that defines the process (\ref{crossdef1}),
we separate $\bar{\O}$ into two regions, $\bar{\O}_c$, $c=1,2$,
containing jet axes, $\hat{n}_c(N)$.  To be specific, we
let $\bar\O_1$ and $\bar\O_2$ be two hemispheres that
cover the entire space.  Region $\bar\O_1$ is centered on $\hat n_1$,
and $\bar\O_2$ is the opposite hemisphere.
We will specify the
method that determines the jet axes $\hat n_1$ and
$\hat n_2$ momentarily. To identify a meaningful jet, of course, the 
total energy
within $\bar\O_1$ should be a large fraction of the available energy, 
of the order of $Q/2$
in dijet events.
In $\rm e^+e^-$ annihilation, if there is a well-collimated
jet in $\bar\O_1$ with nearly half the
total energy, there will automatically be one in $\bar{\O}_2$.

We are now ready to define
the contribution from particles in region $\bar\O_c$ to the $a$-dependent
event shape,
\be
\bar f_{\bar{\O}_c}(N,a) =
\frac{1}{\sqrt{s}}\
\sum_{\hat n_i\in \bar \O_c}\ k_{i,\,\perp}^a\, \o_i^{1-a}\,
\left(1-\hat n_i\cdot \hat n_c\right)^{1-a} \, ,
\label{barfdef}
\ee
where $a$ is any real number less than two. The
sum is over those particles of state $N$  with direction $\hat n_i$ 
that flow into
$\bar\O_c$, and their transverse momenta $k_{i,\perp}$ are measured relative to $\hat{n}_c$.
The jet axis $\hat n_1$ for jet 1 is identified as that axis
that minimizes the specific thrust-related quantity $\bar f_{\bar{\O}_1}(N,a=0)$.
When $\bar{\O}_c$ in Eq.\ (\ref{barfdef}) is extended to all of phase space,
the case $a=0$ is then essentially $1-T$, with $T$ the thrust, while 
$a=1$ is related to the
jet  broadening.

Any choice $a<2$ in (\ref{barfdef}) specifies an infrared safe event shape
variable, because the
contribution of any particle $i$ to the event shape behaves as
$\theta_i^{2-a}$ in the collinear limit, $\theta_i=\cos^{-1} (\hat n_i 
\cdot \hat
n_c ) \rightarrow 0$.  Negative values of $a$ are clearly permissible, and
the limit $a\rightarrow -\infty$
corresponds to the total cross section.
At the other limit, the factorization and resummation
techniques that we discuss below will apply
only to $a<1$. For
$a> 1$, contributions to the event shape (\ref{barfdef}) from energetic
particles near the jet axis are generically larger than
contributions from soft, wide-angle radiation, or equal for
$a=1$.  When this is the case, the
analysis that we present below must be modified, at
least beyond the level of leading logarithm \cite{Dokshitzer:1998kz}.

In summary, once $\hat n_1$ is fixed, we have divided the phase space into
two regions: \begin{samepage}
\begin{itemize}
\item Region $\bar \O_1$, the entire hemisphere centered on
$\hat n_1$, that is, around jet 1.
\item Region $\bar \O_2$, the complementary hemisphere.
\end{itemize} \end{samepage}
In these terms, we define
the complete event shape variable $\bar f(N,a)$ by
\ba
\bar f(N,a) &=& \bar f_{\bar{\O}_1}(N,a)+\bar f_{\bar{\O}_2}(N,a)\, , \label{2jetf}
\ea
with ${\bar f}_{\bar{\O}_c}$, $c=1,2$ given by (\ref{barfdef}) in terms of
the axes $\hat n_1$ of jet 1 and $\hat n_2$ of jet 2.

In Eq.\ (\ref{barfdef}), $a$ is a parameter that allows us to study
various event
shapes within the same formalism; it helps to control the
approach to the two-jet limit.   As noted above,
$a< 2$ for infrared safety, although the factorization
that we will discuss below applies beyond leading logarithm
only to $1>a>-\infty$.  A
similar weight function with a non-integer power has been discussed in
a related context for $2>a>1$ in
\cite{Manohar:1994kq}.

To see how the parameter $a$ affects the shape of the jets, let us
reexpress
the weight function for jet 1 as
\ba
\bar f_{\bar{\O}_1}(N,a) = \frac{1}{\sqrt{s}}\
\sum_{\hat n_i\in \bar \O_1} \o_i \sin^a \theta_i \left( 1 -
\cos \theta_i  \right)^{1-a}, \label{fbarexp}
\ea
where $\theta_i$ is the angle of the momentum of final state
particle $i$ with respect to jet axis $\hat n_1$.
As $a \rightarrow 2$ the weight vanishes only  very slowly for
$\theta_i\rightarrow 0$, and at fixed $\bar f_{\bar{\O}_1}$, the
jet becomes very narrow. On the other hand, as $a \rightarrow -
\infty$, the event
shape vanishes more and more rapidly in the forward direction, and the
cross
section at fixed $\bar f_{\bar{\O}_1}$ becomes more
and more inclusive in the radiation into $\bar{\O}_1$.
The effect of $a$ on the shape
of the radiation is illustrated in Fig.
\ref{shapefig}. In Fig. \ref{shapefig} we compare  the phase
space available to a particle at fixed $\bar f$ for
three different values of $a$. The radial magnitude of each plot
is the maximum energy $\o$ found from Eq. (\ref{fbarexp}):
$r = \bar{f} Q \sin^{-a} \theta (1-\cos \theta)^{a-1}$.
For $a = 1$, as shown in Fig. \ref{shapefig} a), the particle is
restricted to be close to the jet axes, while for $a = 0$,
and $a = -1$, depicted in Figs. \ref{shapefig} b) and c),
respectively, the particle is allowed to be farther away
from the axes.

\begin{figure}[htb]
\begin{center}
a) \hspace*{4.1cm} b) \hspace*{4.1cm} c) \hspace*{3.6cm} \vspace*{-6mm}
\\
\epsfig{file=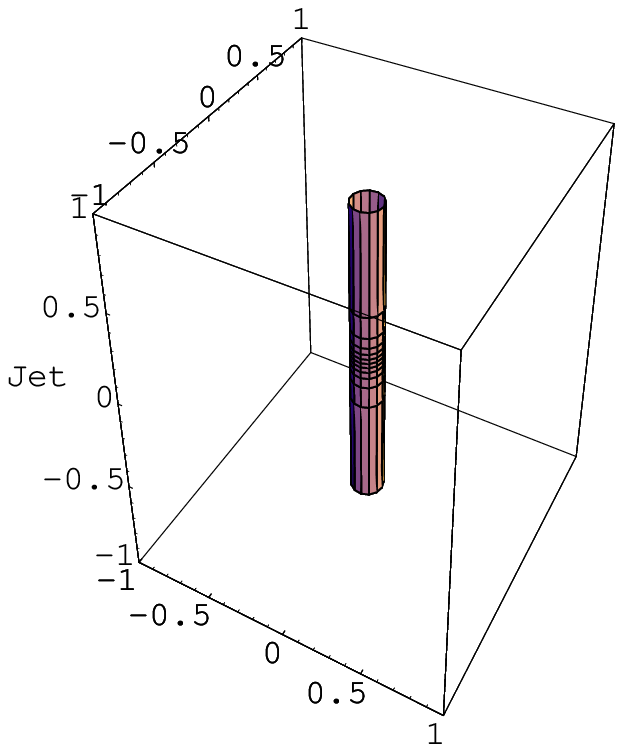,width=3.7cm,clip=0}  \mbox{ }\hspace*{7mm}
\epsfig{file=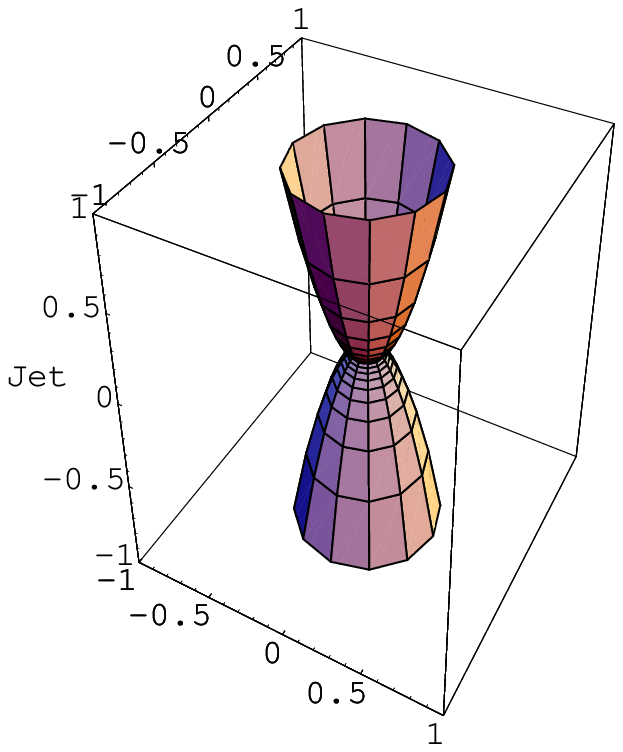,width=3.7cm,clip=0} \mbox{ }\hspace*{7mm}
\epsfig{file=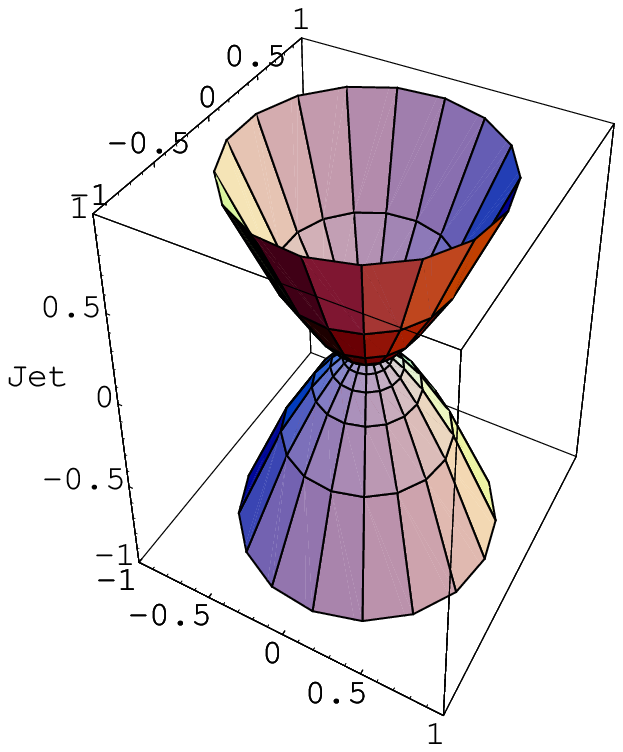,width=3.7cm,clip=0} 
\caption[Illustration of the effect of the parameter $a$ in the
weight (\ref{fbarexp}) on the shape of the event: a) shape for $a
= 1$, b) shape for $a = 0$, c) shape for $a = -1$.]{Illustration of the effect of the parameter $a$ in the
weight (\ref{fbarexp}) on the shape of the event: a) shape for $a
= 1$, b) shape for $a = 0$, c) shape for $a = -1$. The jet axes
are in the vertical direction. The radial normalization
($\bar{f} Q$) is arbitrary, but the same for all three
plots.} \label{shapefig}
\end{center}
\end{figure}

As an aside, the generalized event shape (\ref{barfdef}) goes to zero for all values of $a$ in the limit of two back-to-back jets. The three-jet limit for the thrust $T$ is $2/3$, as is well-known. On the other hand, the three-jet limit for the generalized shape is given by
\be
f^{(\mbox{\tiny 3 jets})}(a) = \frac{1}{3} \sqrt{3}^a, \label{threelim}
\ee
which reduces to $1/3$ for $f(a = 0) = 1 - T$, and to $1/\sqrt{3}$ for the jet broadening, $f(a=1)$. We can also compute the limit for infinitely many homogeneously distributed final-state partons for the generalized event shape (\ref{barfdef}). We find 
\be
f^{(\mbox{\tiny $\infty$ jets})}(a) = \frac{1}{4} \left[2(1+a) - a (2-a) \left(\Psi\left(\frac{3}{2}-\frac{a}{4}\right)-\Psi\left(1-\frac{a}{4}\right)\right) \right] \label{shapelimeq}
,
\ee
 where $\Psi(z) = \Gamma'(z)/\Gamma(z)$ is the digamma function, that is, the logarithmic derivative of the gamma function. This reduces to $1/2$ for the thrust-related shape, and to $\pi/4$ for the jet broadening when $a = 1$. Eq. (\ref{shapelimeq}) is illustrated in Fig. \ref{shapelim}.

\begin{figure}[hbt]
\begin{center}
\epsfig{file=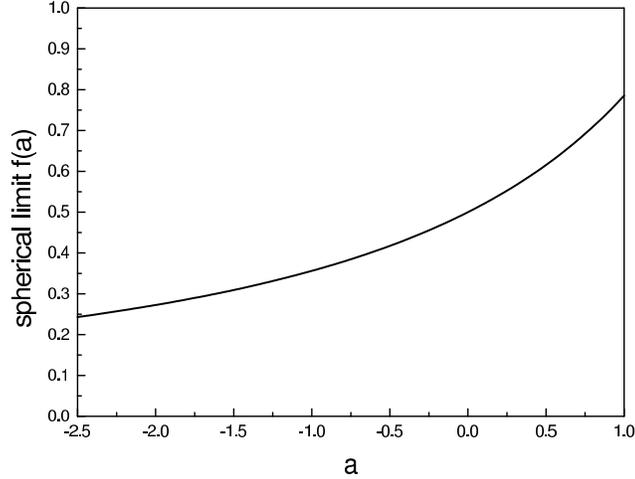,width=7cm,angle=270,clip=0} 
\caption{Spherical limit of the shape (\ref{barfdef}) as a function of $a$, Eq. (\ref{shapelimeq}).} \label{shapelim}
\end{center}
\end{figure}

The  differential cross section
for such dijet events at fixed values of $\bar f$ is now
\ba
{d \bar{\sigma}^{\rm incl}(\bar{\varepsilon},s, a)\over 
d\bar{\varepsilon}\, d\hat n_1}
&=&
{1\over 2s}\ \sum_N\;
|M(N)|^2\, (2\pi)^4\, \delta^4(p_I-p_N) \nonumber\\
&\ & \hspace{10mm} \times \, \delta(\bar{\varepsilon} -\bar f(N,a))\;
\delta^2  (\hat n_1 -\hat n(N))\, ,
\label{eventdef}
\ea
where we sum over all final states $N$ that contribute to the
weighted event, and where $M(N)$ denotes
the corresponding amplitude for ${\rm e^+e^-}\rightarrow N$.
The total momentum is $p_I$, with $p_I^2=s\equiv Q^2$.

 Since we are investigating  two-jet cross sections, we fix the
constant $\bar{\varepsilon}$ to be
 much less than unity:
\be
0 < \bar{\varepsilon} \ll 1.
\label{elasticlim}
\ee
We refer to this as  the elastic limit for the two jets.
In the elastic limit, the dependence of the directions of the
jet axes on soft radiation is weak.  We will return to
this dependence below.
Independent of soft radiation, we can
always choose our coordinate system such
that the
transverse momentum of jet 1 is
zero,
\be
p_{J_1,\, \perp} = 0\, ,
\ee
with $\vec p_{J_1}$ in the $x_3$ direction.  In the limit $\bar
\varepsilon \rightarrow 0$, and in the overall c.m.,
$p_{J_1}$ and $p_{J_2}$ then approach light-like vectors in the plus and
minus directions:
\ba
p_{J_1}^\mu &\rightarrow&  \left(\sqrt{\frac{s}{2} },0^-,0_\perp \right)
\nonumber\\
p_{J_2}^\mu &\rightarrow&  \left(0^+,\sqrt{\frac{s}{2} },0_\perp \right)\, .
\label{lightlike}
\ea
As usual, it is convenient to work in light-cone coordinates (\ref{lccoord}).
For small $\bar{\varepsilon}$, the cross section
(\ref{eventdef}) has
corrections in 
$\ln (1/\bar{\varepsilon})$, analogous to the thrust as discussed in Sec. \ref{sec:resum}, which we will organize in the following.

\section{Factorization of the Cross Section}

\subsection{Leading Regions Near the Two-Jet Limit}

As for the thrust, we identify ``leading regions" in
the momentum integrals of cut diagrams, which can give rise
to logarithmic enhancements of the  
cross section associated with lines approaching the
mass shell.  Within these regions, the lines of a cut diagram
fall into the following subdiagrams:
\begin{itemize}
\item A hard-scattering, or ``short-distance" subdiagram $H$, where all
components of line momenta are far off-shell, by order $Q$.
\item Jet subdiagrams, $J_1$ and $J_2$, where energies are
fixed and momenta are collinear
to the outgoing primary partons and the jet
directions that emerge from the hard scattering. (For
$\bar{\varepsilon}=0$,
the sum of all energies in each jet is one-half the total energy.)
\item A soft subdiagram, $S$ connecting the jet functions $J_1$ and
$J_2$, in which the components of
  momenta $k$ are  small compared to $Q$ in all components.
\end{itemize}

An arbitrary final state $N$ is the
union of substates associated with these subdiagrams:
\be
N=N_s \oplus N_{J_1} \oplus N_{J_2}\, .
\ee
As a result, the  event shape $\bar f$ can
also be written as a sum of contributions from the soft
and jet subdiagrams:
\ba
\bar f(N,a) &=& \bar f^N(N_s,a) + \bar f^N_{\bar{\O}_1}(N_{J_1},a) +
\bar f^N_{\bar{\O}_1}(N_{J_2},a)\, .
\label{fbarf}
\ea
The superscript $N$ reminds us that the contributions of
final-state particles associated with
the soft and jet functions depend implicitly on the
full final state, through the determination of the
jet axes, as discussed in the previous section.

When we sum over all diagrams
that have a fixed final state, the contributions from these leading regions
may be factorized into a set of functions, each of which corresponds to
one of the generic hard, soft and jet subdiagrams, as discussed in Chapter \ref{ch1}.   
The cross section becomes a convolution in
$\bar \varepsilon$, with the sums over states linked
by the delta function which fixes $\hat n_1$, and by momentum
conservation, \newpage
\ba
{d \bar{\sigma}^{\rm incl}( \bar{\varepsilon},s,a)\over d\bar{\varepsilon}\, d\hat n_1}
&=& {d \sigma_0 \over d\hat{n}_1}\
H(s,\hat{n}_1)\
\sum_{N_s,N_{J_c}}\;
\int d\bar{\varepsilon}_s\, {\cal S}(N_s)\, 
\delta(\bar{\varepsilon}_s-\bar{f}^N(N_s,a))\nonumber\\
&\ & \quad \times \prod_{c=1}^2\,
  \int  d\bar{\varepsilon}_{J_c}\, {\cal J}_c (N_{J_c}) \,
\delta(\bar{\varepsilon}_{J_c}-\bar{f}^N_{\bar{\O}_c}(N_{J_c},a))\nonumber\\
&\ & \hspace{15mm} \times\
(2\pi)^4\, \delta^4(p_I-p(N_{J_2})-p(N_{J_1})-p(N_s))
\nonumber\\
&\ & \hspace{15mm} \times\
\delta^2(\hat n_1 -\hat n(N))\;
\delta(\bar{\varepsilon}-\bar{\varepsilon}_{J_1}-\bar{\varepsilon}_{J_2}-\bar{
\varepsilon}_s)
\nonumber\\
&=& {d \sigma_0 \over d\hat{n}_1}\; 
\delta(\bar{\varepsilon})+{\cal O}(\alpha_s)\, .
\label{sigmafact}
\ea
Here  $d\sigma_0/d\hat{n}_1$ is the Born
cross section for the production of a single
particle (quark or antiquark) in direction
$\hat{n}_1$, while the short-distance function
$H(s,\hat{n}_1)=1+{\cal O}(\alpha_s)$, which
   describes corrections to the hard scattering,
is an expansion in $\alpha_s$ with finite coefficients.
The functions ${\cal J}_c (N_{J_c}),\ {\cal S}(N_s)$
describe
the internal dynamics of the jets and wide-angle soft
radiation, respectively.  We will specify these functions below.
We have suppressed their dependence on a factorization scale.

So far, we have specified our sums over states in Eq.\ (\ref{sigmafact})
only when all lines in $N_s$ are
soft, and all lines in $N_{J_c}$ have momenta that are collinear, or
nearly collinear
to $p_{J_c}$.   As $\bar{\varepsilon}$ vanishes, these are the
only final-state momenta that are kinematically possible.
Were we to restrict ourselves to these configurations
only, however, it would not be straightforward to make the individual
sums over $N_s$ and $N_{J_c}$ infrared safe.  Thus, it is necessary to
include soft partons in $N_s$ that are emitted near the jet directions,
and soft partons in the $N_{J_c}$ at wide angles.
We will show below how to define the functions ${\cal J}_c (N_{J_c}),\ {\cal
S}(N_s)$
so that they generate factoring, infrared safe functions that
avoid double counting.
We know on the basis of the arguments in Chapter \ref{ch1}
that corrections to
the factorization of soft from jet functions are suppressed by
powers of the weight function  $\bar \varepsilon$.

\subsection{The Factorization in Convolution Form} \label{sec:approxincl}

Although formally factorized, the jet and soft functions
in Eq.\ (\ref{sigmafact}) are still  linked in a potentially complicated
way through their dependence on the jet
axes.  Our strategy is to simplify this complex dependence
to a simple convolution in contributions to $\bar\varepsilon$,
accurate to leading power in  $\bar\varepsilon$.

First, we note that the cross section of Eq.\ (\ref{sigmafact})
is singular for vanishing $\bar{\varepsilon}$, but is a
smooth function of $s$ and $\hat{n}_1$.  We may therefore make any
approximation that changes $s$ and/or $\hat{n}_1$ by an amount
that vanishes as a power of  $\bar{\varepsilon}$ in the
leading regions.

Correspondingly, the amplitudes for jet $c$ are singular in
$\bar{\varepsilon}_{J_c}$,
but depend smoothly on the jet energy and direction, while
the soft function is singular in 
$\bar{\varepsilon}_s$,
but depends smoothly on the jet directions.  As a result,
at a fixed value of $\bar\varepsilon$ we
may approximate the jet directions and energies by
their values at $\bar{\varepsilon}=0$ in the soft and jet
functions.

Finally, we may make any approximation that affects
the value of  $\bar\varepsilon_{J_c}$ by
amounts that vanish faster than linearly for $\bar\varepsilon\rightarrow 0$.
It is at this stage that we will require that $a<1$, which we will justify in the next subsection. 

Furthermore, we assume that $a$ is not large in absolute value. The event shape at fixed angle decreases exponentially with $a$, and higher-order corrections can be proportional to $a$. We therefore require that $|\ln \bar{\varepsilon}\,| \gg |a|$.

With these observations in mind,
we enumerate the replacements
and approximations by which we
reduce Eq.\ (\ref{sigmafact}), while retaining leading-power accuracy.

\begin{enumerate}

\item  To simplify the definitions  of the jets in Eq.\ (\ref{sigmafact}),
we make the replacements $\bar{f}^N_{\bar{\O}_c}(N_{J_c},a) 
\rightarrow \bar f_c(N_{J_c},a)$ with
\ba
\bar f_c(N_{J_c},a) \equiv
\frac{1}{\sqrt{s}} \sum_{{\rm all}\
\hat n_i
\in N_{J_c} }\
k_{i,\,\perp}^a\, \o_i^{1-a}\, \left(1-\hat n_i\cdot \hat n_c
\right)^{1-a}\, .
\label{fbar2jet1}
\ea
The jet weight function $\bar{f}_c(N_{J_c},a)$ now depends only on particles
associated with $N_{J_c}$.
  The contribution to $\bar{f}_c(N_{J_c},a)$
from particles within region $\bar{\O}_c$,
is exactly the same here as in the weight (\ref{barfdef}), 
but we now include particles
in all other directions.
    In this way, the independent sums over final states of the
jet amplitudes will be naturally infrared
safe.  The value of $\bar f_c(N_{J_c},a)$ differs from
the value of $\bar{f}^N_{\bar{\O}_c}(N_{J_c},a)$, however, due to 
radiation outside
$\bar\O_c$, as indicated by the new subscript.  This radiation is hence at
wide angles to the jet axis.  In the elastic limit (\ref{elasticlim}), it is
also constrained to be soft.  Double counting in contributions
to the total event shape, $\bar f(N,a)$, will be avoided by an appropriate
definition
of the soft function below.
The sums over states are still not yet fully independent,
however, because the jet directions $\hat n_c$ still depend
on the full final state $N$.

\item
Next, we turn our attention to the condition that fixes the jet
direction $\hat n_1$.
Up to corrections in the orientation of $\hat n_1$
that vanish as powers of $\bar{\varepsilon}$,
we may neglect the dependence of $\hat{n}_1$
on $N_s$ and $N_{J_2}$:
\be
\delta(\hat n_1-\hat n(N))  \rightarrow  \delta(\hat n_1 -\hat n(N_{J_1}))\, .
\label{nnJone}
\ee
In Section \ref{sec:approxapp}, we show that
this replacement also leaves the value of
$\bar\varepsilon$ unchanged, up to corrections that vanish as
$\bar\varepsilon^{2-a}$.  Thus, for $a<1$, (\ref{nnJone}) is
acceptable to leading power.
For $a<1$, we can therefore
identify the direction of jet 1 with $\hat{n}_1$.
These approximations simplify Eq.\ (\ref{sigmafact})
by eliminating the implicit dependence of
the jet and soft weights on the full final state.  We may
now treat $\hat n_1$ as an independent vector.

\item  In the leading regions, particles
that make up each final-state jet are associated with states $N_{J_c}$,
while $N_s$  consists of soft particles only.
In the momentum conservation delta function, we
can neglect the four-momenta of lines in $N_s$,
whose energies all vanish as $\bar{\varepsilon}\rightarrow 0$:
\be
\delta^4(p_I-p(N_{J_2})-p(N_{J_1})-p(N_s))
\rightarrow
\delta^4(p_I-p_{J_2}-p_{J_1}).
\ee

\item
Because the cross section is a smooth function of
the jet energies and directions, we may also
neglect the masses of the jets within the
momentum conservation delta function, as in
Eq.\ (\ref{lightlike}).  In this approximation,
we derive in the c.m.,
\ba
\delta^4(p_I-p_{J_2}-p_{J_1})
&\rightarrow&
\delta(\sqrt{s} - \o(N_{J_1})-\o(N_{J_2}))
\, \delta(|\vec p_{J_1}|-|\vec p_{J_2}|)
\, \nonumber \\
& & \qquad \times\, {1\over |\vec p_{J_1}|^2}
\, \delta^2(\hat n_1 + \hat n_2)\nonumber\\
&\rightarrow& {2\over s}\,
\delta\left({\sqrt{s}\over 2} - \o(N_{J_1})\right)
\, \delta\left({\sqrt{s}\over 2} - \o(N_{J_2})\right)
\, \nonumber \\
& & \qquad \times \, \delta^2(\hat n_1 + \hat n_2)\, .
\ea
Our jets are now back-to-back:
\be
\hat n_2 \rightarrow -\hat n_1\, .
\label{fbar2soft}
\ee

\end{enumerate}

Implementing these replacements and approximations for $a<1$,
we rewrite the cross section Eq.\
(\ref{sigmafact}) as
\ba
{d \bar{\sigma}^{\rm incl}(\bar{\varepsilon},s,a)\over 
d\bar{\varepsilon}\, d\hat n_1}
&=&
{d \sigma_0 \over d\hat{n}_1}\
H(s,\hat{n}_1,\mu)\;
\int  d\bar{\varepsilon}_s\,
\bar{S}(\bar{\varepsilon}_s,a,\mu) \,
\nonumber\\
&\ & \times
\prod_{c=1}^2\, \int  d\bar{\varepsilon}_{J_c}\,
\bar{J}_c(\bar{\varepsilon}_{J_c},a,\mu)\,
\delta(\bar{\varepsilon}- \bar{\varepsilon}_{J_1}-\bar{\varepsilon}_{J_2}-
\bar{\varepsilon}_s)\, ,
\label{factoreps}
\ea
with (as above) $H=1+{\cal O}(\alpha_s)$.  Referring to
the notation of Eqs.\ (\ref{sigmafact}) and (\ref{fbar2jet1}),
the functions $\bar{S}$ and $\bar{J}_c$ are:
\ba
\bar{S}(\bar{\varepsilon}_s,a,\mu)
&=&
\sum_{N_s}\; {\cal S}(N_s,\mu)\,
\delta(\bar{\varepsilon}_s-\bar f(N_s,a))
\label{firstSdef}
\\
\bar{J}_c(\bar{\varepsilon}_{J_c},a,\mu)
&=&
\frac{2}{s}  (2\pi)^6\,   \sum_{N_{J_c}}
{\cal J}_c(N_{J_c},\mu) \, \delta(\bar{\varepsilon}_{J_c}-\bar f_c(N_{J_c},a))\, \nonumber \\
& & \qquad \times \,
\delta\left({\sqrt{s}\over 2} - \o(N_{J_c})\right)\,
\delta^2(\hat n_1 \pm \hat n(N_{J_c})),
\nonumber\\
\label{firstJdef}
\ea
with the plus sign in the angular delta function
for jet 2, and the minus for jet 1. The weight functions for the jets are
given by Eq.\
(\ref{fbar2jet1}) and induce dependence on
the parameter $a$.   We have introduced the factorization scale
$\mu$, which we set equal to the
renormalization scale. The factorized cross section (\ref{factoreps}) is of the same form as the thrust cross section, Eq. (\ref{thrustfact}), only the weights differ, with $\bar{\varepsilon}(a = 0) \equiv \tau$. Eq. (\ref{factoreps}) is therefore also illustrated by Fig. \ref{factorized}.  

We note that we must construct the soft functions $\bar{S}(N_s,\mu)$
to cancel the contributions of final-state particles from
each of the $\bar{J}_c(N_{J_c},\mu)$ to $\bar\varepsilon$
from soft radiation outside their respective regions $\bar\O_c$.
Similarly, the jet amplitudes
must be constructed to include collinear enhancements only in their
respective jet directions.  Explicit constructions that satisfy these
requirements will be
specified in the following subsections.

To disentangle the convolution in (\ref{factoreps}), we take Laplace
moments with respect to $\bar{\varepsilon}$, according to Eq. (\ref{thrustmoment}):
\ba
   \frac{d \sigma^{\rm incl} \left(\nu,s,a \right)}{ d \hat{n}_1 }
& = &  \int_0^\infty d\bar{\varepsilon}\, e^{- \nu \,\bar{\varepsilon}}\,
{d \bar{\sigma}(\bar{\varepsilon},a)\over 
d\bar{\varepsilon}\,
d\hat n_1}
\label{trafo} \nonumber
\\
& = & {d \sigma_0 \over d\hat{n}_1}\
H(s,\hat{n}_1,\mu)\;  S(\nu,a,\mu) \,\prod_{c=1}^2\,
J_c(\nu,a,\mu).
\label{trafosig}
\ea
Here and below  unbarred quantities are the transforms in
$\bar{\varepsilon}$,
and barred quantities denote untransformed functions.
\be
S(\nu,a,\mu) =   \int_0^\infty d\bar{\varepsilon}_s \,e^{- \nu
\,\bar{\varepsilon}_s}
\bar{S}(\bar{\varepsilon}_s,a,\mu),
\label{trafodef}
\ee
and similarly for the jet functions. 

Before giving explicit constructions for the hard, jet, and soft functions in Eq. (\ref{factoreps}), we justify the neglect of recoil effects above for $a < 1$.

\subsection{Recoil Effects} \label{sec:approxapp}

We return to the justification
of the technical step represented by Eq.\ (\ref{nnJone}).
According to this approximation, we may
compute the jet functions by identifying
axes that depend only upon particles in the
final states
$N_{J_c}$ associated with those functions, rather
than the full final state $N$.
Intuitively, this is a reasonable estimate, given
that the jet axis should be determined by
a set of energetic, nearly collinear particles.
When we make this replacement, however,  the contributions
to the event shape from energetic particles near the jet axis may
change.  This change is neglected in going from
the original factorization, Eq.\ (\ref{sigmafact}), to the
factorization in convolution form, Eq.\ (\ref{factoreps}),
which is the starting point for the resummation
techniques that we employ in this paper.
The weight functions $\bar f^N(N_i,a)$
in Eq.\ (\ref{sigmafact}) are defined
relative to the unit vector $\hat n_1$ corresponding to
$a=0$, the thrust-like event shape.
The factorization of Eq.\ (\ref{sigmafact})
applies to any $a<2$, but as indicated by the superscript,
individual contributions to $\bar f^N(N_i,a)$ on the
right-hand side continue to depend on the full
final state $N$, through the identification of the jet axis.

To derive the factorization
of Eq.\ (\ref{factoreps}) in a simple convolution
form, we must be able to
treat the thrust axis, $\hat n_1$, as a fixed vector
for each of the states $N_s$, $N_{J_c}$.  This is possible
if we can neglect the effects of recoil from soft,
wide-angle radiation on the direction of the axis.
Specifically, we must be able to make the replacement
\be
\bar f_{\bar{\O}_c}^N(N_{J_c},a) \rightarrow \bar f_c(N_{J_c},a)\, , \label{replace}
\ee
where $\bar f_c(N_{J_c},a)$ is the event shape variable
for jet $c$, in which the axis $\hat n_c$ is
specified by state $N_{J_c}$ {\it only}.  Of course, this
replacement changes the value of the weight, $\bar\varepsilon$,
$\bar f_{\bar{\O}_c}^N(N_{J_c},a) \ne \bar f_c(N_{J_c},a)$.
As we now show, the error induced by this
replacement is suppressed by a power
of $\bar \varepsilon$ so long as $a<1$.  In general,
the error is nonnegligible for $a\ge 1$.
The importance of recoil for jet broadening, at
$a=1$, was pointed out in \cite{Dokshitzer:1998kz}.  We now
discuss how the neglect of such radiation
affects the jet axis
(always determined from $a=0$)
and hence the value of the event shape for arbitrary $a<2$.

The jet axis is
found by minimizing $\bar f(a=0)$
in each state.
The largest influence on the axis ${\hat n}_c$ for jet $c$
is, of course, the set of fast, collinear particles
within the state $N_{J_c}$ associated with the jet function
in Eq.\ (\ref{sigmafact}).
Soft, wide-angle radiation, however,
does affect the precise direction
of the axis.  This is what we mean by `recoil'.

Let us denote by $\o_s$ the energy of the soft wide-angle radiation that is 
neglected in the factorization
(\ref{factoreps}).  Neglecting this soft radiation in
the determination of the jet axis
will result in an axis $\hat n_1(N_{J_c})$, which differs from the
axis $\hat n_1 (N)$
determined from the complete final state $(N)$ by an angle $\Delta_s\phi$:
\be
    \angle\!\!\!) \left(\hat n_1(N), \hat n_1(N_{J_c}) \right) \equiv 
\Delta_s \phi
\sim  {\o_s \over Q}\, .
\label{deltaphi}
\ee
At the same time, the soft, wide-angle radiation also contributes
to the total event shape
$\bar f(N,a) \sim (1/Q)k_\perp^a (k^-)^{1-a}$ at the
level of
\be
\bar\varepsilon_s \sim {\o_s\over Q} \, ,
\ee
because for such wide-angle radiation, we may take $k_s^-\sim k_{s,\perp}\sim\omega_s$.
In summary, the neglect of wide-angle soft radiation rotates the jet axis
by an angle that is of the order of the contribution
of the same soft radiation to the event shape.

In the factorization (\ref{factoreps}), the contribution of
each final-state particle is taken into account,
just as in Eq.\ (\ref{sigmafact}).  The question
we must answer is how the rotation of the jet axis affects
these contributions, and hence the value of the event
shape.

For a wide-angle particle, the rotation of the jet
axis by an angle of order $\Delta_s\phi$
in Eq.\ (\ref{deltaphi}) leads to a
negligible change in its contributions to the event shape, because
its angle to the axis is a number of order unity, and the
jet axis is rotated only
by an angle of order $\bar\varepsilon_s$.
Contributions from soft radiation are therefore stable
under the approximation (\ref{nnJone}).
The only source of large corrections is then associated
with energetic jet radiation,
because these particles are nearly collinear to the jet axis.

It is easy to see from the form
of the shape function in terms of angles, Eq.\ (\ref{fbarexp}),
that for any value of parameter $a$, a particle
of energy $\omega_i$ at a small angle $\theta_i$ to
the jet axis $\hat n_1 (N)$ contributes to the
event shape at the level
\be
\bar\varepsilon_i \sim  {\omega_i \over Q}\theta_i{}^{2-a}\, .
\ee
The rotation of the jet
axis by the angle $\Delta_s\phi$ due to neglect of soft radiation
may be as large as, or larger than,
$\theta_i$. Assuming the latter, we find a shift in the
$\bar\varepsilon_i$ of order
\be
\delta \bar\varepsilon_i \equiv \bar\varepsilon_i \left( \hat n_1(N) \right)
- \bar\varepsilon_i \left( \hat n_1(N_{J_c}) \right)
\sim {\omega_i\over Q}\,
\left(\Delta_s\phi\right)^{2-a} \sim
{\omega_i\over Q}\, \left({\o_s\over Q}\right)^{2-a} \sim
{\omega_i\over Q}\, \bar\varepsilon_s{}^{2-a}\, .
\ee
The change in $\bar\varepsilon_i$
  is thus suppressed by at least a factor $\bar\varepsilon_s{}^{1-a}$
compared to $\bar\varepsilon_s$, which is the
contribution of the wide-angle soft radiation
to the event shape.  The contributions of nearly-collinear,
energetic radiation to the event shape thus change
significantly under the replacement (\ref{nnJone}),
but so long as $a<1$, these contributions are
power-suppressed in the value
of the event shape, both before and after the approximation
that leads to a rotation of the axis.
For this reason, when $a<1$ (and only when $a<1$), the value
of the event shape is stable whether or not
we include soft radiation in the determination
of the jet axes, up to corrections that are suppressed
by a power of the event shape.  In this case, the
transition from Eq.\ (\ref{sigmafact}) to Eq.\ (\ref{factoreps})
is justified.

\subsection{The Short-Distance Function}
\label{sec:sdf}

As we have seen in Sec. \ref{sec:powerthrust},
in  Feynman  gauge
the subdiagrams of Fig.\ \ref{factorized} that contribute to
$H$ in Eq.\ (\ref{factoreps}) at leading
power in 
$\bar \varepsilon$ are connected to
each of the two jet subdiagrams
  by a single on-shell
quark  line,  along with a possible set of on-shell, collinear gluon lines
that carry scalar polarizations.
The hard subdiagram is
not connected directly to the soft subdiagram in any leading region.

The couplings of the scalar-polarized gluons that connect the jets with
short-distance subdiagrams
may be simplified with the help of the Ward
identities Fig. \ref{wardeik}.  At each order of
perturbation theory, the coupling of scalar-polarized gluons
from either jet to the short-distance function is equivalent
to their coupling to a path-ordered exponential of
the gauge field, oriented in any direction that is not
collinear to the jet.  Corrections are infrared safe, and
can be absorbed into the short-distance function.
Let $h(p_{J_c},\hat{n}_1,{\mathcal{A}})$ represent
the set of all short-distance contributions to diagrams
that couple any number of scalar-polarized gluons to the jets,
in the amplitude for the production of any final state.  The argument
  ${\mathcal{A}}$ stands for the fields that create the
scalar-polarized gluons linking the short-distance function
to the jets.
On a diagram-by-diagram basis, $h$ depends on
the momentum of each of the scalar-polarized gluons.
After the sum over all diagrams, however,
  we can make the replacement:
\be
h(p_{J_c},\hat{n}_1,{\mathcal{A}}^{({\rm q,\bar{q}})})
\rightarrow
\Phi^{({\rm \bar{q}})}_{\xi_2} (0,-\infty;0)\, h_2(p_{J_c},\hat{n}_1,\xi_c)\,
\Phi^{({\rm q})}_{\xi_1} (0,-\infty;0) \, ,
\ee
where $h_2$ is a short-distance function that depends only on the
total momenta $p_{J_1}$ and $p_{J_2}$.  It also depends on vectors $\xi_c$
that characterize  the path-ordered
exponentials $\Phi(0,-\infty;0)$:
\be
\Phi^{(\rm f)}_{\xi_c} (0,-\infty;0)  =  P e^{-i g_s \int_{-\infty}^{0} d
\lambda\; \xi_c \cdot {\mathcal{A}}^{(\rm f)} (\lambda \xi_c )}\, ,
\label{patho}
\ee
where the superscript $(\rm f)$ indicates that the vector potential
takes values in
representation $\rm f$, in our case the representation of a quark or
antiquark.
These operators will be associated with
gauge-invariant definitions of the jet functions below.
To avoid spurious collinear singularities,
we choose the vectors $\xi_c$, $c=1,2$, off the light cone.
  In the full cross section (\ref{trafosig}) the
$\xi_c$-dependence cancels, of course.

The dimensionless short-distance function $H=\left|h_2\right|^2$ in 
Eq.\ (\ref{factoreps})
depends on $\sqrt{s}$ and $p_{J_c}\cdot \xi_c$, but not
on any variable that vanishes with $\bar{\varepsilon}$:
\be
H(p_{J_c},\xi_c,\hat{n}_1,\mu) =  H \left(
\frac{\sqrt{s}}{\mu},\frac{p_{J_c} \cdot 
\xic}{\mu},\hat{n}_1,\as(\mu) \right)\, ,
\label{harddef}
\ee
where
\be
\xic \equiv \xi_c / \sqrt{|\xi_c^2|}\, .
\ee
Here we have observed that each diagram is independent of the overall
scale of the eikonal vector $\xi^\mu_c$.

\subsection{The Jet Functions}\label{sec:jets}

The jet functions and the soft functions in Eq.\ (\ref{factoreps})
can be defined in terms of specific matrix elements, which
absorb the relevant contributions to leading regions in
the cross section, and which are infrared safe.
Their perturbative expansions
  specify the functions ${\cal S}$ and ${\cal J}_c$ of
Eq.\ (\ref{firstJdef}).  We begin with
our definition of the jet functions.

The jet functions, which absorb enhancements collinear to the two
outgoing particles produced in the primary hard scattering, can be
defined in terms of matrix elements
in a manner reminiscent of parton distribution or decay functions \cite{Collins:1981uw}.
To be specific, we consider the
quark jet function:
\ba
\bar{J'}_c^\mu (\bar{\varepsilon}_{J_c},a,\mu)
&=&
  {2\over s}\, \frac{(2\pi)^6}{\Ncol} \; \sum\limits_{N_{J_c}}
{\rm Tr} \left[\gamma^\mu
\left<0 \left|
\Phi^{(\rm q)}_{\xi_c}{}^\dagger(0,-\infty;0) q(0) \right| N_{J_c} \right>
 \nonumber \right. \\ 
 & & \quad  \times\, \left. \left< N_{J_c}\left| \bar{q}(0) \Phi^{(\rm q)}_{\xi_c}(0,-\infty;0)
\right| 0 \right>\right] 
\label{jetdef} \\
& & \quad \times \, \delta(\bar{\varepsilon}_{J_c}-\bar
f_c(N_{J_c},a))\,
\delta\left({\sqrt{s}\over 2} - \o(N_{J_c})\right)\,
\delta^2(\hat n_c - \hat n(N_{J_c}))
\, , \nonumber 
\ea
where $\Ncol$ is the number of colors, and
where $\hat n_c$ denotes the direction of the momentum of
  jet $c$, Eq.\ (\ref{firstJdef}),
with $\hat{n}_2 = - \hat n_1$. 
   $q$ is the quark field, $\Phi_{\xi_c}^{(\rm q)}(0,-\infty;0)$
a path-ordered exponential in the notation of (\ref{patho}),
and the trace is taken over color and Dirac indices.
We have chosen the normalization so that the
jet functions $\bar{J}'{}^\mu$ in (\ref{jetdef}) are dimensionless
and begin at lowest order with
\ba
\bar{J'}_c^\mu{}^{(0)} (\bar{\varepsilon}_{J_c},a,\mu) = \beta_{c}^\mu\,
\delta({\bar \varepsilon}_{J_c})\, ,
\label{norm}
\ea
with $\beta_c^\mu$
the lightlike velocities corresponding to the jet momenta in Eq.\ (\ref{lightlike}):
\be
\beta_1^\mu=\delta_{\mu +}\ , \quad \beta_2^\mu = \delta_{\mu -}\, .
\label{betadef}
\ee
The scalar jet functions of Eq.\
(\ref{firstJdef}) are now obtained by projecting out
the component of $J'_c{}^\mu$ in the jet direction:
\be
\bar{J}_c (\bar{\varepsilon}_{J_c},a,\mu) =  \bar{\beta}_c \cdot
\bar{J'}_c
(\bar{\varepsilon}_{J_c},a,\mu) = \delta(\bar{\varepsilon}_{J_c}) +{\cal
O}(\alpha_s)\, ,
\label{Jnorm}
\ee
where
$\bar{\beta}_1=\beta_2$, $\bar{\beta_2}=\beta_1$ are the lightlike vectors
in the directions opposite to $\beta_1$ and $\beta_2$, respectively.
By construction, the
$\bar{J}_c$ are linear in $\bar{\beta}_c$.

To resum the jet functions in the variables $\bar{\varepsilon}_{J_c}$,
it is convenient to reexpress the weight functions
(\ref{fbar2jet1}) in combinations of light-cone momentum
components that are invariant under boosts in the $x_3$ direction,
\ba
\bar{f}_1\left(N_{J_1},a\right) & = & \frac{1}{s^{1-a/2}}
\sum_{\hat n_i \in N_{J_1} }\
k_{i,\,\perp}^a\, \left(2 p_{J_1}^+k_i^-\right)^{1-a},
  \label{fbarLC1}
\nonumber \\
\bar{f}_2\left(N_{J_2},a \right) & = & \frac{1} {s^{1-a/2}}
\sum_{\hat n_i \in N_{J_2} }\
k_{i,\,\perp}^a\, \left(2 p_{J_2}^-k_i^+\right)^{1-a}.
\label{fbarLC2}
\ea
Here we have used the relation $\sqrt{s}/2 = \o_{J_c}$, valid for
both jets in the c.m.  At the same time, we make the identification,
\be
{1\over s} \delta\left({\sqrt{s}\over 2} - \o(N_{J_c})\right)\,
\delta^2(\hat n_c - \hat n(N_{J_c}))
=
{1\over 4}\, \delta^3\left(\vec p_{J_c}-\vec p(N_{J_c})\right)\, ,
\ee
which again holds in the c.m.\ frame.  The spatial components
of each $p_{J_c}$ are thus fixed.  Given that we are at small
$\bar{\varepsilon}_{J_c}$,
the jet functions may be thought of as  functions of
the light-like jet momenta $p_{J_c}^\mu$ of Eq.\ (\ref{lightlike})
and of $\bar{\varepsilon}_{J_c}$.  Because the vector jet function is
constructed
to be dimensionless, $\bar{J}'{}_c^\mu$ in Eq.\ (\ref{jetdef})
is proportional to $\beta_c$
rather than $p_{J_c}$.  Otherwise, it is free of explicit
$\beta_c$-dependence.

The jet functions can now be written in terms of boost-invariant
arguments,
homogeneous of degree zero in $\xi_c$:
\ba
\bar J_c\left(\bar{\varepsilon}_{J_c},a,\mu\right) &=&
\bar{\beta}_c{\,}_\mu \ \Bigg [\
  \beta_c^\mu
\, \bar{J}_c^{(1)} \left(\frac{p_{J_c} \cdot \xic}{\mu},
\bar{\varepsilon}_{J_c} \, \frac{\sqrt{s}}{\mu} \, \left( 
\frac{\sqrt{s}}{2 p_{J_c} \cdot \xic} \right)^{1-a}, a,\as(\mu) 
\right)
\nonumber
\\
&\ & \hspace{-5mm} +\  \, \frac{2\, \xi_c^\mu\ \beta_c\cdot \xi_c
}{\left|\xi_c\right|^2} \,
\bar{J}_c^{(2)} \left(\frac{p_{J_c} \cdot \xic}{\mu},
\bar{\varepsilon}_{J_c} \, \frac{\sqrt{s}}{\mu} \, \left( 
\frac{\sqrt{s}}{2 p_{J_c} \cdot \xic} \right)^{1-a}, a, \as(\mu)
\right) \Bigg ]\, , \nonumber \\
& &
\label{primedef2}
\ea
where ${\bar J}^{(1)}$ and ${\bar J}^{(2)}$ are independent functions, and
where we have suppressed possible dependence on
${\hat \xi}_{c, \, \perp}$.
For jet $c$, the weight $\bar{\varepsilon}_{J_c}$ is fixed by
$\delta(\bar{\varepsilon}_{J_c}-\bar{f}_c(N_{J_c},a))$,
where on the right-hand side of  the expression for the weight (\ref{fbarLC1}),
the sum over each particle's momentum involves the overall factor
$(2 p_{J_c}^\pm/\sqrt{s})^{1-a}$.
After integration over final states at fixed $\bar{\varepsilon}_{J_c}$,
the jet can thus depend on the vector $p_{J_c}^\mu$.
At the same time, it is easy to see from the definition
of the weight that $p_{J_c}^\mu$ can only appear
in the combination $(1/\bar{\varepsilon}_{J_c} \sqrt{s})^{1/(1-a)}\,
(2 p_{J_c}^\mu/\sqrt{s})$.
This vector can combine with $\xi_c$ to form an invariant, and
all $\xi_c$-dependence comes about in this way.

Expression (\ref{primedef2}) can be further simplified by noting that
\be
2\, \bar{\beta}_c \cdot  \xi_c\ {\beta}_c \cdot  \xi_c    =
\xi_c^2 + \xi_{c,\,\perp}^2\,   \, .
\ee
Choosing $\xi_{c,\,\perp} = 0$, we find a single combination,
\ba
\bar J_c\left(\bar{\varepsilon}_{J_c},a,\mu\right)
=
\bar J_c\left( \frac{p_{J_c} \cdot \xic}{\mu},
  \bar{\varepsilon}_{J_c} \, \frac{\sqrt{s}}{\mu} \, \left( \zeta_c 
\right)^{1-a}, a, \as(\mu)
\right)\, ,
\ea
where, in the notation of Eq.\ (\ref{primedef2}), $\bar J_c=\bar
J_c^{(1)}+\bar J_c^{(2)}$, and we have defined
\ba
\zeta_c\equiv {\sqrt{s} \over 2 p_{J_c}\cdot \xic} \, . 
\label{zetadef}
\ea
In these terms, the Laplace  moments of the jet function inherit
dependence on the
moment variable $\nu$ through
\ba
J_c \left(\nu,a,\mu \right)
&=& \int_0^\infty d\bar{\varepsilon}_{J_c} \; {\rm e}^{-\nu \bar{\varepsilon}_{J_c}}\, \bar
J_c\left(\bar{\varepsilon}_{J_c},a,\mu\right)
\nonumber\\
& \equiv &
J_c\left(\frac{p_{J_c} \cdot \xic}{\mu}, \frac{\sqrt{s}}{\mu \nu} \, 
\left(\zeta_c \right)^{1-a},
a,\as(\mu) \right),
\label{primedef}
\ea
where the unbarred and barred quantities denote transformed and
untransformed functions, respectively.  The soft function will be defined below in
a manner that avoids double counting in the cross section.

\subsection{The Soft Function}\label{sec:soft}

Given  the definitions for the jet functions in the
previous subsection, and the factorization (\ref{factoreps}),
we may in principle calculate  the soft function $S$
order by order in perturbation theory.
We can derive a more explicit definition of the soft function,
however, by relating it to an eikonal
analog of Eq.\  (\ref{factoreps}).

As reviewed  in  Refs.\ \cite{Collins:gx,Berger:2001ns} and Sec. \ref{sec:factor},
soft radiation at wide angles from the jets decouples
from the collinear lines within the jet.
As a result, to
compute amplitudes for wide-angle radiation, the jets
may be replaced by nonabelian phases, or Wilson lines.
We therefore construct a dimensionless
quantity, $\sigma^{(\mbox{\tiny eik})}$,
in which gluons are radiated by path-ordered exponentials
$\Phi$, which mimic the color flow of outgoing quarks,
\be
\Phi^{({\mathrm f})}_{\beta_c} (\infty,0;x)  =  P e^{-i g_s \int_{0}^{\infty} d
\lambda \beta_c \cdot {\mathcal{A}}^{({\mathrm f})} (\lambda \beta_c + x )},
\ee
with $\beta_c$ a light-like velocity in either of the jet directions.
For the two-jet cross section at measured
$\bar \varepsilon_{\mbox{\tiny eik}}$, we define
\ba
\bar{\sigma}^{(\mbox{\tiny 
eik})}\left(\bar{\varepsilon}_{\mbox{\tiny eik}}, a ,\mu
\right)\!\!\!
&\!\! \equiv  \!\!\!& \!\!\!{1\over \Ncol}\
\sum_{N_{\mbox{\tiny eik}}} \left< 0
\left| \Phi^{(\bar {\rm q})}_{\beta_2}{}^\dagger(\infty,0;0)
\Phi^{(\rm q)}_{\beta_1}{}^\dagger(\infty,0;0)
\right| {N_{\mbox{\tiny eik}}} \right>
\nonumber \\
& \ & \times\;
\left< N_{\mbox{\tiny eik}}
\left| \Phi^{(\rm q)}_{\beta_1}(\infty,0;0)
\Phi_{\beta_2}^{(\bar{\rm q})}(\infty,0;0)
\right| 0 \right>  \ 
\delta\left(\bar{\varepsilon}_{\mbox{\tiny eik}} - 
\bar{f}(N_{\mbox{\tiny eik}},a) \right)
\nonumber\\
&=&   \delta(\bar{\varepsilon}_{\mbox{\tiny 
eik}}) +{\cal O}(\alpha_s)\,
.
\label{eikdef}
\ea
 The sum is over all final states $N_{\mbox{\tiny eik}}$ in
the eikonal cross section. The renormalization
scale in this cross section, which will also serve as a factorization
scale, is denoted $\mu$.  Here the event shape function
$\bar{\varepsilon}_{\mbox{\tiny eik}}$
is defined by $\bar{f}(N_{\rm eik},a)$ as in Eqs.\  (\ref{barfdef}) and 
(\ref{2jetf}),
distinguishing between the hemispheres around the jets.
As usual, $\Ncol$  is the number of colors,
and a trace over color is understood.

The eikonal cross section (\ref{eikdef}) models the soft
radiation away from the jets, including the radiation into $\O$,
accurately.
It also contains enhancements
for configurations collinear to the jets, which, however,
are  already taken into account by the partonic jet
functions in (\ref{factoreps}).  Indeed, (\ref{eikdef}) does not reproduce
the
partonic cross section accurately for collinear radiation.  
It is also easy to verify at lowest
order that even at fixed $\bar{\varepsilon}_{\mbox{\tiny eik}}$
the eikonal cross section (\ref{eikdef}) is
ultraviolet divergent in dimensional regularization,
unless we also impose a cutoff on
the energy of real gluon emission collinear to $\beta_1$
or $\beta_2$.  

The construction of the soft function $S$
from $\bar{\sigma}^{(\mbox{\tiny eik})}$ is nevertheless possible
  because the eikonal cross
section (\ref{eikdef}) factorizes in the same manner
as the cross section itself, into eikonal jet functions
and a soft function, as in Eq. (\ref{eikfact}).  The essential point \cite{Laenen:2000ij,Kidonakis:1998nf} is that
the soft function in the factorized eikonal cross section
is the same as in the original cross section (\ref{factoreps}).
The eikonal jets organize all collinear enhancements
in (\ref{eikdef}), including the spurious ultraviolet
divergences.  These eikonal jet functions are defined
analogously to their partonic counterparts, Eq.\ (\ref{jetdef}),
but now with ordered exponentials replacing the quark fields,
\ba
\bar{J}_c^{(\mbox{\tiny eik})}\left(\bar{\varepsilon}_c,a,\mu \right)
\!\!& \equiv  & \!\! {1\over \Ncol}\,
\sum_{N_c^{(\mbox{\tiny eik})}}
\left<0 \left| \Phi^{({\mathrm f}_c)}_{\xi_c}{}^\dagger(0,-\infty;0)
\Phi_{\beta_c}^{({\mathrm f}_c)}{}^\dagger(\infty,0;0) \right| 
N_c^{(\mbox{\tiny eik})} \right>
\nonumber \\
& \ &  \left< N_c^{(\mbox{\tiny eik})} \left|
\Phi^{({\mathrm f}_c)}_{\beta_c}(\infty,0;0)
\Phi_{\xi_c}^{({\mathrm f}_c)}(0,-\infty;0) \right|  0 \right> \,
\delta\left(\bar{\varepsilon}_c -  \bar{f}_c(N_c^{(\mbox{\tiny 
eik})},a) \right)
\nonumber\\
&=& \delta(\bar{\varepsilon}_c) +{\cal O}(\alpha_s)\, ,
\label{eikjetdef}
\ea
where ${\mathrm f}_c$ is a quark or antiquark, and where the
trace over color is understood.
The weight functions are given as above, by Eq.\ (\ref{fbar2jet1}),
with the sum over particles in all directions.

In terms of the eikonal jets, the eikonal cross section (\ref{eikdef})
factorizes as
\ba
\bar{\sigma}^{(\mbox{\tiny
eik})}\left(\bar{\varepsilon}_{\mbox{\tiny eik}},a,\mu \right)
& \equiv  &
\int  d \bar{\varepsilon}_s \,
\bar{S}\left(\bar{\varepsilon}_s,a,\mu \right)
\prod\limits_{c = 1}^2 \int d \bar{\varepsilon}_c \,
\bar{J}_c^{(\mbox{\tiny eik})}\left(\bar{\varepsilon}_c,a,\mu \right)\;
   \nonumber \\
   & & \qquad \times \, \delta \left(\bar{\varepsilon}_{\mbox{\tiny eik}} -
\bar{\varepsilon}_s-\bar{\varepsilon}_1-\bar{\varepsilon}_2 \right),
\label{eikfact2}
\ea
where we pick the factorization scale equal to the renormalization scale
$\mu$.  As for the full cross section, the convolution
in (\ref{eikfact2}) is simplified by a Laplace
transformation (\ref{primedef}) with respect to 
$\bar{\varepsilon}_{\mbox{\tiny eik}}$,
which allows us to solve for the soft function as
\be
S \left(\nu,a,\mu\right) =
\frac{\sigma^{(\mbox{\tiny eik})}\left(\nu,a,\mu \right) }
{\prod\limits_{c = 1}^2 J_c^{(\mbox{\tiny eik})}\left(\nu,a,\mu\right) }
=1+{\cal O}(\alpha_s)\, .
\label{s0}
\ee
In this ratio, collinear logarithms
in $\nu$ and the unphysical ultraviolet divergences and their
associated cutoff dependence cancel between the eikonal
cross section and the eikonal jets, leaving a soft
function that is entirely free of collinear enhancements.
The soft function retains $\nu$-dependence through soft
emission.  In addition, because soft radiation
within the eikonal jets can be factored from its collinear
radiation, just as in the partonic jets, all
logarithms in $\nu$ associated with wide-angle radiation
are identical between the partonic and eikonal jets,
and factor from logarithmic corrections associated with
collinear radiation in both cases.
As a result, the inverse eikonal jet
functions cancel contributions from the wide-angle soft radiation of
the partonic jets in the
transformed cross section (\ref{trafosig}). 

 We note that the directions of the non-lightlike eikonals $\Phi_{\xi_c}$ in Eq. (\ref{eikjetdef}) can be inferred from the requirement that the soft function only approximates soft radiation. This can be seen as follows \cite{Collins:1999dz}: At the one-loop level in the frame (\ref{betadef}) the integrand of the eikonal cross section (\ref{eikdef}) is proportional to
\be
I_{\sigma^{(\mbox{\tiny 
eik})}} \sim  \frac{1}{k^2 + i \epsilon} \frac{1}{-k^- +  i \epsilon}  \frac{1}{k^+ +  i \epsilon} .
\ee
This expression is only a good approximation if all components are comparably soft. To avoid collinear overcounting we subtract
\ba
I_{\sigma^{(\mbox{\tiny 
eik})}}  - \sum_c I_{J_c^{(\mbox{\tiny eik})}} &\sim & \frac{1}{k^2 + i \epsilon} \left[\frac{1}{-k^- +  i \epsilon}  \frac{1}{k^+ +  i \epsilon} \right. \nonumber \\
& & \left. \hspace*{-15mm} - \frac{1}{-k^- +  i \epsilon}   \frac{\xi_1^-}{\xi_1^+ k^- + \xi_1^- k^+ +  i \epsilon} -   \frac{\xi_2^+}{-\xi_2^+ k^- - \xi_2^- k^+ +  i \epsilon}  \frac{1}{k^+ +  i \epsilon} \right]\, . \nonumber \\
& & \mbox{ } \label{softo1}
\ea
In order that Eq. (\ref{softo1}) only reproduces the soft region, and in order to avoid spurious Glauber/Coulomb pinches, we have to require that
\ba
\left|\xi_1^+ \right| \gg \left|\xi_1^-\right|, &\quad & \left|\xi_2^-\right| \gg \left|\xi_2^+\right|, \nonumber \\
\xi_1^+ < 0, & & \xi_2^- < 0.
\ea
In other words, the $\xic$ are not light-like, and pointing into the past, $\Phi_{\xi_c}^{({\mathrm f}_c)}(0,-\infty;$ $0)$.

As in the case of the partonic jets, Eq.\ (\ref{primedef}),  we need to
identify
the variable through which $\nu$ appears in the soft
function.
We note that dependence on the velocity vectors $\beta_c$
and the factorization vectors $\xi_c$ must be scale invariant
in each, since they arise only from eikonal lines and vertices.  
The eikonal jet functions cannot depend explicitly on the scale-less, lightlike
eikonal velocities $\beta_c$, and  $\sigma^{\rm (eik)}$
is independent of the $\xi_c$.  Dependence on the factorization
vectors $\xi_c$ enters only
  through the weight functions, (\ref{fbarLC1}) for the eikonal
jets, in a manner analogous to the case of the partonic jets. This results in
a dependence on $(\zeta_c)^{1-a}$, as above, with $\zeta_c$ defined in
Eq. (\ref{zetadef}).  In summary, we may characterize the arguments of the soft function in
transform space as
\be
S\left(\nu,a,\mu \right) =
S\left(
\frac{\sqrt{s}}{\mu \nu} \, \left( \zeta_c \right)^{1-a},
a, \as(\mu)
\right)\, .
\label{Sargs}
\ee

\section{Resummation}

We may summarize the results of the previous
section by rewriting the transform of the factorized cross section
(\ref{trafosig})
in terms of the hard, jet and soft functions identified above,
which depend on the kinematic variables and the moment $\nu$
according to Eqs.\ (\ref{harddef}), (\ref{primedef}) and (\ref{Sargs})
respectively,
\ba
\frac{d \sigma^{\rm incl} \left(\nu,s,a \right)}{ d \hat{n}_1 }
\!\!&=&\!\!
  {d \sigma_0 \over d\hat{n}_1}\ H \left(
\frac{\sqrt{s}}{\mu},\frac{p_{J_c} \cdot \xic}{\mu},\hat{n}_1,\as(\mu)
\right)\, \nonumber \\
&  & \hspace*{-17mm} \times \,
\prod_{c=1}^2\;
J_c\left(\frac{p_{J_c} \cdot \xic}{\mu}, \frac{\sqrt{s}}{\mu \nu} \,
(\zeta_c)^{1-a},
a,\as(\mu) \right)\,
   S\left(
\frac{\sqrt{s}}{\mu \nu}\left(\zeta_c \right)^{1-a},
a, \as(\mu)
\right)\, .\nonumber \\
& &
   \label{factorcom}
\ea
For $a = 0$ this coincides with the expression for the thrust in moment space, Eq. (\ref{thrustmomentb}) with slightly changed notation.
The natural scale for the strong coupling
in the short-distance function $H$ is $\sqrt{s}/2$.
Setting $\mu = \sqrt{s}/2$, however, introduces large logarithms of $\nu$ in both the
soft and jet functions.

The purpose of this section is to control these logarithms by
the identification and solution of  renormalization
group and evolution equations, as in Sec. \ref{sec:resum}. The cross section (\ref{trafosig}) is independent of the factorization scale, and of the choice of the eikonal directions, $\xic$, leading to equations analogous to (\ref{thrustRGE}) and (\ref{thrustxi}). The resummation of single logarithms is straightforward. Following Sec. \ref{sec:single}, we obtain analogously
\ba
   \label{resume}
\frac{d \sigma^{\rm incl} \left(\nu,s,a \right)}{ d \hat{n}_1 }
&=&
{d \sigma_0 \over d\hat{n}_1}\ H \left( \frac{\sqrt{s}}{\mu},\frac{p_{J_c}
\cdot \xic}{\mu},\hat{n}_1,\as(\mu) \right)\,
\nonumber\\
&\ & \hspace{-15mm} \times\; S\left( (\zeta_c)^{1-a}, a
,\as\left( \frac{\sqrt{s}}{\nu} \right)
\right)\,
\exp\left\{ -\int\limits_{\sqrt{s}/\nu}^{\mu} \frac{d
\lambda}{\lambda} \, \gamma_s\left(\as(\lambda)\right)\right\}
\\
&\ & \hspace{-15mm} \times\;
J_c \left( \frac{p_{J_c} \cdot \xic }{\tilde{\mu}_0 },
\frac{\sqrt{s}}{\tilde{\mu}_0 \nu} \, \left(\zeta_c \right)^{1-a} ,
a,\as(\tilde{\mu}_0) \right)
   \,\exp\left\{-\int\limits_{\tilde{\mu}_0}^\mu \frac{d \lambda}{\lambda}
\gamma_{J_c}\left(\as(\lambda)\right)\right\} \, , \nonumber
\ea
where we have left the scale in the jet functions $\tilde{\mu_0}$ free for the moment.

\subsection{Evolution} \label{sec:jetinclevol}

The remaining unorganized ``large" logarithms in (\ref{resume}),
are in the jet functions,
which we will resum by using the consistency equation (\ref{thrustxi}). Analogous to the thrust in Sec. \ref{sec:jetdouble} we obtain the equation satisfied by the
jet functions \cite{Collins:1981uk,Contopanagos:1996nh},
\ba
   \frac{\partial }{\partial \ln \left(p_{J_c} \cdot \xic\right)}
\ln\ J_c \left( \frac{p_{J_c} \cdot \xic}{\mu},
\frac{\sqrt{s}}{\mu \nu} \, (\zeta_c)^{1-a} ,a,\as(\mu)
\right)
& \ & \nonumber
\\
&\ & \hspace{-70mm} =
   K_c\left(\frac{\sqrt{s}}{\mu\,
\nu}(\zeta_c)^{1-a},
a,\as(\mu)
\right)   +  G_c\left(\frac{p_{J_c} \cdot \xic}{\mu},\as(\mu)\right)   \, .
   \label{KGenda}
\ea
The functions $K_c$ and $G_c$
compensate
the $\xi_c$-dependence of the soft and hard functions, respectively,
which determines the kinematic variables upon which they may depend.
In particular, notice the combination of $\nu$- and $\xi_c$-dependence
required by the arguments of the jet function, Eq.\ (\ref{primedef}).

As in Section \ref{sec:jetdouble} $K_c+G_c$ are renormalized
additively, and satisfy \cite{Collins:1981uk}
\ba
\mu \frac{d}{d \mu}\
K_c\left(\frac{\sqrt{s}}{\mu\, \nu}\left(\zeta_c\right)^{1-a},
a,\as(\mu) \right) & = & - \gamma_{K_c}
\left(\as(\mu)\right),
\nonumber\\
\mu \frac{d}{d \mu}G_c\left(\frac{p_{J_c} \cdot \xic }{\mu},\as(\mu)\right)
  & = &  \gamma_{K_c}
\left(\as(\mu)\right) \, .
\label{Gevola}
\ea
Since $G_c$ and hence
$\gamma_{K_c}$, may be computed from
virtual diagrams, they do not depend on $a$, and $\gamma_{K_c}$ is the
universal Sudakov anomalous dimension, computed in Chapter \ref{ch4} \cite{Berger:2002sv,Sen:sd,Collins:1981uk,Korchemsky:wg}.

With the help of these evolution equations, the terms $K_c$ and $G_c$
in Eq. (\ref{KGenda}) can be reexpressed as
\cite{Collins:1984kg}
\ba
   K_c\left(\frac{\sqrt{s}}{\mu\, \nu}\left(\zeta_c\right)^{1-a},a,\as(\mu)
\right)   +  G_c\left(\frac{p_{J_c} \cdot \xic}{\mu},\as(\mu)\right)
&\ &  \nonumber \\
&\ & \hspace{-90mm} =
K_c\left(\frac{1}{c_1},a,\as\left(c_1 \,
\frac{\sqrt{s}}{\nu}\left(\zeta_c\right)^{1-a}\right)
\right)
+  G_c\left(\frac{1}{c_2},\as\left(c_2 \, p_{J_c} \cdot \xic \right) \right)
\nonumber \\
& & \hspace{-60mm} - \int\limits_{c_1 {\sqrt{s}}\, \left(\zeta_c\right)^{1-a}/{\nu} }^{ c_2\,
p_{J_c} \cdot \xic }
\frac{d  \lambda'}{\lambda'} \gamma_{K_c}\left(\as\left(\lambda'\right)
\right)
\nonumber \\
&\ & \hspace{-90mm} =
   - B'_c\left(c_1,c_2,a,  \as\left(c_2 \, p_{J_c} \cdot \xic \right) \right)
-
2 \int\limits_{c_1 {\sqrt{s}}\, \left(\zeta_c\right)^{1-a}/{\nu} }^{ c_2 \,
p_{J_c}
\cdot \xic }
\frac{d  \lambda'}{\lambda'} A'_c\left(c_1, a, \as\left(\lambda'\right)
\right)\, , \nonumber \\ & &
\label{ABabbra}
\ea
where in the second equality we have shifted the argument of
the running coupling in $K_c$, and have introduced the notation
\ba
B'_c\left(c_1,c_2,a, \as\left(\mu \right) \right)
& \equiv & -
K_c\left(\frac{1}{c_1},a, \as\left(\mu \right)  \right) -
G_c\left(\frac{1}{c_2} ,\as\left(\mu \right)\right),
\nonumber \\
2 A'_c\left( c_1, a, \as\left(\mu \right) \right) & \equiv &  \gamma_{K_c}
\left(\as(\mu) \right) + \beta(g(\mu)) \frac{\partial}{\partial
g(\mu)} K_c\left(\frac{1}{c_1},a, \as(\mu)\right). \qquad
\label{ABdefa}
\ea
The primes on the functions $A'_c$ and $B'_c$ are to distinguish
these anomalous dimensions from their somewhat more familiar versions given below. As noted above, $A'_c$ is related to $A$ calculated in the previous chapter.
\newpage

The solution to Eq. (\ref{KGenda}) with $\mu = \tilde{\mu}_0$ is
\ba
J_c \left( \frac{p_{J_c} \cdot \xic }{\tilde{\mu}_0},
  \frac{\sqrt{s}}{\tilde{\mu}_0 \nu} \, \left(\zeta_c
  \right)^{1-a} ,a,\as(\tilde{\mu}_0)
\right)
\!\!&\!=&\!\!
J_c \left( \frac{\sqrt{s}}{2 \,\zeta_0 \,\tilde{\mu}_0},
  \frac{\sqrt{s}}{\tilde{\mu}_0 \nu} \, \left(\zeta_0
  \right)^{1-a} ,a,\as(\tilde{\mu}_0)
\right)   \nonumber \\
&\ & \hspace{-67mm}
   \times \, \exp \left\{ -\int\limits_{\frac{\sqrt{s}}{2\zeta_0} }^{p_{J_c} \cdot
\xic}
   \frac{d \lambda}{\lambda} \Bigg[B'_c\left(c_1,c_2,a,
\as\left(c_2 \lambda \right) \right)  \right.  \nonumber \\
& & \hspace*{-30mm} \left.  +\, 2 \int\limits_{c_1 \frac{s^{1-a/2}
}{ \nu
(2\,\lambda)^{1-a} } }^{c_2\, \lambda}\frac{d \lambda'}{\lambda'} A'_c\left(
c_1,
a,\as\left(\lambda'\right) \right) \Bigg] \right\}\, ,
  \label{orgsola}
\ea
where we evolve from $\sqrt{s}/(2\,\zeta_0)$ to $p_{J_c} \cdot \xic =
\sqrt{s}/(2 \,\zeta_c)$ (see Eq.\ (\ref{zetadef})) with
\be
\zeta_0 = \left(\frac{\nu}{2}\right)^{1/(2-a)}. \label{zeta0}
\ee
After combining Eqs.\ (\ref{resume}) and (\ref{orgsola}),
the choice $\tilde{\mu}_0 = \sqrt{s}/(2\zeta_0) = \frac{\sqrt{s}}{\nu}
(\zeta_0)^{1-a}$
  allows us to control
all large logarithms in
the jet functions:
\ba
J_c \left( \frac{p_{J_c} \cdot \xic}{\mu},
\frac{\sqrt{s}}{\mu \nu} \, (\zeta_c)^{1-a} ,a,\as(\mu)
\right)
\!\!& \!\!=\!\!& \!\!
J_c \left(
1, 1,a,\as\left(\frac{\sqrt{s}}{2 \, \zeta_0} \right) \right)
\,\nonumber \\
& \times  &\!\! \!\! \exp \left\{-\int\limits_{\sqrt{s}/(2 \zeta_0)}^\mu
\frac{d\lambda}{\lambda} \gamma_{J_c} \left(\as(\lambda)\right) \right\}
\, \nonumber \\
&\times & \!\!\!\!
     \exp \left\{ -\int\limits_{\frac{\sqrt{s}}{2\, \zeta_0} }^{p_{J_c}
\cdot \xic}
   \frac{d \lambda}{\lambda} \Bigg[B'_c\left(c_1,c_2,a, \as\left(c_2 \lambda
\right) \right) \right. \nonumber \\
& + & \!\!\!\!\left.  2 \int\limits_{c_1 \frac{s^{1-a/2} }{ \nu
(2\,\lambda)^{1-a} } }^{c_2\, \lambda}\frac{d \lambda'}{\lambda'} A'_c\left(
c_1,
a,\as\left(\lambda'\right) \right) \Bigg] \right\} . \qquad \,
  \label{jetxienda}
\ea
As observed above, we treat $a$ as a fixed parameter, with $|a|$ small compared to
$\ln\nu$.

\subsection{The Resummed Event Shape}

Putting everything together, and setting $\mu = \sqrt{s}/2$, we arrive at
\ba
\frac{d \sigma^{\rm incl} \left(\nu,s,a \right)}{ d \hat{n}_1 }
&=&
{d \sigma_0 \over d\hat{n}_1}\ H \left(\frac{2 p_{J_c}
\cdot \xic}{\sqrt{s} },\hat{n}_1,\as(\sqrt{s}/2) \right)\,
\nonumber\\
&\  &  \hspace*{-2cm} \times\  S \left((\zeta_c)^{1-a},
a,\as\left(\frac{\sqrt{s}}{\nu}
\right) \right) \,
\exp \left\{  - \int\limits_{\sqrt{s} / \nu }^{\sqrt{s}/2} \frac{d
\lambda}{\lambda} \gamma_s\left(\as(\lambda)\right) \right\} \nonumber \\
&\  &  \hspace*{-2cm} \times\ \prod_{c=1}^2\, J_c
\left(1,1,a,\as\left(\frac{\sqrt{s}}{2 \, \zeta_0} \right) \right)
   \exp \left\{ - \int\limits_{\sqrt{s} / (2 \, \zeta_0)}^{\sqrt{s}/2}
\frac{d \lambda}{\lambda}  \gamma_{J_c}\left(\as(\lambda)\right) \right\}
\nonumber \\
& \ & \hspace*{-2cm}
\times \, \exp \left\{ -\int\limits_{\frac{\sqrt{s}}{2\, \zeta_0} }^{p_{J_c} \cdot
\xic}
   \frac{d \lambda}{\lambda} \Bigg[B'_c\left(c_1,c_2,a,
\as\left(c_2 \lambda \right) \right) \right. \nonumber \\
& & + \left. \, 2 \int\limits_{c_1 \frac{s^{1-a/2}
}{ \nu
(2\,\lambda)^{1-a} } }^{c_2\, \lambda}\frac{d \lambda'}{\lambda'} A'_c\left(
c_1,
a,\as\left(\lambda'\right) \right) \Bigg] \right\}\, . \nonumber \\
   \label{globalend}
\ea
$\nu$ appears in up to two logarithms per loop,
characteristic of conventional Sudakov resummation.  We work
 to next-to-leading logarithm in $\nu$,
by which we mean the level $\as^k\, \ln^k\nu$ in the exponent.  As usual, this 
requires one loop in
$B'_c$ and $\gamma_{J_c}$, and two loops in the
Sudakov anomalous dimension $A'_c$, Eq.\ (\ref{ABdefa}).
These functions are straightforward to calculate from their definitions
given in the previous sections.

\section{Results at NLL in Transform Space} \label{sec:resevent}

In this section, we describe the low-order calculations and results that
provide explicit expressions for the resummed  event shape distributions at next-to-leading
logarithm in $\nu$ (NLL).   We go on to verify that the cross section is independent on the choice of the eikonal vectors $\xi_c$, and relate the case $a = 0$ at NLL to the known expressions for the thrust derived with a coherent branching formalism. 

Below we employ the
standard notation,
\ba
\gamma(\as) = \sum_{n=0}^\infty \gamma^{(n)}\ \left({\as\over
\pi}\right)^n
\ea
for any expansion in $\as$.

\subsection{Analytical Results at NLL} \label{sec:resan}

\subsubsection{The Soft Function}

The soft function is normalized to $S^{(0)} =
1$
as can be seen from (\ref{s0}).

The one-loop soft anomalous dimension
is readily calculated in Feynman gauge from
the combination of virtual  diagrams in $\sigma^{\rm (eik)}$, Eq. (\ref{eikdef}), and 
$J^{\rm  (eik)}$, Eq. (\ref{eikjetdef}), 
in  Eq.\ (\ref{s0}).  The calculation and the result are equivalent
to those of Ref.\ \cite{Kidonakis:1998nf}, where the soft function was
formulated in axial gauge,
\be
\gamma_s^{(1)}  = - 2 \, C_F
\left[ \sum_{c=1}^2 \ln \left(\beta_c \cdot \xic \right) - \ln \left(
\frac{\beta_1 \cdot \beta_2}{2} \right) - 1 \right]\, .
\label{softad}
\ee
The first, $\xi_c$-dependent logarithmic term is associated with the eikonal
jets, while the second is a finite remainder from the
combination of $\sigma^{\rm (eik)}$ and $J^{\rm  (eik)}$ in (\ref{s0}).
Whenever $\xi_{c,\,\perp}=0$, the logarithmic terms cancel
identically,  leaving only the final  term, which comes from 
the $\xic$ eikonal self-energy diagrams in the eikonal
jet functions.

\subsubsection{The Jet Functions}

Recall from Eq. (\ref{Jnorm}) that the lowest-order jet function is given
by $J_c^{(0)} = 1.$

The anomalous dimensions of the jet functions are found to be
\be
\gamma_{J_c}^{(1)} = - \frac{3}{2} \, C_F
\, ,
\label{jetAD}
\ee
the same for each of the jets.
The jet anomalous dimensions
are process-independent, but of course flavor-dependent. The same
anomalous dimensions for final-state quark jets appear in three- and
higher-jet cross sections.

\subsubsection{The $K$-$G$-Decomposition}

The anomalous dimension for the $K$-$G$-decomposition
is, as noted above, the Sudakov anomalous dimension (see Eqs. (\ref{a1number}) and (\ref{a2number})),
\ba
\gamma_{K_c}^{(1)} & = & 2 C_F,   \\
\gamma_{K_c}^{(2)} & = & 2 C_F K \equiv  C_F \left[ C_A \left( \frac{67}{18} - \frac{\pi^2}{6} \right) - \frac{10}{9} T_F N_f \right], 
, \ea
also independent of the jet-direction. 

$K_c$ and $G_c$, the functions that describe the evolution
of the jet functions in Eq.\ (\ref{KGenda}), are given at one loop by
\ba
K_c^{(1)} \left(\frac{s^{1-a/2}}{\mu \nu} \left(2 p_{J_c} \cdot
\xic\right)^{a-1}\!\!\!\!\!\!,a\right) \!&\!\! = &\!\! - C_F
\, \ln\left(e^{2 \gamma_E-(1-a)}  \frac{\mu^2
\nu^2}{s^{2-a}} \left(2 p_{J_c} \cdot \xic \right)^{2(1-a)} \right) , \nonumber \\
& & \\
G_c^{(1)} \left(\frac{p_{J_c} \cdot \xic}{\mu} \right) & = & - C_F
\, \ln\left( e^{-1} \frac{\left(2 \, p_{J_c} \cdot \xic \right)^2}{\mu^2}
\right)\, .
\ea
Evolving them to the values of $\mu$ with which they
appear in the functions $A_c'$ and $B'_c$, Eq.\ (\ref{ABdefa}),
they become
\ba
K_c^{(1)} \left(\frac{1}{c_1},a\right) & = & - C_F
\, \ln \left( e^{2 \gamma_E-(1-a)} c_1^2 \right) , \\
G_c^{(1)} \left(\frac{1}{c_2}\right) & = & - C_F
\ln \left( e^{-1} \frac{4}{c_2^2}
\right) .
\ea
Recall that  $G_c$ is computed from virtual diagrams
only, and  thus does not
depend on the weight function.  It therefore agrees with 
the result found in \cite{Collins:1981uk}.
The soft-gluon contribution, $K_c$, which involves
real gluon diagrams, does depend on the cross section
being resummed.

With the definitions  (\ref{ABdefa}) of $A'_c$ and
$B'_c$ we obtain
\ba
A_c^{\prime \,(1)}  & = & C_F  \label{A1prime}
, \\
A_c^{\prime\, (2)} \left(c_1,a\right) & = & \frac{1}{2} C_F
\left[ K + \frac{\beta_0}{2}
\ln \left( e^{2 \gamma_E -1 +a } c_1^2 \right) \right], \\
B_c^{\prime\, (1)} \left(c_1,c_2,a\right) & = & 2 C_F
\ln \left(
e^{ \gamma_E -1 +a/2 } \frac{2 \,c_1}{c_2} \right).
\ea
Here $\beta_0$ is the one-loop coefficient (\ref{beta0}) of the QCD beta-function.

\subsubsection{The Hard Scattering, and the Born Cross Section}

At NLL only the lowest-order hard scattering function contributes, which
is normalized to
\be
H^{(0)}(\alpha_s(\sqrt{s}/2)) = 1\,.
\ee
At this order the hard function is independent of the eikonal vectors
$\xi_c$, although it acquires $\xi_c$-dependence at higher order
through the factorization described in Sec.\ \ref{sec:sdf}.
For completeness,  we also
give the electromagnetic Born cross section $\frac{d\sigma_0}{d \hat n_1}$,
at fixed polar and azimuthal angle:
\be
\frac{d \sigma_0}{d \hat n_1} =
\Ncol \left( \sum_{\rm f} Q_{\rm f}^2 \right) \frac{\alpha_{\rm em}^2}{4 s}
\left( 1 + \cos^2 \theta \right),
\label{bornCross}
\ee
where $\theta$ is the c.m.\ polar angle of $\hat{n}_1$,
$e \, Q_{\rm f}$ is the charge of quark flavor $\rm f$, and $\alpha_{\rm em} = e^2/(4 \pi)$
is the fine
structure constant.

\subsection{Independence of the  Vectors $\xi_c$} \label{sec:gauge}

It is instructive to verify how dependence on
the eikonal vectors $\xi_c$ cancels in the exponents of the
resummed cross section (\ref{globalend}) at the
accuracy at which we work, single and double logarithms of $\nu$.
In these exponents,
$\xi_c$-dependence
enters only through the
combinations
$(\beta_c \cdot \xic)$ and
$(p_{J_c} \cdot \xic)$.

Let us introduce the following notation for the exponents in Eq.
(\ref{globalend}), to
which we will return below:
\ba
E_1 & \equiv &  - \int\limits_{\sqrt{s}/\nu}^{\sqrt{s}/2} \frac{d
\lambda}{\lambda} \gamma_s\left(\as(\lambda)\right) - \sum_{c=1}^2
\int\limits_{\sqrt{s}/(2 \, \zeta_0)}^{\sqrt{s}/2}
  \frac{d \lambda}{\lambda} \gamma_{J_c} \left(\as(\lambda)\right),
  \label{E1} \\
E_2 &  \equiv &  - \sum_{c=1}^2 \int\limits_{\sqrt{s}/(2\, \zeta_0) }^{p_{J_c} \cdot \xic}
   \frac{d \lambda}{\lambda} \Bigg[B'_c\left(c_1,c_2,a,
\as\left(c_2 \lambda \right) \right)  \nonumber \\
& & \qquad \qquad + \, 2 \int\limits_{c_1 \frac{s^{1-a/2}
}{ \nu
(2\,\lambda)^{1-a} } }^{c_2\, \lambda}\frac{d \lambda'}{\lambda'} A'_c\left(
c_1,
a,\as\left(\lambda'\right) \right) \Bigg]. \label{E2}
\ea
At NLL, explicit $\xi_c$ dependence is
found only in $\gamma_s$, Eq. (\ref{softad}), for $E_1$,
and in the upper limit of the $\lambda$ integral of $E_2$.
We then find that
\ba
{\partial \over  \partial \ln\beta_c\cdot\xic}\left(E_1+E_2\right)
& = &
2C_F\, \, \int\limits_{\sqrt{s}/\nu}^{\sqrt{s}/2}
\frac{d
\lambda}{\lambda}\; \frac{\as(\lambda)}{\pi}
-2C_F \int_{c_1{s^{1-a/2}\over
\nu(2p_{J_c}\cdot \xic)^{1-a}}}^{c_2\,p_{J_c}\cdot\xic}
{d\lambda'\over\lambda'}\;
\frac{\as(\lambda')}{\pi} \nonumber \\
& & \qquad  +\,{\rm NNLL}
\, .
\label{gaugevar}
\ea
Here the second term stems entirely from $A^{\prime\,(1)}$, Eq. 
(\ref{A1prime}).
The remaining contributions are of NNLL order,
that is, proportional to $\as^k(\sqrt{s})
  \ln^{k-1} \left (\nu \, \beta_c \cdot \xic \right)$,
as may be verified by expanding the running couplings.
Thus, as required by the factorization procedure,
the relevant $\xi_c$-dependence cancels between
the resummed soft and jet functions, which give rise
to the first and second integrals, respectively, in Eq.\ (\ref{gaugevar}).

As a result, we can choose
\be
p_{J_c} \cdot \xic = \frac{\sqrt{s}}{2}\, .
\label{xiid}
\ee

\subsection{The Inclusive Event Shape at NLL in Transform Space}

We can simplify the differential event shape, Eq.\ (\ref{globalend}),
by absorbing the soft anomalous dimension $\gamma_s$ into 
the remaining terms.  We will find a form that can be compared
directly to the classic  NLL  resummation for the thrust
($a=0$).   This is done by rewriting the integral over
the soft anomalous dimension as
\ba
\int_{\sqrt{s}/\nu}^{\sqrt{s}/2} {d\lambda\over \lambda}\;
\gamma_s\left(\as(\lambda)\right)
&=&
\int_{\sqrt{s}/\left[2(\nu/2)^{1/(2-a)}\right]}^{\sqrt{s}/2}
{d\lambda\over \lambda}\;
\gamma_s\left(\as(\lambda)\right)
\nonumber \\
& & \quad +
\int_{\sqrt{s}/\nu}^{\sqrt{s}/\left[2(\nu/2)^{1/(2-a)}\right]}
{d\lambda\over \lambda}\;
\gamma_s\left(\as(\lambda)\right)
\nonumber\\
& = &
\int_{\sqrt{s}/\left[2(\nu/2)^{1/(2-a)}\right]}^{\sqrt{s}/2}
{d\lambda\over \lambda}\;
\gamma_s\left(\as(\lambda)\right) \nonumber \\
& & \quad + (1-a)
\int_{\sqrt{s}/\left[2(\nu/2)^{1/(2-a)}\right]}^{\sqrt{s}/2}
{d\lambda\over \lambda}\;
\gamma_s\left(\as\left( \frac{s^{1-a/2}}{\nu (2 \lambda)^{1-a}} 
\right)\right) \nonumber \\
& = & 
(2-a)\int_{\sqrt{s}/\left[2(\nu/2)^{1/(2-a)}\right]}^{\sqrt{s}/2}
{d\lambda\over
\lambda}\;
\gamma_s\left(\as(\lambda)\right) \nonumber\\
&\ & \quad 
- (1-a)\int_{\sqrt{s}/\left[2(\nu/2)^{1/(2-a)}\right]}^{\sqrt{s}/2}
{d\lambda\over \lambda}\; \nonumber \\
& & \quad \quad \times \, \int_{s^{1-a/2}/\left[\nu (2 \lambda)^{1-a} 
\right]}^\lambda
\frac{d \lambda'}{\lambda'}
\beta(g(\lambda'))\, {\partial \over \partial g}\,
\gamma_s\left(\as(\lambda')\right)\, . \nonumber \\
& &
\label{split}
\ea
In the first equality we split the $\lambda$ integral so that the limits of 
the first term match those of the $B'_c$ integral of Eq.\ (\ref{globalend}).
In the second equality we have changed variables in the
second term according to
\be
\lambda \rightarrow \left({s^{1-{a\over 2}}\over
2^{1-a}\nu\lambda}\right)^{1\over 1-a}\, ,
\ee
so that the limits of the second integral also match. 
In the third equality of Eq. (\ref{split}),
  we have reexpressed the running coupling at the old
scale
$\lambda$
  in terms of the new scale.
This is a generalization of the procedure of Ref.\ \cite{Catani:1990rp},
applied originally to the threshold-resummed
Drell-Yan cross section \cite{Sterman:1986aj,Catani:ne}.

Using Eq.\ (\ref{split}), and
identifying $p_{J_c} \cdot \xic$ with $\sqrt{s}/2$
(Eq. (\ref{xiid})) in the inclusive event shape
distribution, Eq. (\ref{globalend}),
we can rewrite this distribution at NLL as
\ba
\frac{d \sigma^{\rm incl} \left(\nu,s,a \right)}{ d \hat{n}_1 }
&=&
{d \sigma_0 \over d\hat{n}_1}\
\nonumber \\
&   & \hspace*{-27mm} \times\ \prod_{c=1}^2\,
   \exp \left\{ - \int\limits_{\sqrt{s} / 
\left[2(\nu/2)^{1/(2-a)}\right]}^{\sqrt{s}/2}
\frac{d \lambda}{\lambda}  \Bigg[ B_c
\left(c_1,c_2,a,\as\left(\lambda\right)\right) \right. \nonumber \\
& & \left. \qquad 
+   2 \int\limits_{c_1 \frac{s^{1-a/2} }{ \nu
(2\,\lambda)^{1-a} } }^{c_2\, \lambda}\frac{d \lambda'}{\lambda'} A_c\left(
c_1,
a,\as\left(\lambda'\right) \right) \Bigg] \right\}\, , \nonumber \\
   \label{globalendcat}
\ea
where we have rearranged the contribution of $\gamma_s$ as:
\ba
A_c \left( c_1, a,\as\left(\mu \right) \right)\!\! &\! \equiv \!\!&
A'_c \left( c_1, a,\as\left(\mu \right) \right)
- \frac{1}{4} (1-a)\,  \beta(g(\mu))\, {\partial \over \partial g}\,
\gamma_s\left(\as(\mu)\right) ,  \nonumber \\
  B_c \left(c_1,c_2,a,\as\left(\mu\right)\right) & \equiv &
\gamma_{J_c} \left(\as(\mu) \right)
  + \left( 1 - \frac{a}{2} \right) \gamma_s \left(\as(\mu) \right)  \nonumber \\
  & & \qquad +
  B'_c \left(c_1,c_2,a,\as\left(\mu\right)\right).
  \ea
Next, we replace
  the lower limit of the $\lambda'$-integral
by an explicit $\theta$-function. Then we exchange orders of integration, 
and change variables in the term containing $A$
from the dimensionful variable $\lambda$ to the dimensionless combination
\be
u = {2\lambda\lambda'\over s}\, .
\ee
\newpage
We find
\ba
\frac{d \sigma^{\rm incl} \left(\nu,s,a \right)}{ d \hat{n}_1 }
&=&
{d \sigma_0 \over d\hat{n}_1}\  \prod_{c=1}^2\,
   \exp \left\{ - \int\limits_{\sqrt{s} / 
\left[2(\nu/2)^{1/(2-a)}\right]}^{\sqrt{s}/2}
\frac{d \lambda}{\lambda}
  B_c \left(c_1,c_2,a,\as\left(\lambda\right)\right) \right\}
\nonumber\\
&\ & \hspace*{-35mm} \times\
\prod_{c=1}^2\,
   \exp \left\{ - 2 \int_0^{\sqrt{s}} \frac{d \lambda'}{\lambda'} \;
\int_{\lambda'{}^2/s}^{\lambda'/\sqrt{s}}\, {d u \over u}\;
\theta\left( c_1^{-1}\nu\, {\lambda'{}^a u^{1-a}\over s^{a/2}}-1
\right)\;
A_c\left( c_1,
a,\as\left(\lambda'\right) \right) \right\} \, . \nonumber \\
   \label{globalendcat2}
\ea
Here, the $\theta$-function vanishes for small $\lambda'$, and the remaining effects of 
replacing the lower boundary of the $\lambda'$ integral by 0 are 
next-to-next-to-leading logarithmic.

A further change of variables allows us to write the NLL resummed event shapes
in a form familiar from the NLL resummed thrust.
In the first line of Eq. (\ref{globalendcat2}), we
replace $\lambda^2 \rightarrow u s/4$. In the second line we relabel 
$\lambda' \rightarrow \sqrt{q^2}$,
and exchange orders of integration.
Finally, choosing as in Eq. (\ref{cipick})
\begin{eqnarray*}
c_1 & = & e^{-\gamma_E}, \nonumber \\
c_2 & = & 2, 
\end{eqnarray*}
we find at NLL
\ba
\frac{d \sigma^{\rm incl} \left(\nu,s,a \right)}{ d \hat{n}_1 } \!\!
&\!\!=& \!\!\!
{d \sigma_0 \over d\hat{n}_1}\  \prod_{c=1}^2\,
   \exp \left\{ \int\limits_0^1 \frac{d u}{u} \left[ \,
   \int\limits_{u^2 s}^{us} \frac{d q^2}{q^2} A_c\left(\as(q^2)\right)
   \left( e^{- u^{1-a} \nu \left(q^2/s\right)^{a/2} }-1 \right) 
\right. \right.\nonumber \\
   & & \qquad \qquad \quad \quad
   + \frac{1}{2} B_c\left(\as(u s/4)\right) \left( e^{-u 
\left(\nu/2\right)^{2/(2-a)} e^{-\gamma_E}} -1 \right)
   \bigg] \Bigg\}, \nonumber \\
   & &
\label{thrustcomp}
\ea
and reproduce the well-known coefficients
\ba
A_c^{(1)}  & = & C_F
, \\
A_c^{(2)}  & = & \frac{1}{2} C_F  K, \\
B_c^{(1)} & = & - \frac{3}{2} \, C_F,
\ea
independent of $a$.  In Eq.\ (\ref{thrustcomp}), we have made use of the
relation (\ref{thetarel}).
With these choices, we reproduce the
NLL resummed thrust cross section \cite{Catani:1992ua,Catani:1991kz} when $a = 0$. This also follows directly from eikonal exponentiation, as shown in Section \ref{sec:thrustexp}.

The choices of
the $c_i$ in Eq.\ (\ref{cipick}) cancel all purely soft NLL
components ($\gamma_s$ and $K_c$). The
remaining double logarithms stem from simultaneously soft and collinear
radiation, and single logarithms arise from collinear configurations only.
At NLL, the cross section is determined by the
anomalous dimension $A_c$, which is the coefficient
of the singular $1/[1-x]_+$ term in the
nonsinglet evolution kernel computed in Chapter \ref{ch4} \cite{Berger:2002sv,Korchemsky:1988si,Albino:2000cp}, and the
quark anomalous dimension.
All radiation in dijet events
thus appears to be emitted coherently by the two jets \cite{Catani:1992ua,Catani:1991kz}.
This, however, is not necessarily true
beyond next-to-leading logarithmic accuracy for dijets, and
is certainly not the case for multijet events \cite{Kidonakis:1998nf}.  

\subsubsection{Explicit Expressions}

It is straightforward to perform the integrals in Eq. (\ref{thrustcomp}), using the running coupling in terms of the coupling $\alpha$ evaluated at the hard scale $\sqrt{s}/2$ (compare to (\ref{1as}) and (\ref{2as})):
\ba
\alpha & \equiv & \as\left(\frac{\sqrt{s}}{2} \right)  \label{scaledef} \\
 \as(\mu) \!\!& = \!\!&\!\! \frac{\alpha}{1 + \frac{\beta_0}{2 \pi} \alpha \ln \frac{2 \mu}{\sqrt{s}}} \left[ 1 - \frac{\beta_1}{4 \pi \beta_0} \frac{\alpha}{1 + \frac{\beta_0}{2 \pi} \alpha \ln \frac{2 \mu}{\sqrt{s}}} \ln \left(1 + \frac{\beta_0}{2 \pi} \alpha \ln \frac{2 \mu}{\sqrt{s}} \right) + \dots \right].\nonumber \\
& & 
\ea
The term with $\beta_1$ is only necessary for the integral containing $A_c^{(1)}$. In the following we drop the subscript $c$ and multiply the exponent by 2, since both jets give equal contributions. 

We find 
\ba
\frac{1}{\frac{d \sigma_0}{d \hat n_1}} \frac{d \tilde{\sigma} \left(\nu,s,a \right)}{ d \hat{n}_1 } & = & \exp\Bigg\{2 \, \ln (\nu) \, g_1 \left( \frac{\beta_0}{2 \pi} \frac{\alpha}{2-a} \ln \nu,a\right) \nonumber \\
& & \qquad + \, 2\, g_2 \left( \frac{\beta_0}{2 \pi} \frac{\alpha}{2-a} \ln \nu,a\right) + {\mathcal{O}} \left( \as^n \ln^{n-1} \nu \right) \Bigg\},\quad  \label{expg}
\ea
where the functions $g_1$ and $g_2$ that resum leading and next-to-leading logarithms, respectively, are given by
\ba
g_1(x,a) & = & - \frac{4}{\beta_0} \frac{1}{1-a} \frac{1}{x} A^{(1)} \Bigg[ \left( \frac{1}{2-a} - x \right) \ln (1-(2-a) x) \nonumber \\
& & \qquad \qquad \qquad  - (1-x) \ln (1-x) \Bigg], \label{g1} \\
g_2(x,a) & = & \frac{2}{\beta_0} B^{(1)} \ln (1-x) \nonumber \\
& & \quad  - \, \frac{8}{\beta_0^2} \frac{1}{1-a}  A^{(2)} \left[ (2-a) \ln (1-x) - \ln (1-(2-a)x) \right] \nonumber \\
& & \quad - \, \frac{4}{\beta_0} \gamma_E \frac{1}{1-a}  A^{(1)} \left[ \ln (1-x) - \ln(1-(2-a) x)\right] \nonumber \\
& & \quad + \, \frac{4}{\beta_0} \ln 2 \frac{1}{1-a} \, A^{(1)} \left[ (2-a) \ln (1-x) -  \ln(1-(2-a) x)\right] \nonumber \\
& & \quad - \frac{\beta_1}{\beta_0^3} \frac{1}{1-a} A^{(1)} \left[ 2 \ln(1-(2-a) x) - 2(2-a) \ln(1-x) \right. \nonumber \\
& & \qquad \qquad \qquad \qquad + \left. \, \ln^2 (1-(2-a) x) - (2-a) \ln^2 (1-x) \right]. \nonumber \\
& & \label{g2}
\ea
The factors of 2 in (\ref{expg}) are due to the fact that the two jets give equal contributions. Eq. (\ref{expg}) reduces for $a = 0$ to the form found in \cite{Catani:1992ua}, up to the term proportional to $\ln 2$ in $g_2$ which is due to the fact that we use the hard scale $\sqrt{s}/2$ instead of $\sqrt{s}$ in \cite{Catani:1992ua}.
The cross section at an arbitrary scale $\mu$ is easily found from (\ref{expg}) at NLL:
\ba
\frac{1}{\frac{d \sigma_0}{d \hat n_1}} \frac{d \tilde{\sigma} \left(\nu,s,a \right)}{ d \hat{n}_1 } & \equiv & \left[\,{\mathcal{J}}(\nu,s,a)\right]^2  \nonumber \\
& = & \exp\Bigg\{2 \, \ln (\nu)\, g_1 \left( \frac{\beta_0}{2 \pi} \frac{\as(\mu)}{2-a} \ln \nu,a\right) \nonumber \\
& & \quad + \, 2 \, \left(\frac{\beta_0}{2 \pi} \right)^2 \frac{\as^2(\mu)}{2-a} \ln^2 \nu\, \ln \left( \frac{2 \mu}{\sqrt{s}} \right) g_1' \left( \frac{\beta_0}{2 \pi} \frac{\as(\mu)}{2-a} \ln \nu,a\right) \nonumber \\
& & \quad + \, 2\, g_2 \left( \frac{\beta_0}{2 \pi} \frac{\as(\mu)}{2-a} \ln \nu,a\right) + {\mathcal{O}} \left( \as^n \ln^{n-1} \nu \right) \Bigg\}, \label{expgrun}
\ea
where
\ba
g_1' \left( x,a\right) & = & \frac{\partial }{\partial x} g_1 (x,a) \nonumber \\
& = &  - \frac{4}{\beta_0} \frac{1}{1-a} \frac{1}{x^2}  A^{(1)} \left[ \ln (1-x) -  \frac{1}{2-a}  \ln (1-(2-a) x) \right]. \label{g1prime}
\ea
Setting $\mu = \sqrt{s}$ as in \cite{Catani:1992ua} cancels the term proportional to $\ln 2$ in Eq. (\ref{g2}), and we reproduce for $a = 0$ the form of \cite{Catani:1992ua}.

\section{Inverse Transform}\label{sec:inv}

To perform the inversion we follow the method of \cite{Catani:1992ua}, which is well suited for resummed formulae without soft contributions, like (\ref{thrustcomp}). We also note that there are a variety of other techniques for inverting Laplace transforms, see for example \cite{Contopanagos:1993yq,Catani:1996yz}, which differ in their treatment of ambiguities in the transform due to the asymptotic nature of the perturbative series. These ambiguities manifest themselves in the transformed cross section (\ref{thrustcomp}) as singularities when reaching the Landau pole in the running coupling. In this study we bypass these issues by expanding in terms of a fixed coupling, as in (\ref{expg}). Nevertheless, we have to keep in mind that our result still contains these ambiguities, which manifest themselves as power corrections. We return to this issue below, in Sec. \ref{sec:powerevent}.

First, we consider the integrated cross section
\ba
{d \sigma(\bar{\varepsilon},s, a)\over 
 d\hat n_1} & = &  \frac{d\sigma_0}{d \hat n_1} \, \frac{1}{2 \pi i} \,\int_{C} \frac{d\nu}{\nu} e^{\nu \bar{\varepsilon}} \left[{\mathcal{J}}(\nu,s,a)\right]^2 , \label{intinv}
 \ea
 with $\mathcal{J}$ given in terms of $g_1$, $g_2$, and $g'_1$ in Eq. (\ref{expgrun}).
 The contour lies in the complex plane to the right of all singularities of the integrand.
 Here we have dropped the subscript $c$ because the two jets give equal contributions.

We can now perform a Taylor expansion of the exponent with respect to $\ln \nu$ around $\ln \nu = \ln 1/\bar{\varepsilon}$, because the functions $g_1$ and $g_2$ vary more slowly with $\nu$ than $\nu \bar{\varepsilon}$. At NLL accuracy we can neglect all derivatives of order two and higher in the Taylor expansion of the exponent. Performing the integral is then straightforward, using
\be
\frac{1}{2 \pi i} \int_C du e^{u- (1-\gamma) \ln u} = \frac{1}{\Gamma(1-\gamma)}.
\ee
We find
\ba
\frac{1}{\frac{d\sigma_0}{d \hat n_1} } {d \sigma(\bar{\varepsilon},s, a)\over 
 d\hat n_1} & = & \nonumber \\
 & & \hspace*{-25mm} \frac{\exp \Bigg\{ 2 \ln \frac{1}{\bar{\varepsilon}} \, g_1 (x,a) + 2 g_2 (x,a) + 2 (2-a) x^2 \ln\left(\frac{2 \mu}{\sqrt{s}} \right) g'_1 (x,a) \Bigg\} }{\Gamma\Bigg[1-2 g_1 (x,a) - 2 x g'_1(x,a)\Bigg]}, \quad \label{crossintsol}
\ea
where
\be
x \equiv \frac{\as(\mu)}{\pi} \frac{\beta_0}{2\,(2-a)} \ln \frac{1}{\bar{\varepsilon}}, \label{xdef}
\ee
and the functions $g_i,i = 1,2$ are given in Eqs. (\ref{g1}), (\ref{g2}), and (\ref{g1prime}).

The differential cross section (\ref{eventdef}) is easily obtained from Eq. (\ref{crossintsol}),
\be
\frac{1}{\frac{d \sigma_0}{d \hat n_1}} {d \sigma(\bar{\varepsilon},s, a)\over 
 d \bar{\varepsilon}\, d\hat n_1}  = \frac{1}{\bar{\varepsilon}} \frac{d}{d \ln \bar{\varepsilon}} {d \sigma(\bar{\varepsilon},s, a)\over 
 d\hat n_1} . \label{crossdiff}
 \ee 
 
\subsection{Double and Leading Logarithmic Approximation}

The double logarithmic approximation (DL) which sums only terms proportional to $\as^n \ln^{2n-1} \bar{\varepsilon}$ in the differential cross section  stems entirely from the first order in $\as$ in $g_1$ in the numerator. We find a DL contribution \emph{independent of $a$}
\be 
\frac{\bar{\varepsilon}}{\frac{d \sigma_0}{d \hat n_1}} {d \sigma(\bar{\varepsilon},s, a)\over 
 d \bar{\varepsilon}\, d\hat n_1} = - 2 A^{(1)} \frac{\as}{\pi} \ln \bar{\varepsilon} \, e^{- A^{(1)} \frac{\as}{\pi} \ln^2 \bar{\varepsilon}}. \label{DL}
 \ee
 Since the DL approximation turns out to be $a$-independent, this result is equal to the DL approximation for the thrust cross section \cite{Binetruy:1980hd,Schierholz:1979bt}. 

The leading logarithmic contribution (LL) is given by
\be 
\frac{\bar{\varepsilon}}{\frac{d \sigma_0}{d \hat n_1}} {d \sigma(\bar{\varepsilon},s, a)\over 
 d \bar{\varepsilon}\, d\hat n_1} = 2 \left( g_1(x) + x g'_1(x) \right) e^{-2 g_1(x) \ln \bar{\varepsilon}}, \label{LL}
 \ee
 where $x$ is defined in (\ref{xdef}).
 
The full result at NLL order can be obtained by inserting (\ref{crossintsol}) into (\ref{crossdiff}). 

\section{Numerical Results}

\subsection{Matching with Fixed Order Calculations}

Analytical fixed order expressions for the thrust exist only up to ${\mathcal{O}}(\as)$ \cite{DeRujula:1978yh}, for matching at NLL it is necessary, however, to know the fixed order contributions up to ${\mathcal{O}}(\as^2)$. These can be calculated numerically, not only for the thrust, using, for example, the program EVENT2 \cite{Catani:1996jh}. 

In the following we will use direct matching, that is, the cross section is computed as
\be
\frac{d \sigma \left(\bar{\varepsilon},s,a \right)}{d \bar{\varepsilon}\, d \hat{n}_1 } = \frac{d \sigma_{\mbox{\tiny resum}} \left(\bar{\varepsilon},s,a \right)}{d \bar{\varepsilon}\, d \hat{n}_1 } - \frac{d \sigma_{\mbox{\tiny resum}}^{\mbox{\tiny exp, (2)}} \left(\bar{\varepsilon},s,a \right)}{d \bar{\varepsilon}\, d \hat{n}_1 } + \frac{d \sigma_{\mbox{\tiny fixed}}^{(2)} \left(\bar{\varepsilon},s,a \right)}{d \bar{\varepsilon}\, d \hat{n}_1 }, \label{match}
\ee
where the subscript resum denotes our resummed cross section, the second term on the right, $d \sigma_{\mbox{\tiny resum}}^{\mbox{\tiny exp, (2)}}$ is the resummed cross section expanded up to order $\as^2$, and $d \sigma_{\mbox{\tiny fixed}}^{(2)}$ is the fixed order cross section calculated with EVENT2 up to order $\as^2$. Other matching schemes are possible, they differ from the above formula at order $\as^3$ and NNLL.

\subsubsection{Determination of the Thrust Axis}

The implementation of our generalized weight (\ref{2jetf}) with (\ref{barfdef}) is straightforward, since the weight is defined with respect to the thrust axis, and since the method of event generation in EVENT2 is Lorentz-invariant. The default frame of the program is the partonic center-of-mass frame, which is also convenient for our purposes.

At order ${\mathcal{O}}(\as^2)$, up to only four partons contribute to the final state. For two- or three-parton final states, the thrust axis is given simply by the direction of the parton that carries the biggest energy fraction. For four- or higher parton final states, it is best to rewrite Eq. (\ref{thrust}) as \cite{Brandt:1978zm}
\be
T = \max\limits_{\epsilon_i = \pm 1} \frac{\left|  \sum_i \epsilon_i \vec{p}_i \right| }{\sum_j \left| \vec{p}_j \right|} . \label{thrustre}
\ee
This may be interpreted as applying Eq. (\ref{thrust}) to an event with $2n$ particles, half of which carry the momenta in the original directions $\vec{p}_i$ and the other half in the directions $-\vec{p}_i$, which automatically balances the momentum. Eq. (\ref{thrustre}) then follows from \cite{Brandt:1964sa}
\be
\sum_{i \in C_+} \vec{p}_i \cdot \hat{n} = \left|  \sum_{i \in C_+} \vec{p}_i \right| \cos \theta_{(\mbox{\tiny tot}) n},
\ee
 where $C_+$ is the class of momenta for which $\vec{p}_i \cdot \hat{n} > 0$ with $\vec{p}_i$ in the doubled set of momenta, and $\theta_{(\mbox{\tiny tot})n}$ is the angle between the sum over all momenta in $C_+$, $\sum_{i \in C_+} \vec{p}_i \equiv \vec{p}_{\mbox{\tiny tot}}$, and vector $\hat{n}$. Maximizing this expression gives (\ref{thrustre}), where $\theta_{\mbox{\tiny (tot)}n} = 0$. Geometrically, this can be interpreted as the task of finding the three-dimensional polygon built out of all participating momenta with the largest diameter. The direction of the largest diameter corresponds to the thrust axis. The momenta comprise a polygon due to momentum conservation $\sum_i \vec{p}_i = 0$, and different orderings of the vectors result in different polygons, possibly with crossing of lines.
To find the thrust axis via Eq. (\ref{thrustre}) requires a test of $2^{n-1}$ possibilities (The factor $2^{-1}$ comes from the fact that $T$ is unchanged when all $\epsilon_i \rightarrow (- \epsilon_i)$.). This is quite feasible and fast for four partons in the final state, the maximal number needed at NLO.

\subsection{Matched Results at NLL} \label{sec:matchres}

Here we show some representative results for the matched resummed cross sections for various values of the parameter $a$. The NLL resummed cross section is found from Eq. (\ref{crossintsol}) which is valid up to power corrections for values of $\bar{\varepsilon}$ away from the end-point region $\beta_0/(2 \pi) \as \ln \bar{\varepsilon} < 1$, or $\bar{\varepsilon} > \LQCD/\sqrt{s}$. For $\bar{\varepsilon} \sim \LQCD/\sqrt{s}$ non-perturbative corrections become dominant. We therefore cut off the small values of $\bar{\varepsilon}$ from our plots.

\begin{figure}[hbt]
\begin{center}
\epsfig{file=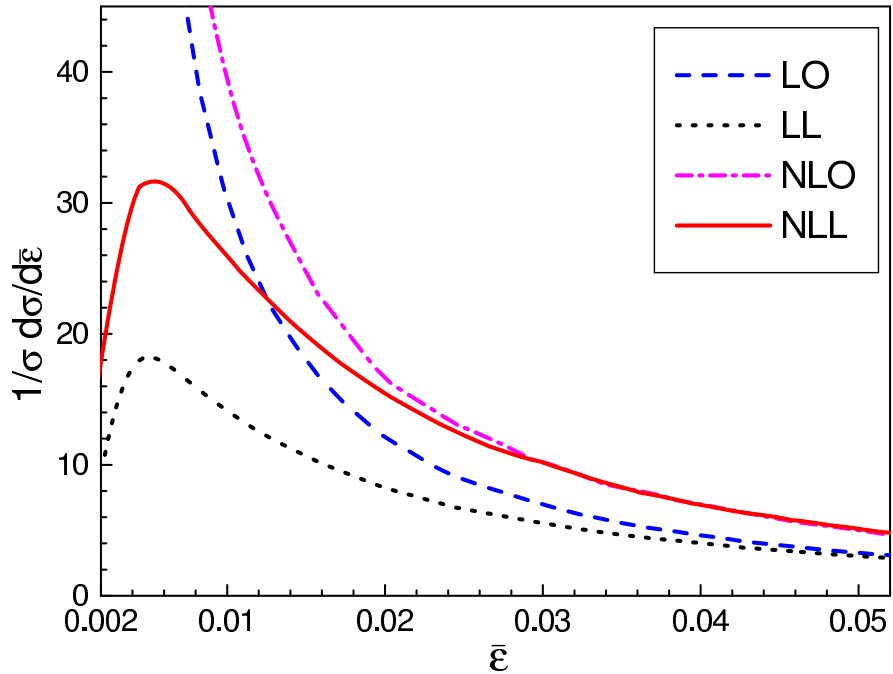,height=12.5cm,angle=270,clip=0}
\caption{Comparison of fixed order and resummed cross sections for $a = -0.5$ at c.m. energy $\sqrt{s} = 91$ GeV as a function of $\bar{\varepsilon}$. Shown are the leading order (LO), matched leading logarithmic (LL), next-to-leading order (NLO) and matched next-to-leading logarithmic (NLL) results.}  \label{logfocomp}
\end{center}
\end{figure}

In Fig. \ref{logfocomp} we compare the cross sections evaluated at fixed order via EVENT2 \cite{Catani:1996jh} with our resummed results matched according to Eq. (\ref{match}). Shown are the differential cross sections $\frac{1}{\sigma} \frac{d \sigma}{d \bar{\varepsilon}}$, normalized by the total cross section, as a function of $\bar{\varepsilon}$ at a center-of-mass energy of $\sqrt{s} = 91$ GeV. As one can see from Fig. \ref{logfocomp}, fixed order calculations are sufficient for not too small values of $\bar{\varepsilon}$. However, the fixed order results are singular at the phase-space boundary $\bar{\varepsilon} \rightarrow 0$, and resummation is necessary for accurate quantitative predictions. 

Furthermore, Fig. \ref{logfocomp} shows that the difference between LL and NLL resummed results is quite sizeable. As mentioned in the introduction to event shapes, Sec. \ref{sec:thrust}, event shapes are used in precision measurements of the running coupling \cite{Bethke:2002rv}.  Although higher logarithmic corrections are generally expected to be not as sizeable, it may be necessary to include NNLL corrections to reduce the theoretical        uncertainty in the determination of $\as$. At NNLL accuracy not only the knowledge of the three-loop coefficient $A^{(3)}$ is necessary, whose fermionic part was computed in the previous chapter, but also the full resummed expression (\ref{globalend}) instead of its simplified version (\ref{thrustcomp}), valid only up to NLL, has to be used in the computation.

In Fig. \ref{PYTH} we plot the matched resummed results at NLL for $a = -1$ and $a = 0$ (the thrust-related shape), normalized as above, $\frac{1}{\sigma} \frac{d \sigma}{d \bar{\varepsilon}}$, as a function of $\bar{\varepsilon}$ for a c.m. energy of $\sqrt{s} = 91$ GeV. For $a = 0$ we reproduce the results of \cite{Catani:1992ua}. We compare our predictions, valid at the partonic level, to the corresponding cross sections computed by \textsc{PYTHIA} \cite{Sjostrand:2000wi}, version 6.215 \cite{Sjostrand:2001yu}, at the hadronic level, using \textsc{PYTHIA}'s implementation of the string fragmentation model \cite{Andersson:ia,Andersson:tv}. We use the default settings of the program. The string picture seems to model the hadronization process fairly well, as the comparison with some recent data for the thrust shows, displayed also in Fig. \ref{PYTH}. We note that our definition of the thrust-related shape, Eq. (\ref{barfdef}) at $a = 0$, is only equivalent to $1- T$ at the partonic level, with $T$ defined in terms of three-momenta as in Eq. (\ref{thrust}). At the partonic level mass-effects are negligible, in contrast to the hadronic level. For the comparison with \textsc{PYTHIA} in Fig. \ref{PYTH} we use the definition of our event shape in terms of energies, (\ref{barfdef}), for both the partonic and the hadronic level. Other prescriptions are of course possible \cite{Salam:2001bd}. 

The difference between partonic and hadronic level, in terms of location of the peak and overall shape at small values of $\bar{\varepsilon}$, is due to hadronization and other non-perturbative effects. These, as briefly mentioned in Sec. \ref{sec:intropower} and \ref{sec:inv}, manifest themselves as power corrections to the perturbatively calculated cross section. We will give some brief qualitative arguments below, but reserve a more quantitative study for future work.\newpage

\begin{figure}[hbt]
\begin{center}
\epsfig{file=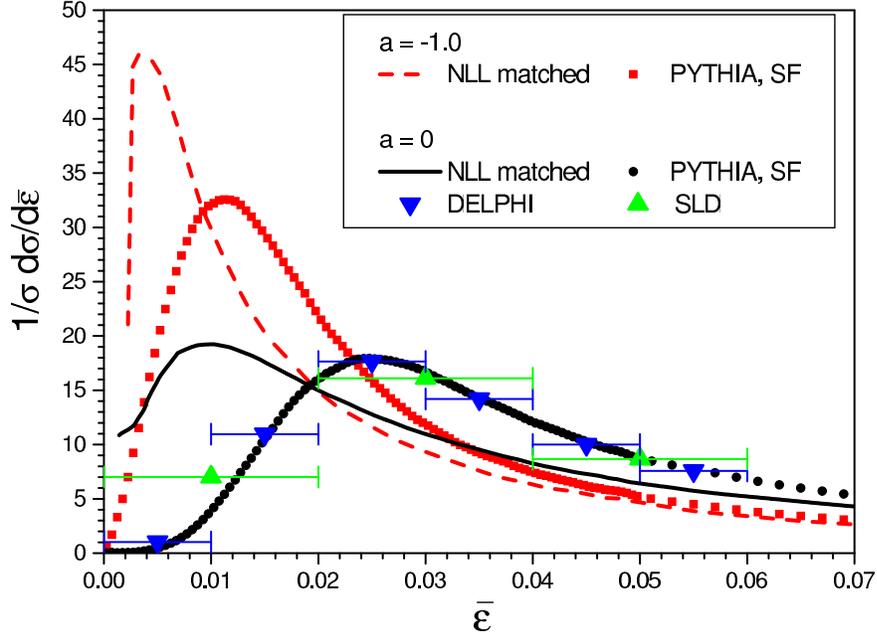,height=12.5cm,angle=270,clip=0}
\caption{Comparison of matched NLL resummed cross sections for $a = -1$ and $a = 0$ to the corresponding cross sections calculated by \textsc{PYTHIA} with string hadronization (SF) at $\sqrt{s} = 91$ GeV. The data for $a = 0$ ($\bar{\varepsilon} = 1 - T$) are taken from  \cite{Abe:1994mf} (SLD) and \cite{Abreu:1996na} (DELPHI).}  \label{PYTH}
\end{center}
\end{figure}

\section{Power Corrections} \label{sec:powerevent}

Starting from Eq. (\ref{thrustcomp}), we can proceed analogously to Sec. \ref{sec:webpower} to infer the form of nonperturbative corrections. Changing orders of integration in (\ref{thrustcomp}), and using Eq. (\ref{asreexp}), we obtain 
\ba
\tilde{\sigma}^{(\mbox{\tiny eik})}_a (\nu) & = & \exp \Bigg\{ 2\, \int\limits_0^{Q^2} \frac{d q^2}{q^2} A\Bigg( \frac{\as(Q^2)}{1+ \frac{\beta_0}{4 \pi} \as(Q^2) \ln \frac{q^2}{Q^2}} \Bigg)  \nonumber \\
& & \qquad \times \,  \int\limits_{q^2/Q^2}^{q/Q} \frac{d u}{u} \left( e^{- \nu  u^{1-a} (q/Q)^a } - 1 \right) \Bigg\}, \label{powera}
\ea
where $\sqrt{s} = Q$. Expanding the exponential in the exponent, changing variables as in Eq. (\ref{varchange}),
\be
t_n \equiv n \, t = \frac{n}{2} \as(Q^2) \ln \frac{Q^2}{q^2},
\ee
and performing the integral over $u$ 
results in
\ba
\tilde{\sigma}^{(\mbox{\tiny eik})}_a (\nu) & = & \exp \Bigg\{ \frac{2}{1-a} \,  \sum_{n = 1}^\infty \frac{1}{n^2\, n!}   \frac{\left(-\nu\right)^n}{\as(Q^2)} \int\limits_0^{\infty} d t_n\, A\Bigg(\frac{\as(Q^2)}{1 -\frac{\beta_0}{2 \pi\,n} t_n} \Bigg)  e^{-\frac{t_n}{\as(Q^2)}} \nonumber \\
& & \hspace*{4cm} \times \, \left[ 1- e^{-\frac{t_n\,(1-a)}{\as(Q^2)}} \right] \Bigg\}. \label{powera2}
\ea

Comparing to the Borel integral, Eq. (\ref{Binv}), we find an ambiguity proportional to $1/Q$, just as for the thrust. Here, however, corrections are suppressed by non-integer powers, by ${\mathcal{O}}(1/Q^{1-a})$, due to the lower limit in the integral over $u$ in (\ref{powera}).
\be 
\ln \tilde{\sigma}_a(\nu,Q)  = \ln \tilde{\sigma}_{a,\,\mbox{\tiny PT}}(\nu,Q) + \ln \tilde{\sigma}_a^{\mbox{\tiny power}} \left(\frac{\nu}{Q} \right)  + {\mathcal{O}}\left( \frac{\nu}{Q^{2-a}} \right)\, ,\label{powershape}
\ee
where
\be
\ln \tilde{\sigma}_a^{\mbox{\tiny power}} \left(\frac{\nu}{Q} \right)  \equiv \frac{1}{1-a} \ln \tilde{f}^{\mbox{\tiny power}} \left( \frac{\nu}{Q} \right) . \label{frel}
\ee
 Moreover, from Eq. (\ref{powershape}) with (\ref{frel}) we find that shape distributions with two different parameters $a,b < 1$ have leading nonperturbative corrections that are related to each other by \cite{Sterman:2003wk}
\be
\ln \tilde{\sigma}_a^{\mbox{\tiny power}} \left(\frac{\nu}{Q} \right) =  \frac{1-b}{1-a} \ln \tilde{\sigma}_b^{\mbox{\tiny power}} \left(\frac{\nu}{Q} \right),\label{powerrel}
\ee
as was observed by G. Sterman. 

As can be seen from Fig. \ref{PYTH}, these power corrections shift the peak of the cross section as well as change its shape. The leading effect is a shift of the distribution \cite{Korchemsky:1994is,Dokshitzer:1997ew,Dokshitzer:1997iz}.  In general, however, the effect of the leading power corrections is also to change the shape of the distribution \cite{Korchemsky:1999kt,Korchemsky:1998ev}. The latter effect is most prominent at small values of the shape, $\bar{\varepsilon} \sim \LQCD/Q$. 

That the leading effect at moderate $\bar{\varepsilon}$ is a shift of the distribution can be seen from the first term in the expansion (\ref{powera2}), $n = 1$. From the above, the ambiguity due to this term, denoted by the superscript (1), is proportional to 
\be
\ln \tilde{\sigma}_a^{\mbox{\tiny power (1)}} = - \frac{\lambda_1}{1-a} \frac{\nu}{Q},
\ee
where $\lambda_1$ is a nonperturbative, constant parameter (compare to Eqs. (\ref{amb}) and (\ref{lamp})). Inserting this expression into (\ref{powershape}), we obtain the cross section in momentum space from (\ref{intinv})
\be
{d \sigma(\bar{\varepsilon},s, a)\over 
 d\hat n_1} =   \frac{d\sigma_0}{d \hat n_1} \, \frac{1}{2 \pi i} \,\int_{C} \frac{d\nu}{\nu} e^{\nu \left(\bar{\varepsilon} - \frac{1}{1-a} \frac{\lambda_1}{Q}\right)} \left[{\mathcal{J}}(\nu,s,a)\right]^2 .
 \ee
Thus the integrated cross section is shifted to the right by an amount 
\be
\Delta \bar{\varepsilon}(a,Q) = \frac{1}{1-a} \frac{\lambda_1}{Q}. \label{shift}
\ee
To first approximation this also holds for the differential cross section (\ref{crossdiff}) for not too small values of $\bar{\varepsilon}$.
As mentioned above, for very small $\bar{\varepsilon} \sim \LQCD/Q$ not only the first term with $n = 1$ in (\ref{powera2}) has to be considered. The result is then a change in the overall shape of the distribution \cite{Korchemsky:1999kt,Korchemsky:1998ev}. 

Here we only study the shift of the distribution (\ref{shift}), more precisely, the shift of the peak $\Delta \bar{\varepsilon}_{p}$. From Eq. (\ref{powerrel}) we infer, that the shifts of the peaks for different values of $a$ multiplied by $(1-a)$ are the same when measured at the same scale $Q$:
\be
(1-a ) \,\Delta \bar{\varepsilon}_{p} (a,Q) = (1-b)\, \Delta\bar{\varepsilon}_{p} (b,Q). \label{scala}
\ee

Fig. \ref{lineshift} shows the shifts of the peaks between our NLL resummed predictions and the corresponding hadronic distributions computed with \textsc{PYTHIA} with the string fragmentation model as in Sec. \ref{sec:matchres}, multiplied by $(1-a)$, at c.m. energy $Q = \sqrt{s} = 91$ GeV. We compare the so-obtained value to the shift of the peak for the thrust  ($a = 0$) determined in \cite{MovillaFernandez:2001ed} between resummed predictions and experimental data. The shifts for different values of $a$ obey the relation Eq. (\ref{scala}) surprisingly well, although the peaks are at fairly small values of $\bar{\varepsilon}$. This seems to confirm the universality of the power corrections within the class of event shapes under consideration, Eq. (\ref{powerrel}).  

\begin{figure}[hbt]
\begin{center}
\epsfig{file=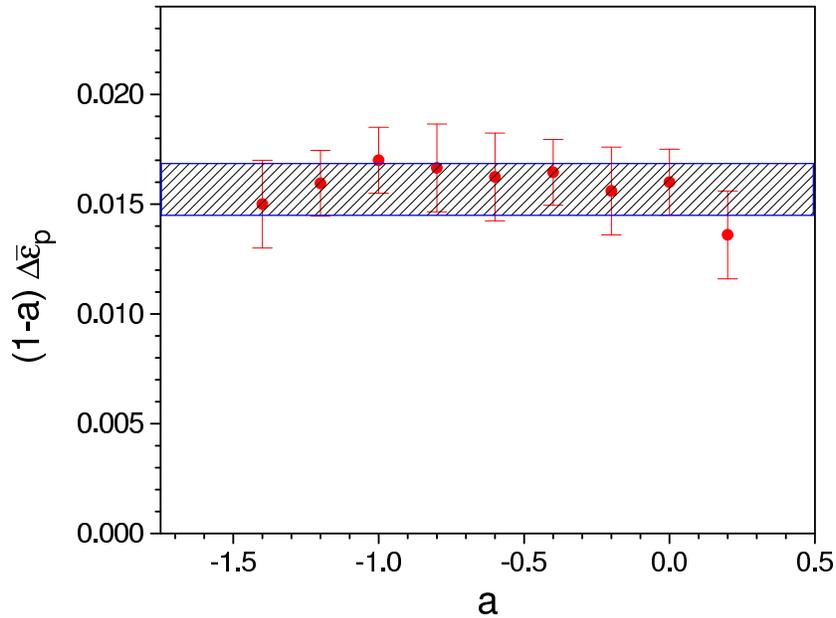,height=11cm,angle=270,clip=0}
\caption{Shifts of the peaks $\Delta \bar{\varepsilon}_{p}(a,Q = \sqrt{s} = 91 \mbox{ GeV})$ of the distributions $\frac{1}{\sigma} \frac{d \sigma}{d \bar{\varepsilon}}$ between NLL partonic resummed predictions and hadronic cross sections computed with \textsc{PYTHIA} with string fragmentation. The result is multiplied by $(1-a)$. The blue band is the shift of the peak for the thrust determined in \cite{MovillaFernandez:2001ed}  between resummed predictions and experimental data.}  \label{lineshift}
\end{center}
\end{figure}

\section{Summary and Outlook}

We have
introduced a general class of inclusive event shapes in $e^+e^-$ dijet events
which
reduce to the thrust and the jet broadening distributions as special
cases.
We have derived analytic expressions in transform
space, and have shown the equivalence of our formalism at NLL 
with the well-known result for the thrust \cite{Catani:1992ua,Catani:1991kz}.
Separate studies of this class of event shapes 
at higher orders, more quantitative studies of power corrections, and a comparison to experiment are certainly of interest.
We reserve these studies for future work.

 Measurements of jet events in general rely on our
determination of the energies of  jets in the presence of
an ``underlying event", consisting of particles not
directly associated with jet production. 
An understanding of the underlying event requires
good control over perturbative bremsstrahlung associated
with the hard scattering itself, so that the two effects
may be separated. In the next chapter we will correlate the general class of event shapes discussed above with the measurement of energy flow into restricted parts of phase space. Such correlations between event shapes and energy flow emphasize radiation directly from the partons that undergo the hard scattering. Studies of these shape/flow correlations may therefore help to disentangle
the underlying event from the bremsstrahlung of the
hard scattering.

\chapter{Interjet Energy Flow/Event Shape Correlations}
\label{ch6}

Energy flow \cite{Sveshnikov:1995vi,Tkachov:1995kk,Korchemsky:1997sy,Berger:2002jt} into angular regions between energetic
jets gives information that is in some ways complementary
to what we learn from event shapes.  In perturbation theory,
the distribution of particles in the final state reflects
interference between radiation from different jets \cite{Dokshitzer:nm},
and there is ample evidence for perturbative
antenna patterns in interjet radiation at both $\rm e^+e^-$ \cite{Bartel:1981kh,Bartel:1983ii,Althoff:1985wt,Akrawy:1991ag,Akers:1995xs}
and hadron colliders \cite{Ellis:1996eu,Abbott:1999cu}.  Energy flow between
jets must also encode the mechanisms that neutralize color in the
hadronization process, and the transition of QCD from weak
to strong coupling.
Knowledge of the interplay
between energy and color flows \cite{Berger:2001ns,Kidonakis:1998nf} may help 
identify the underlying
event in hadron collisions \cite{Huston:zr,Tano:2002hc},
to distinguish QCD bremsstrahlung
from signals of new physics.  Nevertheless, the systematic computation of
energy flow into interjet regions has turned out to be
subtle \cite{Dasgupta:2001sh,Dasgupta:2002bw,Dasgupta:2002dc} for reasons that we will review below,
and requires a careful construction of the class of 
jet events.  

Here we introduce correlations
between event shapes and energy flow, ``shape/flow correlations",
   that are sensitive primarily to radiation from the highest-energy
jets.  So long as
the observed energy is not too small, in a manner to be quantified
below, we may control logarithms of the ratio of energy flow
to jet energy. This chapter is based on our publications \cite{Berger:2001ns,Berger:2002ig,Berger:2003iw}. 

The energy flow observables that we discuss
below are distributions
associated with radiation into a chosen interjet angular region,
$\O$.  Within $\O$ we identify a kinematic quantity $Q_\O\equiv
\varepsilon Q$, at c.m.\ energy $Q$, with
$ \varepsilon\ll 1$.  $Q_\O$ may be the sum of energies, transverse
energies or related
observables for the particles
emitted into $\O$.  Let us denote by $\bar\O$ the complement
of $\O$.  We are interested in the distribution of $Q_\O$
for events with a fixed number of jets in $\bar \O$.
This set of events may be represented  schematically as
\begin{equation}
A + B \rightarrow \mbox{  Jets }  + X_{\bar{\O}}
   + R_\O (Q_\O)\, .
\label{event}
\end{equation}
Here $X_{\bar\O}$ stands for radiation into the regions
between $\O$ and the jet axes, and $R_\O$ for
radiation into $\O$.  

We start by describing the subtleties associated with the computation of interjet energy flow in \ref{sec:nonglobal}, before proposing shape/flow correlations as a means to control secondary effects. After giving a two-loop example, we factorize and resum large logarithmic corrections to the shape/flow correlations in $e^+e^-$ dijet events. In Sections \ref{sec:analyticalshape} and \ref{sec:numerics} we give analytical and numerical results for $e^+e^-$ events at NLL. In Sec. \ref{sec:hadronic} we show the application of our formalism to events with hadrons in the initial state. For hadronic events, the nontrivial color flow can be described via matrices in the space of color exchanges. Explicit color decompositions are listed in Appendix \ref{app3}.

\section{Non-global Logarithms} \label{sec:nonglobal}

The subtlety associated with the computation of energy flow
concerns the origin of logarithms, and is illustrated by
Fig. \ref{eventfig}.
Gluon 1 in Fig.\ \ref{eventfig} is
an example of a primary gluon,
emitted directly from
the hard partons  near a jet axis.
Phase space integrals for primary emissions contribute single logarithms
per loop: $(1/Q_\O)\as^n \ln^{n-1} (Q/Q_\O) = 
(1/\varepsilon Q)\as^n\ln^{n-1}(1/\varepsilon)$, $n\ge 1$, and
these logarithms exponentiate in a straightforward fashion \cite{Berger:2001ns}.
At fixed $Q_\O$
for Eq.\ (\ref{event}), however, there is another source of
potentially large logarithmic
corrections in $Q_\O$.  These are illustrated by gluon 2
in the figure, an example of
secondary radiation in $\O$, originating a parton emitted
by one  of the leading jets that define the event into intermediate region
   $\bar{\O}$.
As observed by Dasgupta and
Salam \cite{Dasgupta:2001sh,Dasgupta:2002bw,Dasgupta:2002dc}, emissions into $\O$ from such secondary
partons   can also result in logarithmic corrections, of the form
$(1/Q_\O)\as^n \ln^{n-1}(\bar{Q}_{\bar{\O}}/Q_\O)$, $n\ge 2$,
where $\bar{Q}_{\bar{\O}}$ is the maximum energy
emitted into $\bar{\O}$.  These logarithms arise
from strong ordering in the energies of the primary
and secondary radiation
because real and virtual enhancements
associated with secondary emissions do not
cancel each other fully at fixed $Q_\O$.

If the cross section is
fully inclusive outside of $\O$, so that no restriction
is placed on the radiation into $\bar{\O}$,
$\bar{Q}_{\bar{\O}}$ can approach $Q$, and
the secondary logarithms can become as important as
the primary logarithms.   Such a cross section, in
which only radiation into a fixed portion of phase 
space ($\O$) is specified, was termed ``non-global" by
Dasgupta and Salam, and the associated logarithms
are also called non-global \cite{Dasgupta:2001sh,Dasgupta:2002bw,Dasgupta:2002dc,Dokshitzer:2003uw,Banfi:2000si,Burby:2001uz,Banfi:2002hw,Appleby:2002ke}. 

In effect, a  non-global definition of energy 
flow is not restrictive
enough to limit final states to a specific set of jets, and
non-global logarithms are produced by jets of intermediate energy,
emitted in directions between region $\O$ and
the leading jets.  
Thus, interjet energy flow 
does not always originate directly from the leading jets, in the absence of
a systematic criterion for suppressing intermediate radiation.
Correspondingly, non-global logarithms
reflect color flow at all scales, and do not
exponentiate in a simple manner.
Our aim in this chapter is to formulate a set of observables
for interjet radiation in which non-global logarithms
are replaced by calculable corrections, and which
reflect the flow of color at short distances.
By restricting the sizes of event shapes,
we will limit radiation in
region $\bar{\O}$, while retaining the chosen jet structure.

An important observation  that
we will employ below is that non-global logarithms are not produced
by secondary emissions that are very close to a jet
direction, because a jet of parallel-moving
particles emits soft radiation coherently.  By
fixing the value of an event shape near the
  limit of narrow jets, we avoid
final states with large energies in $\bar{\O}$ away
from the jet axes.
At the same time, we will identify limits in which non-global logarithms
reemerge as leading corrections, and where the
methods introduced to study nonglobal effects in Refs.\ \cite{Dasgupta:2001sh,Dasgupta:2002bw,Dasgupta:2002dc,Banfi:2000si,Burby:2001uz,Banfi:2002hw,Appleby:2002ke} provide 
important insights.

\begin{figure}[htb]
\begin{center}
\epsfig{file=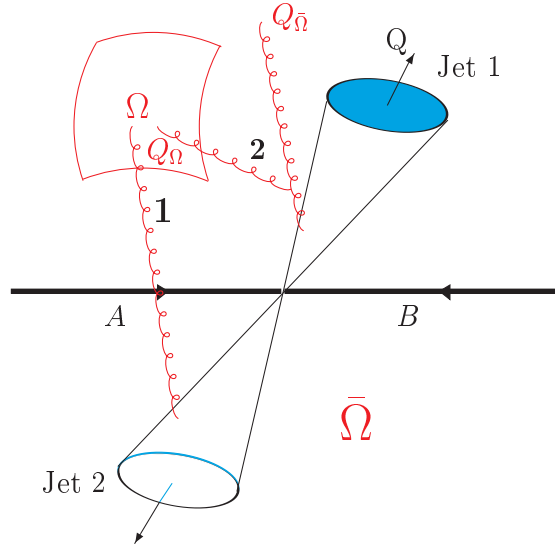,height=8cm,clip=0}
\caption{Sources of global and non-global logarithms in dijet events.
Configuration 1, a
primary emission, is the source of global logarithms.  Configuration 2
can give non-global logarithms.}
\label{eventfig}
\end{center}
\end{figure}

To formalize these observations,
we study below correlated observables for $e^+e^-$
annihilation into two jets.
(In Eq.\ (\ref{event}) $A$ and $B$ denote positron and electron.) In
$e^+e^-$ annihilation dijet
events, the underlying color flow pattern is
simple, which enables us to concentrate  on the
energy flow within the event. We give an outlook on the application of our formalism to cross sections with hadrons in the initial state, where the color flow is non-trivial \cite{Berger:2001ns,Kidonakis:1998nf}.
In this chapter, 
we will correlate the generalized event shape $\bar{f}(a)$, introduced in the previous chapter, with energy flow. 
To avoid large non-global
logarithmic corrections we weigh events by
  $\exp[-\nu \bar{f}]$, with $\nu$
the Laplace transform conjugate variable. The parameter $a$ allows us to dial the amount of energy emitted into the interjet radiation, thus to control non-global logarithmic corrections.

For the restricted set
of events with narrow jets, energy flow is proportional
to the lowest-order cross section for gluon
radiation into the selected region.  The resummed cross section,
however, remains sensitive to color flow at short distances
through anomalous dimensions associated with coherent
interjet soft emission.  In a sense, our results show that
an appropriate selection of jet events automatically
suppresses nonglobal logarithms, and confirms
the observation of coherence in interjet radiation
\cite{Dokshitzer:nm,Ellis:1996eu}.  

\section[Shape/Flow Correlations in $e^+e^-$ Dijet Events]{Shape/Flow Correlations in $e^+e^-$ Dijet \\ Events}

\subsection{Shape/Flow Correlations}

In the notation of Eq.\ (\ref{event}), we will study an event shape
distribution for the process
\be
e^+ + e^- \rightarrow J_1(p_{J_1}) + J_2(p_{J_2})  +
X_{\bar{\O}} \left(\bar{f}\right) + R_\O (Q_\O)\, ,
\label{crossdef}
\ee
at c.m.\ energy $Q \gg Q_\O \gg \LQCD $.
Two jets with momenta $p_{J_c},\, c = 1,\,2$ emit
soft radiation (only) at wide angles.  Again,
$\O$ is a region between the
jets to be specified below, where the total energy or the transverse
energy $Q_\O$ of the soft radiation is measured,
and $\bar{\O}$ denotes the remaining
phase space (see Fig. \ref{event}).  Radiation into $\bar{\O}$
is constrained by event shape $\bar{f}$, Eq. (\ref{2jetf}).  We
refer to cross sections at fixed values (or transforms) of $\bar{f}$ and
$Q_\O$
as shape/flow  correlations.

We find the jet axes as described in Section \ref{sec:eventshape}, minimizing the thrust-related quantity $\bar f_{\bar{\O}_1}(N,a=0)$. In contrast to Sec. \ref{sec:eventshape}, we divide now the phase space into
three regions:
\begin{itemize}
\item Region $\O$, in which we
measure, for example,
the energy flow,
\item Region $\bar \O_1$, the entire hemisphere centered on
$\hat n_1$, that is, around jet 1, except its intersection with $\O$,
\item Region $\bar \O_2$, the complementary hemisphere, except its
intersection with $\O$.
\end{itemize}
We will study the correlations of the  set of event
shapes $\bar{f}(a)$ with the energy flow into $\O$, denoted as
\be
f(N) =  {1\over \sqrt s}\ \sum_{\hat n_i\in\O} \o_i\, .
\label{eflowdef}
\ee

The  differential cross section
for such dijet events at fixed values of $\bar f$ and $f$ is given by
\ba
{d \bar{\sigma}(\varepsilon,\bar{\varepsilon},s, a)\over d \varepsilon
\,d\bar{\varepsilon}\, d\hat n_1}
&=&
{1\over 2s}\ \sum_N\;
|M(N)|^2\, (2\pi)^4\, \delta^4(p_I-p_N) \nonumber\\
&\ & \hspace{8mm} \times
\delta( \varepsilon-f(N))\, \delta(\bar{\varepsilon} -\bar f(N,a))\;
\delta^2  (\hat n_1 -\hat n(N))\, ,
\label{eventdefcorr}
\ea
to be contrasted with the corresponding inclusive event shape, Eq. (\ref{eventdef}). $\varepsilon$, like $\bar{\varepsilon}$, is required to be  
 much less than unity in the elastic limit:
\be
0 < \varepsilon \ll 1.
\label{elasticlimcorr}
\ee
We choose our coordinate system in the same way as in the previous chapter, Eq. (\ref{lightlike}).

Here
we seek to control corrections in the single-logarithmic variable
$\alpha_s(Q) \ln (1/\varepsilon)$,
with $\varepsilon=Q_\O/Q$.  Such a resummation is most
relevant when
\be
\alpha_s(Q) \ln \left({1 \over \varepsilon}\right) \ge 1 \rightarrow
\varepsilon \le  \exp\left({- 1\over \alpha_s(Q) }\right)\, .
\label{einequal}
\ee
Let us compare these logarithms to non-global
effects in shape/flow correlations.
At $\nu=0$ and for $a\rightarrow -\infty$,
the cross section becomes inclusive outside $\O$.  As we show below,
the non-global logarithms discussed in Refs.\ \cite{Dasgupta:2001sh,Dasgupta:2002bw,Dasgupta:2002dc}
appear in shape/flow correlations as logarithms of the form
$\alpha_s(Q)\, \ln(1/(\varepsilon \nu))$, with $\nu$ the moment variable
conjugate to the event shape.  To treat these logarithms
as subleading for small $\varepsilon$ and (relatively)
large $\nu$, we require that
\ba
\alpha_s(Q)\, \ln \left({1 \over \varepsilon \nu}\right) < 1  \rightarrow
\varepsilon >
{1\over\nu}\;
\exp\left({- 1\over \alpha_s(Q) }\right)     \, .
\label{enuinequal}
\ea
For large $\nu$, there is a substantial range of $\varepsilon$
in which both (\ref{einequal}) and (\ref{enuinequal}) can
hold.  When $\nu$ is large, moments of the
correlation are dominated precisely by events with
strongly two-jet energy flows, which is the natural
set of events in which to study the influence of color
flow on interjet radiation.  (The peak of
the thrust cross section is at $(1-T)$ of order one-tenth
at LEP energies, corresponding to $\nu$ of order ten, 
so the requirement of large $\nu$ is not overly restrictive.) 
 In the next subsection, we
show how the logarithms of $(\varepsilon \nu )^{-1}$ emerge in a
low order example. 

\subsection{Low Order Example}
\label{sec:loe}

\begin{figure} \center
\includegraphics*{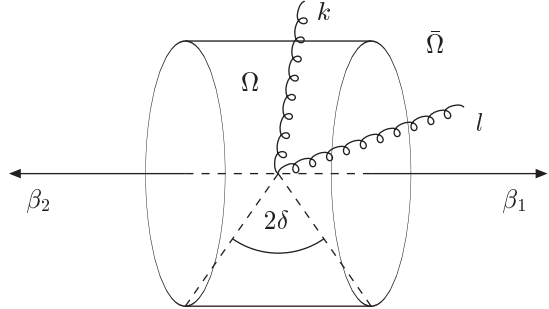}
\caption{\label{kinematics} A kinematic configuration that gives rise to
the
non-global logarithms. A soft gluon with momentum $k$ is radiated
into the region $\Omega$, and an
energetic gluon with momentum $l$ is radiated into $\bar\Omega$.
Four-vectors
$\beta_1$ and $\beta_2$, define the directions of jet 1 and jet 2,
respectively.}
\end{figure}

In this section, we check the general ideas developed above with
the concrete
example of a two-loop cross section for the process
(\ref{crossdef}). This is the lowest order in which a
non-global logarithm occurs, as observed in \cite{Dasgupta:2001sh,Dasgupta:2002bw,Dasgupta:2002dc}. We
normalize this cross section to the Born cross section for
inclusive dijet production. A similar
analysis for the same geometry has been carried out in \cite{Dasgupta:2002bw} and 
\cite{Appleby:2002ke}.

The kinematic configuration we consider is shown in Fig. \ref{kinematics}.
Two fast partons, of velocities
   $\vec{\beta_1}$ and $\vec{\beta_2}$, are treated in eikonal
approximation.
In addition, gluons are emitted into the final state.
A soft gluon with momentum $k$ is radiated into region $\Omega$ and
an energetic gluon with momentum
$l$ is emitted into the region $\bar{\Omega}$.
We consider the cross section at fixed energy,
$\o_k\equiv \varepsilon\sqrt{s}$.
As indicated above, non-global logarithms arise from
strong ordering of the energies of the gluons,
which we choose as $ \ol \gg \ok $.
In this region, the gluon $l$ plays the role of a ``primary"
emission, while $k$ is a ``secondary" emission.

For our calculation, we take the angular region $\Omega$
to be a ``slice" or ``ring'' in polar
angle of width $2 \delta$, or
equivalently, (pseudo) rapidity interval $(-\eta,\eta)$, with
\be \label{rapidity}
\Delta \eta =2\eta= \ln\left(\frac{1+\sd}{1-\sd}\right)\, ,
\label{deltaeta1}
\ee
The lowest-order diagrams for
this process are those shown
in Fig. \ref{diagrams}, including distinguishable diagrams
in which the momenta $k$ and $l$ are interchanged.

\begin{figure}
\hspace*{-10mm}
\includegraphics*{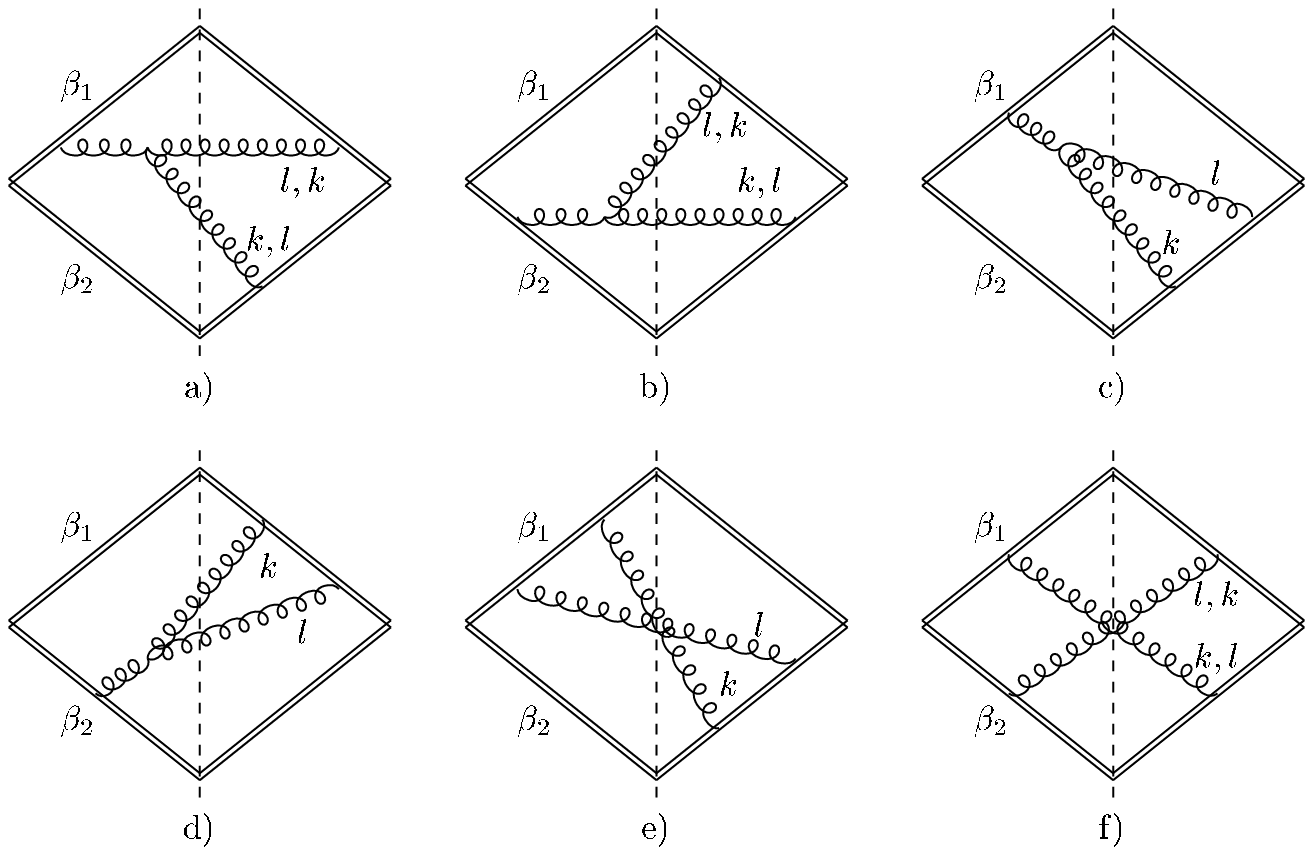}
\caption[The relevant two-loop cut diagrams
corresponding to the emission of
two real gluons in the final state contributing to the eikonal cross
section.]{\label{diagrams} The relevant two-loop cut diagrams
corresponding to the emission of
two real gluons in the final state contributing to the eikonal cross
section.
The dashed line represents the final state, with
contributions to the amplitude
to the left, and to the complex conjugate amplitude to the right.}
\end{figure}

The diagrams of Fig.\ \ref{diagrams}
give rise to color structures $C_F^2$ and $C_FC_A$,
but terms proportional to $C_F^2$  may be associated with a factorized
contribution to the cross section, in which the
gluon $k$ is emitted coherently by the combinations
of the gluon $l$ and the eikonals.
To generate the $C_FC_A$ part, on the other hand, gluon $k$
must ``resolve"
gluon $l$ from the eikonal lines,
giving a result that depends on
the angles between $\vec l$ and the
eikonal directions.

We choose the reference frame such
that the momenta of the final
state particles are given by:
\ba \label{momenta}
{\beta}_1 & = & (1,0,0, 1), \nonumber \\
{\beta}_2 & = & (1,0,0,-1), \nonumber \\
l & = & \ol (1, s_l, 0, c_l), \nonumber \\
k & = & \ok (1, s_k \cfi, s_k \sfi, c_k).
\ea
Here we define $s_{l,k} \equiv \sin
{\theta}_{l,k}$ and $c_{l,k} \equiv
\cos {\theta}_{l,k}$. $\theta_l$ is the angle between the vectors
$\vec{l}$ and $\vec{\beta_1}$,  $\theta_k$ is the
angle between the vectors $\vec{k}$ and $\vec{\beta_1}$
and $\phi$ is the azimuthal angle
of the gluon with momentum $k$ relative to the plane defined by
$\beta_1$, $\beta_2$ and $l$. The
available phase space in polar angle for the radiated gluons is
${\theta}_k \in (\pi/2 - \delta, \,
\pi/2 + \delta)$ and ${\theta}_l \in (0,\, \pi/2 - \delta) \cup
(\pi/2 + \delta, \, \pi)$.

Using the diagrammatic rules for eikonal lines and
vertices, as listed in Appendix \ref{sec:eikfeyn}, we can write down the
expressions corresponding to each diagram separately. For example,
diagram \ref{diagrams} a) gives
\ba \label{a}
a) \, + \, (k \leftrightarrow l)\!\! &\!\! =\!\! &\!\! \left[ f_{abc} \mathrm{Tr}(T^a T^b T^c) \right] \left( -i g_s^4 \,
\beta_1^{\alpha} \beta_2^{\beta} \beta_1^{\gamma} \right) \,
V_{\alpha \beta \gamma}(k+l, -k, -l) \, \nonumber \\
& & \times \frac{1}{\beta_1
\cdot (k+l)} \, \frac{1} {2 k \cdot l} \, \frac{1}{\beta_1 \cdot l}
\, \frac{1}{\beta_2 \cdot k} \nonumber \\
& + &  \, (k \leftrightarrow l).
\ea
$V_{\alpha \beta \gamma}(k+l, -k, -l) =
[ (2k+l)_{\gamma} g_{\alpha \beta} + (l-k)_{\alpha} g_{\beta \gamma}
- (2l + k)_{\beta} g_{\alpha
\gamma}]$ is the
momentum-dependent part of the three gluon vertex.
Using the color identity $f_{abc}$ $\mathrm{Tr}$ $(T^a T^b T^c) = i C_F \Ncol
C_A /2$, and the approximation $\beta_j \cdot l \gg \beta_j \cdot
k$ for $j=1,2$, which is valid due
to the strong ordering of the final state gluon energies, we arrive at
\be
a) + (k \leftrightarrow l) = \frac{1}{4} \, C_F \Ncol C_A \, g_s^4 \,
\frac{\beta_1 \cdot \beta_2}{k
\cdot l}
\left( \frac{1}{\beta_1 \cdot k \, \beta_2 \cdot l} +
\frac{2}{\beta_1 \cdot l \, \beta_2 \cdot k}
\right).
\ee
We proceed in a similar manner for the rest of the diagrams. The
results are:
\ba \label{b-e}
b) + (k \leftrightarrow l) & = & \frac{1}{4} \, C_F \Ncol C_A \, g_s^4 \,
\frac{\beta_1
\cdot \beta_2}{k \cdot l}\left(
\frac{2}{\beta_1 \cdot k \, \beta_2 \cdot l} + \frac{1}{\beta_1 \cdot
l \, \beta_2
\cdot k}\right), \nonumber \\
c) & = & \frac{1}{4} \, C_F \Ncol C_A \, g_s^4 \, \frac{\beta_1 \cdot \beta_2}
{k \cdot l}\frac{1}{\beta_1 \cdot l} \frac{1}{\beta_2 \cdot k}, \nonumber
\\
d) & = & \frac{1}{4} \, C_F \Ncol C_A \, g_s^4 \, \frac{\beta_1 \cdot \beta_2}
{k \cdot l}\frac{1}{\beta_1 \cdot k} \frac{1}{\beta_2 \cdot l}, \nonumber
\\
e) & = & C_F \Ncol (C_F - C_A/2) \, g_s^4 \, \frac{(\beta_1 \cdot \beta_2)^2}
{\beta_1 \cdot l \, \beta_2 \cdot l} \frac{1}{\beta_1 \cdot k \,
\beta_2 \cdot k}, \nonumber \\
f) + (k \leftrightarrow l) & = & C_F \Ncol (C_F - C_A/2) \, g_s^4 \,
\frac{(\beta_1 \cdot \beta_2)^2}{\beta_1 \cdot l \, \beta_2 \cdot l}
\frac{2}{\beta_1 \cdot k \, \beta_2 \cdot k}.
\ea
The color factors in the last two equations of
(\ref{b-e})
are obtained from the identity $\mathrm{Tr}(T^a T^b T^a T^b) = C_F \Ncol
(C_F - C_A/2)$. Combining the terms proportional to
the color factor $C_F \Ncol C_A$, and including the complex conjugate
diagrams, we find for the squared amplitude
\ba \label{mm}
|M|^2 & = & 2 \, g_s^4 \, C_F \Ncol C_A \, \beta_1 \cdot \beta_2 \left(
\frac{1}{k \cdot \l \, \beta _1
\cdot k \, \beta_2 \cdot l} +
\frac{1}{k \cdot \l \, \beta _1 \cdot l \, \beta_2 \cdot k} \right. \nonumber \\
& & - \left.
\frac{\beta_1 \cdot \beta_2}{\beta_1
\cdot l \, \beta_2 \cdot l \, \beta_1 \cdot k \, \beta_2 \cdot k} \right).
\ea

We take, as indicated above, a Laplace transform
with respect to the shape variable, and
identify the logarithm in the conjugate variable $\nu$. In the frame (\ref{momenta})
we find that the logarithmic $C_FC_A$-dependence
of Fig.\ \ref{diagrams} may be written as a dimensionless
eikonal cross section in
terms of one energy and two polar angular integrals as
\ba
\label{ps0}
{d\sigma_{\rm eik}\over d\, \varepsilon}
   & = & C_F C_A \left(\frac{\alpha_s}{\pi}\right)^2 \,
\frac{1}{\varepsilon} \,
\int_{-\sd}^{\sd}
\mathrm{d} c_k \, \int_{\sd}^{1}
\mathrm{d} c_l \, \int_{\varepsilon \sqrt{s}}^{\sqrt{s}} \frac{\mathrm{d}
\ol}{\ol} \, e^{-\nu \, \ol \,
(1-c_l)^{1-a} \, s_l^a / Q}
\nonumber \\
& & \times\ \left[ \frac{1}{c_k + c_l} \, \frac{1}{1+c_k}
\left(\frac{1}{1+c_l} + \frac{1}{1-c_k}\right) - \frac{1}{s_k^2} \,
\frac{1}{1+c_l} \right]\, .
\ea
In this form, the absence of collinear singularities
in the $C_FC_A$ term at $\cos\theta_l=+ 1$ is manifest, independent of $\nu$.
Collinear singularities in the $l$ integral completely
factorize from the $k$ integral, and are proportional
to $C_F^2$.
The logarithmic dependence
on $\varepsilon$ for $\nu > 1$ is readily found to be
\ba
\label{psa}
{d\sigma_{\rm eik}\over d\, \varepsilon} = C_F C_A
\left(\frac{\alpha_s}{\pi}\right)^2 \, \frac{1}{\varepsilon} \,
\ln\left(\frac{1}{\varepsilon \nu}\right)\, C(\Delta \eta)\, ,
\ea
where  $C(\Delta \eta)$ is a finite function of the
angle $\delta$, or equivalently, of the rapidity
width of the region $\Omega$, Eq. (\ref{rapidity}).
\ba \label{ps3}
C(\delta) & = & \frac{\pi^2}{6} +
\ln \left(\frac{\cot\delta \, (1 + \sd)}{4} \right) \, \ln
\left(\frac{1+\sd}{1-\sd} \right)
+ \mathrm{Li}_2\left(\frac{1-\sd}{2}\right) \nonumber \\
& - & \mathrm {Li}_2\left(\frac{1+\sd}{2}\right) -
\mathrm{Li}_2\left(-\frac{2 \sd}{1-\sd}\right) -
\mathrm{Li}_2\left(\frac{1-\sd}{1+\sd}\right) .
\ea
or equivalently,
\ba 
C(\Delta \eta) & = &  \frac{\pi^2}{6} +
\de \left(\frac{\de}{2} - \ln \left(2 \sinh (\de)\right) \right)
+ \mathrm{Li}_2 \left ( \frac{e^{-\de/2}}{2 \, \cosh (\de/2)} \right)
 \nonumber \\
& - &\mathrm {Li}_2 \left( \frac{e^{\de/2}}{2 \, \cosh
(\de/2)} \right) - \mathrm{Li}_2
\left( -2 \sinh (\de/2) \,
e^{\de/2} \right) - \mathrm{Li}_2 (e^{-\de}) . \nonumber \\
& & \label{psRapidity}
\ea
The coefficient $C (\Delta \eta)$
as a
function of $\de$ is shown in
Fig. \ref{crossSec_vs_rapidity}.
Naturally, $C$ is a monotonically increasing function of $\Delta\eta$.
For $\de \rightarrow 0$,
\be
C \sim {\mathcal O}(\de \, \ln \de)\, ,
\ee
and the cross section vanishes, as expected. On the other hand, as the size of
region $\Omega$ increases, $C$ rapidly saturates and reaches its limiting 
value \cite{Dasgupta:2002bw}
\be
\lim_{\de \rightarrow \infty} \, C = \frac{\pi^2}{6}\, .
\ee

\begin{figure} \center
\epsfig{file=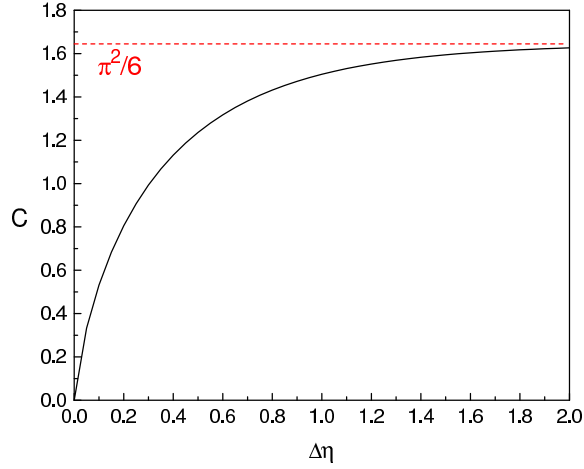,height=8cm,angle=270,clip=0}
\caption{\label{crossSec_vs_rapidity} $C(\Delta \eta)$, as defined in (\ref{psRapidity}),
as a function of rapidity
width
$\de$ of the region $\Omega$. The dashed line is its limiting value,
$C (\de \rightarrow \infty) = \pi^2/6$.}
\end{figure}

We can contrast  the result (\ref{psa}) to what happens when $\nu=0$,
that is, for an inclusive,  non-global cross section.  In this case,
recalling that $\varepsilon=Q_\O/Q$,
we find in place of Eq.\ (\ref{psa}) the non-global logarithm
\ba
\label{psb}
{d\sigma_{\rm eik}\over d\, \varepsilon} = C_F C_A
\left(\frac{\alpha_s}{\pi}\right)^2 \, \frac{1}{\varepsilon} \,
\ln\left(\frac{Q}{Q_\O}\right)\, C(\Delta \eta)\, .
\ea
As anticipated, the effect of the transform is to
replace the non-global logarithm in $Q/Q_\O$, by a logarithm
of $1/( \varepsilon \nu)$.  We are now ready to generalize this
result, starting from the factorization properties of
the cross section near the two-jet limit.

\section{Factorization of the Cross Section for $e^+e^-$}

As in the previous Chapter, an arbitrary final state $N$ is the
union of substates associated with the leading regions:
\be
N=N_s \oplus N_{J_1} \oplus N_{J_2}\, .
\ee
Therefore, the  event shape $\bar f$ can, as above,
be written as a sum of contributions from the soft
and jet subdiagrams, Eq. (\ref{fbarf}).
In contrast, the energy flow weight, $f(N)$, depends only on
particles emitted at wide angles, and is hence insensitive
to collinear radiation:
\ba
f(N) = f(N_s)\, .
\ea

Analogous to Eq. (\ref{factoreps}) for the inclusive event shape we can write the cross section in convolution form, 
\ba
{d \bar{\sigma}(\varepsilon,\bar{\varepsilon},s,a)\over d \varepsilon\,
d\bar{\varepsilon}\, d\hat n_1}
&=&
{d \sigma_0 \over d\hat{n}_1}\
H(s,\hat{n}_1,\mu)\;
\int  d\bar{\varepsilon}_s\,
\bar{S}(\varepsilon,\bar{\varepsilon}_s,a,\mu) \,
\nonumber\\
&\ & \times
\prod_{c=1}^2\, \int  d\bar{\varepsilon}_{J_c}\,
\bar{J}_c(\bar{\varepsilon}_{J_c},a,\mu)\,
\delta(\bar{\varepsilon}- \bar{\varepsilon}_{J_1}-\bar{\varepsilon}_{J_2}-
\bar{\varepsilon}_s)\, ,
\label{factorcorr}
\ea
where we have made the same approximations listed in Sec. \ref{sec:approxincl}. The difference between (\ref{factoreps}) and (\ref{factorcorr}) lies in the soft function, which now depends on the energy flow $\varepsilon$. 
\be
\bar{S}(\varepsilon,\bar{\varepsilon}_s,a,\mu)
=
\sum_{N_s}\; {\cal S}(N_s,\mu)\, \delta(\varepsilon-f(N_s)) \,
\delta(\bar{\varepsilon}_s-\bar f(N_s,a))
\label{firstSdefcorr}
\ee
We will discuss the explicit construction of the new soft function below. 

Upon taking the Laplace
moments with respect to $\bar{\varepsilon}$, (\ref{factorcorr}) becomes a simple product in moment space:
\ba
    {d \sigma(\varepsilon,\nu,s,a)\over d \varepsilon \,d\hat n_1}
& = &  \int_0^\infty d\bar{\varepsilon}\, e^{- \nu \,\bar{\varepsilon}}\,
{d \bar{\sigma}(\varepsilon,\bar{\varepsilon},a)\over d \varepsilon\,
d\bar{\varepsilon}\,
d\hat n_1}
\label{trafocorr} \nonumber
\\
& = & {d \sigma_0 \over d\hat{n}_1}\
H(s,\hat{n}_1,\mu)\;  S(\varepsilon,\nu,a,\mu) \,\prod_{c=1}^2\,
J_c(\nu,a,\mu).
\label{trafosigcorr}
\ea

Since radiation at wide angles decouples from the jet and the hard scattering functions, these are not affected by measuring the energy flow $\varepsilon$. Their construction is therefore the same as discussed in Sections \ref{sec:sdf} and \ref{sec:jets}. The construction of the soft function needs to be amended to include dependence on the energy flow.

\subsection{The Soft Function}

Because wide-angle, soft
radiation is independent of the internal
jet evolution, products of
nonabelian phase operators 
generate the same wide-angle radiation as the full jets. Since $\O$ is at wide angles from the jets, we can define for the two-jet cross section at measured
$\varepsilon$ and $\bar \varepsilon_{\mbox{\tiny eik}}$
\ba
\bar{\sigma}^{(\mbox{\tiny 
eik})}\left(\varepsilon,\bar{\varepsilon}_{\mbox{\tiny eik}}, a ,\mu
\right)\!\!\!
&\!\! \equiv  \!\!\!& \!\!\!{1\over \Ncol}\
\sum_{N_{\mbox{\tiny eik}}} \left< 0
\left| \Phi^{(\bar {\rm q})}_{\beta_2}{}^\dagger(\infty,0;0)
\Phi^{(\rm q)}_{\beta_1}{}^\dagger(\infty,0;0)
\right| {N_{\mbox{\tiny eik}}} \right>
\nonumber \\
& \ & \hspace{5mm} \times\;
\left< N_{\mbox{\tiny eik}}
\left| \Phi^{(\rm q)}_{\beta_1}(\infty,0;0)
\Phi_{\beta_2}^{(\bar{\rm q})}(\infty,0;0)
\right| 0 \right>  \nonumber \\
& &  \hspace{5mm} \times\; \delta\left(\varepsilon - f(N_{\mbox{\tiny eik}})
\right)
\delta\left(\bar{\varepsilon}_{\mbox{\tiny eik}} - 
\bar{f}(N_{\mbox{\tiny eik}},a) \right)
\nonumber\\
&=& \delta(\varepsilon)\,  \delta(\bar{\varepsilon}_{\mbox{\tiny 
eik}}) +{\cal O}(\alpha_s)\,
.
\label{eikdefcorr}
\ea
 The sum is over all final states $N_{\mbox{\tiny eik}}$ in
the eikonal cross section. $\mu$ denotes the renormalization scale which is set equal to the factorization scale. As in the previous chapter, the shape function
$\bar{\varepsilon}_{\mbox{\tiny eik}}$
is defined by $\bar{f}(N_{\rm eik},a)$ as in Eqs.\  (\ref{barfdef}) and 
(\ref{2jetf}),
separately for each of the hemispheres around the jets.

In order to avoid double counting of collinear radiation, we subtract eikonal jets as defined in Eq. (\ref{eikjetdef}). The eikonal cross section (\ref{eikdefcorr}) then
factorizes as
\ba
\bar{\sigma}^{(\mbox{\tiny
eik})}\left(\varepsilon,\bar{\varepsilon}_{\mbox{\tiny eik}},a,\mu \right)
& \equiv  &
\int  d \bar{\varepsilon}_s \,
\bar{S}\left(\varepsilon,\bar{\varepsilon}_s,a,\mu \right)
\prod\limits_{c = 1}^2 \int d \bar{\varepsilon}_c \,
\bar{J}_c^{(\mbox{\tiny eik})}\left(\bar{\varepsilon}_c,a,\mu \right)\; \nonumber \\
& & \qquad \times
   \delta \left(\bar{\varepsilon}_{\mbox{\tiny eik}} -
\bar{\varepsilon}_s-\bar{\varepsilon}_1-\bar{\varepsilon}_2 \right) .
\label{eikfactcorr}
\ea
In  Laplace
transform space (\ref{primedef}) we can solve for the soft function as
\be
S \left(\varepsilon,\nu,a,\mu\right) =
\frac{\sigma^{(\mbox{\tiny eik})}\left(\varepsilon,\nu,a,\mu \right) }
{\prod\limits_{c = 1}^2 J_c^{(\mbox{\tiny eik})}\left(\nu,a,\mu\right) }
=\delta(\varepsilon)+{\cal O}(\alpha_s)\, .
\label{s0corr}
\ee
The soft function retains $\nu$-dependence through soft
emission, which is also restricted by the weight function
$\varepsilon$.  

Following the same argumentation as in Sec. \ref{sec:soft} we can deduce the arguments of the soft function in transform space as
\be
S\left(\varepsilon,\nu,a,\mu \right) =
S\left(\frac{\varepsilon \sqrt{s}}{\mu},\varepsilon\nu,
\frac{\sqrt{s}}{\mu \nu} \, \left( \zeta_c \right)^{1-a},
a, \as(\mu)
\right)\, .
\label{Sargscorr}
\ee

\section{Resummation for $e^+e^-$ Shape/Flow Correlations}

Summarizing the results so far, we can rewrite Eq. (\ref{trafosigcorr})
in terms of the hard, jet and soft functions identified above,
which depend on the kinematic variables and the moment $\nu$
according to Eqs.\ (\ref{harddef}), (\ref{primedef}) and (\ref{Sargscorr})
respectively,
\ba
\frac{d \sigma \left(\varepsilon,\nu,s ,a\right)}{d\varepsilon\,d
\hat{n}_1 }
&=&
  {d \sigma_0 \over d\hat{n}_1}\ H \left(
\frac{\sqrt{s}}{\mu},\frac{p_{J_c} \cdot \xic}{\mu},\hat{n}_1,\as(\mu)
\right)\,\nonumber \\
& \ &\hspace{15mm} \times\
\prod_{c=1}^2\;
J_c\left(\frac{p_{J_c} \cdot \xic}{\mu}, \frac{\sqrt{s}}{\mu \nu} \,
(\zeta_c)^{1-a},
a,\as(\mu) \right)\,
\nonumber \\
& \ &\hspace{15mm} \times\
   S\left(\frac{\varepsilon \sqrt{s}}{\mu}, \varepsilon \nu,
\frac{\sqrt{s}}{\mu \nu}\left(\zeta_c \right)^{1-a},
a, \as(\mu)
\right)\, .
   \label{factorcomcorr}
\ea
As above, the natural scale for the coupling
in the hard scattering $H$ is $\sqrt{s}/2$. Setting $\mu = \sqrt{s}/2$ in Eq. (\ref{factorcomcorr}) introduces large logarithms of $\varepsilon$ in the soft function and large logarithms of $\nu$ in the jet and soft functions.

\subsection{Energy Flow Dependence}

To resum large logarithms of $\varepsilon$ in the soft function, we use the renormalization group equation
\be
\mu \frac{d}{d \mu} \frac{d \sigma \left(\varepsilon,\nu,s ,a\right)}{d\varepsilon\,d
\hat{n}_1 } = 0 \label{muev}
\ee
which follows from the independence of the physical correlation of the factorization scale. 
Applying Eq.\ (\ref{muev}) to the factorized correlation (\ref{factorcomcorr}), we
derive the following consistency conditions, which are themselves 
renormalization
group equations:
\ba
\mu \frac{d}{d \mu}\;
\ln\, S\left(\frac{\varepsilon \sqrt{s}}{\mu}, \varepsilon \nu,
\frac{\sqrt{s}}{\mu \nu} (\zeta_c)^{1-a},
a, \as(\mu) \right)
& = & -
\gamma_s\left(\as(\mu)\right),
\label{softmu}
\\
\mu \frac{d}{d \mu}\;
\ln\, J_c\left(\frac{ p_{J_c} \cdot \xic }{\mu },
  \frac{\sqrt{s}}{\mu \nu} \, (\zeta_c)^{1-a} ,
a,\as(\mu) \right) & = & - \gamma_{J_c}\left(\as(\mu)\right),
\label{jetmu}
\\
\mu \frac{d}{d \mu}\; \ln\,  H\left( \frac{\sqrt{s}}{\mu},\frac{p_{J_c} \cdot
\xic}{\mu},\hat{n}_1,\as(\mu) \right)
&=& \gamma_s\left(\alpha_s(\mu)\right)
+ \sum_{c=1}^2\gamma_{J_c}\left(\alpha_s(\mu)\right)\, . \nonumber \\
& &
\label{Hmu}
\ea
The anomalous dimensions $\gamma_d$, $d=s,\, J_c$ can
depend only on variables held in common between  at least two
of the functions.  Because each function is infrared safe,
while ultraviolet divergences are present only in virtual
diagrams, the anomalous dimensions cannot depend on
the parameters $\nu$, $\varepsilon$ or $a$.  This leaves
as arguments of the $\gamma_d$ only
the  coupling $\as(\mu)$, which we exhibit, and $\zeta_c$, which
we suppress for now.

Solving Eqs. (\ref{softmu})
and (\ref{jetmu}) we find
\ba
S\left(\frac{\varepsilon \sqrt{s}}{\mu}, \varepsilon \nu,
\frac{\sqrt{s}}{\mu \nu} \left(\zeta_c \right)^{1-a},
a, \as(\mu) \right)
& = &
S\left(\frac{\varepsilon \sqrt{s}}{\mu_0}, \varepsilon \nu,
\frac{\sqrt{s}}{\mu_0 \nu} \left( \zeta_c \right)^{1-a},
a, \as(\mu_0) \right) \, \nonumber \\
& & \qquad \times e^{-\int\limits_{\mu_0}^\mu \frac{d
\lambda}{\lambda} \gamma_s\left(\as(\lambda)\right)},
\label{softevol}
\nonumber \\
&\ & \\
J_c \left(\frac{ p_{J_c} \cdot \xic }{\mu},
  \frac{\sqrt{s}}{\mu \nu} \, \left(\zeta_c \right)^{1-a} , a, \as(\mu)
\right)
& = &
J_c \left( \frac{p_{J_c} \cdot \xic }{\tilde{\mu}_0 },
\frac{\sqrt{s}}{\tilde{\mu}_0 \nu} \, \left(\zeta_c \right)^{1-a} ,
a,\as(\tilde{\mu}_0) \right)
   \,
   \nonumber \\
& & \qquad \times e^{-\int\limits_{\tilde{\mu}_0}^\mu \frac{d \lambda}{\lambda}
\gamma_{J_c}\left(\as(\lambda)\right)}\, , \nonumber
\\
&\ &\label{jetevol}
\ea
for the soft and jet functions.  As suggested above, we will eventually pick
$\mu\sim \sqrt{s}$
to avoid large logs in $H$.
Using these expressions in Eq. (\ref{factorcomcorr}) we can avoid
logarithms of $\varepsilon$ or $\nu$ in the soft function, by evolving from
$\mu_0 = \varepsilon \sqrt{s}$ to the factorization scale $\mu \sim \sqrt{s}$.
No choice of $\tilde{\mu}_0$, however, controls all logarithms of $\nu$ in
the jet functions.  Leaving $\tilde \mu_0$ free, we find for the
cross section (\ref{factorcomcorr}) the
intermediate result
\ba
   \label{resumecorr}
\frac{d \sigma \left(\varepsilon, \nu,s ,a\right)}{d\varepsilon \, d
\hat{n}_1 }
&=&
{d \sigma_0 \over d\hat{n}_1}\ H \left( \frac{\sqrt{s}}{\mu},\frac{p_{J_c}
\cdot \xic}{\mu},\hat{n}_1,\as(\mu) \right)\,
\nonumber\\
&\ & \hspace{-15mm} \times\; S\left(1, \varepsilon \nu, (\zeta_c)^{1-a}, a
,\as(\varepsilon \sqrt{s})
\right)\,
\exp\left\{ -\int\limits_{\varepsilon \sqrt{s}}^{\mu} \frac{d
\lambda}{\lambda} \, \gamma_s\left(\as(\lambda)\right)\right\}
\\
&\ & \hspace{-15mm} \times\;
J_c \left( \frac{p_{J_c} \cdot \xic }{\tilde{\mu}_0 },
\frac{\sqrt{s}}{\tilde{\mu}_0 \nu} \, \left(\zeta_c \right)^{1-a} ,
a,\as(\tilde{\mu}_0) \right)
   \,\exp\left\{-\int\limits_{\tilde{\mu}_0}^\mu \frac{d \lambda}{\lambda}
\gamma_{J_c}\left(\as(\lambda)\right)\right\} \, . \nonumber
\ea
We have avoided introducing logarithms of $\varepsilon$ into the jet functions,
which originally only depend on $\nu$, by evolving the soft and the
jet functions independently.
The choice of $\mu_0=\varepsilon\sqrt{s}$ or
$\sqrt{s}/\nu$ for the soft function is to some extent a matter of
convenience,
since the two choices differ by logarithms of $\varepsilon\nu$.
In general, if we choose $\mu_0=\sqrt{s}/\nu$, logarithms
of $\varepsilon\nu$ will appear multiplied by coefficients that reflect
the
size of region $\O$.  An example is Eq.\ (\ref{ps0}) above.
When $\O$ has a small angular size, $\mu_0=\sqrt{s}/\nu$ is generally the
more natural choice, since then logarithms in $\varepsilon\nu$ will
enter with small weights.  In contrast, when $\O$ grows to cover most
angular directions, as in
the study of rapidity gaps \cite{Oderda:1999kr}, it is more
natural to choose $\mu_0 = \varepsilon\sqrt{s}$.

\subsection{The Resummed Correlation}

The remaining unorganized logarithms reside in the jet functions. Since the definition of the jet functions is unchanged compared to the inclusive event shape of the previous chapter, we can immediately use the conclusions of Section \ref{sec:jetinclevol} to resum these logarithms. Using Eq. (\ref{jetxienda}) in Eq. (\ref{resumecorr}), we obtain after setting $\mu = \sqrt{s}/2$
\ba
\frac{d \sigma \left(\varepsilon, \nu,s,a \right)}{d \varepsilon\, d
\hat{n}_1 }
&=&
{d \sigma_0 \over d\hat{n}_1}\ H \left(\frac{2 \, p_{J_c}
\cdot \xic}{\sqrt{s}},\hat{n}_1,\as\left(\frac{\sqrt{s}}{2}\right) \right)\,
  \nonumber \\
& \ & \hspace*{-2cm}
\times \, S\left(1, \varepsilon \nu, (\zeta_c)^{1-a}, a,\as(\varepsilon
\sqrt{s} ) \right)
\, \exp \left\{  - \int\limits_{\varepsilon \sqrt{s}}^{\sqrt{s}/2} \frac{d
\lambda}{\lambda} \gamma_s\left(\as(\lambda)\right) \right\} \nonumber \\
& \ & \hspace*{-2cm}
\times \, \prod_{c=1}^2\, J_c \left(1,1,a,\as\left(\frac{\sqrt{s}}{2 \,
\zeta_0}\right) \right)
   \exp \left\{ - \int\limits_{\sqrt{s}/(2 \, \zeta_0)}^{\sqrt{s}/2}
  \frac{d \lambda}{\lambda} \gamma_{J_c} \left(\as(\lambda)\right) \right\}
\nonumber \\
& \ & \hspace*{-2cm}
\times \, \exp \left\{ -\int\limits_{\sqrt{s}/(2\, \zeta_0) }^{p_{J_c} \cdot
\xic}
   \frac{d \lambda}{\lambda} \Bigg[B'_c\left(c_1,c_2,a,
\as\left(c_2 \lambda \right) \right) \right. \nonumber \\
& & \qquad \left. +  2 \int\limits_{c_1 \frac{s^{1-a/2}
}{ \nu
(2\,\lambda)^{1-a} } }^{c_2\, \lambda}\frac{d \lambda'}{\lambda'} A'_c\left(
c_1,
a,\as\left(\lambda'\right) \right) \Bigg] \right\}\, . \nonumber \\
& & \label{evolend}
\ea

Alternatively, we can combine all jet-related exponents in Eq.
(\ref{evolend}) in the correlation.
As we have verified in Section
\ref{sec:gauge}, the cross section is independent of the choice of $\xi_c$.
As a result, we can choose $
p_{J_c} \cdot \xic = \frac{\sqrt{s}}{2}\,.$
This choice allows us to combine $\gamma_{J_c}$ and
$B'_c$ in Eq. (\ref{evolend}),
\ba
\frac{d \sigma \left(\varepsilon, \nu,s,a \right)}{d \varepsilon\, d
\hat{n}_1 }
\!\!&=&\!\!
{d \sigma_0 \over d\hat{n}_1}\ H \left(1,
\hat{n}_1,\as\left(\frac{\sqrt{s}}{2}\right) \right)\,
  \nonumber \\
& \ & \hspace*{-3.3cm}
\times \, S\left(1, \varepsilon \nu, 1, a,\as(\varepsilon \sqrt{s} ) \right)
\, \exp \left\{  - \int\limits_{\varepsilon \sqrt{s}}^{\sqrt{s}/2} \frac{d
\lambda}{\lambda} \gamma_s\left(\as(\lambda)\right) \right\} \,
\prod_{c=1}^2\, J_c \left(1,1,a,\as\left(\frac{\sqrt{s}}{2 \, \zeta_0}\right)
\right) \nonumber \\
& \ & \hspace*{-3.3cm}
\times \, \exp \left\{ -\int\limits_{\sqrt{s}/(2\, \zeta_0) }^{\sqrt{s}/2}
   \frac{d \lambda}{\lambda} \Bigg[ \gamma_{J_c} \left(\as(\lambda)\right) +
B'_c\left(c_1,c_2,a,
\as\left(c_2 \lambda \right) \right) \right. \nonumber \\
& & \qquad \left. +  2 \int\limits_{c_1 \frac{s^{1-a/2}
}{ \nu
(2\,\lambda)^{1-a} } }^{c_2\, \lambda}\frac{d \lambda'}{\lambda'} A'_c\left(
c_1,
a,\as\left(\lambda'\right) \right) \Bigg] \right\}\, , \nonumber \\
& & \label{evolendnoxi}
\ea
with $\zeta_0$ given by Eq. (\ref{zeta0}).

In Eqs. (\ref{evolend}) and (\ref{evolendnoxi}), the energy flow
$\varepsilon$ appears at the level
of one logarithm per loop, in $S$, in the first exponent.
Leading logarithms of $\varepsilon$ are
therefore resummed by knowledge of $\gamma_s^{(1)}$.
At the same time, $\nu$ appears in up to two logarithms per loop, as in the inclusive event shape. 

Only the soft
function $S$ in Eqs.\ (\ref{evolend}) and (\ref{evolendnoxi})
contains information on the geometry of $\O$. The exponents are
partially process-dependent, but geometry-independent. In other words, in Eq. (\ref{evolendnoxi}) the non-global effects are factored from the global event shape, residing only in the soft function and the corresponding anomalous dimension. Dokshitzer and Marchesini \cite{Dokshitzer:2003uw} have recently obtained the same result, by investigating how parton branching is affected by correlating the non-global measurement of energy flow with the shape of the jets. Their formalism extends the one presented here to include also non-global effects at leading logarithmic accuracy for $\varepsilon$ in the large $\Ncol$-limit. 

In the next 
section we will derive
explicit expressions for the quantities in (\ref{evolendnoxi}), suitable for
resummation to leading logarithm in $\varepsilon$ and next-to-leading
logarithm in $\nu$.

\section[Analytical Results for $e^+e^-$ Shape/Flow Correlations]{Analytical Results for $e^+e^-$ Shape/Flow \\ Correlations} \label{sec:analyticalshape}

Here we summarize the low-order calculations that
provide explicit expressions for the resummed shape/flow correlations and
inclusive event shape distributions at next-to-leading
logarithm in $\nu$ and leading logarithm in $\varepsilon$
(we refer to this level collectively as NLL below).   Furthermore, we exhibit the expressions for the
correlation that we will evaluate numerically in Sec.\ \ref{sec:numerics}.

\subsection{Results for the Soft Function}\label{sec:softrescorr}

As already noted above, the soft function is the only function in the shape/flow correlation that differs from the functions entering the corresponding inclusive event shape because only the soft function depends on the geometry of $\O$. Below we give the results for the lowest order soft function for two choices of $\O$. The remaining functions in (\ref{evolend}) are the same as the ones listed in Sec. \ref{sec:resan} for the inclusive event shape.

The soft function is normalized to $S^{(0)}(\varepsilon) =
\delta(\varepsilon)$
as can be seen from (\ref{s0corr}).
For non-zero $\varepsilon$,
$d\sigma /d \varepsilon$ is given at lowest order  by
\be
S^{(1)} \left( \varepsilon \neq 0, \Omega\right) =
C_F \frac{1}{\varepsilon}
\int\limits_\O  d \mbox{PS}_2\,
\frac{1}{2 \pi} \frac{\beta_1 \cdot \beta_2}{\beta_1
\cdot \hat{k} \,\beta_2 \cdot \hat{k} }\, ,
\label{oneLoopSoft}
\ee
where $\mbox{PS}_2$ denotes the two-dimensional angular phase space
to be integrated over region $\O$, and $\hat k \equiv k / \omega_k$.
We emphasize again that the soft function contains the only
geometry-dependence of the
cross section. Also, $S^{(1)}$ for $\varepsilon \neq 0$ is independent of
$\nu$
and $a$.

As an example, consider a cone with opening angle $2 \delta$,
centered at angle $\alpha$ from jet 1.  In this case,
the lowest-order soft function is given by
\be
S^{(1)} \left( \varepsilon \neq 0, \alpha, \delta \right) =
C_F \frac{1}{\varepsilon}
\ln \left(\frac{1-\cos^2 \alpha }{\cos^2 \alpha - \cos^2 \delta}\right).
\label{softcone}
\ee
Similarly, we may choose $\O$ as a ring
extending angle $\delta_1$ to the right and $\delta_2$ to
the left of the plane perpendicular to the jet directions
in the center-of-mass.  In this case, we obtain
\be
S^{(1)} \left( \varepsilon \neq 0, \delta_1, \delta_2 \right) =
   C_F \frac{1}{\varepsilon}
\ln \left(
\frac{(1+ \sin \delta_1)}{(1-\sin \delta_1)}\frac{(1+ \sin
\delta_2)}{(1-\sin \delta_2)} \right)
= C_F \frac{2}{\varepsilon}\, \Delta\eta\, ,
\label{deltaeta2}
\ee
with $\Delta\eta$ the rapidity spanned by the ring.
For a ring centered around the center-of-mass ($\delta_1 =
\delta_2 = \delta$) the angular integral reduces to the form that we
encountered in the example  of Sec.\ \ref{sec:loe},
and that we will use in our
numerical examples of Sec.\ \ref{sec:numerics}, with $\Delta\eta$
given by Eq.\ (\ref{deltaeta1}).

\subsection{Closed Expressions for the Correlation}

Given the explicit results above and in Sec. \ref{sec:resan}, the integrals in  the
exponents of the resummed correlation, Eq.\ (\ref{evolend}), may
be easily performed in closed form.
We give the analytic results for the exponents
of Eq. (\ref{evolend}), as defined in
Eqs. (\ref{E1})\footnote{Here we have for the correlation a lower limit of $\varepsilon \sqrt{s}$ instead of $\sqrt{s}/\nu$ in the first term.} and  (\ref{E2}). As in Eq. (\ref{xiid}),
we identify $p_{J_c} \cdot \xic$ with
$\sqrt{s}/2$.
\ba
e^{E_1(a)} & = & \left(
\frac{ \as(\sqrt{s}/ 2)}{\as(\varepsilon \sqrt{s})}\right)^{\frac{4
C_F}{\beta_0}}
\left( \frac{\as\left(\frac{\sqrt{s}}{2 \,\zeta_0}\right)}{ \as(\sqrt{s}/2)}
  \right)^{\frac{6
C_F}{\beta_0} } , \label{E1result} \\
e^{E_2(a)} & = &   \left( \frac{ \as(c_2 \, \sqrt{s}/2)}
{\as\left(\frac{c_2 \,\sqrt{s}}{2 \, \zeta_0} \right)} \right)^{\frac{4
C_F}{\beta_0} \kappa_1(a)}
   \left( \frac{  \as\left(\frac{c_1 \, \sqrt{s}}{2 \,\zeta_0 }\right)} {
\as\left(\frac{c_1 \, \sqrt{s}}{\nu}\right)}
\right)^{\frac{1}{a-1} \frac{4 C_F}{\beta_0} \kappa_2(a)} \,
 \nonumber \\
& & \quad \times \,\left( \frac{ \as(c_2 \, \sqrt{s}/2)}{\as\left(\frac{c_1 \, \sqrt{s}}{2 \,
\zeta_0 }\right)} \right)^{
\frac{1}{2-a} \, \frac{8
C_F}{\beta_0} \, \ln (\nu / 2)}, \label{E2result}
\ea
with
\ba
\kappa_1(a) \!\!& = &\!\! \ln \left(\frac{4}{c_2^2 e}\right) +
\frac{4\pi}{\beta_0} \left[\as\left(\frac{c_2 \, \sqrt{s}}{2 \,\zeta_0}\right)
\right]^{-1}
- \frac{2 K}{\beta_0}
\nonumber \\
& & \qquad - \frac{\beta_1}{2 \beta_0^2} \ln \left( \left(\frac{\beta_0}{4 \pi
e}\right)^2
\as\left(\frac{c_2\, \sqrt{s}}{2}\right) \as\left(\frac{c_2\, \sqrt{s}}{2 \,
\zeta_0}\right)\right), \nonumber \\
& & \label{C1} \\
\kappa_2(a) \!\!& = & \!\!(1 - a - 2 \gamma_E) + \frac{4\pi}{\beta_0}
\left[\as\left(\frac{\sqrt{s}}{\nu}\right)\right]^{-1}
- \frac{2 K}{\beta_0}
\nonumber \\
& & \qquad - \frac{\beta_1}{2 \beta_0^2} \ln \left( \left(\frac{\beta_0}{4 \pi
e}\right)^2
\as\left(\frac{c_1 \, \sqrt{s}}{\nu}\right) \as\left(\frac{c_1 \, \sqrt{s}}{2
\, \zeta_0}\right)\right). \nonumber \\
& & \label{C2}
\ea
  We have used the two-loop running coupling, when appropriate,
to derive Eqs.\ (\ref{E1result}) - (\ref{C2}).
The results are expressed in terms of the one-loop running coupling (\ref{1as}),
and the first two coefficients in the expansion of the QCD beta-function,
$\beta_0$ and $\beta_1$, Eqs. (\ref{beta0}) and (\ref{beta1}), respectively.

Combining the expressions for the exponents, Eqs.\ (\ref{E1result}) and
(\ref{E2result}), for the Born cross section,
Eq.\ (\ref{bornCross}), and for the soft function, Eq.\ (\ref{oneLoopSoft}), in
Eq.\ (\ref{evolend}),
the complete differential cross section, at LL in $\varepsilon$ and at NLL
in $\nu$, is given by
\ba
\frac{d \sigma \left(\varepsilon, \nu,s ,a\right)}{d\varepsilon\, d
\hat{n}_1 }
&=&
\Ncol \left( \sum_{\rm f} Q_{\rm f}^2 \right) \frac{\pi \alpha_{\mbox{\tiny
em}}^2}{2 s} \left(1+ \cos^2 \theta \right) C_F \frac{\as(\varepsilon
\sqrt{s})}{\pi} \frac{1}{\varepsilon} \qquad \qquad \qquad \nonumber \\
& & \hspace*{-32mm}\times \, \int\limits_\O  d \mbox{PS}_2\,
\frac{1}{2 \pi} \frac{\beta_1 \cdot \beta_2}{\beta_1
\cdot \hat{k} \, \beta_2 \cdot \hat{k} } 
\left(
\frac{ \as\left(\frac{\sqrt{s}}{2}\right)}{\as(\varepsilon
\sqrt{s})}\right)^{\frac{4 C_F}{\beta_0}}
\left( \frac{\as\left(\frac{\sqrt{s}}{2 \,\zeta_0}\right)}{
\as\left(\frac{\sqrt{s}}{2}\right)}
  \right)^{\frac{6
C_F}{\beta_0} }
  \nonumber \\
& & \hspace*{-32mm}\times \, \left( \frac{ \as\left(c_2 \, \frac{\sqrt{s}}{2}\right)}
{\as\left(\frac{c_2 \,\sqrt{s}}{2 \, \zeta_0} \right)} \right)^{\frac{4
C_F}{\beta_0} \kappa_1(a)}
   \left( \frac{  \as\left(\frac{c_1 \, \sqrt{s}}{2 \,\zeta_0 }\right)} {
\as\left(\frac{c_1 \, \sqrt{s}}{\nu}\right)}
\right)^{\frac{1}{a-1} \frac{4 C_F}{\beta_0} \kappa_2(a)} \,
\left( \frac{ \as\left(c_2 \, \frac{\sqrt{s}}{2}\right)}{\as\left(\frac{c_1
\, \sqrt{s}}{2 \, \zeta_0 }\right)} \right)^{
\frac{1}{2-a} \, \frac{8
C_F}{\beta_0} \, \ln \left(\frac{\nu}{2}\right)}\hspace*{-2.0cm} . \nonumber \\
& &
\label{resulto1}
\ea
These are the expressions that we will numerically evaluate in the
next section.  We note that this is not the only possible closed form for the
resummed correlation at this level of accuracy.  When a full
next-to-leading order calculation for this set of event shapes
is given, the matching procedure of
\cite{Catani:1992ua,Catani:1991kz} may be more convenient.

\newpage
\section{Numerical Results for $e^+e^-$ Shape/Flow Correlations} \label{sec:numerics}

Here we show some representative examples of numerical results for the
correlation, Eq.\ (\ref{resulto1}). We pick the constants $c_i$ as in
Eq.\ (\ref{cipick}),
unless stated otherwise. The effect of different choices
is nonleading, and is numerically small, as we will see below.
In the following we choose the
region $\O$ to be a ring between the jets, centered in their
center-of-mass, with a width of $\Delta \eta = 2$, or equivalently, opening
angle $\delta \approx 50$ degrees (see Eq.\ (\ref{rapidity})).
The analogous cross section for
  a cone centered at 90 degrees from the jets
(Eq.\ (\ref{softcone})) has a similar behavior. 
In the following, the center-of-mass energy
$Q=\sqrt{s}$ is  chosen to be $100$ GeV.

Fig. \ref{num1} shows the dependence of the differential cross
section (\ref{evolend}), multiplied by $\varepsilon$
and normalized by the Born cross section,
$\frac{\varepsilon d \sigma/(d\varepsilon d\hat n_1)}{d
\sigma_0/d \hat n_1}$, on the measured energy $\varepsilon$ and
on the parameter $a$, at fixed $\nu$.
In Fig. \ref{num1} a), we plot $\frac{\varepsilon
d \sigma/(d\varepsilon d\hat n_1)}{d \sigma_0/d \hat n_1}$ for
$\nu = 10$, in Fig. \ref{num1} b) for $\nu  = 50$.
As $\nu$ increases, the radiation into the
complementary region $\bar \O$ is more restricted,
as illustrated by the comparison of Figs.
\ref{num1} a)  and b). Similarly, as $a$ approaches 1, the
cross section falls, because the jets are restricted to
be very narrow.  On the other hand,
as $a$ assumes more and more negative values at
fixed $\varepsilon$, the correlations (\ref{evolend}) approach a
constant value.  For $a$ large and negative, however, non-global
dependence on
$\ln\varepsilon$ and $|a|$ will emerge from higher order corrections
in the soft function, which we do not include in Eq.\ (\ref{resulto1}).

In Fig.\ \ref{numci} we investigate the sensitivity of the resummed
correlation,
Eq.\ (\ref{resulto1}), to our choice of the constants $c_i$. The effect of these constants is
of next-to-next-to-leading logarithmic order in the event shape. We plot the
differential cross section $\varepsilon
\frac{\varepsilon d \sigma/(d\varepsilon d\hat n_1)}{d
\sigma_0/d \hat n_1}$, at $Q = 100$ GeV, for fixed $\varepsilon = 0.05$ and
$\nu = 20$, as a function of $a$. The effects of changes in the $c_i$ are
of the order of a few percent for moderate values of $a$.

\begin{figure}[htb]
\vspace*{7mm}
\begin{center}
a) \hspace*{9.0cm}  \vspace*{-6mm} \\
\epsfig{file=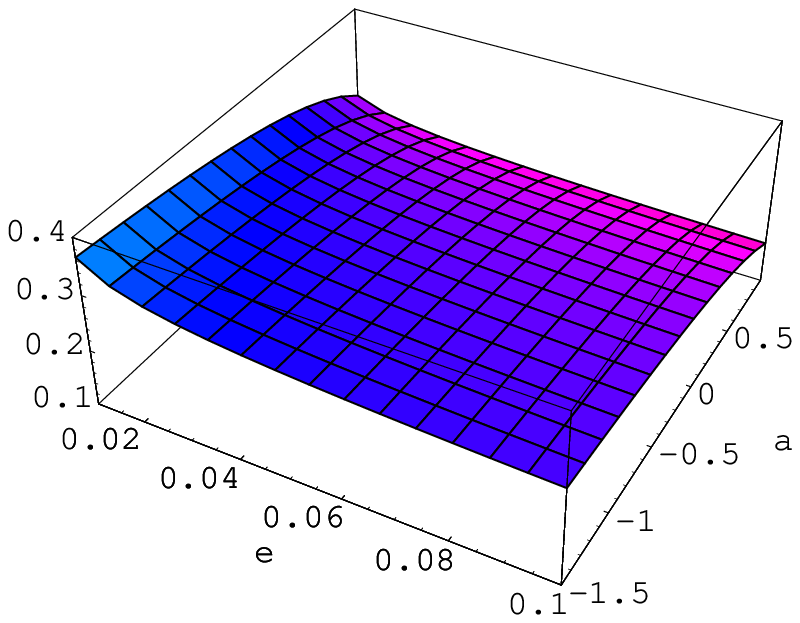,width=8.2cm,clip=0}  \vspace*{5mm}\\
b) \hspace*{9.0cm} \vspace*{-6mm} \\ 
\epsfig{file=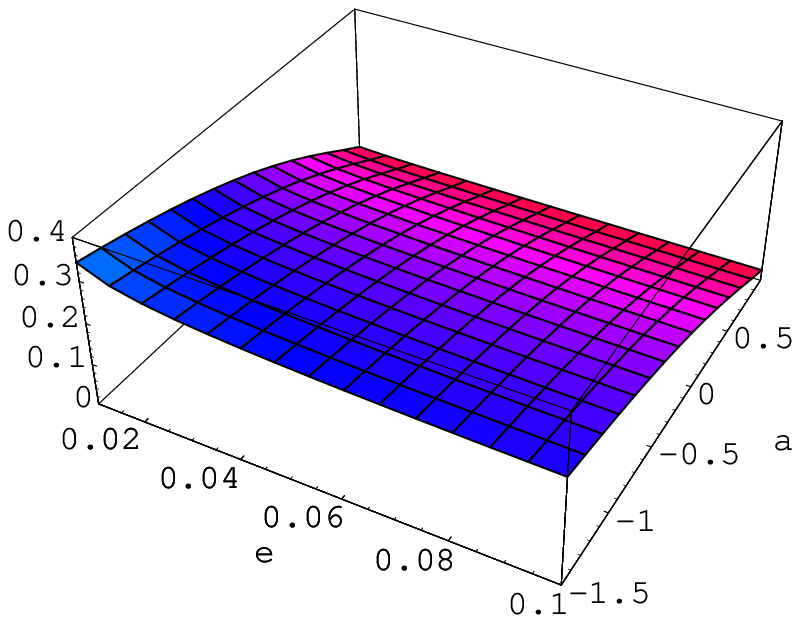,width=8.2cm,clip=0}
\caption{Differential cross section $\frac{\varepsilon d
\sigma/(d\varepsilon d\hat n_1)}{d \sigma_0/d \hat n_1}$,
normalized by the Born cross section, at $Q = 100$ GeV,
as a function of
$\varepsilon$ and $a$ at fixed $\nu$: a) $\nu = 10$, b) $\nu =
50$. $\O$ is a ring (slice) centered around the jets, with a
width of $\Delta \eta = 2$.} \label{num1}
\end{center}
\end{figure}
\clearpage

\begin{figure}[htb]
\begin{center}
\epsfig{file=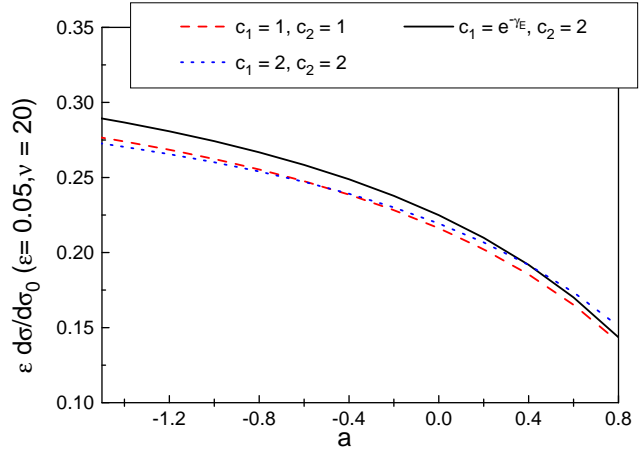,height=9cm,angle=270,clip=0}
\caption{Differential cross section $\frac{\varepsilon d
\sigma/(d\varepsilon d\hat n_1)}{d \sigma_0/d \hat n_1}$,
normalized by the Born cross section, at $Q = 100$ GeV,
as a function of
$a$ at fixed $\nu = 20$ and $\varepsilon = 0.05$.
  $\O$ is chosen as in Fig. \ref{num1}. Solid line: $c_1 = e^{-\gamma_E},\,
c_2 = 2$,
as in Eq.  (\ref{cipick}), dashed line: $c_1 = c_2 = 1$, dotted line:
$c_1 = c_2 = 2$. } \label{numci}
\end{center}
\end{figure}

Finally, we illustrate the sensitivity of these results to the flavor
of
the primary partons. For this purpose we study the corresponding ratio of the shape/flow 
correlation to the cross section for gluon jets produced
by a hypothetical color singlet source. Fig.\ \ref{num2} displays the ratio
of the differential cross section
$d\sigma^q(\varepsilon,a)/(d\varepsilon d \hat n_1)$, Eq.\ (\ref{resulto1}),
normalized
by the lowest-order cross section, to the
analogous quantity with gluons as
primary partons in the outgoing jets, again at
$Q = 100$ GeV.
This ratio is multiplied by $C_A/C_F$ in the figure to compensate for
the difference in the normalizations of the lowest-order soft functions.
Gluon jets have wider angular extent, and hence are
suppressed relative to quark jets with increasing  $\nu$ or $a$,
as can be seen by
comparing Figs. \ref{num2} a) and b). Fig. \ref{num2} a) shows the ratio
at $\nu = 10$, and Fig. \ref{num2} b) at $\nu = 50$.
These results suggest sensitivity to
the more complex color and flavor flow
characteristic of hadronic scattering \cite{Berger:2001ns,Kidonakis:1998nf}. We will discuss the extension of the above formalism to correlations with hadronic initial states in the next section.

\begin{figure}[htb]
\vspace*{6mm}
\begin{center}
a) \hspace*{9.0cm}  \vspace*{-6mm} \\
\epsfig{file=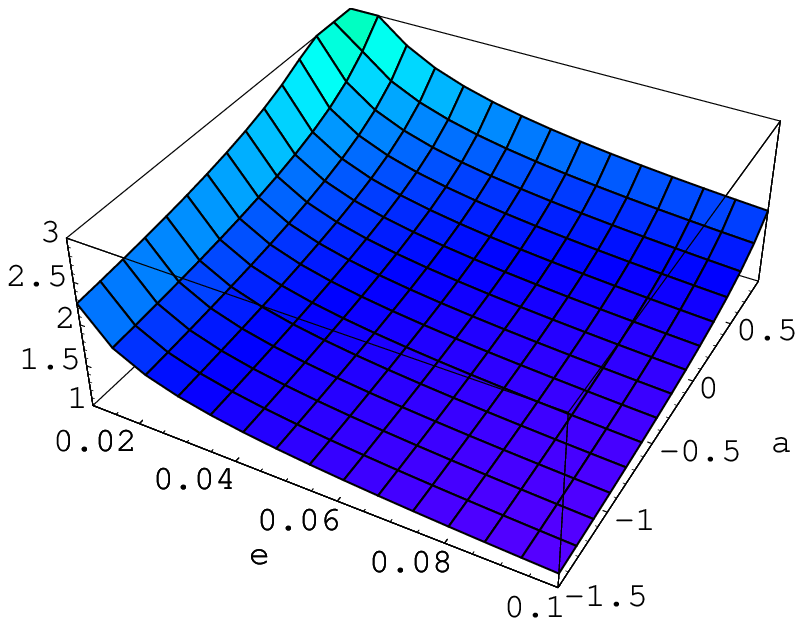,width=8.2cm,clip=0}  \vspace*{5mm}\\
b) \hspace*{9.0cm} \vspace*{-6mm} \\ 
\epsfig{file=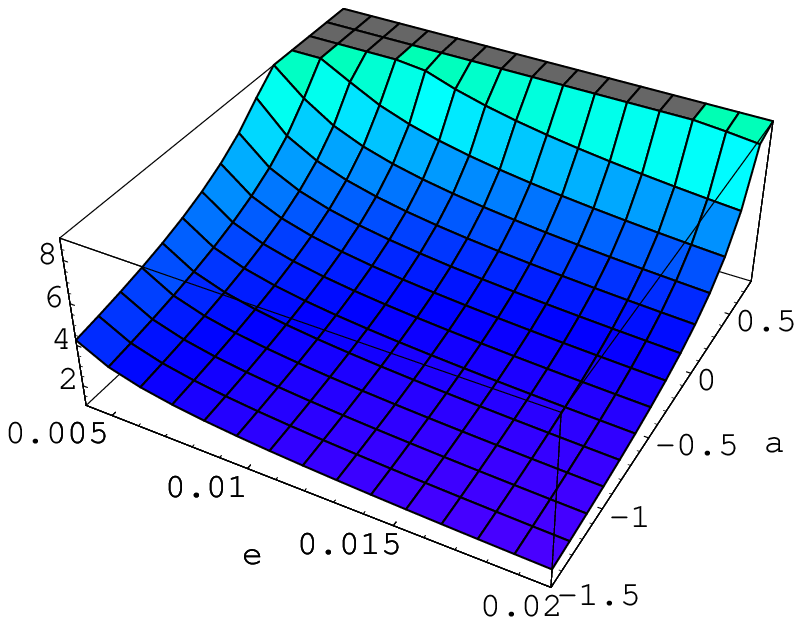,width=8.2cm,clip=0}
\caption[Ratios of differential cross sections for quark to
gluon jets  at
$Q = 100$ GeV
as a
function of $\varepsilon$ and $a$ at fixed $\nu$: a) $\nu = 10$,
b) $\nu = 50$.]{Ratios of differential cross sections  for quark to
gluon jets $\frac{C_A}{C_F} \left(\frac{\varepsilon d
\sigma^q/(d\varepsilon d\hat n_1)}{d \sigma_0^q/d \hat n_1}\right)
\left(\frac{\varepsilon d
\sigma^g/(d\varepsilon d\hat n_1)}{d \sigma_0^g/d \hat n_1}\right)^{-1}$ at
$Q = 100$ GeV
as a
function of $\varepsilon$ and $a$ at fixed $\nu$: a) $\nu = 10$,
b) $\nu = 50$. $\O$ as in Fig. \ref{num1}, $c_1$ and $c_2$ as in Eq.
(\ref{cipick}).}
\label{num2}
\end{center}
\end{figure}
\clearpage

\section{Extension to Hadronic Cross Sections}\label{sec:hadronic}

The study of interjet
radiation in hadronic cross sections may shed light
 on short-distance color and flavor flow,
and on the dynamics of hadronization.  Its use
in the  analysis of jet events in hadronic
scattering, to distinguish
new physics signals from QCD background, was
explored in \cite{Ellis:1996eu,Khoze:1996qb}.  
The discussion below, which was published in \cite{Berger:2001ns}, is 
closely related to
the treatment of rapidity gap
(``color singlet exchange")
events in Refs.\ \cite{Oderda:1999kr,Oderda:1998en},
for jets at high $p_T$ and large rapidity
separations. 

In the following we study hadronic events corresponding to Eq. (\ref{crossdef}), that is, the incoming partons are now hadrons:
\be
A + B \rightarrow J_1(p_{J_1}) + J_2(p_{J_2})  +
X_{\bar{\O}} \left(\bar{f}\right) + R_\O (Q_\O)\, ,
\label{crossdefhad}
\ee
 We refrain from giving an explicit construction of a shape function $\bar{f}$ suitable for accounting for all incoming (beam) and outgoing jets. Such a construction is certainly possible, although nontrivial due to the more involved kinematics. Below we assume the existence of the shape function $\bar{f}$ and instead concentrate on the study of the interrelation of energy and color flow. We reserve a full study for future work. 

\subsection{Refactorization of the Cross Section}
 
Collinear factorization theorems \cite{Collins:1987pm,Collins:gx}
ensure that we may write the cross section for (\ref{crossdefhad}) at fixed $\varepsilon = Q_\O/Q$ and $\bar{\varepsilon}$
as a sum of convolutions of parton
distribution functions $\phi$ (PDFs)
that incorporate long-distance dynamics,
with hard-scattering functions, $\omega$,
          that summarize short-distance dynamics,
\ba
\frac{d\sigma_{AB}}{d\vec p_1\, d \varepsilon d\bar{\varepsilon}}
&=&
\sum_{f_A,f_B} \int dx_A\, dx_B\;
{\phi}_{f_A/A}\left(x_A, \mu_F\right) {\phi}_{f_B/B}
\left(x_B,\mu_F \right)\nonumber\\
&\ & \hspace{15mm} \times\,
\omega_{f_Af_B}\left(x_Ap_A,x_Bp_B,\vec
p_1,  \varepsilon,\bar{\varepsilon},\mu_F,\alpha_s(\mu_F)\right)\, ,
\label{cofact}
\ea
with factorization scale $\mu_F$. Here the hard scale $Q$ is set by $p_T$, the transverse momentum of the observed jet.
The sum is over parton types, $f_A,f_B$.
Corrections to
this relation begin in general with powers
of $\Lambda^2_{\rm QCD}/Q^2_\O$, due to  multiple
scattering of partons.   Our analysis below, which
begins with Eq.\ (\ref{cofact}), is thus accurate
up to such corrections, and requires that the momenta of the outgoing jets are large compared to the hadronization scale $\sim \LQCD$.
We take the renormalization scale
equal to the factorization scale.

Soft gluon emission outside the angular extent of the jets
decouples from the kinematics of the hard scattering,
and from the internal evolution of the jets \cite{Collins:gx,Sterman:2002qn,Dokshitzer:nm,Bassetto:ma}.
As a result, we may express the partonic cross section in Eq.\ (\ref{cofact})
as a sum of terms, each characterized by a definite
number of jets produced at the hard scattering.
To lowest order in $\alpha_s$, only two jets are possible.
Thus, at leading order, the production
of the high-$p_T$ jets is given by the set of Born-level
$2\rightarrow 2$ partonic processes, which we label f,
\be
{\rm f}: \hspace{5mm} f_A+ f_B\rightarrow f_1+ f_2\, .
\label{pprocess}
\ee
We distinguish $q \bar{q} \rightarrow q \bar{q}$ ($f_1 = q$) from
$q \bar{q} \rightarrow \bar{q} q$  ($f_1 = \bar{q}$).
We may now write the single-jet
inclusive cross section at fixed $\varepsilon$ and $\bar{\varepsilon}$ as
\ba
\frac{d\sigma_{AB}}{d\eta dp_T  d \varepsilon d\bar{\varepsilon}}
&= & \sum_{\rm f} \int dx_Adx_B \;
{\phi}_{f_A/A}\left(x_A, \mu_F\right) {\phi}_{f_B/B}
\left(x_B,\mu_F\right)\, \nonumber \\
& & \qquad \times \, \delta\left( p_T-{\sqrt{\hat s}\over
2\cosh\hat\eta}\right)\,
\frac{d \hat{\sigma}^{\rm (f)}}
{d\hat\eta  d \varepsilon d \bar{\varepsilon}}\, ,
\label{hatsigfact}
\ea
with $\hat\eta=\eta-(1/2)\ln(x_A/x_B)$ the jet
rapidity in the partonic center-of-mass, where $\eta$ is defined in Eq. (\ref{rapidity}), and
$\hat s=x_Ax_Bs$.
As above, we neglect the effects
of recoil of the observed jet against relatively soft
radiation. 
 
The partonic cross section, $d \hat{\sigma}^{\rm (f)}/(d\hat\eta  d \varepsilon d \bar{\varepsilon})$ can be refactorized analogous to Eq. (\ref{factorcorr}), but now we have to take the non-trivial color flow into account. The regions that give leading contributions are as above a hard scattering, soft, and jet functions, two for the outgoing jets, and two jet functions for the beam jets. The latter jets must be defined to avoid double counting due to the parton distribution functions. Such a definition is quite non-trivial. Furthermore, the Glauber/Coulomb region discussed in Sec. \ref{sec:Glauber} has to be treated with care, since in shape/flow correlations the phase space for radiation is restricted. Nevertheless, with appropriate definitions of the event shape and the jet functions it is possible to avoid contour pinches in the Glauber/Coulomb region, as discussed in \cite{Kidonakis:1998bk}. Here we do not attempt to give a full treatment of the beam jets or of the construction of suitable shape functions. Instead, we concentrate on the relation between energy and underlying color flow.

As was observed, for example, in \cite{Berger:2001ns,Sterman:2002qn,Kidonakis:1998nf,Kidonakis:1998bk}, there is no unique way of defining color exchange in a finite amount of time since gluons of any energy, including soft gluons, carry octet color charge. We therefore expect the functions from which we construct the refactorized partonic cross section to be described by matrices in the space of possible color exchanges. This is because as the (re)factorization scale changes, gluons that were included in the hard function become soft and vice versa.
Due to intrajet coherence \cite{Dokshitzer:nm}, however, the evolution of the jets themselves is independent of the color exchanges. Once a jet is formed, collinear radiation cannot change its color structure.  Therefore, the refactorized partonic shape/flow correlation can be written in moment space as
\ba
    {d \hat{\sigma}^{\rm (f)}(\varepsilon,\nu,p_T,\mu_F)\over d \varepsilon \,d \hat{\eta}}
\!\!& \!\!= &\!\! 
H^{\rm (f)}_{LI}(p_T,\hat{\eta},\mu,\mu_F)\;  S^{\rm (f)}_{IL}(\varepsilon,\nu,\hat{\eta},\mu) \!\prod_{c=A,B,1,2}\!
J^{\rm (f)}_c(p_T,\hat{\eta},\nu,\mu,\mu_F). \nonumber \\
& &
\label{trafosighad}
\ea
analogous to the corresponding correlation in $e^+e^-$ events, Eq. (\ref{trafosigcorr}), but now the hard and soft functions, $H$ and $S$, respectively, are matrices in the space of color flow. Repeated indices in color space, $L,I,$ are summed over. The superscripts $(\rm f)$ label the underlying partonic process, as in (\ref{pprocess}). $\mu$ is a refactorization scale, not necessarily equal to the factorization scale in Eq. (\ref{hatsigfact}). The large scale in the problem is here, from (\ref{hatsigfact}), the transverse momentum of the observed jet $p_T$.

In Eq.\ (\ref{trafosighad}), the hard scattering function
$H_{LI}$ begins at order $\alpha_s^2$,
while the soft function $S_{IL}$ begins
          at zeroth order. Compared to the $e^+e^-$ correlation, Eq. (\ref{trafosigcorr}), we have absorbed the lowest order Born cross section into the hard scattering function due to the flavor dependence (\ref{pprocess}). In Appendix \ref{app:hard} we list the color bases and the hard scattering matrices for all possible partonic subprocesses for proton-antiproton collisions. Let us now turn to the discussion of the soft function. 

\subsection{The Soft Function}

As we have seen above, wide-angle soft radiation decouples from
jet evolution \cite{Collins:ta}.  Soft
radiation away from jet directions is equally well
described by radiation from a set of path-ordered
exponentials -- nonabelian phase operators
or Wilson lines -- that replace each of the partons
involved  in the hard scattering (four here),
\ba
\Phi_{\beta}^{(f)}(\infty,0;x)
&=&
P\exp\left(-ig_s\int^{\infty}_{0}d{\lambda}\;
{\beta}{\cdot} {\mathcal{A}}^{(f)} ({\lambda}{\beta}+x)\right)\, ,
\nonumber\\
\Phi_{\beta'}^{(f')}(0,-\infty;x)
&=&
P\exp\left(-ig_s\int_{-\infty}^{0}d{\lambda}\;
{\beta'}{\cdot} {\mathcal{A}}^{(f')} ({\lambda}{\beta'}+x)\right)\, ,
\label{inoutPhidef}
\ea
where $P$ denotes path
ordering.  The first line describes an outgoing, and
the second an incoming, parton,
whose flavors and four-velocities are  $f$ and
$\beta$, and $f'$ and $\beta'$, respectively.
The vector potentials ${\mathcal{A}}^{(f)}$
are in the color representation appropriate to flavor $f$,
and similarly for ${\mathcal{A}}^{(f')}$.

Because wide-angle, soft
radiation is independent of the internal
jet evolution, products of
nonabelian phase operators,
linked at the hard scattering
by a tensor in the space of color indices
generate the same wide-angle radiation as the full jets.
The general form for these operators,
exhibiting their color indices, is
\ba
{\mathcal{W}}_{I\, \{c_i\}}^{\rm (f)}(x) &=& \sum_{\{d_i\}}
\Phi_{\beta_2}^{(f_2)}(\infty,0;x)_{c_2,d_2}\;
\Phi_{\beta_1}^{(f_1)}(\infty,0;x)_{c_1,d_1}
\nonumber\\
&\ & \ \times \left(c_I^{\rm (f)}\right)_{d_2,d_1;d_A,d_B}\;
\Phi_{\beta_A}^{(f_A)}(0,-\infty;x)_{d_A,c_A}\;
\Phi_{\beta_B}^{(f_B)}(0,-\infty;x)_{d_B,c_B}
\, . \nonumber \\
& &
\label{Wdef}
\ea
The $c_I$ are color tensors in a convenient basis, listed in Appendix \ref{app:hard}.
Examples will be given below.
The perturbative expansions
for these operators are in terms of standard eikonal
vertices and propagators, and have been given in
detail in Refs.\ \cite{Kidonakis:1998nf,Kidonakis:1998bk}.
       In these terms, we define an eikonal cross section analogous to Eq. (\ref{eikdefcorr})
$\bar{\sigma}_{IL}^{(\mbox{\tiny eik})\,\rm (f)}$ at measured $\varepsilon$ and $\bar{\varepsilon}_{\mbox{\tiny eik}}$ as
\ba
\bar{\sigma}_{IL}^{(\mbox{\tiny eik})\,\rm (f)}\left(
\varepsilon,\bar{\varepsilon},\mu,\hat{\eta},\alpha_s(\mu)\right) &=& \frac{1}{\mbox{Tr} \left(c_I^\dagger c_L \right)}
\sum_{N_{\mbox{\tiny eik}}} \sum_{\{b_i\}}\,  \langle 0|\,
\bar T\left[\left({\mathcal{W}}^{\rm (f)}_I(0)\right)_{\{b_i\}}^\dagger\right]\,
|N_{\mbox{\tiny eik}}\rangle \nonumber \\
&\ & \qquad \times\,
\langle N_{\mbox{\tiny eik}}|\, T\left[{\mathcal{W}}^{\rm (f)}_L(0)_{\{b_i\}}\right]\, |0\rangle\, \nonumber \\
& & \qquad \times\,
\delta\left(\varepsilon - f(N_{\mbox{\tiny eik}})\right) \;
\delta\left(\bar{\varepsilon}_{\mbox{\tiny eik}}-\bar{f}(N_{\mbox{\tiny eik}})\right)
\nonumber\\ & = &  {\mathbf{1}}_{IL} \,\delta(\varepsilon) \delta (\bar{\varepsilon}_{\mbox{\tiny eik}}) +{\cal O}(\alpha_s).
\label{Sdef}
\ea
The indices $L$ and $I$ refer to the color exchange at the
hard scattering between the partons in reaction f,
as built into the definitions of the ${\mathcal{W}}$'s,
Eq.\ (\ref{Wdef}). 
The matrix elements  in Eq.\ (\ref{Sdef})
require renormalization, and we may identify the
corresponding renormalization scale with the
refactorization scale of Eq.\ (\ref{trafosighad}).

The $\varepsilon$-dependence of the
matrix $\bar{\sigma}_{IL}^{\rm (f)}$ in Eq.\ (\ref{Sdef}) is  the same as in the
full
partonic cross section, up to corrections due to differences between
the jets and the nonabelian
phase operators.   
The eikonal cross section (\ref{Sdef}) reproduces the wide-angle radiation accurately, however, the approximation fails for collinear radiation. As in the electron-positron case, Eq. (\ref{eikfactcorr}), we have to subtract collinear emission to avoid double counting in the full partonic cross section. Since we have not specified constructions for the beam jets, we do not give explicit constructions of the corresponding eikonal jets. Here we only note that the eikonal jets, like their partonic counterparts, are independent of the color exchange, which is therefore contained fully in the eikonal cross section:
\be
S_{IL}^{\rm (f)} \left(\varepsilon,\nu,\hat{\eta},\mu\right) =
\frac{\sigma_{IL}^{(\mbox{\tiny eik}),\,\rm (f)}\left(\varepsilon,\nu,\hat{\eta},\mu \right) }
{\prod\limits_{c = A,B,1,2} J_c^{(\mbox{\tiny eik})\,\rm (f)}\left(\nu,\hat{\eta},\mu\right) }
= {\mathbf{1}}_{IL}\, \delta(\varepsilon) +{\cal O}(\alpha_s)\, .
\label{s0had}
\ee

Using (\ref{s0had}) in (\ref{trafosighad}) we obtain
\ba
    {d \hat{\sigma}^{\rm (f)}(\varepsilon,\nu,p_T,\mu_F)\over d \varepsilon \,d \hat{\eta}}
\!\!& \!\!= &\!\! 
H^{\rm (f)}_{LI}(p_T,\hat{\eta},\mu,\mu_F)\;  \sigma^{(\mbox{\tiny eik})\,\rm (f)}_{IL}(\varepsilon,\nu,\hat{\eta},\mu)\nonumber \\
& & \times \, \prod_{c=A,B,1,2}\!
\frac{J^{\rm (f)}_c(p_T,\hat{\eta},\nu,\mu,\mu_F)}{J_c^{(\mbox{\tiny eik})\,\rm (f)}\left(\nu,\hat{\eta},\mu\right) }. 
\label{trafosighad2}
\ea
We observe that all information about color exchange is contained in the hard scattering and in the eikonal cross section,  independent of the details of the partonic and eikonal jets since these are proportional to the identity matrix in color space.

\subsection{Resummation in Color Space}

Without explicitly specifying the definitions of event shape, jet and eikonal jet functions, we can nevertheless proceed in a manner analogous to the $e^+e^-$ case to resum large logarithmic enhancements in $\varepsilon$ and partially in $\nu$. Since the jet functions are color diagonal, their solutions to the renormalization group equations are analogous to Eq. (\ref{orgsola}), where the integration limits may be slightly changed, depending on the exact definition of the shape function. 

The resummation of large logarithms in the soft function now depends on the color exchange. As above, we demand that the following condition be fulfilled
\be
\mu{d\over d\mu}\;
\left[\, \frac{d\hat \sigma^{\rm (f)}}{d\hat\eta d \varepsilon}\, \right] =
0\, .
\label{muindephad}
\ee
This condition applied to the right-hand side
of (\ref{trafosighad2}) leads to a renormalization group equation for the eikonal cross section that encodes all information about the color evolution,
\be
\left(\mu\frac{\partial}{\partial\mu}+\beta(g_s)
\frac{\partial}{\partial g_s}\right) \sigma^{(\mbox{\tiny eik})\,\rm (f)}_{IL}
          = -\left({\Gamma}^{\rm (f)}_{\mbox{\tiny eik}}(\hat\eta)\right)^{\dagger}_{IJ} \sigma^{(\mbox{\tiny eik})\,\rm (f)}_{JL} -
\sigma^{(\mbox{\tiny eik})\,\rm (f)}_{IJ}\left({\Gamma}_{\mbox{\tiny eik}}^{\rm (f)}(\hat\eta)\right)_{JL}\, ,
\label{rgS0}
\ee
with, as usual, $g_s=\sqrt{4\pi\alpha_s}$.
Here $\left({\Gamma}_{\mbox{\tiny eik}}^{\rm (f)}(\hat\eta)\right)_{IL}$
may be thought of as an anomalous
dimension matrix \cite{Kidonakis:1998nf,Kidonakis:1998bk}. This anomalous dimension matrix depends only on the directions of the jets through the directions of the eikonal lines, $\hat\eta$, but is geometry-independent otherwise.

To solve Eq.\ (\ref{rgS0}),
we go to a basis for the color matrices $c_I^{\rm (f)}$
that diagonalizes $\Gamma_{\mbox{\tiny eik}}^{\rm (f)}(\hat\eta)$,
through a transformation matrix $R$,
\be
\left(\Gamma_{\mbox{\tiny eik}}^{\rm (f)}(\hat\eta)
\right)_{\gamma\beta} \equiv
\lambda^{\rm (f)}_\beta(\hat\eta)\delta_{\gamma\beta}=
R^{\rm (f)}_{\gamma I\, }\left(\Gamma_{\mbox{\tiny eik}}^{\rm
(f)}(\hat\eta)\right)_{IJ}\,
R^{\rm (f)}{}^{-1}_{J\beta},
\label{Gamdia}
\ee
where
\be
\lambda^{\rm (f)}_\beta(\hat\eta)= \sum\limits_{n > 0} \left({\alpha_s\over \pi}\right)^n\,
\lambda_\beta^{\rm (f,\; n)}(\hat\eta)
\label{lambdadef}
\ee
are the eigenvalues of $\Gamma_{\mbox{\tiny eik}}^{\rm (f)}(\hat\eta)$.
Here and below, Greek indices $\beta, \; \gamma$ indicate that a matrix
is evaluated in the basis where
the eikonal anomalous dimension has been diagonalized.
Thus, for the eikonal and short-distance functions we also write,
\ba
\sigma^{(\mbox{\tiny eik})\,\rm (f)}_{\gamma\beta} &=&
\left[\left(R^{\rm (f)-1}\right)^{\dagger}\right]_{\gamma L} \;
          \sigma^{(\mbox{\tiny eik})\,\rm (f)}_{LK} \; \left[R^{\rm (f)-1}\right]_{K\beta}
\nonumber \\
H^{\rm (f)}_{\gamma\beta} &=& \left[R^{\rm (f)}\right]_{\gamma K} \;
H^{\rm (f)}_{KL} \; \left[R^{\rm (f)\dagger}\right]_{L\beta}.
\label{colortransform}
\ea
The transformation matrix $R^{\rm (f)-1}$ is given by the eigenvectors
of the anomalous dimension matrix,
\begin{equation}
\left( R^{(\rm f)\, -1} \right)_{K \beta}
\equiv \left( e_\beta^{(\rm f)}
\right)_K\, .
\label{Rdef}
\end{equation}

Analogous to Eq. (\ref{softevol}) we find the solution to Eq.\ (\ref{rgS0}), which resums leading logarithms
of $\varepsilon$.
We introduce a combination of eigenvalues
of $\Gamma_{\mbox{\tiny eik}}^{\rm (f)}$, $E_{\gamma\beta}^{\rm (f,\, n)}(\hat\eta)$, given
by
\be
E^{\rm (f,\, n)}_{\gamma\beta}(\hat\eta) =
{\lambda}^{\rm (f,\, n)\star}_{\gamma}(\hat\eta) +
{\lambda}^{\rm (f,\, n)}_{\beta}(\hat\eta)\, .
\label{Edef}
\ee
The soft eikonal function is then
\ba
\sigma^{(\mbox{\tiny eik})\,\rm (f)}_{\gamma\beta}\left(\hat\eta,\Omega,
\frac{\varepsilon p_T}{\mu},\varepsilon \nu, \frac{p_T}{\mu \nu},\alpha_s(\mu)\right)
&=&
\sigma^{(\mbox{\tiny eik})\,\rm (f)}_{\gamma\beta}\left(\hat\eta,\Omega,1,\varepsilon \nu,\alpha_s(\varepsilon p_T)\right)
\ \nonumber \\
&  & \hspace*{-15mm}  \times \exp \left\{ - \sum\limits_{n>0} E^{\rm (f,\, n)}_{\gamma\beta}(\hat\eta)\; \int\limits_{\varepsilon\, p_T}^{\mu}
{d\mu'\over
\mu'}\;
\left(\frac{\alpha_s(\mu')}{\pi}
\right)^n\right\} , \nonumber \\
& &
\label{softsoln}
\ea
where repeated color indices are summed over.
We now set the refactorization scale, $\mu$, equal to the
transverse momentum of the observed jet, $p_T$, and find with Eqs. (\ref{trafosighad2}) and (\ref{softsoln}),
for the partonic cross section,
\ba
\frac{{d\hat{\sigma}}^{\rm (f)}
\left(p_T,\hat\eta,\mu_F,\varepsilon,\nu,\alpha_s(\mu_F)\right) }
{d\hat\eta d \varepsilon  } 
&=&
\sum_{\beta,\,\gamma}\;
          H^{\rm (f)}_{\beta\gamma}\left(p_T,\hat\eta,
{\alpha}_s({\mu_F})\right) 
\nonumber\\
& &   \hspace*{-50mm} \times \,S^{\rm (f)}_{\gamma\beta}\left(\hat\eta,\Omega,1,\varepsilon \nu, \alpha_s(\varepsilon p_T)\right)\,
\exp \left\{ - \sum\limits_{n>0} E^{\rm (f,\, n)}_{\gamma\beta}(\hat\eta)\; \int\limits_{\varepsilon\, p_T}^{\mu}
{d\mu'\over
\mu'}\;
\left(\frac{\alpha_s(\mu')}{\pi}
\right)^n \right\} \, \nonumber \\
& &   \hspace*{-50mm} \times \, \!\prod_{c=A,B,1,2}\!
J^{\rm (f)}_c(p_T,\hat{\eta},\nu,\mu_F) \nonumber \\
&&  \hspace*{-55mm} =
\sum_{\beta,\,\gamma}\;
          H^{\rm (f)}_{\beta\gamma}\left(p_T,\hat\eta,
{\alpha}_s({\mu_F})\right) 
\nonumber\\
& &   \hspace*{-50mm} \times \,\sigma^{(\mbox{\tiny eik})\,\rm (f)}_{\gamma\beta}\left(\hat\eta,\Omega,1,\varepsilon \nu, \alpha_s(\varepsilon p_T)\right)\,
\exp \left\{ - \sum\limits_{n>0} E^{\rm (f,\, n)}_{\gamma\beta}(\hat\eta)\; \int\limits_{\varepsilon\, p_T}^{\mu}
{d\mu'\over
\mu'}\;
\left(\frac{\alpha_s(\mu')}{\pi}
\right)^n \right\} \, \nonumber \\
& &   \hspace*{-50mm} \times \, \!\prod_{c=A,B,1,2}\!
\frac{J^{\rm (f)}_c(p_T,\hat{\eta},\nu,\mu_F)}{J_c^{(\mbox{\tiny eik})\,\rm (f)}\left(p_T,\hat{\eta},\nu \right) }.
\label{HSresum}
\ea
In the second equality we have expressed the soft function in terms of its decomposition into an eikonal function and eikonal jets, Eq. (\ref{s0had}).
As in (\ref{evolend}) all dependence on $\varepsilon$ is factored into the eikonal function and its anomalous dimension, and all geometry dependence is contained in the eikonal matrix, as shown explicitly by the argument $\O$. 

\subsection{Renormalization and Color Mixing}

We now turn to the calculation of
the anomalous dimensions $\Gamma_{\mbox{\tiny eik}}^{\rm (f)}$,
introduced in Eq.\ (\ref{rgS0}) above. The renormalization of multi-eikonal
vertices has been discussed in  some detail in Ref.\
\cite{Kidonakis:1998nf},
and we will follow the method
outlined there.
In \cite{Kidonakis:1998nf}, the soft anomalous dimensions
$\Gamma_S^{\rm (f)}$ were computed in axial gauge,
after dividing an eikonal scattering amplitude by
eikonal self-energy functions (eikonal jets).  This extra factorization
eliminated double poles in dimensional
regularization, which are associated with collinear
emission by the nonabelian phase operators.  In
axial gauge, these singularities appear only
in self-energy diagrams. In Feynman gauge, the cancellation amounts to dividing out eikonal jet functions, as discussed above. Nevertheless, as already emphasized above, all non-trivial color structure is contained in the eikonal function.
The ultraviolet divergences
in $\sigma^{(\mbox{\tiny eik})\,\rm (f)}_{IL}$ may be compensated
through local counterterms, just as in \cite{Kidonakis:1998nf}.

We consider the diagrams
shown in Fig.\ \ref{examplefig}.
Figs.\ \ref{examplefig} a) and \ref{examplefig} c)
are virtual corrections,
while \ref{examplefig} b) shows the corresponding
diagrams for real gluon
emission.  
We compute the counterterms for Fig. \ref{examplefig} just as in the previous chapters.
From these counterterms,
we may simply read off the
entries of the anomalous dimension matrix
$\Gamma^{\rm (f)}_{\mbox{\tiny eik}}$ in an MS
renormalization scheme:
\begin{equation}
\left(\Gamma_{\mbox{\tiny eik}}^{\rm (f)}(\hat\eta,\alpha_s(\mu))\right)_{IJ}  =
-  \alpha_s
\frac{\partial}{\partial \alpha_s}
\left(
Z_{\mbox{\tiny eik}\, 1}^{\rm (f)}\left(\hat\eta,\alpha_s(\mu),\varepsilon\right)
\right)_{IJ} 
\label{residue}
\, .
\end{equation}
Here the subscript 1 denotes the $1/\varepsilon$-pole of the counterterm (compare to Eq. (\ref{prela})).
The calculation of the $Z$'s is thus essentially
equivalent to the calculation of the anomalous dimensions.
To carry out these calculations, however, we must specify
a basis of color tensors, $c_I$. Convenient bases for various $2 \rightarrow 2$ scattering processes are listed in Appendix \ref{app:colbas}.

\begin{figure}
\centerline{\epsfxsize=12cm \epsffile{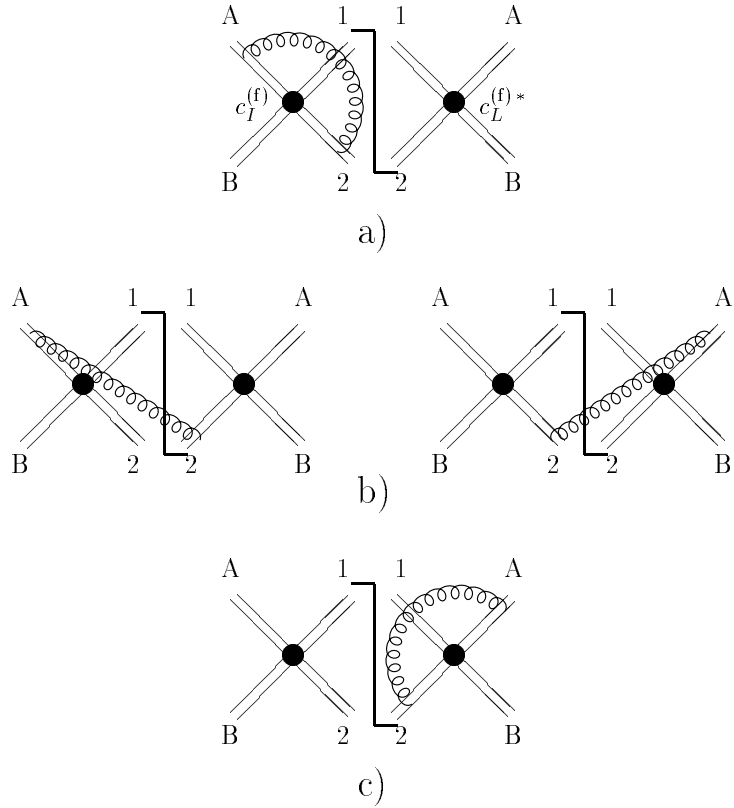}}
\caption[Diagrams for the calculation of the
anomalous dimension ma\-trix.]{Diagrams for the calculation of the
anomalous dimension matrix.
The double lines represent eikonal propagators, linked
by vertices $c_I^{\rm (f)}$ and $c_L^{\rm (f)}{}^*$,
in the amplitude and its complex conjugate.  The
vertical line represents the final state.}
\label{examplefig}
\end{figure}

Let $Z^{(ij)}$ denote the contribution to the
counterterms from all the one-loop graphs in which
the gluon connects eikonal lines $i$ and $j$.  In
this notation, the calculation of Fig.\ \ref{examplefig}
gives us $Z^{(A2)}$.   The $Z$'s are
constructed to give local counterterms that cancel those ultraviolet divergences that are left over
after the real-virtual cancellation has been carried out.

To find the $Z^{(ij)}_{IJ}$'s in
the color bases listed in the appendix,
we rewrite the various one-loop virtual diagrams
in terms of the original color basis, using the identity
shown in Fig. \ref{colorfig},
\begin{equation}
T^a_{ij} T^a_{kl} = \frac{1}{2} \left( \delta_{il} \delta_{jk} -
\frac{1}{\Ncol} \delta_{ij} \delta_{kl} \right)\,
\label{identity}
\end{equation}
for quark processes.  For scattering processes involving gluons,
many useful identities can be found, for example, in \cite{MacFarlane:1968vc}.
This procedure results in a $2\times 2$ matrix decomposition
for scattering
involving only quark
and antiquark eikonal lines,
describing the mixing under renormalization of their
color exchanges.
The annihilation of a pair of quark and antiquark
eikonals into two gluon eikonal
lines gives a $3\times 3$ matrix structure, while for incoming and
outgoing gluonic eikonals,
we get an $8\times 8$ matrix \cite{Kidonakis:1998nf}.

For a given $Z^{(ij)}$,
the momentum-space integral appears as an overall factor.
It is then convenient to introduce the notation
\ba
\left(Z_{\mbox{\tiny eik}\, 1}^{(ij)} - 1\right)_{LI}
=
\zeta^{(ij)}_{LI}\; \omega_1^{(ij)}\, ,
\label{factorZ}
\ea
where the $\zeta_{LI}$ are the coefficients that
result from the color decomposition of the virtual
diagram, and where the $\omega^{(ij)}_1$s include the ultraviolet
   pole part of the
momentum space integral for the eikonal function $\sigma^{(\mbox{\tiny eik})\,\rm (f)}$ and
remaining overall constants.
Defined in this fashion, the sign of each $\omega^{(ij)}_1$
depends on the flow of flavor in the underlying
process (see below). From (\ref{residue}) and (\ref{factorZ})
the relation between the $\omega$'s and the anomalous dimension matrices
is
\begin{equation}
\left( \Gamma_{\mbox{\tiny eik}}^{(\rm f)} \right)_{LI} = -  
\sum_{\{(ij)\}} {\zeta^{(ij)}}_{LI}
   \;\omega^{(ij)}_{\,1} \;, \label{fulladm}
\end{equation}
where, as above, the subscript 1 denotes the $1/\varepsilon$ pole.

\begin{figure}
\centerline{\epsfysize=3cm \epsfbox[71 356 545 478]{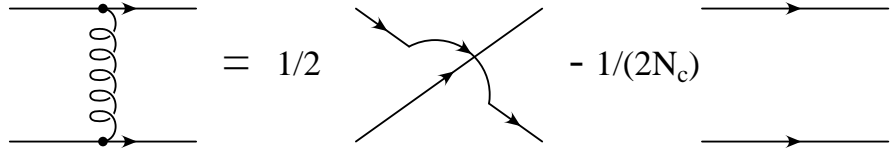}}
\caption{Color identity corresponding to Eq.\ ({\protect \ref{identity})}.}
\label{colorfig}
\end{figure}

We note that the color decompositions $\zeta^{(ij)}$ are the same for the eikonal anomalous dimensions and the corresponding first order contributions to the soft function. The difference lies in the momentum parts, which can always be factored from the color part as in Eq. (\ref{factorZ}).
\begin{equation}
\left( \sigma^{(\rm eik, f)} \right)_{LI} =  
\sum_{\{(ij)\}} {\zeta^{(ij)}}_{LI}
   \;\omega^{(ij)}(\O) \;, \label{softmat}
\end{equation}
 analogous to the results for $e^+e^-$ annihilation listed in Section \ref{sec:softrescorr}, where now $\o^{(ij)}$ without subscript denotes the lowest order contribution after cancellation of UV divergences.  At lowest order for $\varepsilon \neq 0$, the soft function is independent of the details of the eikonal jets, but depends on the geometry of the region $\O$ in which the energy flow is measured. 

As an example, we give the color decomposition of
$\Gamma^{(A2)}$ in $q \bar{q}
\rightarrow q \bar{q}$, for which we find, using Eq.\ (\ref{identity}),
\begin{eqnarray}
\zeta^{(A2)}  = \left( \begin{array}{cc} 0 &
\frac{C_F}{2 \,\Ncol}  \\
1 & - \frac{1}{\Ncol}  \end{array} \right)\, . \label{A2example}
\end{eqnarray}

\subsection{Results for Lowest Order Momentum Parts} \label{sec:momentum}

\subsubsection{Anomalous Dimensions}

 The momentum parts of the anomalous dimensions are given by virtual graphs only, because real emissions are restricted over the whole phase space by the weight functions $f$ and $\bar{f}$:
\be
\o^{(ij)}_{\,1} = - i g_s^2 \int_{P.P.}  \frac{d^n k}{(2 \pi)^{n}} \frac{1}{k^2+i\epsilon}  \frac{\Delta_i \Delta_j \,\beta_i \cdot \beta_j}{(\delta_i\beta_i
\cdot k +i \epsilon) \,(\delta_j \beta_j \cdot k + i \epsilon) }. \label{ompoledef}
\ee
$P.P.$ denotes the pole part, and the integral is over all of phase space, independent of the details of the weight functions.  
In Eq. (\ref{ompoledef}), collinear divergences cancel against contributions from eikonal jets. Here $\delta_i=1(-1)$ for momentum
$k$ flowing in the same (opposite) direction as the momentum
flow of line $i$, and  $\Delta_i=1(-1)$ for $i$ a quark (antiquark)
line. 
For example,
Fig.\ \ref{examplefig}a for $q \bar{q} \rightarrow q \bar{q}$ has
$i$ = A, $j$ = 2, with $\Delta_A=1$,
$\Delta_2=-1$ and with $\delta_A=\delta_2$.

The factorization 
of momentum and color parts as in (\ref{factorZ}) is somewhat ambiguous for gluon eikonal 
lines since here the orientation with which the gluon attaches is important. The sign 
resulting from the way of attachment can either be factored into the momentum or into 
the color parts, their product is of course independent of this choice. 
Below we will attach all gluons in such a way that the momentum parts for processes 
involving gluons 
acquire \emph{the same sign} as the momentum parts for the process 
$q \bar{q} \rightarrow q \bar{q}$. 

After combining Eq. (\ref{ompoledef}) with the eikonal jets, collinear poles cancel. This may not seem obvious from the color decomposition, (\ref{A2example}), since the eikonal jets are diagonal in color space. However, the off-diagonal elements are in such combinations that collinear singularities cancel among themselves, without invoking the eikonal jets. The resulting soft anomalous dimension can then be written as
\begin{equation}
\left( \Gamma_S^{(\rm f)} \right)_{LI} = 
\sum_{\{(ij)\}} {\zeta^{(ij)}}_{LI}
   \;\Gamma^{(ij)} \;, \label{fulladmsoft}
\end{equation}
where the $\Gamma^{(ij)}$ are given at lowest order by (compare to (\ref{softad}) and \cite{Kidonakis:1998nf})
\be
\Gamma^{(ij,\,1)} = - \Delta_i \Delta_j \delta_i \delta_j  \left[ \ln \left(\beta_i \cdot \hat{\xi_i} \right)+ \ln \left(\beta_j \cdot \hat{\xi_j} \right) - \ln \left(
\frac{\beta_i \cdot \beta_j}{2} \right) - 1 \right], \label{oneloopadm}
\ee
where the $\beta_c, \, c = A,B,1,2$ are lightlike vectors in the directions of the jet momenta $p_{J_c}$, and the $\xi_c$ are the corresponding non-lightlike eikonal jet momenta, as in the case for $e^+e^-$ annihilation.

\subsubsection{Lowest Order Soft Function}

On the other hand, the lowest order eikonal function in Eq. (\ref{HSresum}) is given by the integral over the phase space in which the energy flow $\varepsilon$ is measured, analogous to the results for $e^+e^-$ annihilation listed in Section \ref{sec:softrescorr}. As above, only the eikonal function contributes to the lowest order soft function $S^{(1)}$.
\be
\left( S^{(\rm f)} \right)_{LI} = 
\sum_{\{(ij)\}} {\zeta^{(ij)}}_{LI}
   \; S^{(ij)} \;. \label{fullsoft}
\end{equation}
The momentum parts of the lowest order soft contributions $S^{(ij,\,1)}$ are thus, analogous to Eq. (\ref{oneLoopSoft}) given by
\be
S^{(ij,\,1)}(\varepsilon \neq 0,\O,\delta_i\beta_i,\delta_j\beta_j,\Delta_i,\Delta_j) = \frac{1}{\varepsilon}  \int\limits_\O  d \mbox{PS}_2\,
\frac{1}{2 \pi} \frac{\Delta_i \Delta_j \,\beta_i \cdot \beta_j}{\delta_i\beta_i
\cdot \hat{k} \,\delta_j \beta_j \cdot \hat{k} }\, ,
\label{oneLoopSofthad}
\ee
where the notation is as in Eq. (\ref{oneLoopSoft}). 
     
\subsubsection{Color Decompositions}

We may summarize the color decompositions of Eqs. (\ref{fulladmsoft}) and (\ref{fullsoft}) by  
 introducing the following combinations,
\begin{eqnarray}
\alpha^{\rm (f)} & \equiv & \phi^{(AB,\rm f)} +
\phi^{(12,\, \rm f)}, \nonumber \\
\beta^{\rm (f)} & \equiv & \phi^{(A1,\, \rm f)} +
\phi^{(B2,\, \rm f)}, \nonumber \\
\gamma^{\rm (f)} & \equiv & \phi^{(A2,\, \rm f)} +
\phi^{(B1,\, \rm f)},
\label{abgdef}
\end{eqnarray}
where $\phi^{(ij)}$ denotes either $S^{(ij,\,1)}$ (Eq. (\ref{oneLoopSofthad})) or $\Gamma^{(ij,\,1)}$ (Eq. (\ref{oneloopadm})). 

In these terms, both anomalous dimension matrices and lowest order eikonal contributions, in the following collectively denoted by ${\mathcal{M}}^{(\rm f)}$, 
for the
process $q \bar{q} \rightarrow q \bar{q}$ in the basis (\ref{qqbarbas}) are given by
\begin{equation}
{\mathcal{M}}^{(q\bar q \rightarrow q \bar{q})} = \left( \begin{array}{cc}
C_F \beta &
\frac{C_F}{2 \Ncol} \left(
\alpha +
    \gamma \right) \\
\alpha +
    \gamma  \, &
C_F \alpha  - \frac{1}{2 \Ncol}
\left( \alpha  +
\beta 
+ 2 \gamma \right) \end{array} \right)\, 
\label{qqbarGadm}.
\end{equation}

The color decompositions of the anomalous dimension matrices and soft functions at lowest order for all other $2 \rightarrow 2$ scattering processes are listed in Appendix \ref{app:soft}. 

\subsection{Results for Hadronic Shape/Flow Correlations}

From Eq. (\ref{HSresum}) we obtain at LL in the energy flow, $\varepsilon$,
\ba
\frac{{d\hat{\sigma}}^{\rm (f)}
\left(p_T,\hat\eta,\mu_F,\varepsilon,\nu,\alpha_s(\mu_F)\right) }
{d\hat\eta d \varepsilon  } 
&=&
\sum_{\beta,\,\gamma}\;
          H^{\rm (f)}_{\beta\gamma}\left(p_T,\hat\eta,
{\alpha}_s({\mu_F})\right) 
\nonumber\\
& &   \hspace*{-30mm} \times \,S^{\rm (f,1)}_{\gamma\beta}\left(\hat\eta,\Omega,1,\varepsilon \nu, \alpha_s(\varepsilon p_T)\right)\, \left( \frac{\alpha_s(p_T)} {\alpha_s(\varepsilon p_T)}
\right)^{\frac{2}{\beta_0} E^{\rm (f,1)}_{\gamma\beta}(\hat\eta)}
\, \nonumber \\
& &   \hspace*{-30mm} \times \, \!\prod_{c=A,B,1,2}\!
J^{\rm (f)}_c(p_T,\hat{\eta},\nu,\mu_F).  \label{hadronLL}
\ea
Here, the expression for $S^{\rm (f,1)}$ can be found from (\ref{fullsoft}) with (\ref{oneLoopSofthad}) and the color decompositions listed in Appendix \ref{app:soft}, whereas the eigenvalues are determined from (\ref{oneloopadm}) and the same color decompositions with (\ref{fulladmsoft}). The resummation of the remaining large logarithms in the jet functions proceeds analogous to the electron-positron case in Sec. \ref{sec:jetinclevol}, which we will not perform here, since we have not specified the details of the beam jets and the event shape. However, from Eq. (\ref{hadronLL}), or in general, from Eq. (\ref{HSresum}) we observe, that also for hadronic processes all information about interjet energy flow is contained in the soft function and its anomalous dimension matrix.

\section{Summary and Outlook}

As was emphasized in \cite{Dokshitzer:2003uw}, most actual experimental measurements involve phase-space restrictions, and are therefore influenced more or less by non-global effects. For example, most detectors do not provide full acceptance over the whole angular region, measurements are restricted over a finite range of rapidity. It is therefore necessary to provide theoretical predictions that take these non-global effects into account, as in Refs. \cite{Dasgupta:2001sh,Dasgupta:2002bw,Dasgupta:2002dc,Dokshitzer:2003uw,Banfi:2000si,Burby:2001uz,Banfi:2002hw,Appleby:2002ke}, or alternatively, to study observables that are less sensitive to non-global contributions. 

Moreover, as already mentioned in the introductory section to this chapter, the study of energy flow into only part of phase space may shed light on the underlying event, and may help to distinguish  multiple scatterings effects, perturbative bremsstrahlung emerging from the primary hard scattering, and possible new physics signals. 

Above, we have introduced a set of correlations of interjet energy
flow for the general class of  event shapes discussed in Chapter \ref{ch5}, and have shown
that for these quantities it is possible to control
the influence of secondary radiation and non-global logarithms.
These correlations are sensitive mainly to radiation
emitted directly from the primary hard scattering, through transforms in
the weight functions that suppress secondary, or non-global, radiation.  
 We have presented analytic  and
numerical studies of these shape/flow correlations in $e^+e^-$ dijet events at leading
logarithmic order in the
flow variable and at next-to-leading-logarithmic order in the
event shape. Within our shape/flow correlations the function that encodes the non-global energy flow factorizes from the remaining functions that describe the shape of the event. A similar conclusion was reached in a recent study of such shape/flow correlations (associated distributions) based on coherent parton branching \cite{Dokshitzer:2003uw}.

  The application of our formalism
to multijet events and to scattering with initial state hadrons is
certainly possible, and may shed light on the relationship
between color and energy flow in hard scattering processes with non-trivial
color exchange. We have illustrated above how non-trivial color flow in events with incoming hadrons is treatable within our formalism. We have shown that, analogous to $e^+e^-$ shape/flow correlations, the information about the energy flow factors from the remainder of the cross section. In the hadronic case, however, this factorization is in terms of matrix multiplication in the space of color exchanges. We have provided complete expressions for hard and soft matrices and soft anomalous dimension matrices for $2 \rightarrow 2$ scattering in $p\bar{p}$ collisions. A full treatment of the hadronic case, however, requires an amended definition of the corresponding event shape, which is nontrivial due to the more involved kinematics, and a careful treatment of the incoming jet functions. We postpone the study of these remaining few, but subtle issues. We now summarize the work presented in this thesis.

\chapter{Epilogue: Conclusions and Perspectives}
\label{ch7}

In this thesis we have studied soft gluon radiation which arises in (semi-) inclusive cross sections at the edge of phase space. Soft gluons are responsible for large logarithmic corrections that need to be resummed in order to provide reliable quantitative predictions within perturbation theory. We have found that soft gluon radiation at wide angles from the hard scattering decouples, and is well described by emission from path ordered exponentials or eikonal lines that replace the partons involved in the hard scattering. 

Cross sections built out of these eikonal lines exponentiate directly by reordering of graphs \cite{Sterman:jc,Gatheral:cz,Frenkel:pz}. This leads to important consequences for physical observables in which soft radiation is not inclusively summed over. The leading logarithmic behavior in such events, which is due to emissions that are simultaneously soft and collinear, is controlled by the anomalous dimension of these eikonal cross sections. The exponentiation of these cross sections has lead us to the development of a simplified method to compute their anomalous dimensions, which we have illustrated with the computation of the fermionic three-loop contribution to the singular coefficients of partonic splitting functions \cite{Berger:2002sv}.

As noted above, the same anomalous dimensions arise in the resummation of large logarithms in jet events. In the second part of this thesis we have introduced a generalized class of dijet event shapes, and have studied the correlation of these jet shapes with energy flow into the interjet region \cite{Berger:2002ig,Berger:2003iw}. Studies of interjet energy flow may help to disentangle the underlying event from soft bremsstrahlung effects and from effects due to new physics (beyond the Standard Model). The study of correlations of energy flow with event shapes is sensitive mainly to emissions directly from the underlying hard scattering and allows us to control the influence of secondary radiation, which, if inclusively summed over, masks the underlying scattering in a highly nontrivial way \cite{Dasgupta:2001sh,Dasgupta:2002bw}.  We have resummed large logarithms due to soft bremsstrahlung in the interjet region by solving corresponding evolution equations. The solutions to these evolution equations which follow from factorization contain all large logarithmic corrections in exponentiated form. The leading logarithmic corrections due to soft-collinear emissions can also be obtained by exponentiation of eikonal cross sections on the graphical level, as emphasized above. The remaining exponents are due to only soft or only collinear radiation. At the NLL level, the same result for event shapes and shape/flow correlations can be obtained by studying coherent parton branching \cite{Catani:1992ua,Dokshitzer:2003uw}. Up to NLL level in dijet events, all radiation appears to be emitted coherently. This, however, is not necessarily true beyond next-to-leading logarithm, and certainly not the case in multi-jet configurations. Nevertheless, our formalism is valid beyond NLL and can in principle be extended to multi-jet events. We have illustrated how it can be applied to events with hadrons in the initial state, for example to $p\bar{p}$ collisions \cite{Berger:2001ns}. 

To summarize, we have attempted to extend the application of perturbative QCD to studies of observables which are sensitive to the radiation into only part of phase space, and we have developed a method which considerably simplifies the computation of the leading effects at higher orders in perturbation theory. These topics are tied together by the mechanisms of soft-gluon factorization which leads to quantities built out solely of eikonal lines. The latter exponentiate directly at the level of Feynman graphs. 
The possible applications are numerous, for example the quantitative study of power corrections to the cross sections mentioned above, or the complete study of hadronic shape/flow correlations which may shed light on the interplay of color and energy flow. Further studies of these topics will certainly
provide more  insights into the mechanisms of QCD itself, and facilitate to distinguish QCD effects
from signals of new physics in experimental measurements. 

To conclude with Aristotle - in this thesis we have attempted to explore some aspects of nature (however, infinitely far from settling questions about principles), starting from QCD which is more knowable to us, and which provides the background to new physics that may be more knowable to nature.

\appendix

\chapter{QCD Conventions and Eikonal Feynman Rules}
\label{app1}

\section{The Running Coupling}\label{sec:alpha}

The effective running coupling in QCD 
\begin{equation} 
\alpha_s = \frac{g_s^2}{4 \pi} 
\end{equation}
 obeys the renormalization group equation
\begin{equation} 
\mu \frac{\partial}{\partial \mu} \alpha_s = \beta\left(\alpha_s \right) =  - \frac{\beta_0}{4 \pi} \alpha_s^2 - \frac{\beta_1}{(4 \pi)^2} \alpha_s^3 - \dots.\label{betaf} 
\end{equation}
The expansion of the beta-function is currently known to four loops \cite{vanRitbergen:1997va}. In this thesis we only need the one- and two-loop coefficients, given here for SU(N),
\begin{eqnarray}
\beta_0 & = & \frac{11}{3} C_A - \frac{4}{3} T_F N_f, \label{beta0} \\ 
\beta_1 & = & \frac{34}{3} C_A^2 - \frac{20}{3} C_A T_F N_f - 4 C_F T_F N_f. \label{beta1} 
\end{eqnarray}
$C_F,\,C_A$ are the Casimirs of the fundamental and the adjoint representation, 
\ba
C_F & = & \frac{\Ncol^2 - 1}{2\,\Ncol},\nonumber \\
C_A & = & \Ncol = 3,
\ea
for SU(3). 
$N_f$ denotes the number of flavors, and $T_F = 1/2$ is the standard normalization of the generators of the fundamental representation. Beginning with $\beta_2$, the coefficients of the expansion of the beta-functions are scheme-dependent at higher orders.

The solution of Eq. (\ref{betaf}) at one and two loops, respectively, is given by
\ba
\as(\mu) & = & \frac{2 \pi}{\beta_0 \ln \left(\frac{\mu}{\LQCD} \right)} + \dots,  \label{1as}  \\
\as(\mu) & = & \frac{2 \pi}{\beta_0 \ln \left(\frac{\mu}{\LQCD} \right)} \left[ 1 - \frac{\beta_1}{2 \beta_0^2} 
\frac{\ln \ln \left(\frac{\mu}{\LQCD} \right)^2}{\ln \left(\frac{\mu}{\LQCD} \right)} + \dots \right]. \label{2as}
\ea 
 We have introduced the dimensionful scale parameter $\LQCD$ which parameterizes the initial condition to the differential equation (\ref{betaf}). Conventionally it is determined by the value of $\as$ at the scale equalling the mass of the $Z$ boson, whose world average is currently determined as \cite{Bethke:2002rv,Hagiwara:fs}
\be
\as(M_Z) = 0.1183 \pm 0.0027. \label{asMZ}
\ee
$\LQCD$ depends on the number of active flavors, as can be seen from (\ref{beta0}) and (\ref{beta1}).

\begin{figure}[hbt]
\begin{center}
\epsfig{file=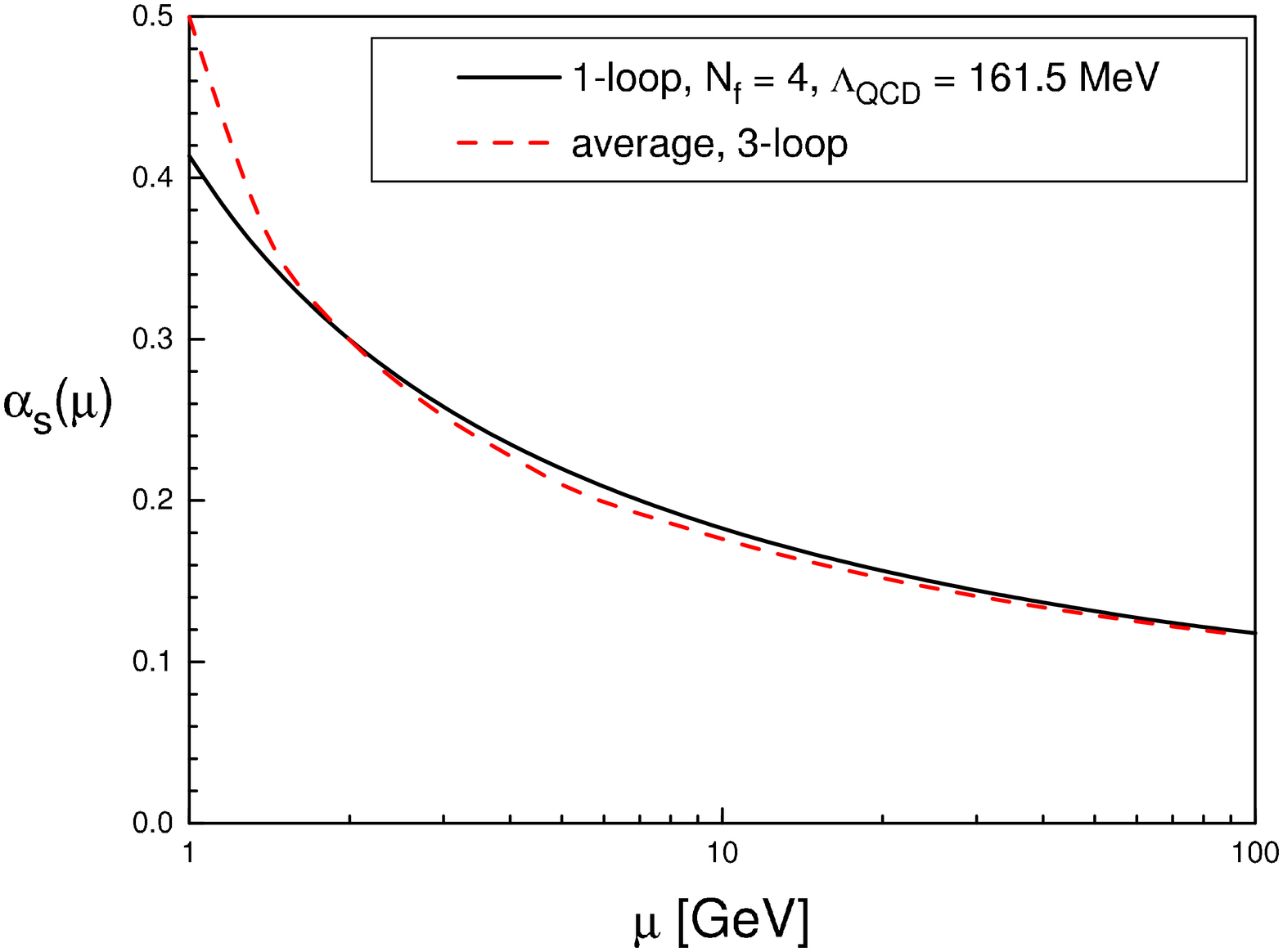,height=12.2cm,angle=270,clip=0}
\caption[One-loop (solid line) and three-loop (dashed red line) strong coupling as a function of scale $\mu$ (in GeV).]{One-loop (solid line) and three-loop (dashed red line) strong coupling as a function of scale $\mu$ (in GeV). The parameters of the one-loop coupling are given in (\ref{asnum}), the values for the three-loop coupling are taken from \cite{asweb}.} \label{asplot}
\end{center}
\end{figure}

For numerical studies in this thesis we use a value of
\be
\LQCD = 161.5 \mbox{ MeV} \quad \mbox{ for } N_f = 4 \label{asnum}
\ee
active flavors, determined by fitting the one-loop term (\ref{1as}) to (\ref{asMZ}).
Fig. \ref{asplot} shows the one-loop coupling with (\ref{asnum}) as a function of the scale $\mu$ (solid black line). For comparison (dashed red line) we plot the coupling at three loops as determined in \cite{Hagiwara:fs} (the corresponding numerical values can be found in \cite{asweb}).

At the one-loop level, we can reexpress the coupling given at scale $\mu$ in terms of the coupling at scale $\tilde{\mu}$,
\be
\as(\mu) = \frac{\as(\tilde{\mu})}{1+ \frac{\beta_0}{2 \pi} \as(\tilde{\mu}) \ln \frac{\mu}{\tilde{\mu}}}. \label{asreexp}
\ee 

\clearpage

\section{Eikonal Feynman Rules} \label{sec:eikfeyn}

\begin{center}
\begin{figure}[htb]
\begin{center}
\epsfig{file=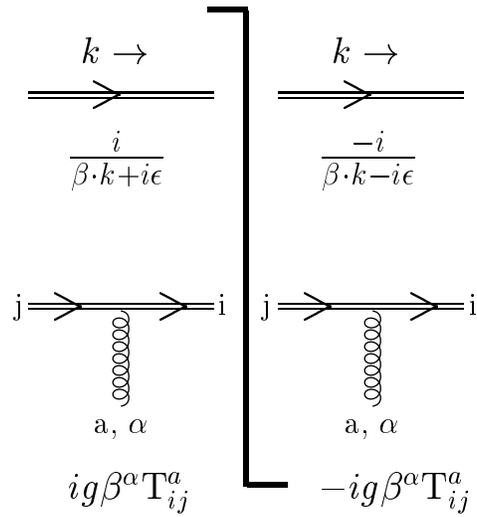,height=8cm,clip=0}
\caption[Feynman rules for eikonal lines in the fundamental representation with velocities $\beta^\mu$, represented by the double lines.]{Feynman rules for eikonal lines in the fundamental representation with velocities $\beta^\mu$, represented by the double lines. The vertical line represents the cut separating the amplitude and its complex conjugate. For an eikonal line in the adjoint representation one has to replace $T^a_{ij}$ with $i f_{ija}$. \label{Frules} }
\end{center}
\end{figure}
\end{center} 



\def\SUNC{
\setlength{\unitlength}{1.2cm}
\mbox{\parbox{2.1em}{\hspace{0.25em}
\begin{picture}(2,1)
\thicklines
\put(0.1,0.4){\line(1,0){1.8}}
\put(1,0.4){\circle{0.9}}
\end{picture}
}}
\hfill}

\def\smallSUNC{
\setlength{\unitlength}{1.5em}
\mbox{\parbox{2.1em}{\hspace{0.25em}
\begin{picture}(2,1)
\thicklines
\put(0.1,0.5){\line(1,0){1.8}}
\put(1,0.5){\circle{0.9}}
\end{picture}
}}
\hfill}


\def\GLASS{
\setlength{\unitlength}{1.2cm}
\mbox{\parbox{3.1em}{\hspace{0.25em}
\begin{picture}(3.1,1)
\thicklines
\put(0.15,0.4){\line(1,0){0.5}}
\put(2.4,0.4){\line(1,0){0.5}}
\put(1.1,0.4){\circle{0.9}}
\put(1.95,0.4){\circle{0.9}}
\end{picture}
}}
\hfill}

\def\smallGLASS{
\setlength{\unitlength}{1.5em}
\mbox{\parbox{3.1em}{\hspace{0.25em}
\begin{picture}(3.1,1)
\thicklines
\put(0.15,0.5){\line(1,0){0.5}}
\put(2.4,0.5){\line(1,0){0.5}}
\put(1.1,0.5){\circle{0.9}}
\put(1.95,0.5){\circle{0.9}}
\end{picture}
}}
\hfill}


\def\BUB{
\setlength{\unitlength}{1.5em}
\mbox{\parbox{2.1em}{\hspace{0.25em}
\begin{picture}(2,1)
\thicklines
\put(0.1,0.5){\line(1,0){0.45}}
\put(1.4,0.5){\line(1,0){0.45}}
\put(1,0.5){\circle{0.9}}
\end{picture}
}}
\hfill}


\def\trig{
\setlength{\unitlength}{1.2cm}
\mbox{\parbox{2.4em}{\hspace{0.25em}
\begin{picture}(2,1.4)
\thicklines
\put(0.05,1){\line(1,0){0.5}}
\put(1,1.44){\line(0,-1){0.88}}
\put(1.45,0.98){\line(1,0){0.5}}
\put(1,1){\circle{0.9}}
\end{picture}
}}
\hfill}

\def\smalltrig{
\setlength{\unitlength}{1.5em}
\mbox{\parbox{2.4em}{\hspace{0.25em}
\begin{picture}(2,1.4)
\thicklines
\put(0.05,0.6){\line(1,0){0.5}}
\put(1,1.04){\line(0,-1){0.88}}
\put(1.45,0.58){\line(1,0){0.5}}
\put(1,0.6){\circle{0.9}}
\end{picture}
}}
\hfill}

\chapter{Multiloop Techniques}
\label{app2}

In this appendix we collect a few items useful for evaluating multi-loop integrals.

\section{Feynman Parametrization} \label{app:feynpar}

Two denominators can be combined by introducing the integral representation:
\be
\frac{1}{A} \,\frac{1}{B} = \int\limits_0^1 dx \frac{1}{(x A + (1-x) B)^2}.
\ee

This can be generalized to $N$ factors raised to arbitrary inverse powers $-\eta_i,\,i = 1,\dots,N$ with the help of integrals over $N$ Feynman parameters $x_i$:
\ba
\prod\limits_{i = 1}^N A_i^{-\eta_i} & = & \left[ \prod\limits_{i=1}^N \Gamma\left(\eta_i \right) \right]^{-1} 
\Gamma\left(\sum\limits_{i = 1}^N \eta_i \right) \int\limits_0^1 d x_1\, x_1^{\eta_1 -1} \dots d x_N\, x_N^{\eta_N -1}
\nonumber \\
& & \times \delta \left(1- \sum\limits_{i = 1}^N x_i \right) \left( \sum\limits_{i = 1}^N x_i\, A_i \right)^{- \sum\limits_{i = 1}^N \eta_i}. \label{feynpar}
\ea

\section{Reduction of Loop Integrals} \label{app:reduce}

In the following we will discuss specifically two-loop integrals, but the methods collected below can be generalized to higher orders. Furthermore, we will only discuss the techniques that are needed for the evaluation of the two-loop gluon self-energy, as shown in Table \ref{graphtab}, graph g)\footnote{The presented methods apply of course also to the corresponding gluonic part.}. A more complete overview and a list of further references can be found, for example, in Ref. \cite{Tejeda-Yeomans:2002eh}. 

We consider two-loop integrals in dimensional regularization of the form
\be
I^n\left[\eta_1,\dots,\eta_N\right] \equiv \int \frac{d^n k}{(2 \pi)^n} \int \frac{d^n l}{(2\pi)^n} \frac{{\mathcal{N}}(k,l)}{A_1^{\eta_1} \dots A_N^{\eta_N}} , \label{general2}
\ee 
where the numerator ${\mathcal{N}}$ is a function of the loop-momenta $k_i = k,l$, and the external momentum $p$ (or momenta, for graphs with more external legs). The massless propagators are of the form $A_i \sim (k_i \pm f(p,k_i))^2 + i \epsilon$, with $f(p,k_i)$ a function of internal and/or external momenta. The remaining notation is as in Eq. (\ref{feynpar}). In the case of graph g), Table \ref{graphtab}, the maximal number of different propagators is $n = 5$, for the graphs with denominator structure
\be
\trig \label{graphg} \qquad \quad.
\ee

After applying Feynman parametrization, (\ref{feynpar}), to the scalar version of Eq. (\ref{general2}) (we consider the case ${\mathcal{N}}(k,l) = 1$) we obtain a new, single denominator with the structure
\be
 \sum\limits_{i = 1}^N x_i\, A_i = a \,k^2 + b \,l^2 + 2 \,c\, k \cdot l + 2\, d \cdot k + 2 \,e \cdot l + f, \label{newdenom}
 \ee
 where $a,b,c$ are linear combinations of the Feynman parameters $x_i$, $d^\mu$ and $e^\mu$ are new vectors that depend on the $x_i$ and the external momentum (more generally, momenta), and $f$ is a scalar, also depending on $x_i$ and the external momentum.

The integral over loop momenta $l$ and $k$ can now be performed via standard techniques, by  a change of variables to complete the square:
\ba
k^\mu & \rightarrow & K^\mu - \frac{c}{a} L^\mu + \frac{c\, e^\mu - b\, d^\mu}{P}\, , \nonumber \\
l^\mu & \rightarrow & L^\mu + \frac{c \,d^\mu - a\, e^\mu}{P}\, , \label{klshift}
\ea
with
\be
P \equiv a\, b - c^2.
\ee
(\ref{newdenom}) becomes
\be
a K^2 + \frac{P}{a} L^2 + \frac{Q}{P},
\ee
where
\be
Q \equiv -a\, e^2 - b \, d^2 + 2\, c\, d \cdot e + f (a\,b-c^2).
\ee

With (\ref{Minkint}) we obtain
\ba
I^n\left[\eta_1,\dots,\eta_N\right] & = & - \frac{1}{(4\pi)^n}\left[ \prod\limits_{i=1}^N \frac{(-1)^{\eta_i}}{\Gamma\left(\eta_i \right)} \int\limits_0^1 d x_i\, x_i^{\eta_i -1} \right]
 \delta \left(1- \sum\limits_{i = 1}^N x_i \right) \nonumber \\
& & \times \Gamma\left( \nu - n\right) P^{\,\nu - 3/2 n} Q^{n - \nu}, \label{general2b}
\ea
where we have introduced the abbreviation
\be
\sum\limits_{i = 1}^N \eta_i = \nu.
\ee
The integrals over the Feynman parameters in (\ref{general2b}) can be expressed in terms of Gamma-, Beta- and (generalized) hypergeometric functions (see, for example, \cite{gradshteyn}). The latter are, however, not always easily expanded into polynomials of $\varepsilon$ when working in $n = 4 -2 \varepsilon$ dimensions. In addition, for a higher number of propagators ($N >$ 5), it is not easy to find closed expressions. It is advantageous to express all possible integrals in terms of a few, simpler, \emph{Master integrals}. 

\subsection{Integration by Parts}

The technique that we use for the calculation of (\ref{graphg}) to reduce it to Master integrals is termed \emph{integration by parts}  (IBP) \cite{'tHooft:fi,Tkachov:wb,Chetyrkin:qh}. IBP allows us to find relations between integrals of the form (\ref{general2}) with different powers of $\eta_i$, so that we can eventually express all such integrals in terms of those with many $\eta_i = 0$, that is, the corresponding propagators are absent.  

IBP identities follow from the fact that the integral over the total derivative with respect to any loop momentum vanishes in dimensional regularization
\be
\int \frac{d^n k}{(2 \pi)^n} \frac{\partial}{\partial k^\mu} J(k,\,\dots) = 0,
\ee
where $J$ is a loop integrand, and can be a scalar or a tensor. For two-loop integrals (\ref{general2}) these identities can be cast into the form
\be
 \int \frac{d^n k}{(2 \pi)^n} \int \frac{d^n l}{(2\pi)^n} \frac{\partial}{\partial k_i^\mu} \left[ \frac{{\mathcal{N}^\mu}(k_i)}{A_1^{\eta_1} \dots A_N^{\eta_N}} \right]  = 0.
\ee

Applying the derivative to the numerator of (\ref{general2}), we obtain
\be
\frac{\partial {\mathcal{N}}^\mu}{\partial k_i^\mu} = n,
\ee
if the numerator depends on $k_i$ and 0 otherwise. Similarly, taking the derivative of one of the terms in the  denominator results in
\be
{\mathcal{N}}^\mu \left[ \frac{\partial}{\partial k_i^\mu} \frac{1}{A_j^{\eta_i}} \right] = - 2 \eta_j \frac{{\mathcal{N}} \cdot A_j}{A_i^{\eta_j+1}}, \label{IBP}
\ee
if $A_j$ is $k_i$-dependent, and 0 otherwise.

The system of equations generated in this manner can then be used to solve for the integrals that have the lowest number of different propagators. We will illustrate this below for (\ref{graphg}). In (\ref{IBP}) we have additional numerators of the form
\be
2 {\mathcal{N}} \cdot A_j \sim 2\, (k_i + g) \cdot (k_i + h) = (k_i + g)^2 + (k_i + h)^2 - (g - h)^2.
\ee
 \emph{Reducible} numerators can be cancelled against corresponding denominators, and result in simplified integrals. We will deal with \emph{irreducible} numerators in the following.

\subsection{Numerators}

Integrals with irreducible numerators (vectors or tensors) can be dealt with by using the following identities from symmetry considerations
\ba
\int \frac{d^n k}{(2 \pi)^n} k^{\mu_1} \dots k^{\mu_{2m+1}} f(k^2) & = & 0, \quad m \mbox{ integer},\nonumber \\
\int \frac{d^n k}{(2 \pi)^n} k^\mu k^\nu f(k^2) & = & \frac{g^{\mu \nu}}{n} \int \frac{d^n k}{(2 \pi)^n} k^2 f(k^2), \label{symmetry}
\ea
and similarly for higher tensors. 

Reexpressing numerators involving the momenta $k_i$ in terms of the new variables (\ref{klshift}) and using Eqs. (\ref{symmetry}), we obtain, for example
\ba
I^n\left[\eta_1,\dots,\eta_N\right]\left[k^\mu\right] & \equiv & \int \frac{d^n k}{(2 \pi)^n} \int \frac{d^n l}{(2\pi)^n} \frac{k^\mu}{A_1^{\eta_1} \dots A_N^{\eta_N}} \nonumber \\
& = & \int \frac{d^n K}{(2 \pi)^n} \int \frac{d^n L}{(2\pi)^n} \frac{c\, e^\mu - b\, d^\mu}{P} \frac{1}{A_1^{\eta_1} \dots A_N^{\eta_N}}. \label{numint1}
\ea
In this expression $d^\mu$ and $e^\mu$ are functions of the external momentum (momenta), independent of the internal loop variables $K$ and $L$. Comparing this to Eq. (\ref{general2b}), we see that factors of $x_i$ can be absorbed into the integrals over Feynman parameters, by increasing the power to which the $i$th propagator is raised
\be
\frac{(-1)^{\eta_i}\,x_i^{\eta_i-1}}{\Gamma(\eta_i)} x_i = - \eta_i \frac{(-1)^{\eta_i+1} \,x_i^{\eta_i}}{\Gamma(\eta_i + 1)},
\ee
whereas the factor $1/P$ can be absorbed by increasing the dimension of the integral
\be
P^{\,\nu - 3/2 n} \frac{1}{P} = P^{\,\nu - 1/2 ( 3n +2) }.
\ee

We can proceed in a similar fashion for more complicated tensors in the numerator. We always obtain integrals that are of the form (\ref{general2}), only with increased dimensions and increased powers of numerators $\eta_i$.

\subsubsection{Example - Reduction of Graph (\ref{graphg})}

As an example for IBP identities, we express graph (\ref{graphg}) in terms of simpler Master integrals:
\be
\,\smalltrig\, \quad = \frac{2 (3 n -8)(3 n - 10)}{(n-4)^2} \frac{1}{(p^2)^2} \,\smallSUNC\, \quad - \frac{2(n-3)}{n-4} \frac{1}{p^2} \,\smallGLASS\, \quad.
\ee
The expressions for the ``sunset'' and the ``glass'' Master integrals are given below.

\subsection{Collection of a Few Integrals} \label{app:intcoll}

Here we collect the results for a few standard integrals in Minkowski and Euclidean space, the former can be obtained by performing a Wick rotation to Euclidean space (see, for example, \cite{book}), expressing the $n$-dimensional Euclidean phase space in terms of polar coordinates
\be
\int d^n k_E \rightarrow \int d \Omega_{n-1} \int\limits_0^\infty d \kappa \kappa^{n-1}
\ee
and integrating over polar angles:
\be
\int d \Omega_m = 2 \pi^{m/2} \frac{\Gamma(m/2)}{\Gamma(m)}.
\ee
A general $n$-dimensional scalar integral in Euclidean space  (denoted by the subscript $E$) is then computed to give
\be
\int d^n k_E \left(k^2_E + M^2 \right)^{-s}  = \pi^{n/2} \frac{\Gamma(s-n/2)}{\Gamma(s)} \left(M^2\right)^{n/2-s}.
\ee
The result for its  counterpart in Minkowski space is after Wick rotation given by
\be
\int d^n k \left(k^2 - M^2 + i \epsilon \right)^{-s}  = (-1)^s i \pi^{n/2} \frac{\Gamma(s-n/2)}{\Gamma(s)} \left(M^2\right)^{n/2-s}. \label{Minkint}
\ee

\subsubsection{A Few Simple Master Integrals} 

The calculation of Eq. (\ref{graphg}), and thus the calculation of graph g) in Table \ref{graphtab}, is reduced to the calculation of three Master integrals after using the IBP techniques and a bit of algebra to reduce tensor integrals. These loop integrals with two or three propagators are textbook exercises. $p$ denotes the external momentum. They are given in terms of Gamma- and Beta-functions by
\ba
\BUB\quad [\eta_1,\eta_2] & \equiv & \int \frac{d^n k}{(2 \pi)^n} \frac{1}{(k^2)^{\eta_1}} \frac{1}{(k^2+p^2)^{\eta_2}}\nonumber \\
&  & \hspace*{-2.5cm} =\, \frac{i}{(4 \pi)^{n/2}} (-1)^{\eta_1+\eta_2}  \frac{\Gamma\left( \eta_1+\eta_2 - \frac{n}{2} \right)}{\Gamma\left(\eta_1\right)\Gamma\left(\eta_2\right)} B\left(\frac{n}{2}-\eta_1,\frac{n}{2}-\eta_2 \right) \left(-p^2\right)^{n/2-\eta_1-\eta_2}, \nonumber \\
& & \\
\smallSUNC \quad & = & \frac{i}{(4 \pi)^{n/2}} (-1)^{n/2} \Gamma\left(2-\frac{n}{2}\right) B\left(\frac{n}{2}-1,\frac{n}{2}-1 \right) \nonumber \\
& & \times \,\, \BUB\quad \left[1,2-\frac{n}{2}\right], \\
\smallGLASS \quad & = & \left\{\,\, \BUB\quad \left[1,1\right] \right\}^2.
\ea
We have listed the sunset and glass graphs for propagators raised to the power 1, that is $\eta_i = 1$.

\section{Rules for LCOPT} \label{app:LCOPT}

Throughout this thesis our convention for light-cone coordinates is as follows:
\ba
k^+ & = & \frac{1}{\sqrt{2}}\left(k^0 + k^3\right), \nonumber \\
k^- & = & \frac{1}{\sqrt{2}}\left(k^0 - k^3 \right), \label{lccoord} \\
k_\perp & =  & \left(k^1,k^2\right). \nonumber
\ea

Light-cone ordered perturbation theory is equivalent to the expressions obtained after performing all minus integrals of all loops in a given graph 
\cite{Chang:1968bh, Kogut:1969xa, Brodsky:1973kb}. 
LCOPT is similar to old-fashioned, or time-ordered, perturbation theory, but ordered along the light-cone, $x^+$, rather than in $x^0$. In a LCOPT diagram all internal lines are on-shell, in contrast to a covariant Feynman diagram, which allows us to identify UV divergent loops in eikonal diagrams more easily. This property and the fact that the rules are equivalent to performing all minus integrals make LCOPT especially suited for multi-loop integrals of eikonal diagrams in Feynman gauge. A covariant Feynman diagram is comprised of one or more LCOPT diagrams.

The rules for LCOPT can be summarized as follows:
\begin{itemize}
\item We start with forming all possible light-cone orderings of a given covariant diagram.
\item Only those configurations are kept which describe possible physical processes once the flow of plus-momenta is specified.
\item For every loop we have a factor
\be
\frac{d l^+\,d^{2-2\varepsilon} l_{\perp}}{(2 \pi)^{3 -2 \varepsilon}}
\ee
 if we work in $n = 4 - 2 \varepsilon$ dimensions.
\item For every internal line we have a factor
\be
\frac{\theta(l_i^+)}{2 l_i^+},
\ee
 corresponding to the flow of plus momentum through the graph.
\item Every intermediate virtual state contributes a factor
\be
\frac{i}{q^- - \sum_j \frac{l^2_{j,\,\perp}}{2 l_j^+} + i \epsilon},  \label{denom}
\ee
where we sum over all momenta comprising that virtual state, and where $q^-$ is the external minus-momentum of the incoming state(s).
\item Every intermediate real state gives a momentum conserving delta-function:
\be
2 \pi \delta \left(q^- - \sum_j \frac{l^2_{j,\,\perp}}{2 l_j^+} \right),
\ee
where the sum is over all momenta in that real state.
\item Since all lines are on-shell, we replace for every Fermion numerator its minus component by its on-shell value:
\be
l^- = \frac{l_\perp^2}{2 l^+}.
\ee
\end{itemize}

\chapter{Color Space Decomposition}
\label{app3}
 
Below we list the color bases, and the decompositions of the lowest order hard matrices and soft   functions in these bases for various $2 \rightarrow 2$ scattering processes relevant for $p\bar{p}$ collisions. We employ various identities 
from \cite{MacFarlane:1968vc} in the calculation of the color decompositions.  $r_i$ labels the color of parton $i$ in (\ref{pprocess}). $\Ncol$ is the number of colors.

\section{Color Bases} \label{app:colbas}

\subsection{Basis for $q\bar{q} \rightarrow q\bar{q}$}  
 
We use the $t$-channel singlet-octet basis  
\begin{eqnarray} 
c_1 & = & \delta_{r_A,\,r_1} \delta_{r_B,\,r_2}, \nonumber \\ 
c_2 & = & - \frac{1}{2 \Ncol} \delta_{r_A,\,r_1} \delta_{r_B,\,r_2} + \frac{1}{2}
 \delta_{r_A,\,r_B} \delta_{r_1,\,r_2}. \label{qqbarbas} 
\end{eqnarray} 
 
\subsection{Basis for $q q \rightarrow q q$}  
 
The natural $t$-channel basis for this process is 
\begin{eqnarray} 
c_1 & = & \delta_{r_A,\,r_1} \delta_{r_B,\,r_2}, \nonumber \\ 
c_2 & = & - \frac{1}{2 \Ncol} \delta_{r_A,\,r_1} \delta_{r_B,\,r_2} + \frac{1}{2}
 \delta_{r_A,\,r_2} \delta_{r_B,\,r_1}. \label{qqbas} 
\end{eqnarray}

\subsection{Basis for $q g \rightarrow q g$} 
 
Here we use the basis 
\begin{eqnarray} 
c_1 & = & \delta_{r_A,\,r_1} \delta_{r_B,\,r_2}, \nonumber \\ 
c_2 & = & d_{r_Br_2  c} \left(T^c_F\right)_{r_1r_A}, \label{qgbas}\\ 
c_3 & = & i f_{r_Br_2  c} \left(T^c_F\right)_{r_1r_A}, \nonumber 
\end{eqnarray} 
where $c_1$ is the $t$-channel singlet tensor, $c_2$ and $c_3$ are the 
symmetric and antisymmetric octet tensors, respectively.

\subsection{Basis for  
$g g \rightarrow q \bar{q}$ and $q \bar{q} \rightarrow g g$} 
 
For these processes it is convenient to use the $s$-channel color basis 
\begin{eqnarray} 
c_1 & = & \delta_{r_A,\,r_B} \delta_{r_1,\,r_2}, \nonumber \\ 
c_2 & = & d_{r_Ar_B  c} \left(T^c_F\right)_{r_1r_2}, \label{ggqbas} \\ 
c_3 & = & i f_{r_Ar_B  c} \left(T^c_F\right)_{r_1r_2}. \nonumber 
\end{eqnarray}

\subsection{Bases for $gg \rightarrow gg$} \label{basgg}
 
 The most convenient procedure to calculate the color structure for this process is 
to start with an overcomplete basis of nine color tensors, then to find a new basis
 consisting of nine tensors in which the anomalous dimension matrix that we will list below is block diagonal, 
and finally to reduce the matrix to the minimal basis consisting of eight tensors. 
 
We start with the following set of color tensors  
\begin{eqnarray} 
c_1 & = & \Tr \left(T_F^{r_A} T_F^{r_B} T_F^{r_2} T_F^{r_1} \right),\nonumber \\ 
c_2 & = & \Tr \left(T_F^{r_A} T_F^{r_B} T_F^{r_1} T_F^{r_2} \right),\nonumber \\ 
c_3 & = & \Tr \left(T_F^{r_A} T_F^{r_1} T_F^{r_2} T_F^{r_B} \right),\nonumber \\ 
c_4 & = & \Tr \left(T_F^{r_A} T_F^{r_1} T_F^{r_B} T_F^{r_2} \right),\nonumber \\ 
c_5 & = & \Tr \left(T_F^{r_A} T_F^{r_2} T_F^{r_1} T_F^{r_B} \right),\nonumber \\ 
c_6 & = & \Tr \left(T_F^{r_A} T_F^{r_2} T_F^{r_B} T_F^{r_1} \right), \label{col9} \\ 
c_7 & = & \frac{1}{4} \delta_{r_A,\,r_1} \delta_{r_B,\,r_2}, \nonumber \\ 
c_8 & = & \frac{1}{4} \delta_{r_A,\,r_B} \delta_{r_1,\,r_2}, \nonumber \\ 
c_9 & = & \frac{1}{4} \delta_{r_A,\,r_2} \delta_{r_B,\,r_1}. \nonumber \\ 
\end{eqnarray} 
We transform to a new basis of 
color tensors. This basis can be rewritten in terms of the tensors $f$ and $d$ with 
the help of the product formula for generators of $SU(\Ncol)$ in the fundamental 
representation, 
\begin{equation} 
T_F^i T_F^j = \frac{1}{2 \Ncol} \delta_{ij} 1 + \frac{1}{2} \left( d_{ijk} + i f_{ijk} \right) T_F^k. 
\end{equation} 
After setting $\Ncol = 3$, the new basis is given by 
\begin{eqnarray} 
c_1' & = & c_1 - c_3 = \frac{i}{4} \left[ f_{r_A r_B c} d_{r_1 r_2 c} - 
d_{r_A r_B c} f_{r_1 r_2 c} \right],\nonumber \\ 
c_2' & = & c_2 - c_5 = \frac{i}{4} \left[ f_{r_A r_B c} d_{r_1 r_2 c} + 
d_{r_A r_B c} f_{r_1 r_2 c} \right],\nonumber \\ 
c_3' & = & c_4 - c_6 = \frac{i}{4} \left[ f_{r_A r_1 c} d_{r_B r_2 c} + 
d_{r_A r_1 c} f_{r_B r_2 c} \right],\nonumber \\ 
c_4' & = & c_1 + c_3 = \frac{1}{6} \delta_{r_A r_1} \delta_{r_B r_2} + 
\frac{1}{4} \left[ d_{r_A r_1 c} d_{r_B r_2 c} + f_{r_A r_1 c} f_{r_B r_2 c} 
\right],\nonumber \\ 
c_5' & = & c_2 + c_5 = \frac{1}{6} \delta_{r_A r_B} \delta_{r_1 r_2} +
 \frac{1}{4} \left[ d_{r_A r_B c} d_{r_1 r_2 c} - f_{r_A r_B c} f_{r_1 r_2 c} 
\right], \label{block}\\ 
c_6' & = & c_4 + c_6 = \frac{1}{6} \delta_{r_A r_1} \delta_{r_B r_2} +
 \frac{1}{4} \left[ d_{r_A r_1 c} d_{r_B r_2 c} - f_{r_A r_1 c} f_{r_B r_2 c} 
\right],\nonumber \\ 
c_7' & = & c_7,\nonumber \\ 
c_8' & = & c_8,\nonumber \\ 
c_9' & = & c_9.\nonumber \\ 
\end{eqnarray} 
The overcomplete basis (\ref{block}) can be reduced to a  
basis of 8 $SU(3)$ color tensors. This basis can be partly expressed in terms  of 
$t$-channel $SU(3)$ projectors for the decomposition into irreducible 
representations of the direct product $8 \otimes 8$ which  corresponds to the color 
content of a set of two gluons: 
\begin{equation} 
\left\{ c_1', c_2', c_3', P_1, P_{8_S}, P_{8_A}, P_{10 \oplus \overline{10}},
P_{27} \right\}, \label{ggbas} 
\end{equation} 
where,
\begin{eqnarray} 
P_1(r_A,r_B;r_1,r_2) & = & \frac{1}{8} \delta_{r_A r_1} \delta_{r_B r_2}, \nonumber \\ 
P_{8_S}(r_A,r_B;r_1,r_2) & = & \frac{3}{5} d_{r_A r_1 c} d_{r_B r_2 c}, \nonumber \\ 
P_{8_A}(r_A,r_B;r_1,r_2) & = & \frac{1}{3} f_{r_A r_1 c} f_{r_B r_2 c}, \nonumber \\ 
P_{10 \oplus \bar{10}}(r_A,r_B;r_1,r_2) & = & \frac{1}{2} \left( \delta_{r_A r_B} 
\delta_{r_1 r_2} - \delta_{r_A r_2} \delta_{r_B r_1} \right)  - \frac{1}{3}
 f_{r_A r_1 c} f_{r_B r_2 c}, \nonumber \\  
P_{27}(r_A,r_B;r_1,r_2) & = & \frac{1}{2} \left( \delta_{r_A r_B} \delta_{r_1 r_2}
 + \delta_{r_A r_2} \delta_{r_B r_1} \right) \nonumber \\
& & \quad   - \frac{1}{8}\delta_{r_A r_1} 
\delta_{r_B r_2} - \frac{3}{5} d_{r_A r_1 c} d_{r_B r_2 c}. \nonumber  \\
& & 
\end{eqnarray}

\section{Hard Scattering Matrices} \label{app:hard}

Explicit hard matrices in color
space at LO  
have been given in Ref.\ \cite{Oderda:1999kr}. 
We exhibit them here for the sake of completeness
with a trivial change in overall normalization relative to \cite{Oderda:1999kr}.

We will express the hard matrices in terms of the partonic Mandelstam variables
\ba  
\hat s &=& x_A x_B s, \nonumber \\  
\hat t &=& -p_T^2  \left(1+e^{-2\eh}\right), \nonumber \\  
\hat u &=& -(\hat s + \hat t).  
\ea  
We also recall that we have from Eq. (\ref{hatsigfact})
\be
p_T = \frac{\sqrt{\hat{s}}}{2 \cosh \hat\eta}.
\ee

\subsection{Hard Matrices for $q\bar{q} \rightarrow q\bar{q}$}  
  
In the basis (\ref{qqbarbas}) the decomposition of  
the Born level hard scattering for same-flavor $q \bar{q} \rightarrow q \bar{q}$  
in color space is given by the matrix 
\begin{equation} 
H^{(1)}\left(\hat{t},\hat{s},\alpha_s(\mu) \right)=\frac{1}{\Ncol^2} \,  
\frac{\alpha^2_s(\mu) \, \pi}{\hat{s}} \left( 
\begin{array}{cc} 
\left(\frac{C_F}{\Ncol}\right)^2 \chi_1  & \frac{C_F}{\Ncol^2} \chi_2 \vspace{2mm} \\ 
\frac{C_F}{\Ncol^2} \chi_2 & \chi_3 
\end{array} 
\right) \,, \label{hardqqb} 
\end{equation} 
where $\chi_1$,  
$\chi_2$, and $\chi_3$  are 
defined by 
\begin{eqnarray} 
\chi_1&=&\frac{\hat{t}^2+\hat{u}^2}{\hat{s}^2}      \nonumber \\ 
\chi_2&=&\Ncol \frac{\hat{u}^2}{\hat{s}\hat{t}}- 
\frac{\hat{t}^2+\hat{u}^2}{\hat{s}^2} \nonumber \\ 
\chi_3&=&\frac{\hat{s}^2+\hat{u}^2}{\hat{t}^2}+ 
\frac{1}{\Ncol^2}\frac{\hat{t}^2+\hat{u}^2}{\hat{s}^2} 
-\frac{2}{\Ncol}\frac{\hat{u}^2}{\hat{s}\hat{t}}\,. 
\end{eqnarray} 
The unequal-flavor cases,
$q\bar q'\rightarrow q\bar q'$ and $q\bar q\rightarrow q'\bar q'$, are found from
(\ref{hardqqb}) by
  dropping the $s$-channel terms for
the former, and the $t$-channel contributions for the latter. The matrix
for $q \bar{q} \rightarrow \bar{q} q$ can be found from (\ref{hardqqb}) via
the transformation $\hat{t} \leftrightarrow \hat{u}$.

\subsection{Hard Matrix for $q q \rightarrow q q$}  
  
For this process we obtain in the basis Eq. (\ref{qqbas})  
a hard matrix which is related to the one for  
$ q \bar{q} \rightarrow q \bar{q}$, Eq. (\ref{hardqqb}),  by 
the crossing transformation $\hat{s} \leftrightarrow \hat{u}$.  
So the $\chi_1$, $\chi_2$ and $\chi_3$    
are given by 
\begin{eqnarray} 
\chi_1&=&\frac{\hat{t}^2+\hat{s}^2}{\hat{u}^2}      \nonumber \\ 
\chi_2&=&\Ncol \frac{\hat{s}^2}{\hat{t}\hat{u}}- 
\frac{\hat{s}^2+\hat{t}^2}{\hat{u}^2}  \nonumber \\ 
\chi_3&=&\frac{\hat{s}^2+\hat{u}^2}{\hat{t}^2}+ 
\frac{1}{\Ncol^2}\frac{\hat{s}^2+\hat{t}^2}{\hat{u}^2} 
-\frac{2}{\Ncol}\frac{\hat{s}^2}{\hat{t}\hat{u}}\,. 
\end{eqnarray} 
As above, for $qq'\rightarrow qq'$  only the
$t$-channel terms are kept.

\subsection{Hard Matrices for $q g \rightarrow q g$} 
 
In the basis Eq. (\ref{qgbas}) we arrive at a hard matrix of the form 
\begin{equation} 
H^{(1)}\left(\hat{t}, 
\hat{s},\alpha_s(\mu) \right)=\frac{1}{\Ncol (\Ncol^2 -1)} \,  
\frac{\alpha^2_s(\mu) \, \pi}{2 \hat{s}} 
\left( 
\begin{array}{ccc} 
\frac{1}{2 \Ncol^2}\chi_1    & 
\frac{1}{2 \Ncol}\chi_1   & 
\frac{1}{\Ncol}\chi_2  \vspace{2mm} \\ 
  \frac{1}{2 \Ncol}\chi_1  &  \frac{1}{2} \chi_1 & \chi_2 \vspace{2mm} \\ 
  \frac{1}{\Ncol} \chi_2 & \chi_2 & \chi_3 
\end{array} 
\right) \, ,  \label{hardqg}
\end{equation} 
with $\chi_1$, $\chi_2$, and $\chi_3$ given by 
\begin{eqnarray} 
\chi_1&=&- \frac{\hat{s}^2 + \hat{u}^2}{\hat{s}\hat{u}}      \nonumber \\ 
\chi_2&=&1-\frac{1}{2} \frac{\hat{t}^2}{\hat{s} \hat{u}}-\frac{\hat{u}^2}{\hat{s}\hat{t}}  
- \frac{\hat{s}}{\hat{t}}   \nonumber \\ 
\chi_3&=&3-4\frac{\hat{s}  
\hat{u}}{\hat{t}^2}- \frac{1}{2} \frac{\hat{t}^2}{\hat{s}\hat{u}} \,. 
\end{eqnarray} 
Again, the matrix for $q g \rightarrow g q$ can be found from the above matrix 
(\ref{hardqg}) by exchanging $\hat{t} \leftrightarrow \hat{u}$.

\subsection{Hard Matrices for  
$g g \rightarrow q \bar{q}$ and $q \bar{q} \rightarrow g g$}

 In the basis (\ref{ggqbas}) we calculated the Born level hard scattering matrix to be 
\begin{equation} 
H^{(1)}\left(\hat{t}, 
\hat{s},\alpha_s(\mu) \right)= \frac{1}{\bar{N}} \,  
\frac{\alpha^2_s(\mu) \, \pi}{2 \hat{s}} 
\left( 
\begin{array}{ccc} 
\frac{1}{2 \Ncol^2}\chi_1    & 
\frac{1}{2 \Ncol}\chi_1   & 
\frac{1}{\Ncol}\chi_2  \vspace{2mm} \\ 
  \frac{1}{2 \Ncol}\chi_1  &  \frac{1}{2} \chi_1 & \chi_2 \vspace{2mm} \\ 
  \frac{1}{\Ncol} \chi_2 & \chi_2 & \chi_3 
\end{array} 
\right) \, ,  \label{hardqqgg}
\end{equation} 
where $\chi_1$, $\chi_2$, and $\chi_3$ are given by 
\begin{eqnarray} 
\chi_1&=&\frac{\hat{t}^2 + \hat{u}^2}{\hat{t}\hat{u}}      \nonumber \\ 
\chi_2&=&1-\frac{1}{2} \frac{\hat{u}^2 - \hat{t}^2}{\hat{t} \hat{u}}    \nonumber \\ 
\chi_3&=&6-4\frac{\hat{t}^2}{\hat{s}^2}+  
\frac{1}{2} \frac{\hat{t}^2 + \hat{u}^2}{\hat{t}\hat{u}} \,, 
\end{eqnarray} 
and the averaging factor is $\bar{N} = \Ncol^2$ for the  
process $q \bar{q} \rightarrow g g$, and  
$\left( \Ncol^2 - 1\right)^2$ for $g g \rightarrow q \bar{q}$. The matrix for 
$g g \rightarrow \bar{q} q$ is once again found by letting 
$\hat{t} \leftrightarrow \hat{u}$.

\subsection{Hard Matrix for $gg \rightarrow gg$} \label{hardsoftgg}
 
In the minimal basis consisting of 8 color tensors, (\ref{ggbas}), 
 the hard matrix is found to be block-diagonal 
\begin{equation} 
  H^{(1)}\left(\hat{t}, 
\hat{s},\alpha_s(\mu) \right)=\left(\begin{array}{cc} 
         0_{3 \times 3}        & 0_{3 \times 5} \\ 
          0_{5 \times 3}      &  H^{(1)}_{5 \times 5} 
\end{array} \right)\, , 
\label{hardblock} 
\end{equation} 
with the $5 \times 5$ block $H^{(1)}_{5 \times 5}$ 
\begin{equation} 
H^{(1)}_{5 \times 5}\left(\hat{t}, 
\hat{s},\alpha_s(\mu) \right)=\frac{1}{16} \, \frac{\alpha^2_s(\mu) \pi}{2 \hat{s}}  
\left(\begin{array}{ccccc} 
9\chi_1 & \frac{9}{2}\chi_1 & \frac{9}{2}\chi_2 & 0 & -3\chi_1 \vspace{2mm}\\ 
\frac{9}{2}\chi_1 & \frac{9}{4}\chi_1 & \frac{9}{4}\chi_2 & 0 &  
-\frac{3}{2}\chi_1 \vspace{2mm}\\ 
\frac{9}{2}\chi_2 & \frac{9}{4}\chi_2 & \chi_3 & 0 &  
-\frac{3}{2}\chi_2 \vspace{2mm}\\ 
0 & 0 & 0 & 0 & 0 \vspace{2mm}\\ 
-3\chi_1 & -\frac{3}{2}\chi_1 & -\frac{3}{2}\chi_2 &0 &\chi_1 
\end{array}\right) \, , 
\end{equation} 
with $\chi_1$, $\chi_2$, and $\chi_3$  defined by 
\begin{eqnarray} 
\chi_1&=&1-\frac{\hat{t}\hat{u}}{\hat{s}^2}-\frac{\hat{s}\hat{t}}{\hat{u}^2}+ 
\frac{\hat{t}^2}{\hat{s}\hat{u}}      \nonumber \\ 
\chi_2&=&\frac{\hat{s}\hat{t}}{\hat{u}^2}-\frac{\hat{t}\hat{u}}{\hat{s}^2}+ 
\frac{\hat{u}^2}{\hat{s}\hat{t}}- 
\frac{\hat{s}^2}{\hat{t}\hat{u}}    \nonumber \\ 
\chi_3&=&\frac{27}{4}-9\left(\frac{\hat{s}\hat{u}}{\hat{t}^2}+\frac{1}{4} 
\frac{\hat{t}\hat{u}}{\hat{s}^2}+\frac{1}{4}\frac{\hat{s}\hat{t}}{\hat{u}^2}\right) 
+\frac{9}{2}\left(\frac{\hat{u}^2}{\hat{s}\hat{t}}+\frac{\hat{s}^2}{\hat{t}\hat{u}}- 
\frac{1}{2}\frac{\hat{t}^2}{\hat{s}\hat{u}}\right) \,. 
\end{eqnarray} 
For this process we have set explicitly $\Ncol = 3$.

\section{Soft Functions for $2 \rightarrow 2$ Scattering} \label{app:soft}

Below we list the anomalous dimension matrices and soft functions at lowest order, Eqs. (\ref{fulladmsoft}) and (\ref{fullsoft}),  in terms of the combinations (\ref{abgdef}). $\phi^{(ij,\,1)}$ is either $S^{(ij,\,1)}$ (Eq. (\ref{oneLoopSofthad})) or $\Gamma^{(ij,\,1)}$ (Eq. (\ref{oneloopadm})). Furthermore, we collectively denote
  anomalous dimension matrices and lowest order eikonal contributions by ${\mathcal{M}}^{(\rm f)}$,

\subsection{Soft Matrices for $q q \rightarrow q q$} 
 
We obtain the color decomposition for $q q \rightarrow q q$ in the basis (\ref{qqbas})
\begin{equation} 
{\mathcal{M}}^{(qq\rightarrow q q)} = \left( \begin{array}{cc} 
C_F \beta^{(qq)} &
 \frac{C_F}{2 \Ncol} \left( \alpha^{(qq)} +
 \gamma^{(qq)} \right) \\ 
\alpha^{(qq)} + 
\gamma^{(qq)} \quad & C_F \gamma^{(qq)} - \frac{1}{2 \Ncol}
 \left( 2 \alpha^{(qq)} + \beta^{(qq)} +
 \gamma^{(qq)} \right) \end{array} \right), 
\end{equation}
where we have abbreviated $(q q \rightarrow q q)$ by $(qq)$, and
where the sign changes compared to $q \bar{q} \rightarrow q \bar{q}$ due to
 the factors of $\Delta_i$ are as follows: 
\begin{eqnarray}
\alpha^{(qq\rightarrow q q)} & = & -\alpha^{(q\bar{q}\rightarrow q \bar q)} \nonumber \\
\beta^{(qq\rightarrow q q)} & = & \beta^{(q\bar{q}\rightarrow q \bar q)} \\
\gamma^{(qq\rightarrow q q)} & = & -\gamma^{(q\bar{q}\rightarrow q \bar q)}.\nonumber
\end{eqnarray}
 
\subsection{Soft Matrix for $q g \rightarrow q g$ and
 $\bar{q} g \rightarrow \bar{q} g$}

In terms of the $\phi^{(ij,\,1)}$s for $q \bar{q} \rightarrow q \bar{q}$,
 as discussed in Sec. \ref{sec:momentum}, the anomalous dimension matrix for $q g \rightarrow q g$ reads in the basis (\ref{qgbas})
\begin{equation} 
{\mathcal{M}}^{(qg \rightarrow q g)} = \left( \begin{array}{ccc}  
C_F \phi^{(A1,\,1)} + C_A \phi^{(B2,\,1)} \,\, & 0 & - \frac{1}{2}
 \left( \alpha +\gamma \right) \\ 
 0 & \chi \,\, & 
- \frac{\Ncol}{4} \left( \alpha + \gamma \right) \\ 
-\left(\alpha + \gamma \right)  & - \frac{\Ncol^2 - 4}{4 \Ncol}
 \left( \alpha + \gamma \right) & \chi \end{array} \right), 
\end{equation} 
where $\chi \equiv  \frac{\Ncol}{4} \left( \alpha - \gamma \right) - 
\frac{1}{2 \Ncol} \phi^{(A1,\,1)} + \frac{\Ncol}{2}  \phi^{(B2,\,1)}$. 
Here and everywhere below we drop the superscripts $(q \bar{q}\rightarrow q \bar q)$.
 
 The matrix for  
\begin{displaymath} 
\bar{q}(p_1, r_1) + g (p_2, r_2) \rightarrow \bar{q} (p_A, r_A)+ g (p_B,r_B) 
\end{displaymath} 
 is the same as for 
\begin{displaymath} 
q (p_A, r_A) + g(p_B, r_B) \rightarrow q (p_1, r_1) + g (p_2, r_2) 
\end{displaymath} 
if we use the same basis and take into account all sign changes due 
to the eikonal Feynman rules and the relabelling of the color indices. 
 
\subsection{Soft Matrices for $g g \rightarrow q \bar{q}$ 
and $q \bar{q} \rightarrow g  g$}

In the basis (\ref{ggqbas}) the color decomposition for $g g \rightarrow q \bar{q}$ is given by 
\begin{equation} 
{\mathcal{M}}^{(gg\rightarrow q\bar{q})} = \left( \begin{array}{ccc}  
C_F \phi^{(12,\,1)} + C_A \phi^{(AB,\,1)} \,\, & 0 & \frac{1}{2} 
\left( \beta +\gamma \right) \\ 
 0 & \xi \,\, & 
 \frac{\Ncol}{4} \left( \beta + \gamma \right) \\ 
\beta + \gamma  &  \frac{\Ncol^2 - 4}{4 \Ncol} \left( \beta + \gamma \right) 
& \xi \end{array} \right), 
\end{equation} 
where $\xi  \equiv  \frac{\Ncol}{4} \left( \beta - \gamma \right) -
 \frac{1}{2 \Ncol} \phi^{(12,\,1)} + \frac{\Ncol}{2}  \phi^{(AB,\,1)}$, and the
 $\phi^{(ij,\,1)}$ again denote those calculated above for $q \bar{q} \rightarrow q \bar{q}$. 
 
The process 
\begin{displaymath} 
q (p_2, r_2) +  \bar{q} (p_1, r_1)  \rightarrow g (p_B, r_B) + g (p_A, r_A) 
\end{displaymath} 
is  related to the  process 
\begin{displaymath} 
g (p_A, r_A) + g (p_B, r_B)  \rightarrow q (p_1, r_2) + \bar{q} (p_2, r_2) 
\end{displaymath} 
by time reversal. After relabelling the color indices appropriately we 
arrive at the following anomalous dimension matrix where only the diagonal 
elements are changed compared to the one stated above: 
\begin{equation} 
{\mathcal{M}}^{(q \bar{q} \rightarrow gg)} = \left( \begin{array}{ccc}  
C_F \phi^{(AB,\,1)} + C_A \phi^{(12,\,1)} \,\, & 0 & \frac{1}{2} \left(
 \beta +\gamma \right) \\ 
 0 & \xi' \,\, & 
 \frac{\Ncol}{4} \left( \beta + \gamma \right) \\ 
\beta + \gamma   &  \frac{\Ncol^2 - 4}{4 \Ncol} \left( \beta + \gamma \right) &
 \xi' \end{array} \right), 
\end{equation} 
where $\xi'  \equiv  \frac{\Ncol}{4} \left( \beta - \gamma \right) - \frac{1}{2 \Ncol}
 \phi^{(AB,\,1)} + \frac{\Ncol}{2}  \phi^{(12,\,1)}$. 
 
\subsection{Soft Matrix for $g g \rightarrow g g$} \label{ggADM} 
 
The soft matrix for 
$g g \rightarrow g g$ is in the minimal basis (\ref{ggbas}) given by 
\begin{equation} 
{\mathcal{M}}^{(gg\rightarrow gg)}
 = \left( \begin{array}{cc} {\mathcal{M}}_{3 \times 3} \, & \, 0_{3\times5} \\ 
0_{5 \times 3} & {\mathcal{M}}_{5 \times 5} \end{array} \right), \label{sadgg} 
\end{equation} 
where the matrix ${\mathcal{M}}_{3 \times 3}$ is given by
\begin{equation} 
{\mathcal{M}}_{3 \times 3} = \left( \begin{array}{ccc} 
 \frac{\Ncol}{2} \left( \alpha + \beta \right) \, & 0 & 0 \\ 
0 & \frac{\Ncol}{2} \left( \alpha - \gamma \right) \, & 0 \\ 
0 & 0 & \frac{\Ncol}{2} \left(\beta - \gamma \right) \end{array} \right), \label{33} 
\end{equation} 
 and ${\mathcal{M}}_{5 \times 5}$ reads
 { \small 
\ba
{\mathcal{M}}_{5 \times 5} & = &
\nonumber \\
& & \hspace*{-18mm} \left( \begin{array}{ccccc} 
3 \beta & 0 & 3 \left( \alpha + \gamma \right) & 0 & 0 \\ 
0 & \frac{3}{4} \left( \alpha + 2 \beta - \gamma \right) & \frac{3}{4} 
\left( \alpha + \gamma \right) & \frac{3}{2} \left( \alpha + \gamma \right) & 0 \\ 
\frac{3}{8} \left( \alpha + \gamma \right) & \frac{3}{4} \left( \alpha + 
\gamma \right) & \frac{3}{4} \left( \alpha + 2 \beta - \gamma \right) & 0 &  
\frac{9}{8} \left( \alpha + \gamma \right) \\ 
0 & \frac{3}{5} \left( \alpha + \gamma \right) & 0 & \frac{3}{2} \left( \alpha -
 \gamma \right) & \frac{9}{10} \left( \alpha + \gamma \right) \\ 
0 & 0 & \frac{1}{3} \left( \alpha + \gamma \right) & \frac{2}{3} \left( \alpha + 
\gamma \right) & 2 \alpha - \beta - 2 \gamma 
\end{array} \right). \nonumber \\
& & \label{finalggADM}
\ea 
 }

\end{singlespace}

\end{document}